\documentclass[11pt,fleqn]{article}
\usepackage{etoc}
\usepackage{amsfonts}
\usepackage{algorithm}
\usepackage{algpseudocode}
\usepackage{amsmath}
\usepackage{amssymb}
\usepackage{a4wide}
\usepackage{hyperref}
\usepackage{float}
\usepackage{multirow}
\usepackage{threeparttable}
\usepackage{lscape}
\usepackage{mathpazo}
\usepackage{chngcntr}
\usepackage{booktabs}
\usepackage[shortlabels]{enumitem}
\counterwithin{table}{section}
\usepackage[round]{natbib}
\usepackage{rotating}
\usepackage{xcolor}
\numberwithin{equation}{section}

\newcommand{\add}[1]{{\textcolor{purple}{#1}}}

\newcommand{\bbeta}{\boldsymbol{\beta}}

\newtheorem{proposition}{Proposition}

\newtheorem{example}{Example}

\newtheorem{remark}{Remark}

\newtheorem{theorem}{Theorem}
\newtheorem{lemma}{Lemma}

\def\*#1{\mathbf{#1}}
\def\+#1{\boldsymbol{#1}}
\usepackage{authblk}
\DeclareMathOperator*{\argmin}{arg\,min}
\newcommand{\norm}[1]{\left\lVert#1\right\rVert}
\usepackage[margin=0.75in,paper=letterpaper]{geometry}
\title{\bf New Tests of Equal Forecast Accuracy for Factor-Augmented Regressions with Weaker Loadings}
\author[1]{Luca Margaritella}
\author[2]{Ovidijus Stauskas}
\affil[1]{Lund University, Department of Economics}
\affil[2]{BI Norwegian Business School, Department of Economics}
\date{}
\begin{document}
\maketitle 
\setlength{\baselineskip}{0.7cm}
    \begin{abstract}\setlength{\baselineskip}{0.7cm}
        We provide the theoretical foundation for the recent tests of equal forecast accuracy and encompassing by \citet{pitarakis2023direct} and \citet{pitarakis2025novel}, when the competing forecast specification is that of a factor-augmented regression model. This should be of interest for practitioners, as there is no theory justifying the use of these simple and powerful tests in such context. In pursuit of this, we employ a novel theory to incorporate the empirically well-documented fact of homogeneously/heterogeneously weak factor loadings, and track their effect on the forecast comparison problem.   \\ 
        \textbf{Keywords}: \textit{Forecast Accuracy}, \textit{Factor-Augmented Regressions}, \textit{Weak Loadings}

    \end{abstract}
\etocdepthtag.toc{mtchapter}
\etocsettagdepth{mtchapter}{subsection}
\etocsettagdepth{mtappendix}{none}
    \section{Introduction} 
    Assessing the \emph{out-of-sample} forecast performance of different models is fundamental for practitioners deciding which specification to employ. Central banks, investment banks, and government economic planning agencies routinely employ forecasting models to make informed decisions and implement effective policies. Therefore, evaluating the \emph{population} predictive ability of various alternative models is for them essential.\footnote{We stress the \emph{out-of-sample} and \emph{in-population} because ``in-sample" mean square error (MSE) based forecasts, and horse races thereof, are void of statistical grounds, see a.o., \cite{diebold2015comparing}.} Competing forecasts may differ in setup (e.g., forecasting inflation via linear regression or exponential smoothing), but the focus is often on augmenting a linear model with additional predictors. In such cases, models are \emph{nested} under the null hypothesis of equal forecast accuracy. Nesting comes naturally in the validation of economic theories. For instance, inflation ($\pi_t$) can be forecast using an AR(1) model, or alternatively, using an ARX(1) model as $\pi_t=\gamma_1+\gamma_2 \pi_{t-1}+\gamma_3\iota_{t-1}+\nu_t$, where interest rates ($\iota_t$) serve as exogenous input. Under the null hypothesis then, $\gamma_3=0$.  Another example of this is forecast accuracy comparison against random walk models in the field of exchange rate considered by \cite{rossi2005testing} (see \citealp{pitarakis2025novel} for an overview).

    While natural, nowadays nested comparisons go hand in hand with a high-dimensional setting. With the availability of contemporary large datasets, researchers and practitioners face an abundance of potential predictors that may or may not improve forecasts beyond standard autoregressive specifications. By far the most popular approach to 'condense' the predictive information is employing the \textit{factor structure} (see a survey in \citealp{eickmeier2008successful}). Here, a large number of potential predictors load on a small number of latent series (factors) that drive their co-movement.\footnote{To be precise, factor models are not a dimension-reduction technique per-se; it is only the reduced rank assumption on the common component, combined with white noise-like assumptions on the idiosyncratics, which have the desired reduction effect \citep[see e.g.,][for an excellent discussion]{barigozzi2024dynamic}. However, for macroeconomic applications, these are often reasonable assumptions.} Therefore, \textit{factor-augmented forecast} is the hallmark example of a nested setup. It also constitutes a whole research industry with applications in macroeconomics and finance, where Principal Components (PC) is the predominant method of factor estimation which extracts eigenvectors corresponding to the largest eigenvalues of the sample covariance matrix (see \citealp{bai2002determining}; or \citealp{bai2006confidence}). Hence, it is not surprising to see two parallel strands of econometric literature. The first focuses on the robustness of the PC procedure itself, while the second develops tests to evaluate out-of-sample forecasts with \textit{estimated} factors. An important example of the former is empirically well-documented and it concerns risks when the loadings through which the factors weight on the high-dimensional set of observables are weak (see examples in \citealp{stock2002macroeconomic}, or \citealp{ludvigson2009macro}). Technically, the eigenvalues diverge at a sub-linear rate  (see e.g. \citealp{uematsu2022inference}; \citealp{bai2023approximate}). Practically, it limits the informativeness of a data set and may cloud forecast comparisons. Indeed, the key examples of the second strand are built only on the strong loadings assumption, and they are the seminal work of \citet{gonccalves2017tests} and \citet{stauskas2022tests}\footnote{This study, however, exploits the Common Correlated Effects (CCE) estimator by \cite{pesaran2006estimation}. While this method has elegant properties when the data admit a specific structure (e.g. blocks), we focus on PC due to its versatility.}. They derive conditions under which (highly non-standard) asymptotic distributions of the tests of \cite{clark2001tests} continue to hold when the competing model is augmented with estimated factors. In the current study, we merge both strands. Firstly, we re-establish standard normal inference by borrowing from the recent developments in the theory of nested environments by \citet{pitarakis2023direct,pitarakis2025novel}. Next, we introduce the possibility of weaker loadings to quantify their effect on the forecast comparison problem. 
    
     Statistical comparison of nested models typically results in a non-standard inference. Particularly, under the null, the population errors of the two competing specifications are identical, leading to zero out-of-sample mean square error (MSE) differentials in the limit, as well as zero asymptotic variances. As a result, the test statistics become asymptotically ill-defined. On the contrary, these problems are bypassed in the non-nested case, and asymptotic normality is relatively easy to establish \citep[see][]{diebold2015comparing}. Such challenges motivated early works by \citet{clark2001tests} and \citet{mccracken2007asymptotics} who introduced adaptive normalizations of the MSEs to recover well-defined asymptotic distributions of the test statistics. However, these distributions are highly non-standard and follow stochastic integrals of Brownian motion that depend on the relative growth rate of in-sample versus out-of-sample observations.  Simulation-based approaches for estimating asymptotically valid critical values exist (see \citealp{clark2012reality}; or \citealp{hansen2015equivalence}), but their practical implementation is very challenging, which makes the results of \cite{gonccalves2017tests} elegant, but impractical.

    More recently, Jean-Yves Pitarakis (JYP) proposed a set of tests for predictive accuracy and encompassing in nested models, avoiding the non-standardness of the asymptotic distribution of the statistics under the null, as well as the variance degeneracy issue of existing procedures. \citet{pitarakis2025novel} proposes a \emph{forecast accuracy} test, still based on the MSE comparison across two nested models. However, in comparing MSEs, it uses partially overlapping out-of-sample segments to compute them, rather than the whole, same out-of-sample span. The intuition goes that as long as the fractions of out-of-sample squared forecast errors associated with the two competing specifications are different, the variance of a suitably normalized test statistic involving the MSEs cannot be degenerate. JYP then proves asymptotic normality of two types of tests under a set of general, nonrestrictive assumptions (more details in Section \ref{sec_econsetup_tests}).

    \citet{pitarakis2023direct} instead, is an \emph{encompassing} test. Meaning, it is based on the forecast encompassing principle \citep[see][]{hendry1982formulation} for which if one forecast offers no additional value over another, then an optimal convex combination of the two cannot yield a lower squared error loss. Defining $\widetilde{u}_{1,t+1}, \widetilde{u}_{2,t+1}$ as the one-step-ahead forecast errors associated with two alternative forecasts, this principle boils down to testing the population moment restriction: $\mathbb{E}({u}_{1,t+1}({u}_{1,t+1}-{u}_{2,t+1}))=0$. However, suitably normalized sample statistics to test such restriction are also plagued by the same issues that plague the equal predictive ability tests: variance degeneracy and nonstandardness of asymptotics. To circumvent these, the simple and brilliant idea in \citet{pitarakis2023direct} is that of considering the linear combination of two subsample means, in place of a unique sample mean, as additive sample counterparts of the population quantities. Simply enough, if $[1:k_0]$ is the set of in-sample observations and $[k_0+1:T]$ the out-of-sample's, this means that the sample counterpart of $\mathbb{E}[{u}_{1,t+1}{u}_{2,t+1}]$ is not going to be $(T-k_0)^{-1} \sum_{t=k_0}^{T-1} \widetilde{u}_{1,t+1}\widetilde{u}_{2,t+1}$ but $1/2 ( m_0^{-1}\sum_{t=k_0}^{m_0+k_0-1}\widetilde{u}_{1,t+1}\widetilde{u}_{2,t+1}+(T-k_0-m_0)^{-1} \sum_{t=m_0+k_0}^{T-1} \widetilde{u}_{1,t+1}\widetilde{u}_{2,t+1})$, for $m_0$ being a split point that should be different from $(T-k_0)/2$. The two subgroups of units will have slightly different variances, and that is the key to circumvent the variance degeneracy issue. As for the previous tests, JYP proves asymptotic normality of his proposed encompassing test statistics under a set of general, high-level assumptions.
    
JYP's tests, for their simplicity, robustness and generality, should be considered the new standard procedures to test for forecast accuracy and encompassing. Crucially, they have potential to make the out-of-sample evaluation of factor-augmented forecasts much more practical. However, as of now, there is no theory formally justifying their appealing properties in this setting as his high-level assumptions accommodate only observed, but not estimated factors. \citet{pitarakis2023direct} only considers a factor-augmented data generating process (cf. DGP2) as a robustness check in the simulations. \footnote{Also, upon reviewing of his simulations, it appears that the factors are not re-estimated at every roll of the out-of-sample window, but are just estimated once, prior to the recursion, over the whole out-of-sample span. This neglects the very out-of-sample nature of the setting (\citet{pitarakis2023direct} simulation scripts are available on the author's GitHub page at \url{https://github.com/jpitarakis/Multi-Step_Encompassing}).} Moreover, given the potential for a wide applicability, it is only natural to provide the justification by reflecting the risks that occur in practice. Indeed, various routes have been taken to accommodate the weakness of factor loadings, which include local factors (see \citealp{freyaldenhoven2022factor}) or sparsity  which prescribes a \textit{sparse PC} as an alternative estimator (see \citealp{uematsu2022inference}). Only recently \cite{bai2023approximate} provided a comprehensive theory for the usual PC to determine the impact of the loading weakness. It is also more general than in the latter studies. Particularly, if assumptions of \cite{uematsu2022inference} were applied to the usual PC setting, factor loadings would need to be stronger for the factors to be identified. 

As the key contribution, we provide the theoretical framework that justifies the use of the whole set of JYP's tests of forecast accuracy and encompassing, when the competing specification is that of a factor augmented autoregression. The factors are estimated at every roll of the out-of-sample window (\textit{recursively}), and the factor loadings are allowed to be strong, as well as homogeneously or heterogeneously weak. This leads to a technical contribution, where we re-work the weaker loadings theory in \citet{bai2023approximate} to the recursive estimation setting, which can be of independent interest. We spell and discuss the assumptions required to show how in all these settings JYP's test statistics maintain Gaussian asymptotic distribution and appealing power properties, as long as the loadings are at most moderately weak. This extra informativeness condition is required to take into account the recursive estimation of factors.

Finally, few words on the notation we are going to use throughout. Firstly, $a$ is a scalar, $\*a$ is a vector and $\*A_t$ is a matrix with $t$ rows. For any generic matrix $\*A$, spectral norm is $\left\|\*A \right\|_{sp}=\sqrt{\lambda_{\mathrm{max}}(\*A'\*A)}$, while $\|\*A \|=\sqrt{\mathrm{tr}(\*A'\*A)}$ is the Frobenius norm with $\mathrm{tr}(.)$ being the trace operator. Vectorization of a matrix $\*A$ is denoted by $\mathrm{vec}(\*A)$, and $\lambda_{\min(\max)}(\*A)$ denote the smallest (largest) eigenvalue. Next, $\lfloor x \rfloor$ represents the integer part of $x$, and $M$ is a positive constant, while $\sup_{a\leq t \leq b}$ ($\inf_{a\leq t \leq b}$) is supremum (infimum). Moreover, $k_0$ represents the in-sample observations, while $T-k_0=n$ are the out-of-sample observations. Convergence in distribution and probability are given by $\to_d$ and $\to_p$, respectively, while weak convergence is given by $\Rightarrow$. Ultimately, $\widetilde{a}$ and $\widehat{a}$ are quantities estimated under the observed and estimated $\*f_t$.
\section{Econometric Setup \& Tests}\label{sec_econsetup_tests}
For $t=1,\ldots,T,$ let us consider the following forecasting model 
\begin{align}
    y_{t+1}&=\+\theta'\*w_t+\+\beta'\*f_t+u_{t+1}=\+\delta'\*z_t+u_{t+1},
\end{align}
where $\*w_t\in \mathbb{R}^k$, $\*f_t\in \mathbb{R}^r$, which are stacked into $\*z_t\in\mathbb{R}^{k+r}$ with conformable parameter vectors: $\+\theta$, $\+\beta$ and $\+\delta$. We consider $\*w_t$ being the ``known factors", this can contain both lags of $y_t$ as well as an intercept, seasonal dummies or time period dummies. Instead, $\*f_t$ is a vector of ``unknown factors".
In particular, we assume that there exist a panel of $N$ series (which excludes $y_t$) whose components $x_{i,t},\; i=1,\ldots,N,$ can be decomposed into two unobservable and mutually orthogonal components: a common component $\chi_{i,t}$ and an idiosyncratic component $e_{i,t}$. Respectively, they represent the comovements and the individual features of the series. For $\chi_{i,t}$, we assume it being low-rank, i.e., to be driven linearly by an $r$-dimensional vector of common \emph{static} factors $\*f_t$, such that: $\chi_{i,t}=\+\lambda_i'\*f_t$, with $\+\lambda_i$ being, for every $i$, an $r$ dimensional vector of factor loadings. Thus, the decomposition takes the following form
\begin{align}
    x_{i,t}=\+\lambda_i'\*f_t+e_{i,t}, \hspace{2mm}\text{for $i=1,\ldots,N$ and $t=1,\ldots,T$,}
\end{align}
 or in matrix notation 
 \begin{align}\label{matrix_not}
     \*X=\*F\+\Lambda'+\*E,
 \end{align}
 where $\*F=(\*f_1,\ldots,\*f_T)'\in \mathbb{R}^{T\times r}$, $\+\Lambda=(\+\lambda_1,\ldots,\+\lambda_N)'\in\mathbb{R}^{N\times r}$ is the matrix of individual factor loadings and $\*E \in \mathbb{R}^{T\times N}$ is the matrix of idiosyncratic components.

We are interested in understanding whether the factor-augmented model ($\+\beta\neq \*0_r$, ``unrestricted") is on average better in terms of out-of-sample forecast accuracy than a simple, possibly autoregressive model ($\+\beta = \*0_r$, ``restricted"). For this purpose, we will split the sample $T$ into $T=k_0+n$, where $k_0$ and $n$ are in- and out-of-sample periods, respectively. Conveniently, we let $k_0=\lfloor T\pi_0 \rfloor$ for $\pi_0\in (0,1)$. Then, we produce \textit{recursive} pseudo out-of-sample forecasts for the restricted and unrestricted models, for $t=k_0,\ldots,T-1$, and compare their errors of the form $\widetilde{u}_{1,t+1}=y_{t+1}-\widetilde{\+\theta}_t'\*w_t$, against  $\widetilde{u}_{2,t+1}=y_{t+1}-\widetilde{\+\theta}'\*w_t-\widetilde{\+\beta}'{\*f_t}=y_{t+1}-\widetilde{\+\delta}_t'\*z_t$ for $t=k_0,\ldots,T-1$. Here and henceforth, the notation $\widetilde{a}$ $(\widehat{a})$ indicates an \textit{infeasible} (\textit{feasible}) estimator; as such, the least squares (LS) estimator
\begin{equation} \label{fls}
\widetilde{\+\delta}_t=\left(\sum_{s=1}^{t-1}\*z_s\*z_s' \right)^{-1}\sum_{s=1}^{t-1}\*z_sy_{s+1},
\end{equation}
is indeed infeasible, given that the factors $\*f_t$ are unobserved. However, we are first going to proceed as if $\*f_t$ was given.  
Reason being that the expansions of the \emph{feasible} test statistics reveal how the key component in this analysis is indeed the difference between the infeasible and feasible forecast error: $\widetilde{u}_{2,t+1}- \widehat{u}_{2,t+1}$, similarly to \citet{gonccalves2017tests} and \citet{stauskas2022tests}. Boundedness of functions of this quantity is indeed the key to the main theoretical results in this paper.  

\subsection{Tests for Forecast Encompassing and Accuracy:  Observed factors}
\par To begin with, we consider in total 3 tests of equal forecast accuracy and encompassing. The test for \emph{forecast encompassing} is given by \cite{pitarakis2023direct} and it has the following form:
\begin{align}\label{arxiv_test_main}
g_{f,1}=\frac{1}{\widetilde{\omega}_1}\Bigg(\frac{1}{\sqrt{n}}\sum_{t=k_0}^{T-1}\widetilde{u}^2_{1,t+1}&-\frac{1}{2}\Bigg[\frac{n}{m_0}\frac{1}{\sqrt{n}}\sum_{t=k_0}^{k_0+m_0-1}\widetilde{u}_{1,t+1}\widetilde{u}_{2,t+1}+ \frac{n}{n-m_0}\frac{1}{\sqrt{n}}\sum_{t=k_0+m_0}^{T-1}\widetilde{u}_{1,t+1}\widetilde{u}_{2,t+1}\Bigg] \Bigg),
 \end{align}
 where $m_0=\lfloor n \mu_0\rfloor = \lfloor (T-k_0) \mu_0\rfloor$ is a cut-off point to split the average for $\mu_0\in (0,1)$, with $\mu_0\neq 1/2$; $\widetilde{\omega}_1$ is the estimated standard deviation of the limiting distribution of the test statistic. 
 \par Two further tests of \emph{forecast accuracy} come from \cite{pitarakis2025novel}: 
 \begin{align}\label{ET1_main}
     g_{f,2}=\frac{1}{\widetilde{\omega}}_2\frac{n}{l^0_1}\left(\frac{1}{\sqrt{n}}\sum_{t=k_0}^{k_0+l^0_1-1}\widetilde{u}^2_{1,t+1}-\frac{l^0_1}{l^0_2}\frac{1}{\sqrt{n}}\sum_{t=k_0}^{k_0+l^0_2-1}\widetilde{u}^2_{2,t+1} \right)
 \end{align}
 and 
 \begin{align}\label{ET2_main}
 g_{f,3}=\frac{1}{\widetilde{\omega}}_{3}\frac{1}{n(1-\tau_0)}\sum_{l_1=\lfloor n\tau_0 \rfloor+1}^n  g_{f,2}(l_1,\lfloor n\lambda^0_2 \rfloor ).
 \end{align}
 Here, $l^0_j=\lfloor n\lambda_j^0 \rfloor$ for $j=1,2$ and $\lambda _j^0\in (0,1)$, controls the two portions of the out-of-sample period over which the forecast errors of both models are compared. One can have $l^0_1>l_2^0$ or vice-versa, which means that both portions are overlapping. However, an equality is ruled out in order to avoid the asymptotic degeneracy of variance. This test has a slight disadvantage when compared to the \cite{pitarakis2023direct}, because of the overlapping evaluation over the effective sample size,
some data, though in principal minimal, is lost in the MSEs comparison. Note also how (\ref{ET2_main}) is simply an average of (\ref{ET1_main}) over some chosen feasible set of $l_1$ for the fixed $l_2^0$ (or, fixed $\lambda_2^0$). The tuning parameter $\tau_0\in (0,1)$ helps to pick that set.\footnote{For a further discussion on how to choose the tuning parameters, we refer to \citet{pitarakis2025novel}, Section 3.} Intuition behind the averaging is the following: if $l_2^0$ is fixed and $l_1$ changes, then the MSE of the \emph{restricted} model accumulates. In effect, the uncertainty over the possible choices of $l_1$ is integrated.\\
\indent  There are several suggestions for different avenues of averaging in \cite{pitarakis2025novel}. One that is significant both practically and theoretically is averaging over $l_2$ while $l_1^0$ is fixed, such that the MSE of the \emph{unrestricted} model accumulates. Because in practice the feasible version of the unrestricted model will contain a factor estimation error, we can track if it interferes with an integration of the uncertainty around $l_2$. This results in a new statistic 
 \begin{align}\label{ET_new_main}
     g_{f,4}=\frac{1}{\widetilde{\omega}}_{4}\frac{1}{n(1-\tau_0)}\sum_{l_2=\lfloor n\tau_0 \rfloor+1}^n   g_{f,2}(\lfloor n\lambda^0_1 \rfloor, l_2  ).
 \end{align}
 This statistic is not present in \cite{pitarakis2025novel}, therefore, as an additional contribution, we provide its full analysis. Lastly, $\widetilde{\omega}^2_j$ for $j=1,\ldots,4$ represent the variance estimators of the four statistics.\\
\indent Algorithm \ref{alg1} below summarizes $g_{f,1}$ - $g_{f,4}$ and their respective variance estimators $\widetilde{\omega}^2_1,\ldots, \widetilde{\omega}_4^2$. It also provides recommendations for the tuning parameter values, which serve to boost statistical power of the tests. The recommendations stem from a theoretical and simulation-based investigation in \cite{pitarakis2025novel}, while in our simulations we experiment with different values to explore the balance size and power. 

 \begin{algorithm}[H]
\caption{Operationalizing Test Statistics}\label{alg1}
\begin{algorithmic}
\State \textbf{Select} in-sample portion $\pi_0 \in (0,1)$ and $k_0=\lfloor T\pi_0\rfloor$
\State \textbf{Obtain} out-of-sample: $n=T-k_0$
\State \textbf{Initialize} with $\widetilde{\+\theta}_{k_0}$ and $\widetilde{\+\delta}_{k_0}$ using $s=1,\ldots k_0$
\For{$t=k_0:T-1$} 
\State $\widetilde{u}_{1,t+1}=y_{t+1}-\widetilde{\+\theta}_{t}'\*w_t, $ and $ \widetilde{u}_{2,t+1}=y_{t+1}-\widetilde{\+\delta}_t'\*z_t $
\EndFor
\State \textbf{Estimate} ${\phi}^2$ as
 $\widetilde{\phi}^2=
   \frac{1}{n}\sum_{t=k_0}^{T-1}\left(\widetilde{u}_{2,t+1}^2-\frac{1}{n}\sum_{t=k_0}^{T-1}\widetilde{u}_{2,t+1}^2 \right)^2$
\State \textbf{Select} $Test\in \{g_{f,1},g_{f,2} ,g_{f,3} ,g_{f,4}  \}$
\If{$Test=g_{f,1}$}
\State $\widetilde{\omega}_1^2=\widetilde{\phi}^2\frac{(1-2\mu_0)^2}{4\mu_0(1-\mu_0)}$;\hspace{2mm}\textbf{recommend}: $\mu_0\neq 0$;\hspace{2mm}  $\mu_0\to 0.5$;
\ElsIf{$Test=g_{f,2}$}
\State$\widetilde{\omega}_2^2=\widetilde{\phi}^2\frac{|\lambda_1^0-\lambda_2^0 |}{\lambda_1^0\lambda_2^0}$; \textbf{recommend}: $\lambda_1^0 \neq \lambda_2^0$;\hspace{2mm} $\lambda_1^0,\lambda_2^0$\hspace{1mm} both close to 1;
\ElsIf{$Test=g_{f,3}$}
\State $\widetilde{\omega}_3^2=\begin{cases}
        \widetilde{\phi}^2\frac{(1-\tau_0)^2+2\lambda_2^0(1-\tau_0+\ln(\tau_0))}{\lambda_2^0(1-\tau_0)^2} \hspace{2mm}\text{if} \hspace{2mm} \lambda_2^0\leq \tau_0,\\
       \widetilde{\phi}^2 \frac{1-\tau_0^2+2\lambda_2^0((1-\tau_0)\ln(\lambda_2^0)+\tau_0\ln(\tau_0))}{\lambda_2^0(1-\tau_0)^2} \hspace{2mm}\text{if} \hspace{2mm} \lambda_2^0> \tau_0;
        \end{cases}$ \textbf{recommend}: $\tau_0\to 1$; \quad$\lambda_2^0=0.5\tau_0+0.5$;
\ElsIf{$Test=g_{f,4}$}
\State$\widetilde{\omega}_4^2=\begin{cases}
    \widetilde{\phi}^2\frac{(1-\tau_0)^2+2\lambda_1^0(1-\tau_0+\ln(\tau_0))}{\lambda_1^0(1-\tau_0)^2} \hspace{2mm}\text{if} \hspace{2mm} \lambda_1^0\leq \tau_0,\\
       \widetilde{\phi}^2 \frac{1-\tau_0^2+2\lambda_1^0((1-\tau_0)\ln(\lambda_1^0)+\tau_0\ln(\tau_0))}{\lambda_1^0(1-\tau_0)^2} \hspace{2mm}\text{if} \hspace{2mm} \lambda_1^0> \tau_0;
        \end{cases}$\hspace{1mm}\textbf{recommend}:  $\tau_0\to 1$;\hspace{2mm} $\lambda_1^0=0.5\tau_0+0.5$; 
\EndIf
\end{algorithmic}
\end{algorithm}
\noindent Note that for $j=1,\ldots, 4$, we have $\omega_j^2=\phi^2\gamma^2_j$, where $\gamma_j^2$ is known. We only need to estimate $\phi^2$ with
\begin{align}\label{infeas_omega}
 \widetilde{\phi}^2=
   \frac{1}{n}\sum_{t=k_0}^{T-1}\left(\widetilde{u}_{2,t+1}^2-\frac{1}{n}\sum_{t=k_0}^{T-1}\widetilde{u}_{2,t+1}^2 \right)^2,
\end{align}
 where in practice we use the feasible estimator $\widehat{\phi}^2$. Similar sample variance estimators have been widely used in different exercises for factor-augmented regressions (see e.g. \citealp{bai2006confidence}, \citealp{gonccalves2017tests}, or \citealp{yan2022factor})\\
 \indent For further exposition it is convenient  to split (\ref{arxiv_test_main}) - (\ref{ET_new_main}) into two components, such that for $j=1,\ldots,4$ we have
\begin{align}
    g_{f,j}= \widetilde{\omega}_j^{-1}(g_{f,j,1}+g_{f,j,2}),
\end{align}
where $g_{f,j,1}$ generates the distribution under the null, $g_{f,j,2}$ generates power under the alternative. Then, as $T\to \infty$, we shall have that, under the null, the $\widetilde{\omega}_{j}^{-1}$-scaled first component converges in distribution to a standard Gaussian density function. To see its key difference from the tests in \cite{clark2001tests}, let $\sigma^2=\mathbb{V}ar(u_{t+1})$ (unconditional variance). Then it is the process $\{u_{t+1}^2-\sigma^2 \}$ that generates the distribution, instead of the predictors. For the local power analysis in $g_{f,j,2}$, we impose $\+\beta=\+\beta^0T^{-1/4}$ as the local alternative similarly to \cite{pitarakis2023direct, pitarakis2025novel}. Thus, we require the following assumptions on the forecast error (A.1) and on the factors and predictors (A.2). They are slightly more primitive than in the original studies in order to make them compatible with and sufficient for the feasible setup explored later.  
\medskip 

\noindent \textbf{A.1 (Forecast Error)}\label{A1}
\begin{enumerate}[label=\roman*)]
    \item $\{u_{t}\}$ is a martingale difference sequence with respect to the filtration $\mathcal{F}_{t-1}=\sigma(\*z_{t-1},\ldots,\allowbreak \*X_{t-1},\ldots, y_{t-1},\ldots)$.
    \item  $\mathbb{E}(u_{t}^2|\mathcal{F}_{t-1})=\sigma_t^2$, $\sup_{k_0\leq t\leq T-1}\sigma^2_t=O_p(1)$, $\sigma^2=\mathbb{E}(\sigma_t^2)$.  
    \item $\sup_{k_0\leq t \leq T-1}\mathbb{E}(u_t^4)<M$.
    \item $u_t$ is independent of all other primitives of the model $\forall i, t, s$.
\end{enumerate}

 \noindent \textbf{A.2 (Factors and Predictors)}\label{A3}
\begin{enumerate}[label=\roman*)]
\item $\sup_{k_0\leq t\leq T-1}\mathbb{E}(\left\|\*z_t \right\|^8)<M$, where $\*z_t=(\*w_t',\*f_t')'\in \mathbb{R}^{k+r}$. 
\item $\sup_{k_0\leq t\leq T-1}\left\|\frac{1}{T}\sum_{s=1}^t(\*z_s\*z_s'-\+\Sigma_\*z) \right\|=O_p(T^{-1/2})$ and $\+\Sigma_\*z=\mathbb{E}(\*z_t\*z_t')=\begin{bmatrix}
    \+\Sigma_\*w & \+\Sigma_{\*w\*f}\\
    \+\Sigma_{\*w\*f}' & \*I_r
\end{bmatrix}$.
\item $\frac{1}{\sqrt{T}}\sum_{s=1}^t\mathrm{vec}(\*z_s\*z_s'-\+\Sigma_\*z)$ converges weakly to a Brownian motion.
\end{enumerate}
A.1 i) considers the filtration at $t-1$, hence this implicitly means that we are looking at the one-step-ahead forecast. As we provide a unified theory for all JYP's tests, and those in \citet{pitarakis2025novel} are only for one-step-ahead, this choice allows us to treat them all at once. Martingale difference sequence (MDS) assumption is natural in this setting (see the same assumption in \citealp{stauskas2022tests} or an equivalent one in \citealp{bai2006confidence}, \citealp{cheng2015forecasting}, or \citealp{karabiyik2021forecasting}). However, we shall note that this is merely for convenience, as further steps ahead and serial correlation can be considered. In particular, we will comment on how to relax the MDS requirement in Remark \ref{Remark1}. As it is likely to occur in practice, we can also allow for conditional heteroskedasticity (e.g. $u_t=\sigma_t\varepsilon_t$, were $\varepsilon_t$ is $\mathrm{IID}(0,1)$, and $\sigma_t^2$ represents ARCH/GARCH effect) in the forecast error if (\ref{infeas_omega}) is replaced by a consistent HAC. 
A.1 ii) and iv) are absent in \cite{gonccalves2017tests}, but they are fairly standard and allow to bring down moment requirements on factors and idiosyncratics. Indeed, in the equivalent assumption of A.2 in \cite{gonccalves2017tests}, 16th moments of factors are required. In our case, it is sufficient to have 8th moments. Generally, 
 parts i) - iii) of A.2 ensure that the tests in \cite{pitarakis2023direct, pitarakis2025novel} are asymptotically normal as required, but under lower-level conditions. For example, part iii) is similar to the one in \cite{clark2001tests} and \cite{stauskas2022tests}, where $\{ \mathrm{vec}(\*z_t\*z_t'-\+\Sigma_\*z)\}_{t=k_0}^{T-1}$ follows a mixing sequence of specific size (see \citealp{hansen1992convergence}).

Proposition \ref{Prop1_main} below gives the said results for $g_{f,j,1}, g_{f,j,2}$ under the above assumptions, whereas in \cite{pitarakis2023direct, pitarakis2025novel} they are obtained under high-level conditions.
\begin{proposition} \label{Prop1_main} Under Assumptions A.1 and A.2, for $j=1,\ldots,4$ as $T\to \infty$ one has
    \begin{align*}
        \widetilde{\omega}_j^{-1}g_{f,j,1}\to_d \mathcal{N}(0,1).
    \end{align*}
For the local power generating terms $g_{f,j,2}$, for $j=1,\ldots,4$ one has
    \begin{align*}
        g_{f,j,2}\to_p \sqrt{1-\pi_0}\+\beta^{0\prime}(\*I_r-\+\Sigma_{\*w\*f}'\+\Sigma^{-1}_\*w\+\Sigma_{\*w\*f})\+\beta^0.
    \end{align*}
    \textbf{Proof:} Online Supplement (Section 2). 
\end{proposition}
Recall the nulls for the encompassing test: $\mathbb{E}(u_{1,t+1}(u_{1,t+1}-u_{2,t+1}))=0$, and for the forecast accuracy tests: $\mathbb{E}(u_{1,t+1}^2-u_{2,t+1}^2)=0$. Both imply that $\+\beta=\*0_r$, and thus that $g_{f,j,2}\to_p 0$. Under the alternatives instead, $g_{f,j,2}$ converges to an expression that explicitly depends on the difference between the diagonal matrix $\+I_r$ and the quadratic form of the covariances between known and unknown factors with the precision matrix of the known factors: $\+\Sigma_{\*w\*f}'\+\Sigma^{-1}_\*w\+\Sigma_{\*w\*f}$. Note that by denoting limit of $g_{f,j,2}$ by $\psi$, we can give credence to the suggested values in Algorithm \ref{alg1}. For a standard normal CDF $\Phi(.)$ and a quantile related to size $\alpha$, the power function for each test is $1-\Phi(q_\alpha-\omega_j^{-1}\psi)$. As $\omega_j^{-1}\propto\gamma_j^{-1}$ for $\gamma_j$ that depends on the tuning parameters, we see that suggestions in Algorithm \ref{alg1} increase precision and boost the power. 
 

\indent Before we move on to the case of the estimated factors, a word on the further compact notation. We note that the components of (\ref{arxiv_test_main}) - (\ref{ET_new_main}) can be expressed as $\sum_{t=\lfloor f_1(T)\rfloor }^{\lfloor f_2(T)\rfloor}d_{u}(\widetilde{u}_{1,t+1}, \widetilde{u}_{2,t+1})$ for a loss-differential $d_{u}(.)$, where $\frac{\lfloor f_j(T)\rfloor}{T}\to q_j$ for $j=1,2$, such that $q_2>q_1$. This helps conduct analysis uniformly over different out-of-sample paths. Therefore, we will compactly formulate the results in terms of $\frac{1}{\sqrt{d_T}}\sum_{t=\lfloor f_1(T)\rfloor }^{\lfloor f_2(T)\rfloor}d_u(.)$, where $d_T=(\lfloor f_2(T)\rfloor-\lfloor f_1(T)\rfloor+1)$. They will then apply to (\ref{arxiv_test_main}) - (\ref{ET_new_main}) simultaneously.
 \begin{example}
     In the case of (\ref{arxiv_test_main}), the first component has $\lfloor f_1(T)\rfloor=\lfloor T\pi_0 \rfloor=k_0$ and $\lfloor f_2(T)\rfloor=T-1$ with $q_1=\pi_0$ and $q_2=1$. The second component has $\lfloor f_1(T)\rfloor=\lfloor T\pi_0 \rfloor=k_0$, $q_1=\pi_0$, while $\lfloor f_2(T)\rfloor=\lfloor T\pi_0 \rfloor+\lfloor (T-k_0)\mu_0 \rfloor-1=k_0+m_0-1$, $q_2=\pi_0+(1-\pi_0)\mu_0$, and so $d_T=m_0$. Then let $M_T=\sqrt{\frac{m_0}{n}}$, such that we can analyze 
     \begin{align*}
       \frac{1}{\sqrt{n}}\sum_{t=k_0}^{k_0+m_0-1}\widetilde{u}_{1,t+1}\widetilde{u}_{2,t+1} =M_T\times \frac{1}{\sqrt{d_T}}\sum_{t=\lfloor f_1(T)\rfloor}^{\lfloor f_2(T)\rfloor}\widetilde{u}_{1,t+1}\widetilde{u}_{2,t+1},
     \end{align*}
     since $M_T=O(1)$. Similar compact expression applies to (\ref{ET1_main}), (\ref{ET2_main}) and (\ref{ET_new_main}). 
 \end{example} 


\subsection{Tests for Forecast Encompassing and Accuracy: Estimated factors}
Clearly, the estimator $\widetilde{\+\delta}_t$ in \eqref{fls} is unavailable as the factors are unobserved. Its feasible counterpart is
\begin{equation}\label{feasible_deltahat}
\widehat{\+\delta}_t=\left(\sum_{s=1}^{t-1}\widehat{\*z}_s\widehat{\*z}_s' \right)^{-1}\sum_{s=1}^{t-1}\widehat{\*z}_sy_{s+1},
\end{equation}
where $\widehat{\*z}_t=(\*w_t', \widehat{\*f_t}')'$. Therefore, to use (\ref{feasible_deltahat}) in the ``for'' loop of Algorithm \ref{alg1}, we additionally obtain $\widehat{\*F}_t$ by PC in the same loop, whose procedure is defined as 
 \begin{align}\label{PC_est}
     (\widehat{\+\Lambda}_t,\widehat{\*F}_t )=\argmin_{\+\Lambda, \*F_t}\frac{1}{Nt}\sum_{i=1}^N\sum_{s=1}^t(x_{i,s}-\+\lambda_i'\*f_s)^2&=(t^{-1}\*X_t'\widehat{\*F}_t, \sqrt{t}\*U_{Nt,r}),
 \end{align}
 where $\*U_{Nt,r}$ is a $t\times r$ matrix of eigenvectors of $(Nt)^{-1}\*X_t\*X_t'$. For the minimization, the usual required normalizations are that ${t}^{-1}\widehat{\*F}_t'\widehat{\*F}_t=\*I_r$ and $\widehat{\+\Lambda}_t'\widehat{\+\Lambda}_t$  diagonal, as $\+\Lambda$ and $\*F$ are not identified separately.
 We stress that $\widehat{\+\Lambda}_t$ is indexed by $t$, because we estimate a different loading matrix for each recursion. This implies that the forecast error for the unrestricted model is given by $\widehat{u}_{2,t+1}=y_{t+1}-\widehat{\delta}'_t\widehat{\*z}_t=y_{t+1}-\widehat{\+\theta}_t'\*w_t-\widehat{\+\beta}_t'\widehat{\*f_t}$. This estimated forecast error is used in the feasible versions of (\ref{arxiv_test_main}) - (\ref{ET_new_main}). 
 \begin{example}
 The feasible version of (\ref{ET1_main}) is given by
 \begin{align}
      g_{\widehat{f},2}&=\frac{1}{\widehat{\omega}}_2\frac{n}{l^0_1}\left(\frac{1}{\sqrt{n}}\sum_{t=k_0}^{k_0+l^0_1-1}\widetilde{u}^2_{1,t+1}-\frac{l^0_1}{l^0_2}\frac{1}{\sqrt{n}}\sum_{t=k_0}^{k_0+l^0_2-1}\widehat{u}^2_{2,t+1} \right),
 \end{align}     
 where $\widehat{\omega}^2_2$ is the feasible variance estimator. 
 \end{example}
\indent Our goal is now to show that (\ref{arxiv_test_main}) - (\ref{ET_new_main}) have the same null asymptotic distribution and power properties as in the original references of \citet{pitarakis2023direct,pitarakis2025novel}. This applies both when $\*f_t$ is observed (infeasible setting), and when the factors are estimated (feasible setting) where the loadings are allowed to be strong or weak. This means, the $r$ eigenvalues of the common component covariance matrix can diverge at a sublinear rate in $N$, i.e., at rate $N^{\alpha}$, for $\alpha\in (0,1]$.  Accommodation of weaker loadings will require specific assumptions, and we follow \citealp{bai2023approximate}, who require that $N^{-\alpha}\+\Lambda'\+\Lambda$ has a positive definite limit. To provide intuition, $\alpha=1$ gives the usual strong loading case. Within this setting, factors and loadings can be consistently estimated with PC yielding the usual rate of $O_p(\max(1/\sqrt{N}, 1/\sqrt{T}))$. Instead, $\alpha=0$ leads to absolutely uninformative loadings similarly to \cite{onatski2012asymptotics}. For instance, let $r=1$, then if $N^{-\alpha}\+\Lambda'\+\Lambda =\sum_{i=1}^N\lambda_i^2 <\infty$, we have square-summable loadings, which implies that individual loadings are practically zero for highly indexed individuals. All situations in-between give weaker (or ``weakly influential") loadings \citep[see][]{de2008forecasting, onatski2012asymptotics}. This means that as $N\to \infty$ the loadings are too small or too sparse for the corresponding eigenvalues to diverge at rate $N$ \citep[see][]{barigozzi2024dynamic}. Because of this sublinear divergence, consistent estimation of the common component with PC is less straightforward. As we also prove later, in the recursive estimation setup, PC only allows recovering the factors associated with eigenvalues that diverge at least at rate equal to ${N}^{\alpha}$, for $\alpha >1/2$, which is a new result in the PC literature \citep[see also:][]{ freyaldenhoven2022factor, bai2023approximate}. \\
 \indent Recall that $\widehat{\*F}_t$ is the matrix of $r$ eigenvectors corresponding to the $r$ largest eigenvalues of $(Nt)^{-1} \*X_t\*X_t'$ and $\*D_{Nt,r}$ ($\*D_{Nt,r}^{2}$) is a matrix with its singular values (eigenvalues) arranged in a decreasing order.\footnote{Let us note that the SVD (and the subsequent decomposition) is here conducted on the $t\times t$ covariance for $\*X'$ i.e., $\frac{1}{Nt}\*X\*X'$, following the work of \citet{bai2023approximate}. The same decomposition and asymptotic expansions below can be done in the (more traditional) case of taking eigenvectors of the $N\times N$ covariance $\frac{1}{Nt}\*X'\*X$  \citep[see][]{stock2002forecasting, barigozzi2023fnets}. Both SVDs return the same set of singular values and nothing changes as long as the interest is in modeling static principal eigenvectors.}  Similar decompositions as in \citet{bai2023approximate} can therefore be obtained to get an explicit expression of the difference between the estimated factors and a rotated version of the true factors, i.e., $\widehat{\*F}_t-\*F_t\*H_{Nt,r}'$. This is important, as one never estimates the true factors but always a rotated version of them. Note that for the rotation matrix we have 
 \begin{align}\label{weak_rotation}
     \*H_{Nt,r}=\*D_{Nt,r}^{-2}(t^{-1}\widehat{\*F}_t'\*F_t)(N^{-1}\+\Lambda'\+\Lambda )=\left(\frac{N}{N^\alpha}\*D_{Nt,r}^{2}\right)^{-1}(t^{-1}\widehat{\*F}_t'\*F_t)(N^{-\alpha}\+\Lambda'\+\Lambda ),
 \end{align}
 which accommodates the fact that we employ the usual PC procedure when the loadings can be weaker. The next assumption imposes structure on the weaker loadings.
 \medskip 

  \noindent \textbf{A.3 (Loadings)}\label{A3}
\begin{enumerate}[label=\roman*)]
\item  $\left\|N^{-\alpha}\+\Lambda' \+\Lambda-\+\Sigma_{\+\Lambda}\right\| =O_p(N^{-\alpha/2})$ for a positive definite $\+\Sigma_{\+\Lambda}$ as $N\to \infty$. The loadings are independent from $\*F_t$ for all $t$.
\item  $\mathbb{E}\left[\left(\frac{1}{N^{\alpha/2}}\sum_{i=1}^N\*f_s'\+\lambda_ie_{i,t} \right)^2\right]<M$ and $\left\|\frac{1}{\sqrt{N^{\alpha}T}}\*F_t'\*E_t\+\Lambda\right\|=O_p(1)$, $\forall t,s$.\\ 
\item $\mathbb{E}\left(\left\|\frac{1}{N^{\alpha/2}}\sum_{i=1}^N\+\lambda_i'e_{i,t} \right\|^8 \right)<M,$ $\forall t$.\\ 
\item $\left\|\frac{1}{N^\alpha T }\*e_i'\*E_t\+\Lambda\right\|=O_p(N^{-\alpha})+O_p(N^{-\alpha/2}T^{-1/2}),$ $\forall i=1,\ldots,N$.\\ 
\item $\sup_{k_0\leq t\leq T-1} \|T^{-1/2}\sum_{s=1}^t\*f_s e_{i,s}\|^2\leq M,$ $\forall i=1,\ldots,N$,\\ $\left\|(NT)^{-1}\*e_s'\*E_t'\*F_t\right\|=O_p(N^{-1})+O_p(T^{-1}),$ $\forall t,s$. 
\end{enumerate}
We treat the loadings as random and make them independent from $\*F_t$ in order to simplify some arguments. Alternatively, we can impose $\mathbb{E}(\*f_t\*f_t'|\+\Lambda)=\*I_r$. They can also be treated as fixed, similarly to \cite{gonccalves2017tests}. Moreover, as in the latter study, we impose a convergence rate to facilitate quantification of our analysis (under $\alpha=1$, it coincides with the natural $N^{-1/2}$ rate) in the recursive setup. Naturally, this slightly strengthens the original assumption in \cite{bai2023approximate}. In general, i) is one of our central assumptions that allows us to conduct an asymptotic analysis uniformly in loading strength. A similar formulation is used by \cite{uematsu2022inference}, but in the context of the sparse PC estimator, and by \cite{he2025huber}, who exploit the minimization of Huber loss function to estimate the loadings. However, neither of these studies deals with hypothesis testing in a forecasting setting. Recently, \cite{boot2025diffusion} used the assumptions of \cite{bai2023approximate} to compare PC forecasts with ridge regression and random projections. The remaining assumptions are the higher-level conditions from both \citet{gonccalves2017tests} (e.g. iii) and \citet{bai2023approximate} (iv and v). The latter are also utilized in \cite{boot2025diffusion}, but they would be directly implied if $\*f_t$ and $e_{i,s}$ were independent for all $i,s,t$. The equivalent of iii) in \cite{gonccalves2017tests} requires 16-th moment, whereas we deem it very strong and bring it down due to A.1 iv). 
\subsubsection{Homogeneous Weak Loadings}
 We obtain an expansion of $\widehat{\*F}_t-\*F_t\*H_{Nt,r}'$ by implementing further modifications from both \cite{bai2023approximate} and \cite{gonccalves2017tests}. We start with the case of homogeneous loadings and move in the next section to the case of heterogeneous loadings, i.e., when the loadings are allowed to be weaker to different degrees. We define the following scalar quantities: 
\begin{align}
    &\gamma_{l,s,\alpha}=\frac{1}{N^\alpha}\sum_{i=1}^N\mathbb{E}(e_{i,l}e_{i,s})=\frac{N}{N^\alpha}\frac{1}{N}\sum_{i=1}^N\mathbb{E}(e_{i,l}e_{i,s})=\frac{N}{N^\alpha}\gamma_{l,s},\label{gamma_fact_main}\\
    & \xi_{l,s,\alpha}=\frac{N}{N^\alpha}\frac{1}{N}\sum_{i=1}^N( e_{i,l}e_{i,s}-\mathbb{E}(e_{i,l}e_{i,s}))=\frac{N}{N^\alpha}\xi_{l,s}\label{xi_fact_main}, \\
    &\eta_{l,s,\alpha}=\frac{1}{N^\alpha}\sum_{i=1}^N\*f_l'\+\lambda_ie_{i,s},\quad \nu_{l,s,\alpha}=\frac{1}{N^\alpha}\sum_{i=1}^N\*f_s'\+\lambda_ie_{i,l}\label{eta_nu_fact_main}.
\end{align}
Note that only the terms that depend on $\+\lambda_i$ are directly scaled by $N^{-\alpha}$. Since (\ref{gamma_fact_main}) and (\ref{xi_fact_main}) are functions of the idiosyncratics only, $N/N^{\alpha}$ can be seen as a ``penalty'' term on the overall rate, as we estimate potentially weaker loadings with PC. Therefore, we obtain the following decomposition for a row $s$: 
\begin{align}\label{row_fact_space_main}
    \widehat{\*f}_s-\*H_{Nt,r}\*f_s=\left(\frac{N}{N^\alpha}\*D_{Nt,r}^{2}\right)^{-1}\left(\frac{1}{t}\sum_{l=1}^t\widehat{\*f}_l\gamma_{l,s,\alpha}+ \frac{1}{t}\sum_{l=1}^t\widehat{\*f}_l\xi_{l,s,\alpha}+\frac{1}{t}\sum_{l=1}^t\widehat{\*f}_l\eta_{l,s,\alpha}+\frac{1}{t}\sum_{l=1}^t\widehat{\*f}_l \nu_{l,s,\alpha}\right).
\end{align}
This is an important term appearing when expanding $\widetilde{u}_{2,t+1}- \widehat{u}_{2,t+1}$, that is, the difference between the infeasible and feasible estimated forecast errors. Specifically, let $\+\Phi_{Nt,r}=\mathrm{diag}(\*I_k, \*H_{Nt,r})\in \mathbb{R}^{(k+r)\times (k+r)}$, where $\*H_{Nt,r}$ is the rotation matrix defined in (\ref{weak_rotation}). Then, in line with \cite{gonccalves2017tests} and \cite{stauskas2022tests}, we obtain 
    \begin{align}\label{u_diff_exp_main}
    \widetilde{u}_{2,t+1}- \widehat{u}_{2,t+1}
    & =(\widehat{\*z}_t-\+\Phi_{Nt,r} \*z_t)'(\+\Phi_{Nt,r}^{-1})'(\widetilde{\+\delta}_t-\+\delta)\notag\\
    &+\widehat{\*z}_t'(\widehat{\+\delta}_t-(\+\Phi_{Nt,r}^{-1})'\widetilde{\+\delta}_t)+(\widehat{\*z}_t-\+\Phi_{Nt,r} \*z_t)'(\+\Phi_{Nt,r}^{-1})'\+\delta= I + II + III.
\end{align}
Because it is a scalar, we have that $III=(\widehat{\*f}_t-\*H_{Nt,r}\*f_t)'(\*H_{Nt,r}^{-1})'\+\beta$, and $III$ is therefore absent when $\+\beta=\*0_r$ (under the null). The component $I$, instead, reveals that the asymptotic equivalence is ensured when the factors are estimated consistently and the infeasible forecasting model is well-specified ($\|\widetilde{\+\delta}_t-\+\delta\|=o_p(1)$). Term $II$ additionally requires that the feasible OLS estimator of the parameters is asymptotically equivalent to the infeasible one. Because $\widehat{\+\delta}_t$ employs the factors which are only identified up to a rotation, the former is naturally rotated, as well. \footnote{Indeed, by the Frisch-Waugh-Lovell argument, the second component is $(\*H_{Nt,r}^{-1})'\widetilde{\+\alpha}_t=(\*H_{Nt,r}'\*F_t'\*M_\*W\*F_t\*H_{Nt,r})^{-1}\*H_{Nt,r}'\*F_t'\*M_\*W\*y_t$, where $\*M_\*W$ is the projection matrix onto the orthogonal complement of the observed predictors.}

The expression (\ref{u_diff_exp_main}) is the integral part of the feasible versions of (\ref{arxiv_test_main}) - (\ref{ET_new_main}). To demonstrate their asymptotic equivalence, it is useful to introduce the following quantities:
\begin{align}
&A=\frac{1}{\sqrt{d_T}}\sum_{t=\lfloor f_1(T) \rfloor}^{\lfloor f_2(T)\rfloor}(\widetilde{u}_{2,t+1}- \widehat{u}_{2,t+1})^2, \label{A_comp} \\ 
  &B=\frac{1}{\sqrt{d_T}}\sum_{t=\lfloor f_1(T)\rfloor }^{\lfloor f_2(T) \rfloor}\widetilde{u}_{1,t+1}(\widetilde{u}_{2,t+1}- \widehat{u}_{2,t+1}), \label{B_comp} \\ 
  & C=\frac{1}{\sqrt{d_T}}\sum_{t=\lfloor f_1(T)\rfloor}^{\lfloor f_2(T) \rfloor}\widetilde{u}_{2,t+1}(\widetilde{u}_{2,t+1}- \widehat{u}_{2,t+1}), \label{C_comp}\\
  & D=|\widetilde{\omega}^2_j-\widehat{\omega}^2_j | \quad \text{for} \quad j=1,\ldots,4, \label{D_comp}
\end{align}
which are expressed in the compact notation in order to be applied to every statistic under consideration. Then, for $j=1,\ldots,4$
\begin{align}\label{general_equivalence}
g_{\widehat{f},j}=g_{f,j}+g_{f,j}\left(\frac{\widetilde{\omega}_j}{\widehat{\omega}_j}-1 \right)+q_j(A,B,C,D),
\end{align}
where $q_j(.)$ is a function, such that $|q_j(A,B,C,D) |=o_p(1) $ if $A,B$ and $C$ are negligible and $\widehat{\omega}_j=\widetilde{\omega}_j+o_p(1)$. Therefore, the asymptotic equivalence holds for $j=1,\ldots,4$ if (\ref{A_comp}) - (\ref{D_comp}) are negligible. Expansion of all the infeasible statistics can be found in the Online Supplementary material.
\begin{example}\label{Example2}
Let us examine the feasible versions of (\ref{ET2_main}) and (\ref{ET_new_main}): 
\begin{align*}
   & g_{\widehat{f},3}=\frac{1}{\widehat{\omega}}_{3}\frac{1}{n(1-\tau_0)}\sum_{l_1=\lfloor n\tau_0 \rfloor+1}^n  g_{\widehat{f},2}(\lfloor n\lambda^0_2 \rfloor, l_1  ),\\
  & g_{\widehat{f},4}=\frac{1}{\widehat{\omega}}_{4}\frac{1}{n(1-\tau_0)}\sum_{l_2=\lfloor n\tau_0 \rfloor+1}^n   g_{\widehat{f},2}(\lfloor n\lambda^0_1 \rfloor, l_2  ).
 \end{align*}
 Then, their expansions in spirit of (\ref{general_equivalence}) are given by
    \begin{align*}
g_{\widehat{f},j}=\begin{cases}
(j=3) \hspace{2mm}g_{f,3}+g_{f,3}\left(\frac{\widetilde{\omega}_{3}}{\widehat{\omega}_{3}} -1\right)\notag\\
       +\underbrace{\frac{1}{\widehat{\omega}}_{3}\frac{n}{l_2^0}\left(\frac{n-\lfloor n\tau_0 \rfloor}{n(1-\tau_0)} \right)\left(\frac{2}{\sqrt{n}} \sum_{t=k_0}^{k_0+l_2^0-1}\widetilde{u}_{2,t+1}(\widetilde{u}_{2,t+1}-\widehat{u}_{2,t+1})-  \frac{1}{\sqrt{n}}\sum_{t=k_0}^{k_0+l_2^0-1}(\widetilde{u}_{2,t+1}-\widehat{u}_{2,t+1})^2\right)}_{q_3(A,C,D)},\\
(j=4) \hspace{2mm} g_{f,4}+g_{f,4}\left(\frac{\widetilde{\omega}_{4}}{\widehat{\omega}_{4}} -1\right)\notag\\
       +\underbrace{\frac{1}{n(1-\tau_0)}\frac{1}{\widehat{\omega}_4}\sum_{l_2=\lfloor n\tau_0 \rfloor}^n\frac{n}{l_2}\left( \frac{2}{\sqrt{n}}\sum_{t=k_0}^{k_0+l_2-1}\widetilde{u}_{2,t+1}(\widetilde{u}_{2,t+1}-\widehat{u}_{2,t+1})- \frac{1}{\widehat{\omega}}_{4} \frac{1}{\sqrt{n}}\sum_{t=k_0}^{k_0+l_2-1}(\widetilde{u}_{2,t+1}-\widehat{u}_{2,t+1})^2\right)}_{q_4(A,C,D)}.
\end{cases}
 \end{align*}
 In contrary to $j=3$, the factor estimation error accumulates over the choices of $l_2$ under $j=4$. Clearly, in both cases for $|q_j(A,C,D)|=o_p(1)$, it is sufficient to have $|A|=o_p(1)$ and $|C|=o_p(1)$. Additionally, if $|\widetilde{\omega}_{j}^2-\widehat{\omega}_{j}^2|=o_p(1)$, then $  g_{\widehat{f},j}=g_{f,j}+o_p(1)$ for $j=3,4$. 
\end{example}
The structure of (\ref{A_comp}) - (\ref{C_comp}) reveals how the behavior of the \textit{out-of-sample} average of (\ref{u_diff_exp_main}) dictates the overall asymptotic analysis. For this, we state more assumptions regarding the idiosyncratic components. 
\medskip 

\noindent \textbf{A.4 (Idiosyncratics)}\label{A4}
\begin{enumerate}[label=\roman*)]
    \item \textit{$\mathbb{E}(\*E_t|\*F_t, \+\Lambda)=\*0_{t\times N}.$}
    \item $\sup_{k_0\leq t\leq T-1}\mathbb{E}(|e_{i,t}|^4)<M,$ $\forall i,t$. 
    \item $\mathbb{E}\left[\left(\frac{1}{\sqrt{N}}\sum_{i=1}^N(e_{i,t}e_{i,s}-\mathbb{E}(e_{i,t}e_{i,s})\right)^4 \right]<M$. 
    \item $\frac{1}{T}\sum_{t=1}^T\sum_{s=1}^T[\mathbb{E}(e_{i,t}e_{i,s})]^2<M$ .
    \item $\frac{1}{N}\sum_{i=1}^N\sum_{j=1}^N\frac{1}{T} \sum_{s_1=1}^{T}\sum_{s_2=1}^{T}\left|\mathrm{C}ov\left( e_{i,s_1}e_{j,s_1}, e_{i,s_2}e_{j,s_2}\right)\right|=O_p(1).$
    \item Let $\mathbb{E}(\*e_t\*e_t')=\+\Sigma_{e}\in \mathbb{R}^{N\times N}$. There exists $m>0$ and some positive $M$, such that $\lambda_{\mathrm{min}}(\+\Sigma_{e})>m$ and $\left\|\+\Sigma_{e} \right\|_{sp}<M.$
    \item $\sup_{k_0\leq t\leq T-1}\left\|\*E_t \right\|_{sp}=\max\{\sqrt{N},\sqrt{T}\}$.
\end{enumerate}
A.4 is similar to conditions applied in \cite{gonccalves2017tests}, except for the lower moment requirement. It allows the idiosyncratics to be weakly dependent over time and cross-sectionally. Part vii) is the same as in \cite{bai2023approximate}, but it is required to hold uniformly in $t$. We formulate it in terms of $T$, because $t=\lfloor s T\rfloor$ for some $s\in (0,1)$ due to the recursive setup. It is primarily used to simplify proofs when the loading weakness is heterogeneous, but it also helps to improve convergence rates. 
\medskip 

\noindent Lemma \ref{Lemma2_main} below formalizes the asymptotic behavior of the out-of-sample average of (\ref{u_diff_exp_main}) by providing the rate of the average square factor approximation error over the recursive samples, when the loadings are weaker. In addition, it provides the uniform rate of the in-sample factor approximation error. 
\noindent \begin{lemma} \label{Lemma2_main} Under A.1 - A.4, as $(N,T)\to \infty$, 
\begin{align*}
    &\mathrm{(i.)}\hspace{2mm}\frac{1}{d_T}\sum_{t=\lfloor f_1(T)\rfloor}^{\lfloor f_2(T)\rfloor}\left\|\widehat{\*f}_t - \*H_{Nt,r}\*f_t \right\|^2= O_p\left(\frac{N^2}{N^{2\alpha}}\frac{1}{T} \right) + O_p(N^{1-2\alpha}),\\
    & \mathrm{(ii.)}\hspace{2mm} \sup_{k_0\leq t\leq T-1} \frac{1}{t}\left\|\widehat{\*F}_t-\*F_t\*H'_{Nt,r} \right\|^2= O_p\left(\frac{N^2}{N^{2\alpha}}\frac{1}{k_0} \right) + O_p(N^{1-2\alpha}).
\end{align*}
\textbf{Proof}: Online Supplement (Section 3.2).
\end{lemma}
Lemma \ref{Lemma2_main} can be seen as an extension of Theorem 4.1 in \cite{gonccalves2017tests} to the case of weaker loadings and different alternative out-of-sample paths. If $\alpha=1$ (strong loadings), we return to the usual rate of $O_p(\max\{1/N, 1/T\})$, which coincides with the result in the latter study. Note that the result immediately implies that we must have $\frac{N}{N^\alpha}\frac{1}{\sqrt{T}}\to 0$ and $\alpha>0.5$ to consistently estimate the factor space in both cases of in- and out-of-sample. In contrast, the theory in \cite{bai2023approximate} requires a lower bound of $\alpha$ different from 0 only for inference exercises, but not consistency. The difference arises, because in part (i) we consider a recursive setup, where the rotation matrix in (\ref{weak_rotation}) changes for every $t=k_0,\ldots,T-1$, and therefore this extra informativeness condition needs to be satisfied. Part (ii.) of Lemma \ref{Lemma2_main} deals with the average square \textit{in-sample} factor estimation error, where we average over $s=1,\ldots, t$ as $ \frac{1}{t}\left\|\widehat{\*F}_t-\*F_t\*H'_{Nt,r} \right\|^2= \frac{1}{t}\sum_{s=1}^t\left\| \widehat{\*f}_s - \*H_{Nt,r}\*f_s\right\|^2$. We use $\widehat{\*F}_{t-1}$ when obtaining the feasible $\widehat{\+\delta}_{t}$ for each recursion. Interestingly, $\alpha>0.5$ is sufficient, but not necessary for part (ii) to hold, since we average for a \textit{given} $\*H_{Nt,r}$. Hence, the approximation rate can be improved with higher-level conditions in both homogeneous and heterogeneous cases, as we point out in Remark \ref{Remark2}. However, part (i) is responsible for the general out-of-sample approximation and will determine the behavior of $g_{\widehat{f},j}$ for $j=1,\ldots, 4$.  \\
\indent The following result employs (ii.) of Lemma \ref{Lemma2_main} to demonstrate the uniform equivalence of feasible and infeasible OLS estimators. 
\begin{lemma}\label{Lemma6_main}
Under Assumptions A.1 - A.4, as $(N,T)\to \infty$ we have \begin{align*}
   & \sup_{k_0\leq t\leq T-1}\left\|T^{1/4}(\widehat{\+\delta}_t-(\+\Phi_{Nt,r}^{-1})'\widetilde{\+\delta}_t)\right\|=o_p(1)
\end{align*}
\textbf{Proof:} Online Supplement (Section 4.2).
\end{lemma}
Similarly to Lemma \ref{Lemma2_main}, Lemma \ref{Lemma6_main} can be seen as a generalization of Lemma 4.1 of \cite{gonccalves2017tests} to weaker loadings. This follows from the direct application of our Lemma \ref{Lemma2_main} (ii) when establishing the uniform consistency of $\widehat{\+\delta}_t$ for (the rotated) $\widetilde{\+\delta}_t$. Hence, $\alpha>0.5$ plays a role, as well. The implication of the lemma is that the rate of consistency is $o_p(T^{-1/4})$, unlike in the latter study, where it is $o_p(T^{-1/2})$. The difference arises through two related channels. Firstly, the tests of \cite{pitarakis2023direct, pitarakis2025novel} have different local power properties. Indeed, we specify $\+\beta =\+\beta^0T^{-1/4}$, while the tests of \cite{clark2001tests} explored in \cite{gonccalves2017tests} require $\+\beta =\+\beta^0T^{-1/2}$. Secondly, we chose a different proving technique, since we only needed to demonstrate consistency, but the exact rate is less important due to different asymptotic properties of our tests.   \\
\indent In order to move on to the main results, we introduce the last assumption. 
\medskip 

\noindent \textbf{A.5 ($N,T$ Expansion Rates)}\label{A5}
\medskip

 i)\;$\frac{N}{N^{\alpha}}\frac{1}{T^{1/4}}\to c>0$, as $(N,T)\to \infty$,\quad ii)\;$\sqrt{T}N^{-\alpha}\to 0$, for $\alpha\in (0.5, 1)$.

 \medskip
 A.5 i) should be seen as a device to make convergence rates more transparent when the loadings are indeed weaker. Specifically, to prove that $B$ and $C$ in (\ref{B_comp}) and (\ref{C_comp}) are negligible, we will need $\frac{N}{N^{\alpha}}\frac{1}{\sqrt{T}}=o(1)$ as before, but at the same time $T^{1/4}\frac{N}{N^{\alpha}}\frac{1}{\sqrt{T}}=O(1)$. Thus, under A.4 i), $\frac{N}{N^{\alpha}}\frac{1}{\sqrt{T}}=O_p(T^{-1/4})$, which is the rate that can be incorporated in the subsequent analysis. Requirements in spirit of A.5 ii) are often met in the PC literature, and it helps to ensure that factor estimation error does not accumulate too fast with the expansion of $N$ and $T$. Here, it is identical to the assumption needed for inference in Lemma 4 of \cite{bai2023approximate}. Under $\alpha=1$ (strong loadings), this naturally coincides with the requirement of $\sqrt{T}N^{-1}=o(1)$ in \cite{gonccalves2014bootstrapping} or \cite{gonccalves2017tests}. Both here and in the latter study this requirement targets sums over an MDS process $\{ (\widehat{\*f}_t-\*H_{Nt,r} \*f_t)u_{t+1}\}_{t=k_0}^{T-1}$ and makes sure that they remain asymptotically negligible. Such terms appear by applying (\ref{u_diff_exp_main}) to components $B$ and $C$ in \eqref{B_comp} and \eqref{C_comp}, respectively.\\
\indent Lemma \ref{Lemma7_main} below is the central outcome that utilizes the interim results discussed above. 
\begin{lemma}\label{Lemma7_main}
Under Assumption A.1 - A.5 as $(N,T)\to \infty$ we have A - D  are asymptotically negligible.
\medskip 

 \noindent \textbf{Proof:} Online Supplement (Section 4.2).
\end{lemma}
To our knowledge, this is the first result that controls factor estimation error in the \textit{out-of-sample} context uniformly in loading strength and alternative out-of-sample paths. Apart from the desired results on $A,B$ and $C$, we can also see that the feasible variance estimator is asymptotically equivalent to the infeasible one, because
\begin{align}\label{feas_omega}
   \widehat{\phi}^2= \frac{1}{n}\sum_{t=k_0}^{T-1}\left(\widehat{u}_{2,t+1}^2-\frac{1}{n}\sum_{t=k_0}^{T-1}\widehat{u}_{2,t+1}^2 \right)^2=\widetilde{\phi}^2+o_p(1),
\end{align}
under our assumptions. This provides the last missing piece to establish the equivalence result. 

\begin{theorem} \label{Theorem1_main} Under Assumption A.1 - A.5, as $(N,T)\to \infty$ we have \begin{align*}
&  g_{\widehat{f},j}=  g_{f,j}+o_p(1), \quad \text{for $j=1,\ldots,4$}
\end{align*}
\textbf{Proof}: Follows from the application of Lemma \ref{Lemma7_main}.
\end{theorem}
The main message of Theorem \ref{Theorem1_main} is that we are able not only to use the battery of new statistics (\ref{arxiv_test_main}) - (\ref{ET_new_main}) in the popular context of factor-augmented forecasts, but that they are also robust to weaker factor loadings, as long as $\alpha>0.5$. This result is a companion to Theorem 4.2 in \cite{gonccalves2017tests}. While it does not necessarily nest their results due to different statistics and their local power properties, we have the same approximation rates under strong loadings ($\alpha=1$).
\medskip 

\noindent Before we generalize our results to the heterogeneously weak loadings, it is important to illustrate how Theorem \ref{Theorem1_main} can be used to improve the statistics. The tests of equal forecast accuracy in (\ref{ET1_main}) and (\ref{ET2_main}) depend on the tuning parameters $\lambda_1^0$ and $\lambda_2^0$. Naturally, while suggestions on their values are provided in Algorithm \ref{alg1}, your choice still alters the power properties of the tests, as argued in \cite{pitarakis2025novel}. To bypass this issue, power-enhanced versions of the statistics are presented, e.g.
    \begin{align*}
       g_{f,2}^{adj}=g_{f,2}+\widetilde{\zeta}(\lambda_1^0,\lambda_2^0),
    \end{align*}
    where $\widetilde{\zeta}(\lambda_1^0,\lambda_2^0)=\frac{1}{\widetilde{\omega}_2}\frac{1}{\lambda_2^0}\frac{1}{\sqrt{n}}\sum_{t=k_0}^{k_0+l_2^0-1}(\widetilde{u}_{1,t+1}-\widetilde{u}_{2,t+1})^2$ in spirit of \cite{fan2015power}. Clearly, the power-adjustment term is infeasible and we must replace it with $\widehat{\zeta}(\lambda_1^0,\lambda_2^0)$. Proposition \ref{prop_2_power_enhc} below demonstrates that the power enhancement procedures remain valid.
    \begin{proposition}\label{prop_2_power_enhc}
    Under Assumption A.1 - A.5, as $(N,T)\to \infty$ we have
        \begin{align*}
g^{adj}_{\widehat{f},2}=g_{\widehat{f},2}+\widehat{\zeta}(\lambda_1^0,\lambda_2^0)=g_{f,2}+\widetilde{\zeta}(\lambda_1^0,\lambda_2^0)+o_p(1),
    \end{align*}
     and the same holds for $g^{adj}_{\widehat{f},3}$ and $g^{adj}_{\widehat{f},4}$, where the adjustment term is an appropriate average of $\widehat{\zeta}(\lambda_1^0,\lambda_2^0)$.
     \medskip 
     
\noindent \textbf{Proof}: Online Supplement (Section 4.4).
    \end{proposition}
    The adjustment term $g^{adj}_{\widehat{f},4}$ concerns our new statistic $g_{\widehat{f},4}$ in (\ref{ET_new_main}). We relegate its asymptotic analysis together with a broader discussion on how the power adjustment terms are constructed to Section 4.4 in the Online Supplementary material. 
    \medskip 
    
\begin{remark}\label{Remark1}
      Assumption A.1 implies that the forecast errors are uncorrelated, and so the model is dynamically correctly specified. In practice, we may encounter e.g. measurement errors, which force $\{u_t \}$ to be correlated over time. Moreover, if we consider $h$-step-ahead forecasts, $u_{t+h}$ typically follows a moving average (MA) process of order $h-1$ (see Assumption R in \citealp{cheng2015forecasting}). To account for such possibilities for any $h\geq 1$, we can introduce three high-level conditions instead of i) of Assumption A.1 that go beyond MA processes:
     \begin{enumerate}[label=(\roman*)]
         \item $\frac{1}{\sqrt{T}}\sum_{j=1}^{\lfloor s (T-h) \rfloor}(u_{j+h}^2-\sigma^2)\Rightarrow \phi B(s)$ as $T\to \infty$, where $B(s)$ is a standard Brownian motion on $s\in [0,1]$ and $\phi^2$ is now the long-run variance.
         \item $\sup_{q\in (0,1)}\left\|\frac{1}{\sqrt{T}}\sum_{s=1}^{\lfloor q (T-h) \rfloor}\*z_su_{s+h}\right\|=O_p(1)$. 
         \item $\left\|\frac{1}{\sqrt{T}}\sum_{t=k_0}^{T-h}\left(\*A_{\lfloor q T \rfloor}\frac{1}{\sqrt{T}}\sum_{s=1}^{t-h}\*z_su_{s+h} \right)\*z_tu_{t+h} \right\|=O_p(1)$, for $\*A_{\lfloor q T \rfloor}$ independent from $u_t$ for all $t$ and such that $\*A_{\lfloor q T \rfloor}\to_p \*A$ positive definite uniformly.
     \end{enumerate}
     Conditions (i) and (ii) often appear in the literature when the forecast errors are dependent over time (see e.g. \citealp{pitarakis2025novel}). For instance, they can be implied if $\{\*z_su_{s+h} \}_{s=1}^{t-h}$ is a mixing of a relevant size. Part $(iii)$ mainly concerns the term analyzed in Lemma (\ref{Lemma6_main}). 
      Note that part (iii) resembles an object that converges weakly to a stochastic integral (see \citealp{hansen1992convergence}). Similar terms are uncovered when sums over an MDS process $\{ (\widehat{\+\delta}_t-(\+\Phi_{Nt,r}^{-1})'\widetilde{\+\delta}_t)'\*z_tu_{t+1}\}_{t=k_0}^{T-1}$ are analyzed in Lemma (\ref{Lemma7_main}) in a one-step-ahead context. For our purposes, it is sufficient for them to remain bounded to accommodate some weak dependence in $\{u_{t+h} \}$.\\
     \indent Parts (i) - (iii) above are enough for $A$ - $C$ of Lemma (\ref{Lemma7_main}). However, the MDS assumption legitimizes the use of (\ref{feas_omega}) as the estimator of $\phi^2$ that enters $D$. The full analysis of HAC-type estimators is beyond the scope of this study. However, under serial dependence, and because $\widehat{u}_{2,t+h}^2=\widetilde{u}_{2,t+h}^2+o_p(1)$ under our conditions, we conjecture that HAC estimator of $\phi^2$ as proposed in both \cite{pitarakis2023direct} and \cite{pitarakis2025novel} will remain consistent (see also discussions in \cite{fosten2016forecast} or \cite{su2025estimation}, where HAC is applied after the first-step factor estimation). 
\end{remark}

\bigbreak

\subsubsection{Heterogeneous Weak Loadings}
 What done in the previous section imposed that the loadings are weaker to the same degree, or ``homogeneously". Alternatively, we  can allow for ``heterogeneously" weak loadings by employing the normalizing matrix $\*B_N=\mathrm{diag}(N^{\alpha_1/2},\ldots, N^{\alpha_r/2})$, where $1\geq \alpha_1>\alpha_2>\ldots >\alpha_r>0$, where the weakest loading cannot still be absolutely uninformative. This means that as $N\to \infty$ some loadings are too small or too sparse for the corresponding eigenvalues to diverge at rate $N$, while others are relatively stronger or right-out strong such that the corresponding eigenvalues diverge at slighlty sub-linear rate or linear rate in $N$. Clearly, the heterogenous case nests the homogeneous one when $\alpha_1=\cdots=\alpha_r.$ 
 Note that $\|\*B_N \|\leq MN^{\alpha_1/2}$ and $\|\*B_N^{-1} \|\leq m N^{-\alpha_r/2}$ for some positive constants $M$ and $m$, which means that the order of (the inverse of) this normalization matrix is dominated by the (weakest) strongest factor loading. Again, recall that $ \widehat{\*F}_t\*D_{Nt,r}^2=\frac{1}{Nt}\*X_t\*X_t'\widehat{\*F}_t$ by the eigenvalue-eigenvector relationship. Then, by using the fact that both $\*B_N$ and $\*D_{Nt,r}^2$ are diagonal, we obtain
$\widehat{\*F}_t\*D_{Nt,r}^2\*B_N^{-1}=\widehat{\*F}_t\*B_N^{-1}\*D_{Nt,r}^2 =\widehat{\*F}_t\*B_N(\*B_N^{-2}\*D_{Nt,r}^2)=\frac{1}{Nt}\*X_t\*X_t'\widehat{\*F}_t\*B_N^{-1}.$ 

As for the above, we are after an expansion of $\widehat{\*F}_t-\*F_t\*H_{Nt,r}'$.\footnote{To be precise, we are after the expansion of $(\widehat{\*F}_t\*B_N-\*F_t\*B_N\overline{\*H}_{Nt,r}')\*B_N^{-1}$, for $\overline{\*H}_{Nt,r}:=(N\*B_N^{-2}\*D_{Nt,r}^2)^{-1}\*B_N^{-1}t^{-1}\widehat{\*F}_t'\*F_t\+\Lambda'\+\Lambda \*B_N^{-1}.$ However, as we explain in the Online Supplement, the component $\*B_N^{-1}t^{-1}\widehat{\*F}_t'\*F_t\+\Lambda'\+\Lambda \*B_N^{-1}$ is bounded in probability for $t=T$ as argued in \cite{bai2023approximate}, and we also show that it is uniformly bounded in Lemma 1 of the mathematical Online Supplement. Furthermore, using the fact that the product of $\*B_N$ and $\*D_{Nt,r}^2$ commutes we can show how $\*B_N^{-1}\overline{\*H}_{Nt,r}\*B_N=\*H_{Nt,r}.$} Hence, we will use the same definitions of the scalar quantities in (\ref{gamma_fact_main}) - (\ref{eta_nu_fact_main}). However, we replace $\alpha$ with $\alpha_r$ (in the interim, absence of a subscript $\alpha$ ($\alpha_r$) means that there is no scaling in terms of $N$). Next, to accommodate heterogeneity in loading weakness, we re-define 
\begin{align}
    &\eta_{l,s,\alpha_r}^D=\frac{\sqrt{N}}{N^{\alpha_r}}\sum_{i=1}^N(N^{-1/2}\*B_N\*f_l)'\*B_N^{-1}\+\lambda_ie_{i,s}=\frac{\sqrt{N}}{N^{\alpha_r}}\eta_{l,s}^D,\\
    &\nu_{l,s,\alpha_r}^D=\frac{\sqrt{N}}{N^{\alpha_r}}\sum_{i=1}^N(N^{-1/2}\*B_N\*f_s)'\*B_N^{-1}\+\lambda_ie_{i,l}=\frac{\sqrt{N}}{N^{\alpha_r}}\nu_{l,s}^D,
\end{align}
 where $\left\|N^{-1/2}\*B_N \right\|=O(N^{(\alpha_1-1)/2})$ that is $O(1)$ when $\alpha_1=1$. Additionally, we introduce $\*Q_{N,\alpha_r}=\*B_N^{-1}\sqrt{N^{\alpha_r}}$, which is $O(1)$, as well. Then, by again using the commutative product, we get 
\begin{align}\label{het_expand}
    & \widehat{\*f}_s-\*H_{Nt,r}\*f_s\\
     &=(N\*B_N^{-2}\*D_{Nt,r}^2)^{-1}\*Q^2_{N,\alpha_r}\left(\frac{1}{t}\sum_{l=1}^t\widehat{\*f}_l\gamma_{l,s,\alpha_r}+ \frac{1}{t}\sum_{l=1}^t\widehat{\*f}_l\xi_{l,s,\alpha_r}+\frac{1}{t}\sum_{l=1}^t\widehat{\*f}_l\eta_{l,s,\alpha_r}^D+\frac{1}{t}\sum_{l=1}^t\widehat{\*f}_l \nu_{l,s,\alpha_r}^D\right).\notag
\end{align}

\bigbreak

To handle heterogeneous weak loadings we require adaptations of the former assumptions A.3, A.5. More specifically:
\medskip

  \noindent \textbf{A.3* (Heterogeneously weak loadings)}\label{A3}
\begin{enumerate}[label=\roman*)]
\item  $\left\|\*B_N^{-1}\+\Lambda' \+\Lambda\*B_N^{-1}- \+\Sigma_{\+\Lambda}\right\|=O_p(N^{-\alpha_r/2})$ for a positive definite $\+\Sigma_{\+\Lambda}$ as $N\to \infty$. Both $N^{-\alpha}\+\Lambda' \+\Lambda$ and $\*B_N^{-1}\+\Lambda' \+\Lambda\*B_N^{-1}$ are diagonal and positive definite with eigenvalues ordered in a decreasing fashion.
\item  $\mathbb{E}\left[\left(\sum_{i=1}^N(N^{-1/2}\*B_N\*f_s)'\*B_N^{-1}\+\lambda_ie_{i,t} \right)^2\right]<M$ and $\left\|\frac{1}{\sqrt{N^{\alpha_1}T}}\*F_t'\*E_t\+\Lambda\right\|=O_p(1)$ for all $t,s$.\\ 
\item $\mathbb{E}\left(\left\|\*B_N^{-1}\sum_{i=1}^N\+\lambda_i'e_{i,t} \right\|^8 \right)<M$ for all $t$.\\ 
\item $\left\|T^{-1}\*e_i'\*E_t\+\Lambda\*B_N^{-1}\right\|=O_p(N^{-\alpha_r})+O_p\left(\sqrt{\frac{N^{\alpha_1}}{N^{\alpha_r}}}\frac{1}{\sqrt{T}} \right)$ for all $i=1,\ldots,N$.
\end{enumerate}

\noindent \textbf{A.5* ($N,T$ Expansion Rates for heterogeneously weak loadings)}\label{A4}\\
  i)\;$\frac{N}{N^{\alpha_r}}\frac{1}{T^{1/4}}\to c>0$, as $(N,T)\to \infty$, \quad ii)\;$\sqrt{T}N^{-\alpha_r}\to 0$, for $\alpha_r\in (0.5,1)$.
  \\
  
Finally, Lemma \ref{Lemma6h_main}, whose proofs are in the Online Supplement 3.2.2 \& 4.2, links the results of Lemma \ref{Lemma2_main}, Lemma \ref{Lemma6_main} and Lemma \ref{Lemma7_main} to the heterogeneous loadings context. With this, results in Theorem \ref{Theorem1_main} follow directly.
\noindent \begin{lemma} \label{Lemma6h_main} Under A.1, A.2, A.3*, A.4, A.5* and with heterogeneous loadings ($1\geq\alpha_1\geq\cdots\geq\alpha_r>1/2$), as $(N,T)\to \infty$, the results of Lemma \ref{Lemma2_main}, Lemma  \ref{Lemma6_main}, Lemma \ref{Lemma7_main} and Proposition \ref{prop_2_power_enhc} continue to hold with $\alpha_r$ in place of $\alpha$. 
\medskip 

\noindent \textbf{Proof:} Online Supplement (Section 3.2 and 4.2 as counterparts of the homogeneous case).
\end{lemma}

\begin{remark}\label{Remark2}
    It is possible to improve the rate of convergence of the in-sample estimated factors in Lemma \ref{Lemma2_main} (ii.). As Lemma 8 in the supplementary material reveals, we have for homogeneous and heterogeneous cases, respectively 
    \begin{align*}
   &(a)\hspace{2mm}\sup_{k_0\leq t\leq T-1} \frac{1}{t}\left\|\widehat{\*F}_t-\*F_t\*H'_{Nt,r} \right\|^2= O_p\left(\frac{N^2}{N^{2\alpha}}\frac{1}{T^2} \right) + O_p(N^{-\alpha}),\\
   &(b) \hspace{2mm}\sup_{k_0\leq t\leq T-1} \frac{1}{t}\left\|\widehat{\*F}_t-\*F_t\*H'_{Nt,r} \right\|^2= O_p\left(\frac{N^2}{N^{2\alpha_r}}\frac{1}{T^2} \right) + O_p(N^{-\alpha_r})
    \end{align*}
   which does not immediately restrict $\alpha$ ($\alpha_r)$. Consequently, the rates in Lemma \ref{Lemma6_main} are improved, as well. This can be achieved by following \cite{bai2023approximate} and employing a high-level assumption in A.4 vii). In this case, we do not need to analyze the sums of (\ref{row_fact_space_main}) and employ summability conditions on the idiosyncratics (the usual approach as in e.g., \citealp{bai2002determining}). This technique is unavoidable in Lemma \ref{Lemma2_main} (i.) due to the out-of-sample design. For (ii.), instead, we can focus on the norm of the whole time stack, and this gives the same rates as in Proposition 1 of \cite{bai2023approximate}. Nevertheless, the out-of-sample rates (part (i.) of Lemma \ref{Lemma2_main}) dictate the behavior of $A$ - $D$ of Lemma \ref{Lemma7_main}, therefore, $\alpha>0.5$ cannot be dispensed with. \\
   \indent Importantly, this improvement could not be achieved if the assumptions of \cite{uematsu2022inference} were applied to the usual PC. This occurs due to their unspecified restrictions between $\*F_t$ and $\*H_{Nt,r}$ that strongly moderate the relationship between $\alpha_1$ and $\alpha_r$ (see also footnote 2 in \citealp{bai2023approximate}).
\end{remark}
\section{Monte Carlo Summary}
We design a DGP similar to DGP2 in \citet{pitarakis2025novel} but where factors are specified in the same way as DGP2 of \citet{bai2023approximate} with strong/ weak homogeneous/ heterogeneous loadings. Throughout, we set the number of factors $r=3$.
\begin{align*}
  &y_{t+1}=c+\theta_1 y_t+\+\beta'\*f_t+u_{t+1}, \quad \*f_t\in\mathbb{R}^{r=3}, \\ &u_{t+1}=\begin{cases}
      NID(0,1) \hspace{2mm} (\textit{Baseline}),\\
\sigma_{t+1}\varepsilon_{t+1},\quad \varepsilon_{t+1}\sim NID(0,1),\quad  \sigma^2_{t+1}=\omega+\alpha u_t^2+\eta\sigma^2_t,\quad \omega,\alpha=0.1,\quad  \eta=0.2,  
  \end{cases} 
  \\ 
&x_{i,t}=\+\lambda_i'\*f_t+e_{i,t},\\
  &e_{i,t}=\rho_ie_{i,t-1}+\sqrt{1-\rho_i^2}v_{i,t},\quad   v_{i,t} =\begin{cases}
  NID(0,1)\hspace{2mm} (\textit{Baseline}),\\
      \epsilon_{i,t}+ \sum_{k = i+1}^{K} \xi( \epsilon_{i-k,t} + \epsilon_{i+k,t}), \quad \epsilon_{i,0} =0, \quad K = 5,\quad \xi=0.4,
  \end{cases}\\
&\*f_t\sim N(0,\*I_3),\quad \+\lambda_i\sim \*G_i\*D\*B_N/\sqrt{N}+\*G_i\pi/\sqrt{N}, \quad  \*F'\*F/T\approx\*I_3, \quad \*B_N^{-1}\*\Lambda'\*\Lambda\*B_N^{-1}\approx \*D^2,
\end{align*}
where: $\*G_i\sim N(0,\*I_3)$, $\*D^2=\operatorname{diag}(3\;2\;1);$ $ \*B_N=\operatorname{diag}(N^{\alpha_1/2}\;N^{\alpha_2/2}\;N^{\alpha_3/2})$, $(\alpha_1,\alpha_2,\alpha_3)=(1, 1, 1)$ for strong homogeneous loadings, $(0.51, 0.51, 0.51)$ for weak homogeneous loadings, $(0.51, 0.7, 1)$ for mixed strong/weak heterogeneous loadings. \footnote{Note that $\+\lambda_i$ is simulated differently than in \cite{bai2023approximate}. The reason is that the rate of $\|\*B_N^{-1}\+\Lambda'\+\Lambda\*B_N^{-1}- \+\Sigma_{\+\Lambda} \|$ plays an important role in our asymptotic analysis, whereas it did not matter for \cite{bai2023approximate}. This simulation method mimics A.3 i), because we can show that under such design 
   $\*B_N^{-1}\+\Lambda'\+\Lambda\*B_N^{-1}= \*D^2 + O_p(N^{-\alpha_r/2})$ as desired. 
} Cross-sectional and time dimensions are: $(N,T)=(800,500)$. For the practical implementations, following \citet{pitarakis2025novel}, we set $c=1.25$, $\theta_1=0.5$, $\rho_i=0.3+N(0,1)_i\times 0.5$, $\+\beta=(0,0,0)'$, $\pi=24$, for size, and $\+\beta=(j,j,j)'$ for $j\in \{0.1, 0.2, 0.3, 0.35, 0.4, 0.45, 0.5, 0.55, 0.6\}$ for power. Before we go to the results, let us mention how we performed the same simulations using \citet{bai2002determining} criterion ``$IC_{p1}$" (see their eq. 9) to actively select the number of factors. We select this number once during the in-sample period to reflect the assumption that $r$ is fixed over time. Since there is no recursion in the selection of the number of factors, this choice is justified by the results in \citet{bai2023approximate} who find that in order to estimate factors with weakly convergent loadings ($\alpha>0$), the criteria in \citet{bai2002determining} remain valid. This turned out to be identical to the results presented here as $IC_{p1}$ always correctly estimates the number of factors.\footnote{We shall mention, however, that others of the \citet{bai2002determining} criteria did not perform as well as $IC_{p1}$ in selecting the factors; we reported $IC_{p1}$ as it is the most remarkable.}
\subsection{Baseline Results}
\begin{figure}[H]
    \centering
    \begin{minipage}{0.48\textwidth}
        \centering
        \includegraphics[width=\textwidth]{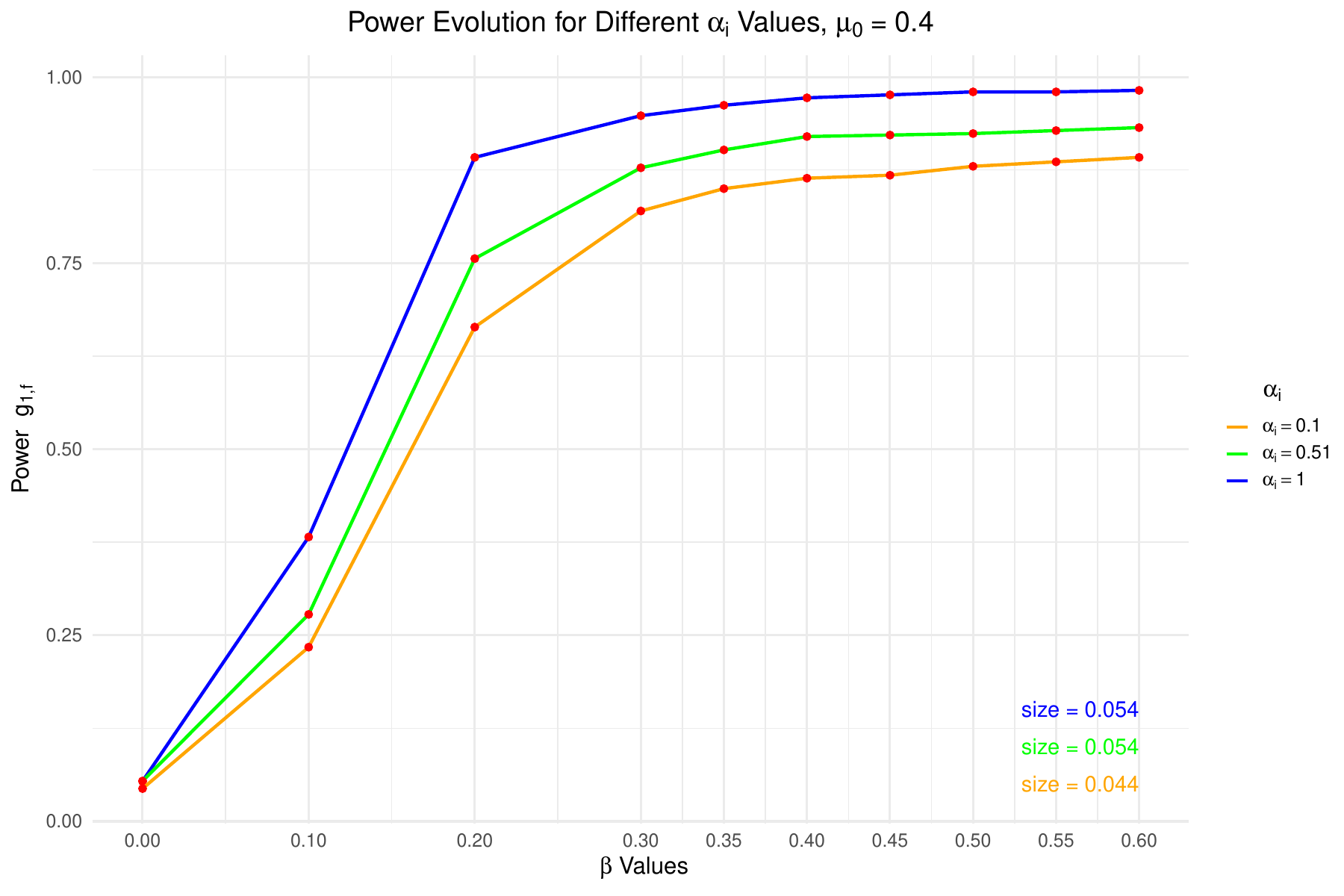}
        \caption{\small $g_{\widehat{f},1}$, Encompassing}
        \label{fig:your_label1}
    \end{minipage}
    \hfill
    \begin{minipage}{0.48\textwidth}
        \centering
        \includegraphics[width=\textwidth]{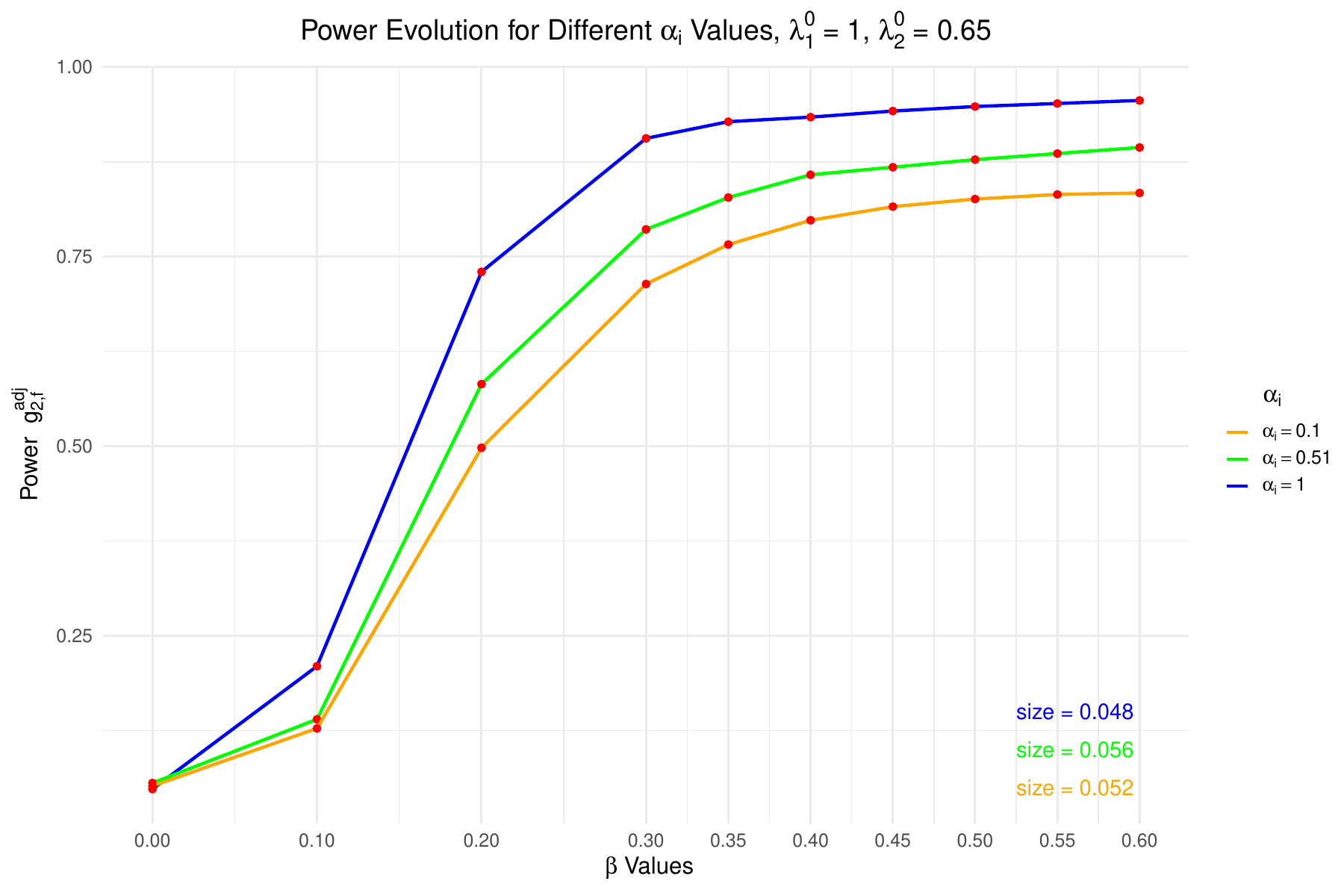}
        \caption{\small $g_{\widehat{f},2}^{adj}$, Forecast Accuracy}
        \label{fig:your_label2}
    \end{minipage}
\end{figure}

\begin{figure}[H]
    \centering
    \begin{minipage}{0.48\textwidth}
        \centering
        \includegraphics[width=\textwidth]{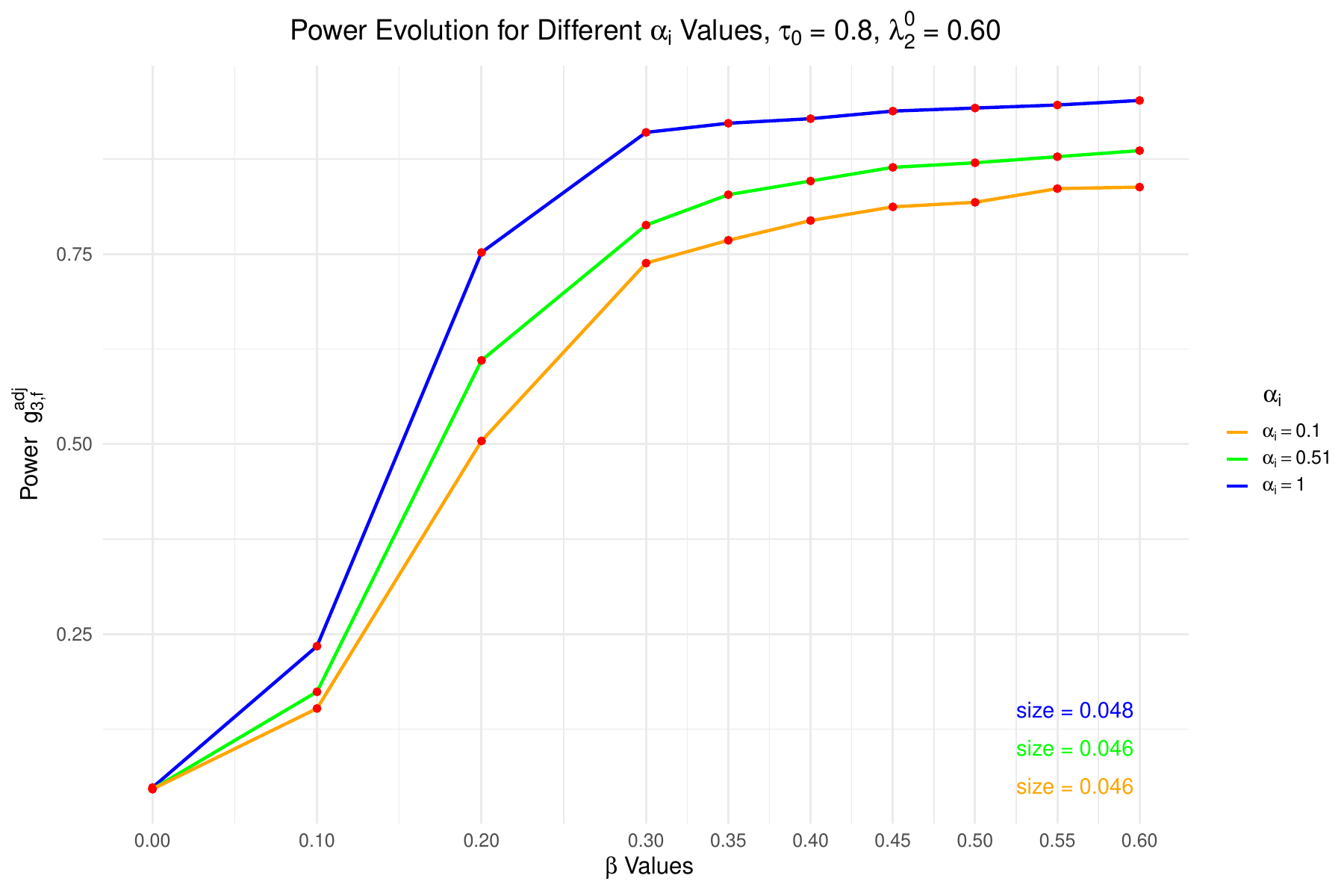}
        \caption{\small $g_{\widehat{f},3}^{adj}$, Forecast Accuracy}
        \label{fig:your_label3}
    \end{minipage}
    \hfill
    \begin{minipage}{0.48\textwidth}
        \centering
        \includegraphics[width=\textwidth]{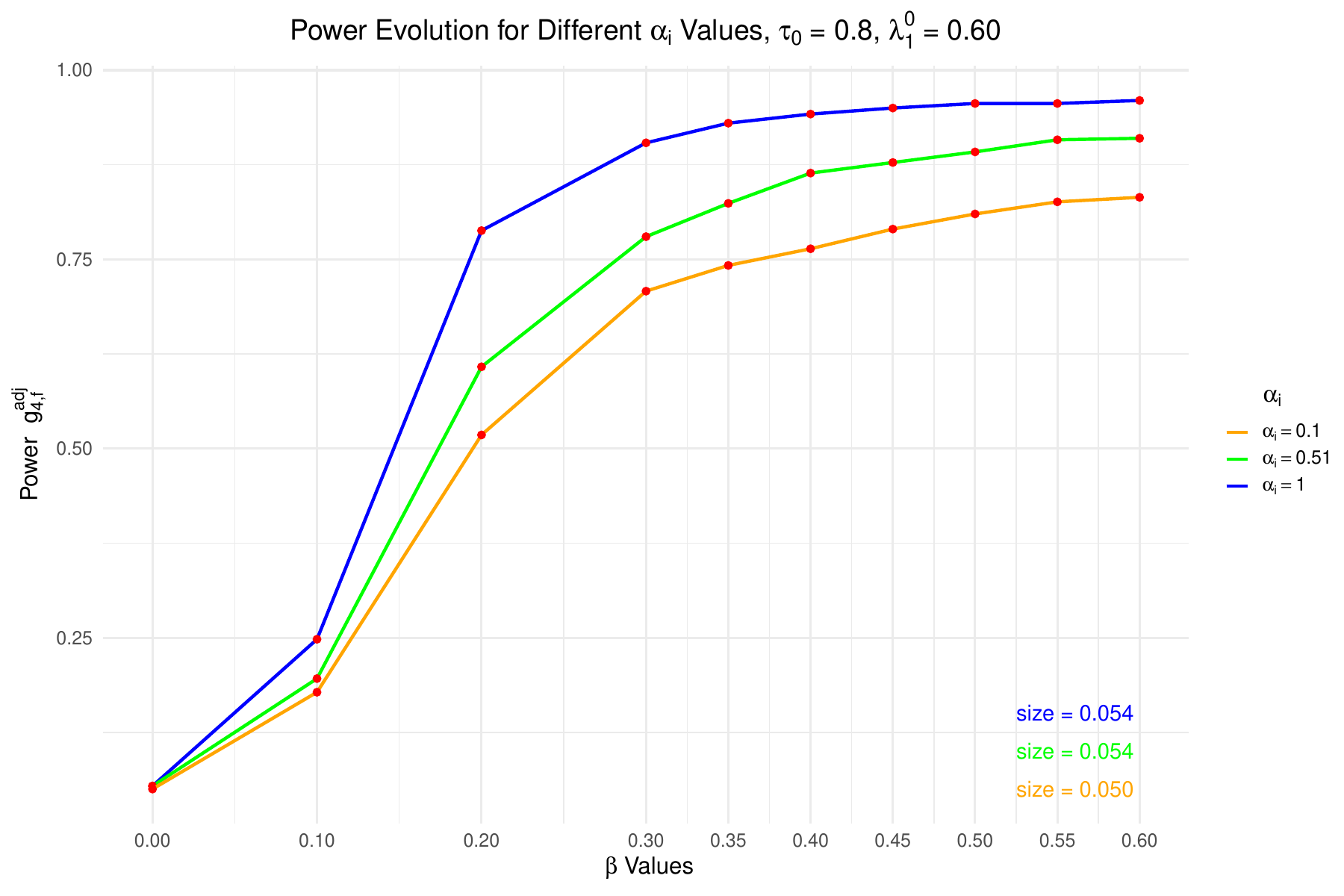}
        \caption{\small $g_{\widehat{f},4}^{adj}$, Forecast Accuracy}
        \label{fig:your_label4}
    \end{minipage}
\end{figure}

We report here a summary of the Monte Carlo in the form of power curves for the different test statistics considered. The setting is baseline, meaning that $u_{t+1}$ is uncorrelated over time and $e_{i,t}$ is uncorrelated cross-sectionally. 
 In Figure \ref{fig:your_label1} - Figure \ref{fig:your_label4}, we find both Encompassing and Forecast Accuracy tests to display satisfactory sizes and powers when factors are included in the alternative forecasting model. In terms of choice of the parameters $\mu_0, \tau_0, \lambda_1^0, \lambda_2^0$ we report here the best performing size-wise (i.e., closest to nominal level $5\%$, see Online Supplement Section 5 for the extended results). As expected, weaker loadings have a dampening effect on power, especially, of course, when the signal ($\+\beta$ value) is low. This effect is more pronounced when looking at the Forecast Accuracy tests $g_{\widehat{f},2}, g_{\widehat{f},2}^{adj}, g_{\widehat{f},3}, g_{\widehat{f},3}^{adj}$, $g_{\widehat{f},4}$, $g_{\widehat{f},4}^{adj}$, which are already affected --power-wise-- by the data-loss due to the sample overlapping discussed earlier. The power adjustment for $g_{\widehat{f},2}, g_{\widehat{f},3}, g_{\widehat{f},4}$ is paramount as the unadjusted versions can lead to severely undersized tests, as can be seen in the tables in Online Supplement Section 5. The $g_{\widehat{f},4}^{adj}$, which averages over the second coordinate thus to let the MSE of the unrestricted model accumulating, has an entirely analogous behavior as $g_{\widehat{f},3}^{adj}$ with some slightly higher power for lower $\+\beta$'s. Overall, we show how all the tests have good finite sample performances when factors are included in the alternative model specification, and loadings can be either strong or weak, homogeneous or heterogeneous. We refer to Online Supplement Section 5 for all the results, including the heterogeneous ones.

\subsection{Cross-Section Dependence}
\begin{figure}[H]
    \centering
    \begin{minipage}{0.48\textwidth}
        \centering
        \includegraphics[width=\textwidth]{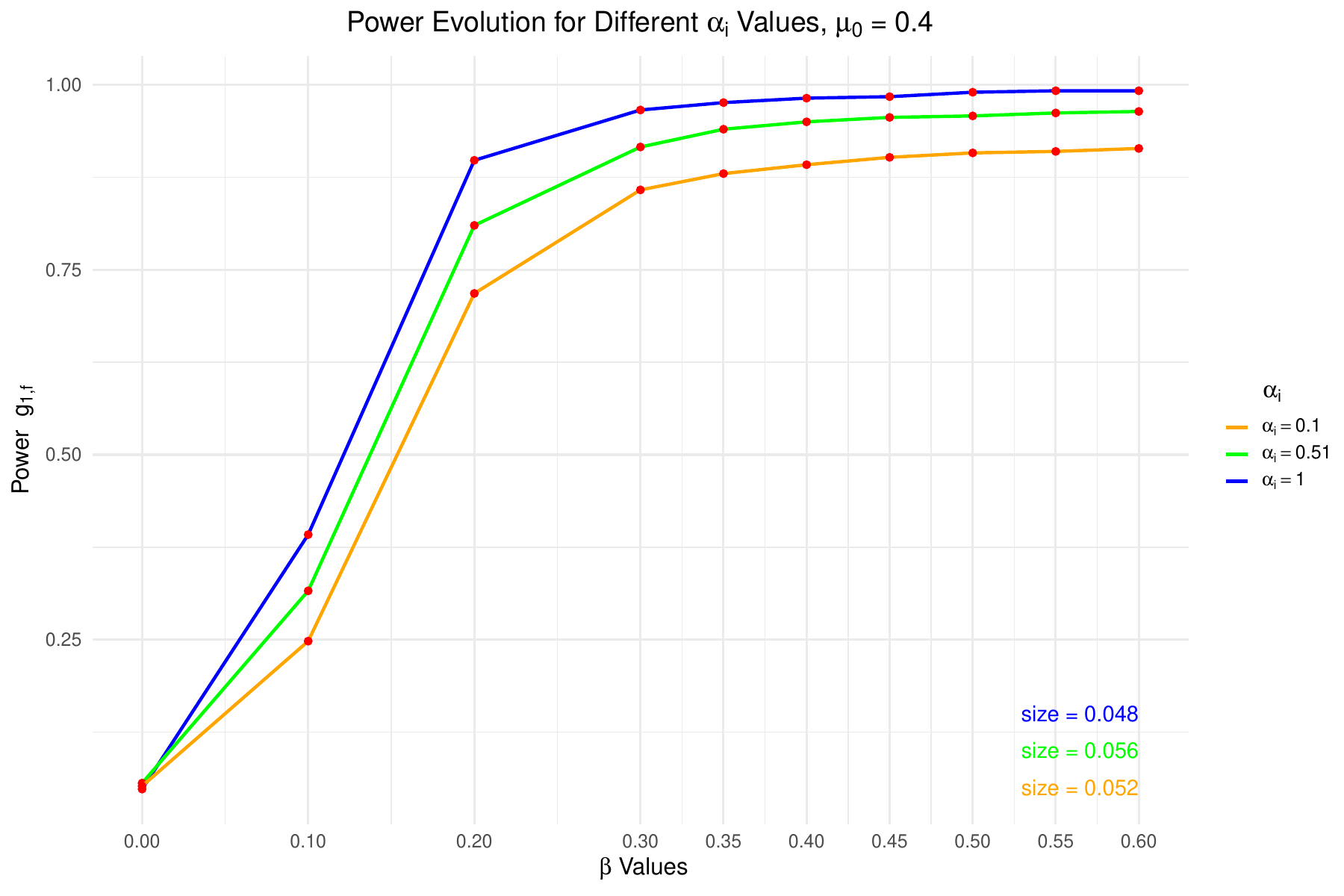}
        \caption{\small $g_{\widehat{f},1}$, Encompassing}
        \label{fig:your_label5}
    \end{minipage}
    \hfill
    \begin{minipage}{0.48\textwidth}
        \centering
        \includegraphics[width=\textwidth]{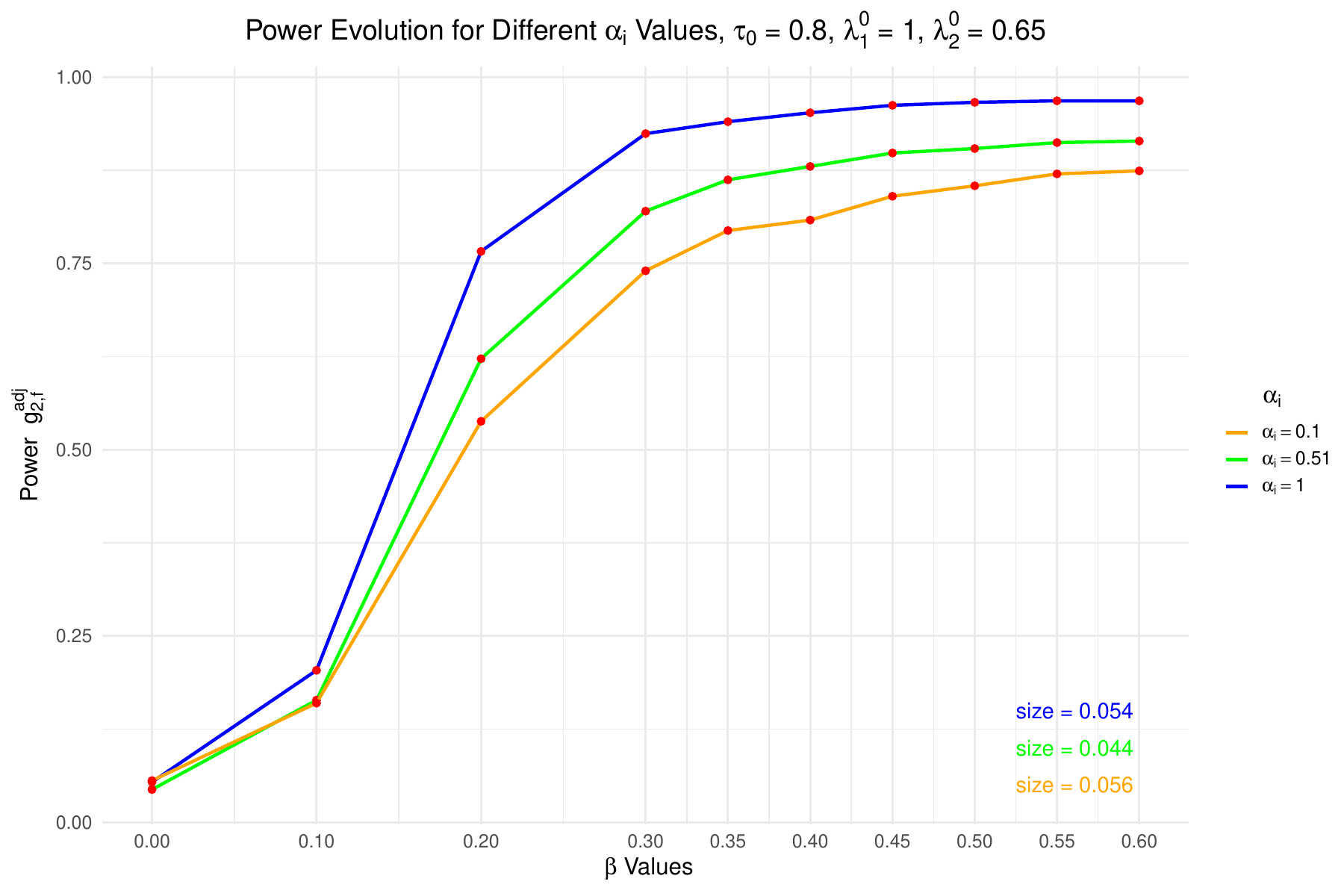}
        \caption{\small $g_{\widehat{f},2}^{adj}$, Forecast Accuracy}
        \label{fig:your_label6}
    \end{minipage}
\end{figure}

\begin{figure}[H]
    \centering
    \begin{minipage}{0.48\textwidth}
        \centering
        \includegraphics[width=\textwidth]{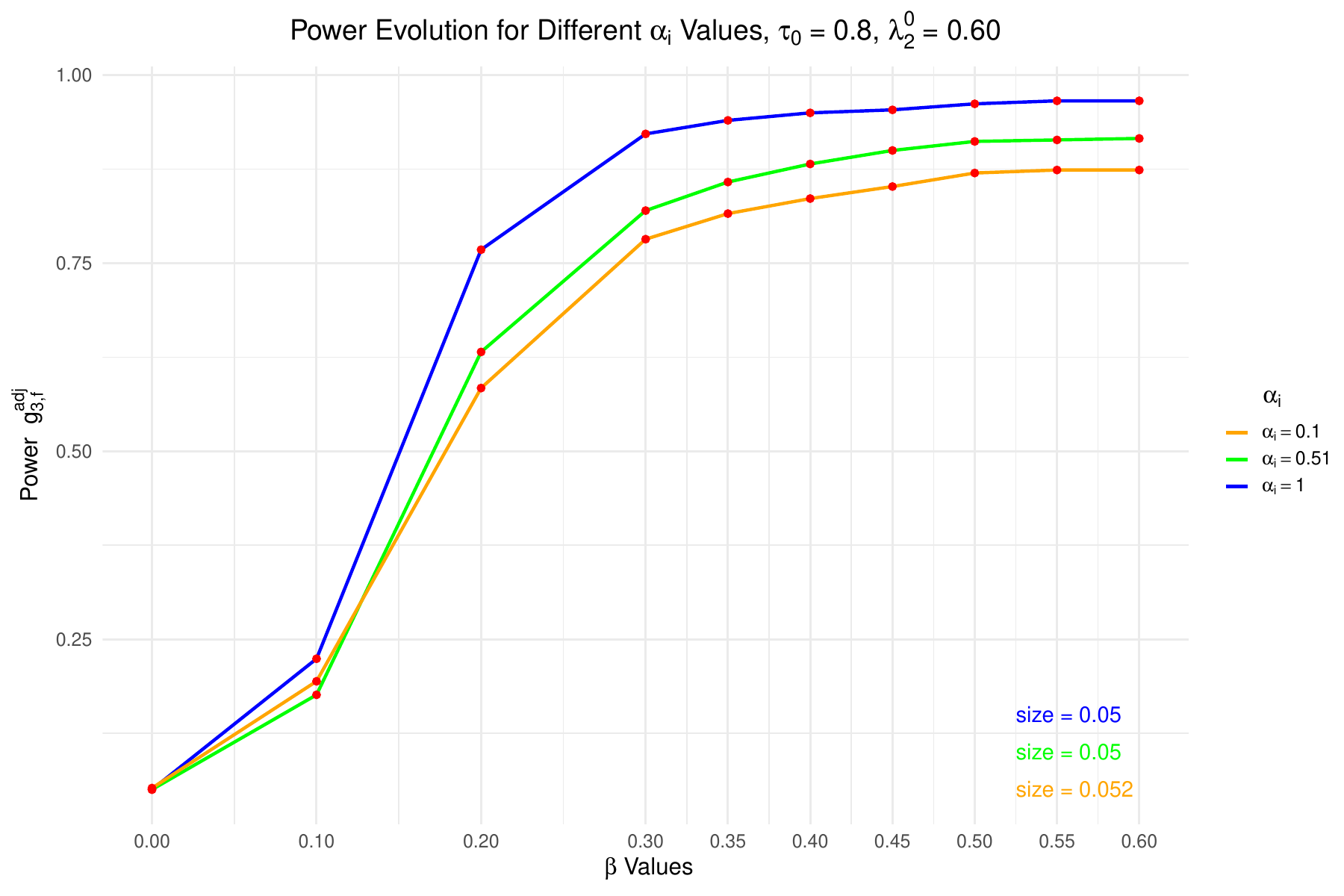}
        \caption{\small $g_{\widehat{f},3}^{adj}$, Forecast Accuracy}
        \label{fig:your_label7}
    \end{minipage}
    \hfill
    \begin{minipage}{0.48\textwidth}
        \centering
        \includegraphics[width=\textwidth]{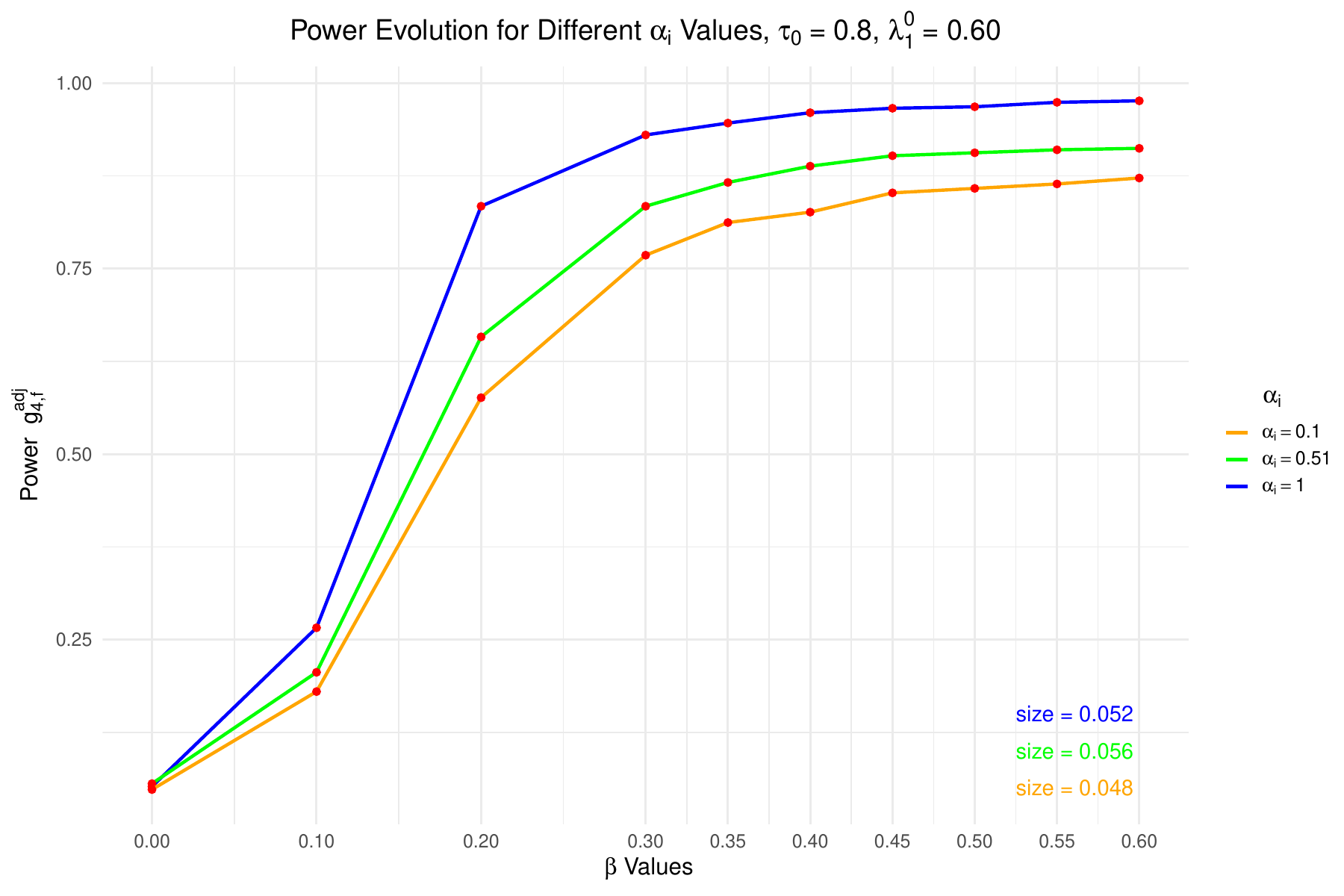}
        \caption{\small $g_{\widehat{f},4}^{adj}$, Forecast Accuracy}
        \label{fig:your_label8}
    \end{minipage}
\end{figure}

In Figure \ref{fig:your_label5} - Figure \ref{fig:your_label8} we introduce weak cross-section dependence in $e_{i,t}$ similarly to \cite{stauskas2022tests}. We see that the results are virtually the same as in the baseline scenario. This signals robustness to both temporal and, for example, spatial dependence structures in idiosyncratics.


\subsection{Conditional Heteroskedasticity}
\begin{figure}[H]
    \centering
    \begin{minipage}{0.48\textwidth}
        \centering
        \includegraphics[width=\textwidth]{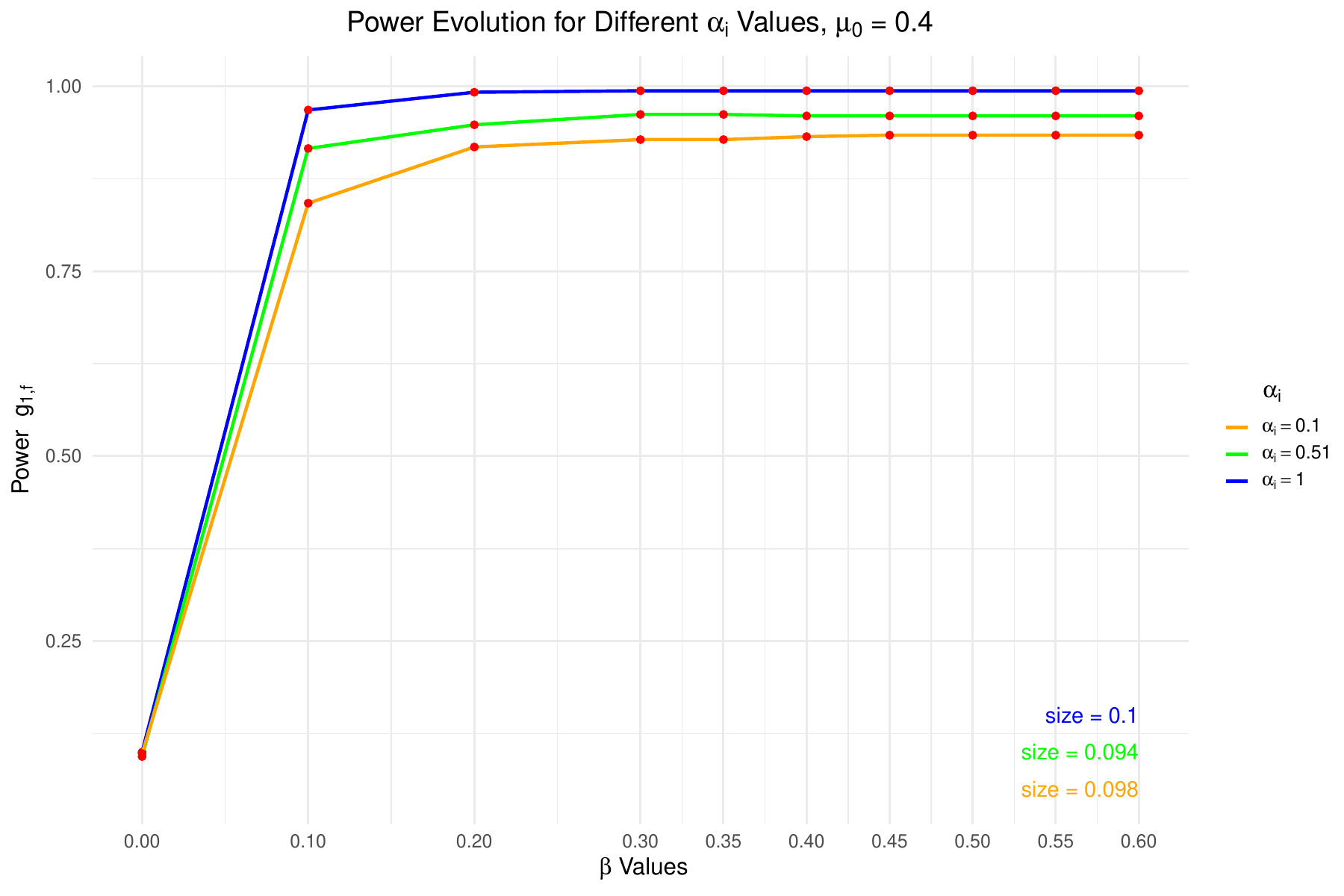}
        \caption{\small $g_{\widehat{f},1}$, Encompassing}
        \label{fig:your_label9}
    \end{minipage}
    \hfill
    \begin{minipage}{0.48\textwidth}
        \centering
        \includegraphics[width=\textwidth]{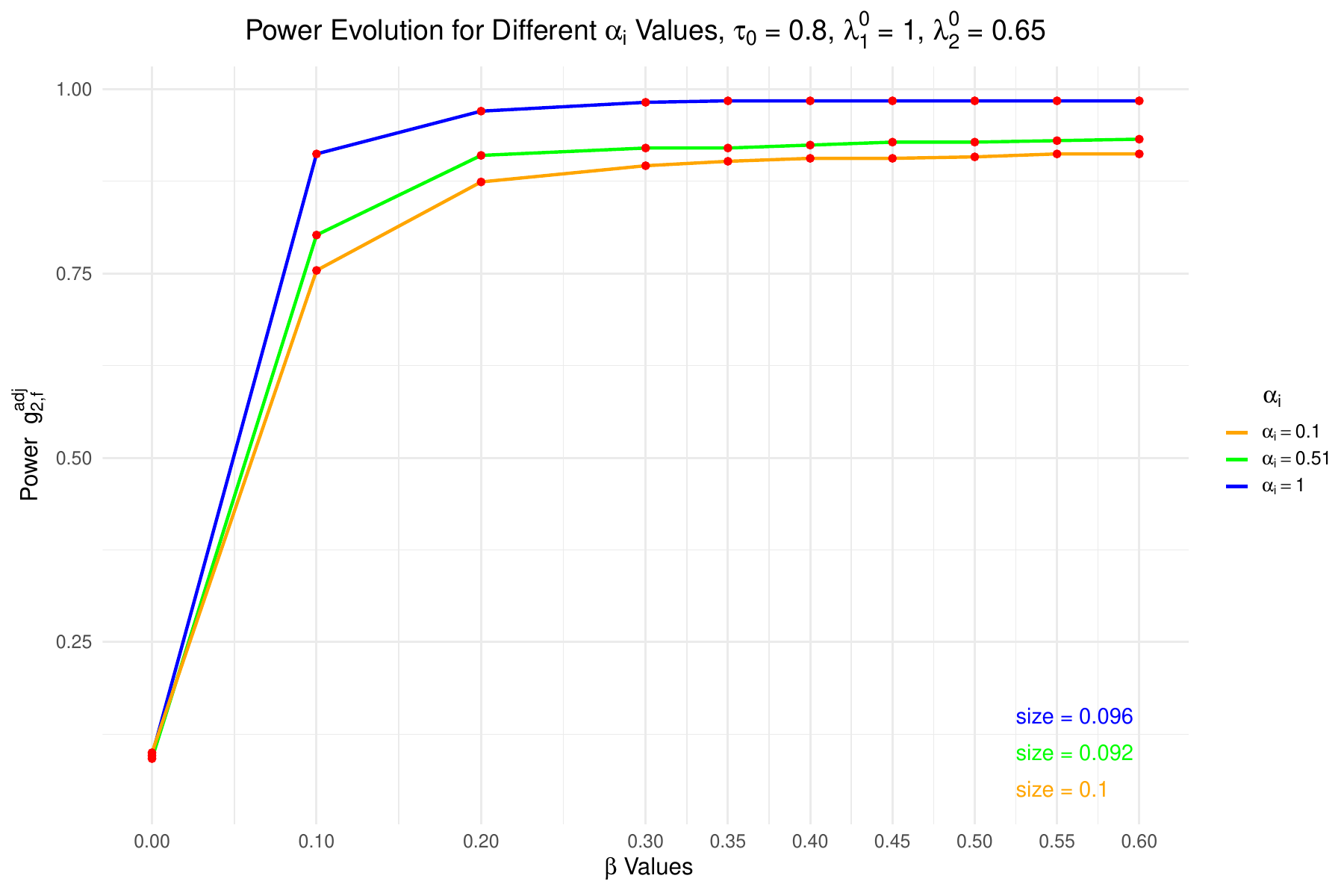}
        \caption{\small $g_{\widehat{f},2}^{adj}$, Forecast Accuracy}
        \label{fig:your_label10}
    \end{minipage}
\end{figure}

\begin{figure}[H]
    \centering
    \begin{minipage}{0.48\textwidth}
        \centering
        \includegraphics[width=\textwidth]{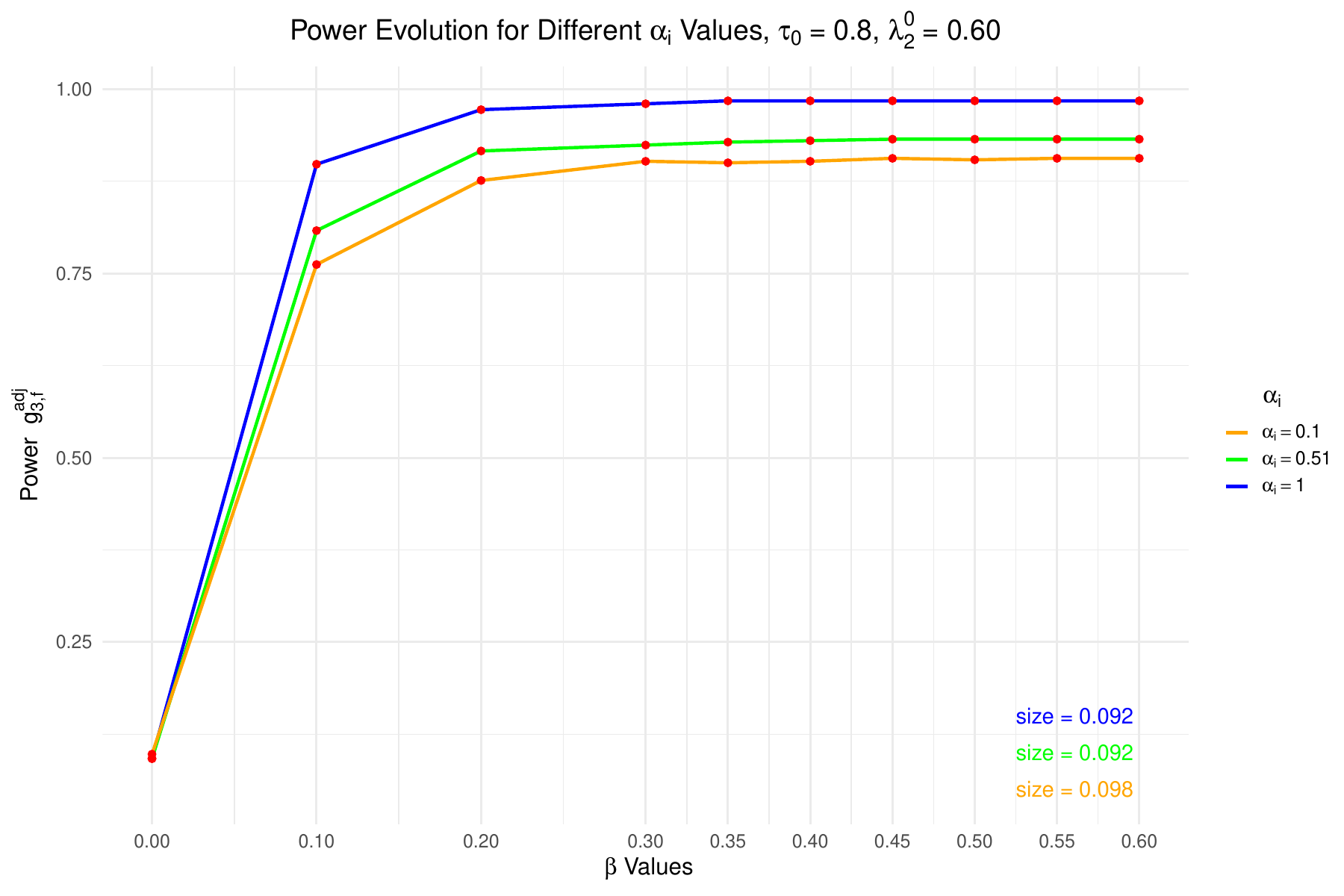}
        \caption{\small $g_{\widehat{f},3}^{adj}$, Forecast Accuracy}
        \label{fig:your_label11}
    \end{minipage}
    \hfill
    \begin{minipage}{0.48\textwidth}
        \centering
        \includegraphics[width=\textwidth]{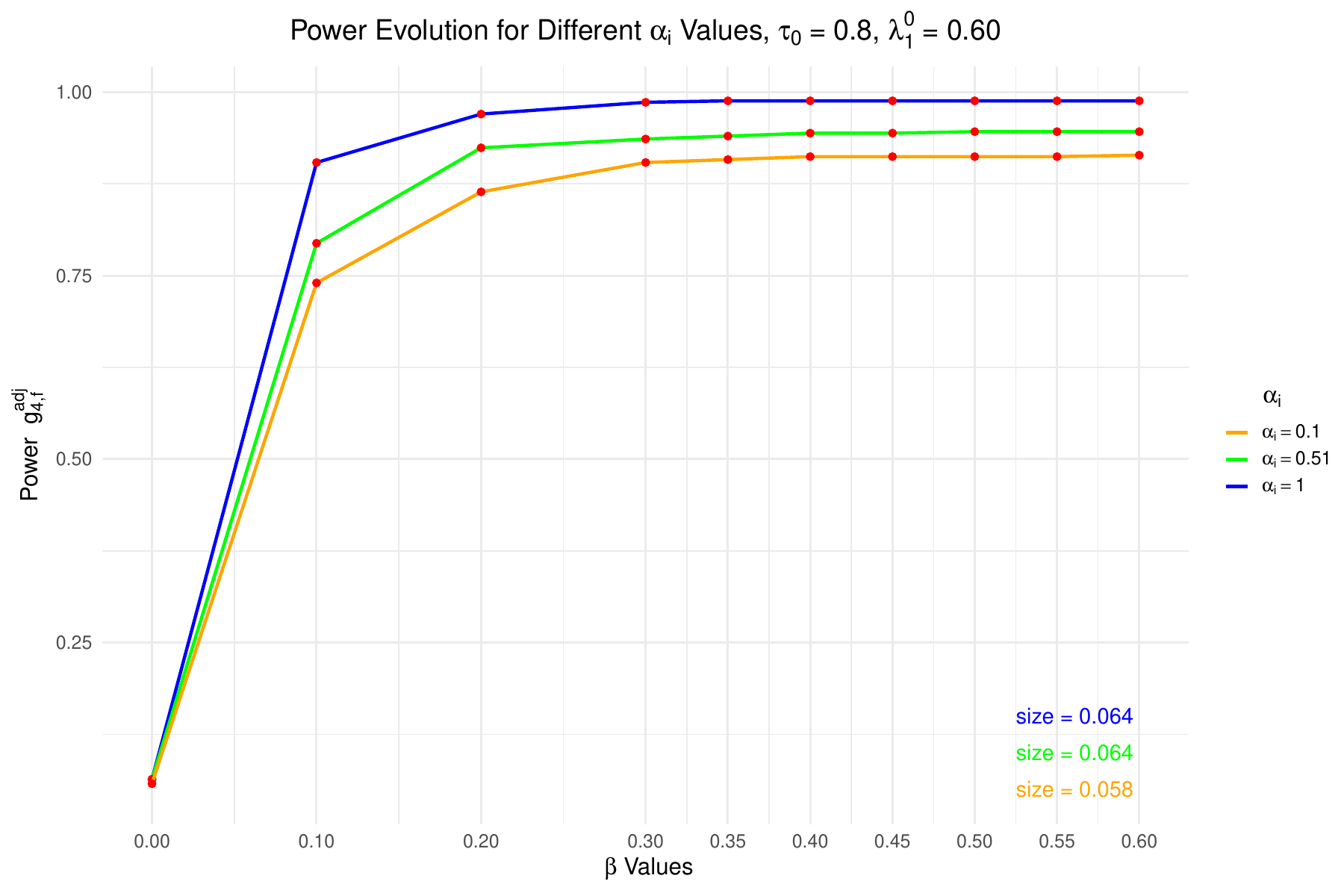}
        \caption{\small $g_{\widehat{f},4}^{adj}$, Forecast Accuracy}
        \label{fig:your_label12}
    \end{minipage}
\end{figure}

In Figure \ref{fig:your_label9} - Figure \ref{fig:your_label12} we introduce, on top of the cross-section dependence of the idiosyncratics, conditional heteroskedasticity in the forecast errors $u_{t+1}$, in the form of a GARCH$(1,1)$ as specified above. Throughout the simulations, we still use (\ref{feas_omega}) to estimate $\phi^2$. The main effect for all tests is a mild inflation of the size\footnote{As one would expect, if the persistence of the past conditional variance of $u_t$ is high(er), i.e., the $\eta$ of the GARCH$(1,1)$ is large(r), the size would suffer more. In such cases HAC-type corrections of the variance are recommended.}, which is however milder for $g_{\widehat{f},4}^{adj}$ compared to $g_{\widehat{f},1}$, $g_{\widehat{f},2}^{adj}$ and $g_{\widehat{f},3}^{adj}$. The power is instead higher for lower values of the coefficient $\boldsymbol{\beta}$ (e.g., $\bbeta=[0.10,\ldots,0.30]$) if compared to the previous results, though this might just be a byproduct of the size increase or of the GARCH time varying variance structure, which potentially makes certain periods in the sample more informative. At the same time, even though (\ref{feas_omega}) estimates unconditional variance, it can be sensitive to conditional heteroskedasticity, especially as it utilizes 4th moments. While a theoretical justification for HAC-type corrections in our testing framework is left for future research, we have experimented with the practical implementations proposed by \citet{pitarakis2023direct} for $g_{\widehat{f},1}$, specifically using Newey-West and Andrews standard errors. These methods demonstrate good empirical performance in our simulations and are thus recommended for practitioners, particularly in settings where conditional heteroskedasticity—such as GARCH-type volatility—is likely present in the forecast errors. The Online Supplement contains heterogeneous weakness results, as well.


\noindent \begin{remark}
    The large combinations of $(N,T)$ in our experiments are selected to demonstrate the asymptotic properties of the tests and the effect of weaker loadings. In Section 5.2 of the Online Supplement, we provide a full set of simulations for smaller samples (e.g., $N=100, T=200, 350$) to reflect more practical scenarios. While natural, the loss in power is not substantial. Nevertheless, small sample performance can be improved by bootstrap, because Assumptions 1-5 in \cite{gonccalves2014bootstrapping} are analogous to ours. Also, their requirement $\sqrt{T}N^{-1}\to c$ for $0\leq c < \infty$ is analogous to, for $c=0$, our current Assumption A.5, where $\sqrt{T}N^{-\alpha}\to 0$ for $\alpha \in (0.5,1)$. As such, their best-case (i.e., without any asymptotic bias term) bootstrap can be used in our context, too. The bootstrap DGP in this case can be constructed using the wild bootstrap scheme described in Section 4 of their study. Specifically, before Algorithm \ref{alg1} starts, we generate $\*X_t^*=\widehat{\+\Lambda}\widehat{\*f}_t+\*e_t^*$ and $y_{t+1}^*=\widehat{\+\delta}'\widehat{\*z}_t+u_{t+1}^*$, where $\widehat{\*z}_t=(\*w_t',\widehat{\*f}_t')'$, while $\widehat{\*f}_t$, $\widehat{\+\Lambda}$ and $\widehat{\+\delta}=(\widehat{\+\theta}', \widehat{\+\beta}')'$ are the initial estimates obtained using the whole time series sample $t=1,\ldots T$. Next, $\*e_t^*=(\nu_{1,t}\widehat{e}_{1,t}, \ldots, \nu_{N,t}\widehat{e}_{N,t})'$ and $u_{t+1}^*=\eta_{t+1}\widehat{u}_{t+1}$, where $\nu_{i,t}$ and $\eta_t$ are IID$(0,1)$ mutually independent variables that scale the residuals from the initial regression. Eventually, Algorithm \ref{alg1} is implemented for $b=1,\ldots, B$ bootstrap samples generated according to the wild scheme.
\end{remark}

\section{Empirical Applications}
\subsection{Inflation Forecasting}\label{sec_emp_appl}
\setlength{\baselineskip}{0.7cm}
We partially replicate the inflation forecast exercise conducted by \citet{pitarakis2023direct}. 
Our focus here is to explore whether \emph{global} inflation can enhance the accuracy of country-level inflation forecasts. This debate is not new; previous studies have provided evidences that global inflation trends can significantly improve domestic inflation forecasts \citep[a.o.,][]{monacelli2009international, ciccarelli2010global}. However, other research suggests that the relevance of global inflation in forecasting domestic rates may stem solely from its ability to capture slow-moving trends in inflation \citep[][]{mikolajun2016advanced}, or that a global inflation factor improves forecasting accuracy primarily at longer horizons \citep[][]{gillitzer2019does}. The question is, also, how to measure global inflation. One approach is to calculate a grand average of country-level inflation rates, represented as $\Bar{\pi}_t=N^{-1}\sum_{i=1}^N \pi_{i,t}$. Another method involves treating global inflation as a few latent factors that can be estimated using PC from the pool of country-level inflation rates. The former approach is employed by \citet{pitarakis2023direct} to illustrate his encompassing test, while we will utilize the latter method to demonstrate how JYP's encompassing tests work in an empirically relevant context where PC factors are considered. There are several reasons why PC global inflation serves as a more accurate measure than a simple grand average of inflation rates. A straightforward sample mean does not consider differences in economic size, inflation volatility, or other factors that may make some countries' inflation rates more indicative of global trends than others. In contrast, PCs can uncover patterns of co-movement in inflation rates that might not be evident from the raw data. Additionally, it is more robust to potential outliers and can adapt to time-varying relationships among countries' inflation rates. However, one could argue that if we assume the existence of only one factor, then all the information contained in that factor is effectively the same as a cross-sectional average of the countries' inflation rates. This would be true, upon essentially three assumptions: (i) the existence of an \emph{exact} factor model underlying the data, i.e., $\pi_{i,t}=\+\lambda_i'\+f_t+e_{i,t}$ with $Cov(e_{i,t}, e_{j,s})=0$, $t,s \in \mathbb{Z}$, $i,j=1,\ldots,N$, $i\neq j$; (ii) a large cross-sectional dimension $N$ (in principle $N\to \infty$); (iii) \emph{all} or \emph{most of} loadings being non-zero (i.e., pervasiveness of factors). Assume that the loadings are fixed. If all these are satisfied, it is clear how $Var(\Bar{\pi}_t)=N^{-2}(\sum_i \lambda_i)^2 Var(f_t)+N^{-2} Var(\sum_i e_{i,t})\to \Bar{\lambda}^2=Var(\Bar{\chi}_t),$ as $N\to \infty,$ meaning how the aggregation of the observed data recovers the same information contained in the factor. Now, (i) is clearly too strong (see also our A.4, v), (ii) is what is referred to as the ``blessing of dimensionality" but in practice it clearly depends on the available data, (iii) is precisely what is challenged by the weaker loadings treated in Section 2. Hence, there are good reasons to re-run this exercise using factors and employing the JYP's tests to check if global inflation computed by means of PC factors improves the country-level inflation forecast. We employ the same dataset provided in \citet{pitarakis2023direct}\footnote{Freely available on JYP's GitHub page: \url{https://github.com/jpitarakis/Multi-Step_Encompassing}}, based on the World Bank \emph{global inflation database} covering the period 1970-2023 for 23 countries at quarterly frequency. We choose the tuning parameters in line with the simulations to balance the power and size. \footnote{For the exact treatment and transformations of the raw data we refer to \citet{pitarakis2023direct}, Section 7.}  
\begin{table}[H]
\centering
\caption{$p$-values: 1-quarter-ahead}
\begin{threeparttable}
\begin{tabular}{l c c c c}
\toprule
Country & $p$-value $g_{\widehat{f},1}$ & $p$-value $g_{\widehat{f},2}^{adj}$ & $p$-value $g_{\widehat{f},3}^{adj}$ & $p$-value $g_{\widehat{f},4}^{adj}$ \\
\midrule
United States of America (USA)  & \textbf{0.007} & 0.469 & \textbf{0.008} & 0.983 \\
United Kingdom (GBR)            & 0.408 & 0.767 & 0.742 & \textbf{0.003} \\
Japan (JPN)                     & 0.284 & 0.388 & 0.094 & \textbf{0.007} \\
France (FRA)                    & \textbf{0.003} & \textbf{0.023} & \textbf{0.010} & 0.877 \\
Germany (DEU)                   & 0.863 & 0.186 & 0.539 & 0.540 \\
Spain (ESP)                     & \textbf{0.000} & \textbf{0.000} & \textbf{0.000} & 0.936 \\
Italy (ITA)                     & 0.859 & 0.306 & 0.845 & 0.392 \\
Netherlands (NLD)              & \textbf{0.001} & \textbf{0.000} & \textbf{0.002} & 0.192 \\
Luxembourg (LUX)               & \textbf{0.000} & 0.194 & \textbf{0.044} & 0.372 \\
Canada (CAN)                   & 0.070 & 0.711 & 0.142 & 0.736 \\
Ireland (IRL)                  & 0.196 & 0.574 & 0.067 & 0.110 \\
Finland (FIN)                  & 0.898 & 0.972 & 0.956 & 0.121 \\
New Zealand (NZL)              & 0.998 & 0.971 & 0.969 & 0.336 \\
Greece (GRC)                   & \textbf{0.004} & 0.359 & 0.525 & \textbf{0.013} \\
Portugal (PRT)                 & 0.241 & 0.355 & 0.329 & \textbf{0.000} \\
Norway (NOR)                   & 0.257 & 0.905 & 0.866 & 0.168 \\
South Korea (KOR)              & 0.459 & 0.714 & 0.569 & \textbf{0.004} \\
Denmark (DNK)                  & 0.414 & 0.473 & 0.527 & \textbf{0.000} \\
Sweden (SWE)                   & 0.496 & 0.855 & 0.787 & \textbf{0.000} \\
Australia (AUS)                & 0.635 & 0.793 & 0.869 & \textbf{0.022} \\
Austria (AUT)                  & \textbf{0.002} & \textbf{0.003} & \textbf{0.009} & \textbf{0.000} \\
Belgium (BEL)                  & 0.233 & 0.173 & 0.485 & 0.348 \\
Switzerland (CHE)             & \textbf{0.000} & 0.138 & \textbf{0.000} & 0.312 \\
\bottomrule
\end{tabular}
\begin{tablenotes}
\footnotesize
\item Notes: $AR(1)$ vs factor augmented $AR(1)$; nr of factors selected with \citet{bai2002determining} $IC_{p1}$, $\mu_0=0.40$, $\tau_0=0.8$, for $g_{\widehat{f},2}^{adj}:$ $\lambda_1^0=1, \lambda_2^0=0.65$; for $g_{\widehat{f},3}^{adj}:$ $\lambda_2^0=0.6$; for $g_{\widehat{f},4}^{adj}:$ $\lambda_1^0=0.6$.
\end{tablenotes}
\end{threeparttable}
\end{table}

\begin{figure}[h]
  \centering
  \begin{minipage}[b]{0.49\textwidth}
    \includegraphics[width=\textwidth]{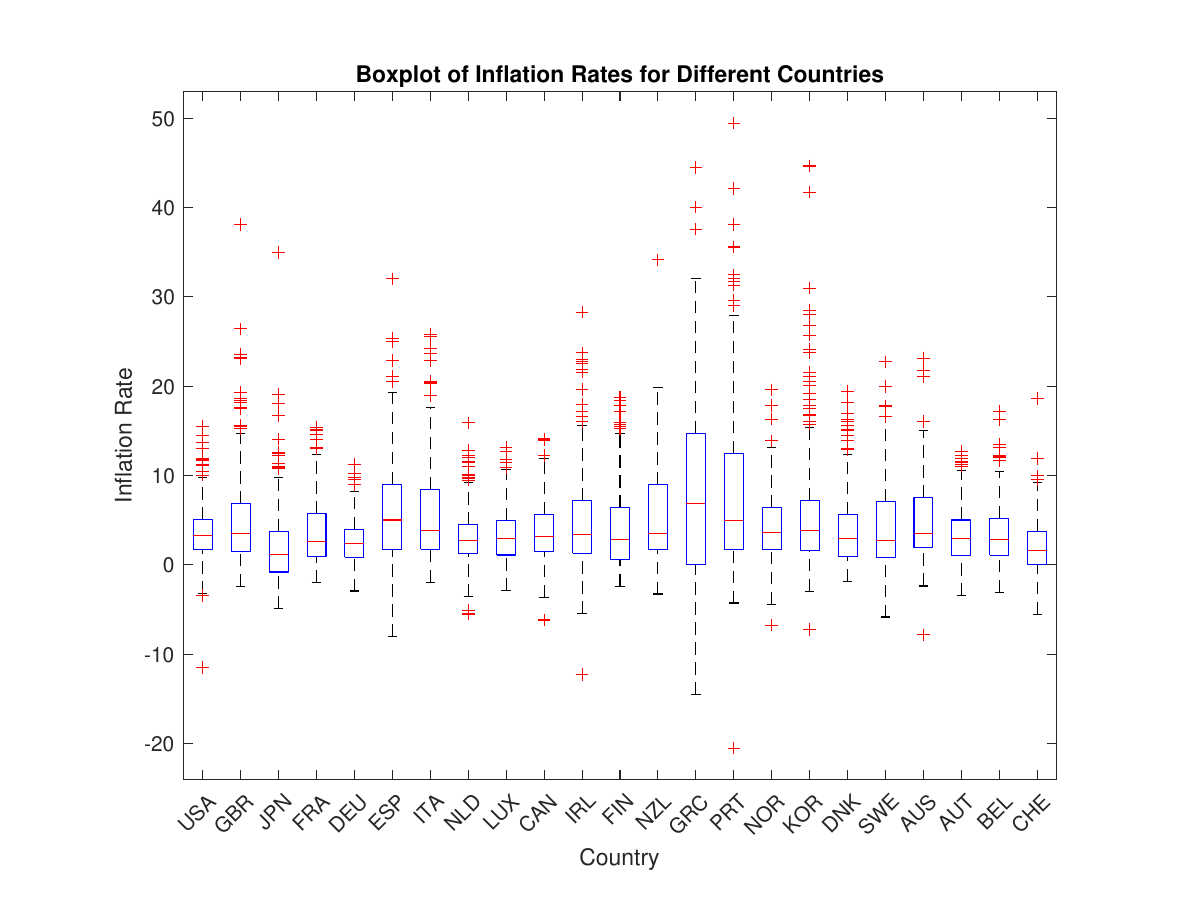}
    \caption{Inflation boxplot}
    \label{boxplot}
  \end{minipage}
  \hfill
  \begin{minipage}[b]{0.49\textwidth}
    \includegraphics[width=\textwidth]{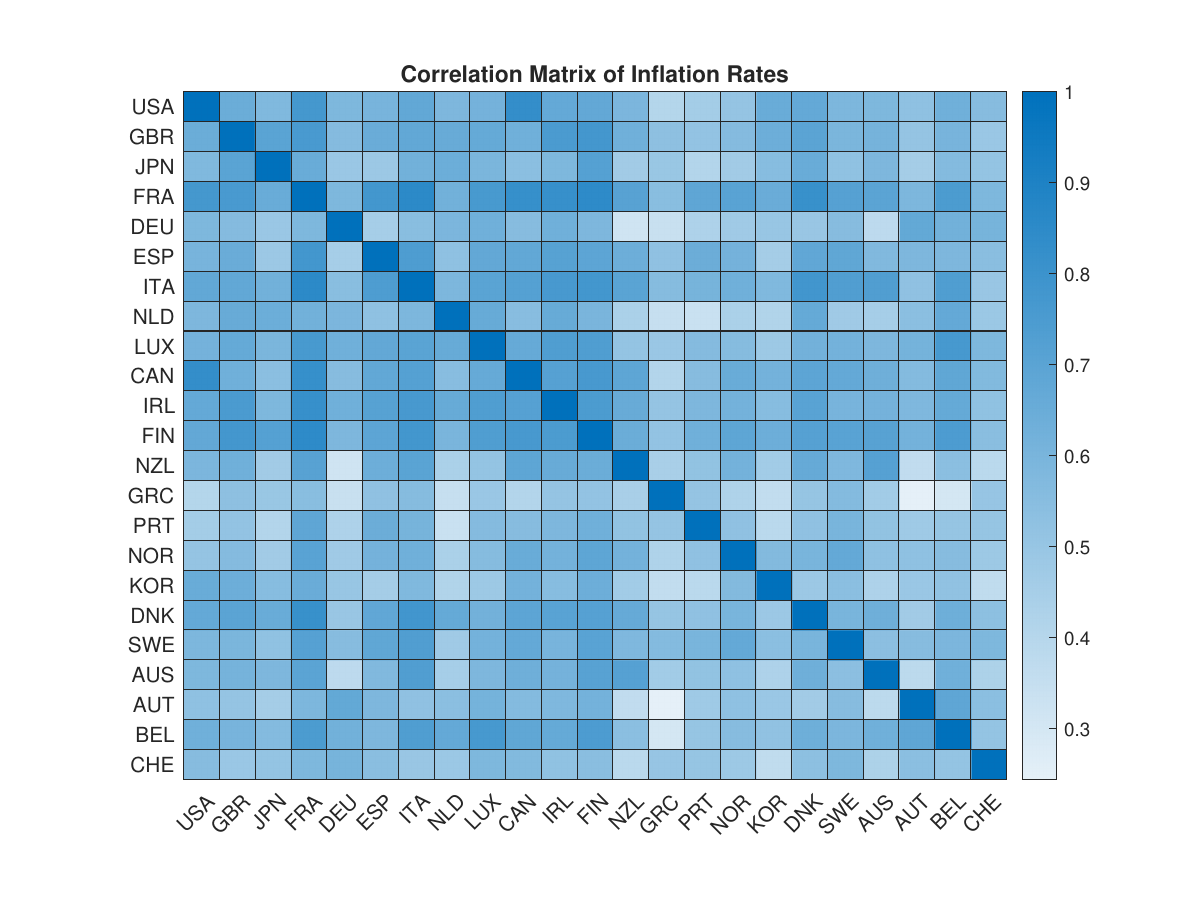}
    \caption{Correlation heatmap}
    \label{heatmap}
  \end{minipage}
\end{figure}

 The base-line model is an $AR(1)$ while the alternative model is a factor-augmented $AR(1)$ where factors are recursively estimated and their number is determined via \citet{bai2002determining} information criterion ($IC_{p1}$, max number$=10$).\footnote{It is well known how the \citet{bai2002determining} criteria depend quite substantially on the maximum number of factors as selected by the practitioner, as well as the relative magnitude of $N,T$ \citep[see e.g.,][]{forni2009opening}. In this case, we find on average that all criteria, including $IC_{p1}$ but excluding $AIC_3$, return the maximum as estimated number of factors. We experimented using $AIC_3$ and the ABC criterion of \citet{alessi2010improved} too, which both on average estimate 4/5 common factors, but we found a completely similar picture in terms of significance, with only slightly higher p-values.} We find patterns of significance across the four test statistics, though rarely all of them at once (only AUT). $g_{\widehat{f},4}^{adj}$ finds significant better forecast accuracy one quarter ahead when using PC global inflation in $9$ of the $23$ countries. It is followed by the encompassing test $g_{\widehat{f},1}$ with $8$ countries ($9$ with CAN if considering $10\%$ nominal level), $g_{\widehat{f},3}^{adj}$ with $7$ and $g_{\widehat{f},2}^{adj}$ with $4$. Noticeably, for some large economies such as USA, FRA, ESP, at least two test statistics of the four are found significant (JPN too if considering $10\%$ nominal level). Some other large/medium-to-large economies like GBR, DEU, ITA have only one or no significance at all. Overall, this paints a mixed picture with regard to the use of global inflation to better forecast country-level inflation, which seems to be clearly a country-specific issue. These tests can therefore be used to gain a sense as to whether PC global inflation might help beyond the simple AR specification. In Online Supplement Section 6 we repeat the exercise without factors but using a grand average of country level inflations (which excludes the country of reference each time) as global inflation. What comes out is an overly suspicious abundant significance across most countries and all tests. Outliers and multicollinearity are likely to affect these results, as evident from the boxplots and pairwise correlation heatmap in Figure \ref{boxplot} and Figure \ref{heatmap}, respectively. It is therefore safe to say that the analysis including PC global inflation is much more trustworthy.

\begin{remark}
We suggest the potential macroeconomics practitioners to interpret the outcomes of the 4 tests as some sort of ``battery of checks", to be used jointly in understanding whether the factor augmentation is worth to improve the quality of the forecast. Acknowledging that, as outlined above, each test has its own nuance, $g_{\widehat{f},1}$ should be given more weight in the decision. This is because, as explained above, due the overlapping evaluation over the effective sample size for the case of $g_{\widehat{f},2}^{adj}$, $g_{\widehat{f},3}^{adj}$, $g_{\widehat{f},4}^{adj}$, some data is lost in the MSEs comparison, which does not occur for $g_{\widehat{f},1}$. As such, a joint significance of  $g_{\widehat{f},1}$ and at least one of $g_{\widehat{f},2}^{adj}$, $g_{\widehat{f},3}^{adj}$, $g_{\widehat{f},4}^{adj}$ should possibly be taken as a solid indication that considering a factor-augmented specification can increase the forecast accuracy.    
\end{remark}
 \subsection{FRED-MD Forecasting}\label{extra_app} We continue the empirical exercise by considering more macroeconomic series, this time from an even larger pool of series. Table \ref{tab:FREDMD} collects the p-values for the tests of encompassing and forecast accuracy described in Section \ref{sec_econsetup_tests}, using the same tunings as in the previous application in Section \ref{sec_emp_appl}, for 14 important macroeconomic series taken from the FRED-MD dataset \citet{mccracken2016fred} (July 2024 vintage, starting January 1960). The FRED-MD dataset includes a wide range of U.S. macroeconomic series such as output (e.g., industrial production), income, labor market indicators (e.g., unemployment rate, payrolls), consumer and producer prices, money supply, interest rates, exchange rates, and financial variables like stock indices and spreads. These series capture economic activity, inflation dynamics, monetary policy, and financial market conditions. All series are cleaned and stationary transformed as prescribed by the Matlab routines provided with FRED-MD. After the necessary cleaning of NAs and outliers, the final dataset contains $117$ series for $772$ data points. Again we choose the tuning parameters in line with the simulations to balance the power and size (see our extensive simulations in the Online Supplement).

We forecast one-month-ahead these $14$ macroeconomic series using the PC factors in the alternative forecasting specification (i.e., $AR(1)$ vs factor augmented $AR(1)$, where factors are estimated via \citet{bai2002determining} $IC_{p1}$ with max number$=15$).\footnote{We choose to follow a common convention in the factor modeling literature by setting a relatively high maximum number of factors, from which the optimal number is selected using IC. The goal of this is to ensure that the true number of factors is not underestimated due to an arbitrarily restrictive upper bound.}\footnote{We collect the full names of the $14$ FRED-MD macroeconomic time series in the Online Supplement (Section 6, Table 6.2)} 
With an average of $7$ estimated common factors, the results are overwhelmingly significant across the four tests.
Apart from ``Real personal consumption expenditures" (DPCERA3M086SBEA), all the other series have at least one significant (at nominal level 5\%) test statistic. This shows how even in a high(er) dimensional dataset such as the FRED-MD, these test statistics are very useful in assessing whether PC factors could be employed to improve the forecast of macroeconomic series.  

\begin{table}[H]
    \centering
    \caption{FRED-MD forecast accuracy \& encompassing}
    \label{tab:FREDMD}
    \begin{threeparttable}
        \begin{tabular}{lcccc}
            \toprule
            Series & p-value $g_{\widehat{f},1}$ & p-value $g_{\widehat{f},2}^{adj}$ & p-value $g_{\widehat{f},3}^{adj}$ & p-value $g_{\widehat{f},4}^{adj}$ \\
            \midrule
            UNRATE & \textbf{0.000} & \textbf{0.000} & \textbf{0.000} & 0.628 \\
            CPIAUCSL & \textbf{0.004} & \textbf{0.029} & \textbf{0.000} & 1.000 \\
            DPCERA3M086SBEA & 0.168 & 0.112 & 0.510 & 0.219 \\
            S\&P 500 & \textbf{0.030} & \textbf{0.021} & 0.118 & 0.898 \\
            PAYEMS & \textbf{0.000} & \textbf{0.000} & \textbf{0.014} & 0.566 \\
            INDPRO & \textbf{0.000} & \textbf{0.000} & \textbf{0.000} & 0.919 \\
            TB3SMFFM & 0.986 & 0.885 & 0.885 & \textbf{0.000} \\
            HOUST & \textbf{0.002} & \textbf{0.000} & \textbf{0.000} & 0.999 \\
            M1SL & \textbf{0.000} & \textbf{0.000} & \textbf{0.000} & 0.999 \\
            M2SL & \textbf{0.000} & \textbf{0.003} & \textbf{0.018} & 0.796 \\
            OILPRICEx & \textbf{0.000} & \textbf{0.000} & \textbf{0.003} & 0.893 \\
            GS10 & 0.999 & 0.963 & 0.985 & \textbf{0.000} \\
            RPI & 0.066 & \textbf{0.012} & \textbf{0.034} & 0.836 \\
            BUSLOANS & 0.117 & \textbf{0.037} & 0.109 & 0.881 \\
            \bottomrule
        \end{tabular}
        \begin{tablenotes}
            \footnotesize
            \item Notes: 1-month-ahead, $AR(1)$ vs factor augmented $AR(1)$; number of factors selected with \citet{bai2002determining} $IC_{p1}$, $\mu_0=0.40$, $\tau_0=0.8$, for $g_{\widehat{f},2}^{adj}$: $\lambda_1^0=1, \lambda_2^0=0.65$; for $g_{\widehat{f},3}^{adj}$: $\lambda_2^0=0.6$; for $g_{\widehat{f},4}^{adj}$: $\lambda_1^0=0.6$.
        \end{tablenotes}
    \end{threeparttable}
\end{table}

\setlength{\baselineskip}{0.7cm}
\section{Conclusion}
We develop the theoretical framework that allows the forecast accuracy and encompassing tests proposed by \citet{pitarakis2023direct, pitarakis2025novel} to be applied when the alternative forecasting model incorporates estimated PC factors. These factors can have loadings that are either strong or weak, whether homogeneously or heterogeneously distributed. Our theoretical findings are supported by both a Monte Carlo simulation and two empirical applications in macroeconomic forecasting.
\newpage
\appendix
\allowdisplaybreaks
\renewcommand{\thesection}{\arabic{section}}
\setcounter{page}{1}

\section*{\Huge \centering Mathematical Supplement}
\etocdepthtag.toc{mtappendix}
\etocsettagdepth{mtchapter}{none}
\etocsettagdepth{mtappendix}{subsection}
\etocsettagdepth{mtappendix}{subsubsection}
\tableofcontents

\section{Notation and Assumptions}
\noindent Throughout this supplement, we will use the following notation. Firstly, $a$ is a scalar, $\*a$ is a vector and $\*A_t$ is a matrix with $t$ rows. For any generic matrix $\*A$, spectral norm is $\left\|\*A \right\|_{sp}=\sqrt{\lambda_{\mathrm{max}}(\*A'\*A)}$, while $\|\*A \|=\sqrt{\mathrm{tr}(\*A'\*A)}$ is the Frobenius (Euclidean) norm with $\mathrm{tr}(.)$ being the trace operator. Vectorization of a matrix $\*A$ is denoted by $\mathrm{vec}(\*A)$, and $\otimes$ represents Kronecker product. By $\lambda_{\min}(\*A)$ and  $\lambda_{\max}(\*A)$ we denote the smallest and largest eigenvalues, respectively. Next, $\lfloor x \rfloor$ represents the integer part of $x$, and $M$ is a positive constant, which does not necessarily have the same value for various statements, while $\sup_{a\leq t \leq b}$ ($\inf_{a\leq t \leq b}$) is supremum (infimum) over a discrete set $\{a,\ldots, b \}$. Moreover, $k_0$ represents the in-sample observations, while $T-k_0=n$ are the out-of-sample observations. Weak convergence and equality in distribution are denoted by $\Rightarrow$ and $\overset{D}{=}$, respectively, while convergence in distribution and probability are given by $\to_d$ and $\to_p$, respectively. Ultimately, $\widetilde{a}$ and $\widehat{a}$ are quantities estimated under the observed and estimated $\*f_t$.\\

\noindent \textbf{A.1 (Forecast Error)}\label{A1}
\begin{enumerate}[label=\roman*)]
    \item $\{u_{t}\}$ is a martingale difference sequence with respect to the filtration $\mathcal{F}_{t-1}=\sigma(\*z_{t-1},\ldots,\allowbreak \*X_{t-1},\ldots, y_{t-1},\ldots)$.
    \item   $\mathbb{E}(u_{t}^2|\mathcal{F}_{t-1})=\sigma_t^2$, $\sup_{k_0\leq t\leq T-1}\sigma^2_t=O_p(1)$, $\sigma^2=\mathbb{E}(\sigma_t^2)$.  
    \item $\sup_{k_0\leq t \leq T-1}\mathbb{E}(u_t^4)<M$.
    \item $u_t$ is independent from all other model primitives for all $i, t, s$.
\end{enumerate}

\noindent \textbf{A.2 (Idiosyncratics)}\label{A2}
\begin{enumerate}[label=\roman*)]
    \item \textit{$\mathbb{E}(\*E_t|\*F_t, \+\Lambda)=\*0_{t\times N}$}
    \item $\sup_{k_0\leq t\leq T-1}\mathbb{E}(|e_{i,t}|^4)<M$ for all $i,t$. 
    \item $\mathbb{E}\left[\left(\frac{1}{\sqrt{N}}\sum_{i=1}^N(e_{i,t}e_{i,s}-\mathbb{E}(e_{i,t}e_{i,s})\right)^4 \right]<M$. 
    \item $\frac{1}{T}\sum_{t=1}^T\sum_{s=1}^T[\mathbb{E}(e_{i,t}e_{i,s})]^2<M$ (square summability).
    \item $\frac{1}{N}\sum_{i=1}^N\sum_{j=1}^N\frac{1}{T} \sum_{s_1=1}^{T}\sum_{s_2=1}^{T}\left|\mathrm{C}ov\left( e_{i,s_1}e_{j,s_1}, e_{i,s_2}e_{j,s_2}\right)\right|=O_p(1)$
    \item Let $\mathbb{E}(\*e_t\*e_t'|\*F_t,\+\Lambda)=\+\Sigma_{e}\in \mathbb{R}^{N\times N}$, which is deterministic. There exists $m>0$ and some positive $M$, such that $\lambda_{\mathrm{min}}(\+\Sigma_{e})>m$ and $\left\|\+\Sigma_{e} \right\|_{sp}<M$
    \item $\sup_{k_0\leq t\leq T-1}\left\|\*E_t \right\|_{sp}=\max\{\sqrt{N},\sqrt{T}\}$.
\end{enumerate}

\noindent \textbf{A.3 (Factors, Loadings and Predictors)}\label{A3}
\begin{enumerate}[label=\roman*)]
\item $\sup_{k_0\leq t\leq T-1}\mathbb{E}(\left\|\*z_t \right\|^8)<M$, where $\*z_t=(\*w_t',\*f_t')'\in \mathbb{R}^{k+r}$. 
\item $\sup_{k_0\leq t\leq T-1}\left\|\frac{1}{T}\sum_{s=1}^t(\*z_s\*z_s'-\+\Sigma_\*z) \right\|=O_p(T^{-1/2})$ and $\+\Sigma_\*z=\mathbb{E}(\*z_t\*z_t')=\begin{bmatrix}
    \+\Sigma_\*w & \+\Sigma_{\*w\*f}\\
    \+\Sigma_{\*w\*f}' & \*I_r
\end{bmatrix}$.
\item $\frac{1}{\sqrt{T}}\sum_{s=1}^t\mathrm{vec}(\*z_s\*z_s'-\+\Sigma_\*z)$ converges weakly. 
\item (\textit{Homogeneous}) $\left\|N^{-\alpha}\+\Lambda' \+\Lambda-\+\Sigma_{\+\Lambda}\right\| =O_p(N^{-\alpha/2})$ for a positive definite $\+\Sigma_{\+\Lambda}$ as $N\to \infty$. (\textit{Heterogeneous}) $\left\|\*B_N^{-1}\+\Lambda' \+\Lambda\*B_N^{-1}- \+\Sigma_{\+\Lambda}\right\|=O_p(N^{-\alpha_r/2})$ for a positive definite $\+\Sigma_{\+\Lambda}$ as $N\to \infty$. Both $N^{-\alpha}\+\Lambda' \+\Lambda$ and $\*B_N^{-1}\+\Lambda' \+\Lambda\*B_N^{-1}$ are diagonal and positive definite with eigenvalues ordered in a decreasing fashion. In both cases, the loadings are independent from $\*F_t$ for all $t$. 
\item  (\textit{Homogeneous}) $\mathbb{E}\left[\left(\frac{1}{N^{\alpha/2}}\sum_{i=1}^N\*f_s'\+\lambda_ie_{i,t} \right)^2\right]<M$ and $\left\|\frac{1}{\sqrt{N^{\alpha}T}}\*F_t'\*E_t\+\Lambda\right\|=O_p(1)$ for all $t,s$. (\textit{Heterogeneous}) $\mathbb{E}\left[\left(\sum_{i=1}^N(N^{-1/2}\*B_N\*f_s)'\*B_N^{-1}\+\lambda_ie_{i,t} \right)^2\right]<M$ and $\left\|\frac{1}{\sqrt{N^{\alpha_1}T}}\*F_t'\*E_t\+\Lambda\right\|=O_p(1)$ for all $t,s$.
\item (\textit{Homogeneous}) $\mathbb{E}\left(\left\|\frac{1}{N^{\alpha/2}}\sum_{i=1}^N\+\lambda_i'e_{i,t} \right\|^8 \right)<M$ for all $t$. (\textit{Heterogeneous}) $\mathbb{E}\left(\left\|\*B_N^{-1}\sum_{i=1}^N\+\lambda_i'e_{i,t} \right\|^8 \right)<M$ for all $t$.
\item (\textit{Homogeneous}) $\left\|\frac{1}{N^\alpha T}\*e_i'\*E_t\+\Lambda\right\|=O_p(N^{-\alpha})+O_p(N^{-\alpha/2}T^{-1/2})$ for all $i=1,\ldots,N$. (\textit{Heterogeneous}) $\left\|\frac{1}{T}\*e_i'\*E_t\+\Lambda\*B_N^{-1}\right\|=O_p(N^{-\alpha_r})+O_p\left(\sqrt{\frac{N^{\alpha_1}}{N^{\alpha_r}}}\frac{1}{\sqrt{T}} \right)$ for all $i=1,\ldots,N$. 
\item $\sup_{k_0\leq t\leq T-1} \|T^{-1/2}\sum_{s=1}^t\*f_s e_{i,s}\|^2\leq M$ for all $i=1,\ldots,N$ and $\left\|\frac{1}{NT}\*e_s'\*E_t'\*F_t\right\|=O_p(N^{-1})+O_p(T^{-1})$ for all $t,s$. 
\end{enumerate}
\noindent \textbf{A.4 ($N,T$ Expansion Rates)}\label{A4}
\begin{enumerate}[label=\roman*)]
\item \textit{(Homogeneous)} $\frac{N}{N^{\alpha}}\frac{1}{T^{1/4}}\to c>0$ as $(N,T)\to \infty$. \textit{(Heterogeneous)} $\frac{N}{N^{\alpha_r}}\frac{1}{T^{1/4}}\to c>0$ as $(N,T)\to \infty$ for $\alpha$ ($\alpha_r$) $\in (0.5,1)$.
\item \textit{(Homogeneous)} $\sqrt{T}N^{-\alpha}\to 0$ for $\alpha>0.5$. \textit{(Heterogeneous)} $\sqrt{T}N^{-\alpha_r}\to 0$ for $\alpha_r>0.5$.
\end{enumerate}
\bigskip 

\noindent The assumptions are similar to the ones in \cite{gonccalves2017tests} and \cite{bai2023approximate} as they are needed to accommodate recursive factor estimation and weaker loadings, respectively. Discussion of differences is in order. For instance, \hyperref[A1]{A.1} ii) and iv) are absent in \cite{gonccalves2017tests}, but they are fairly standard and allow to bring down moment requirements on factors and idiosynratics. Indeed, in the equivalent of \hyperref[A3]{A.3} in \cite{gonccalves2017tests}, 16th moments are required in parts i) and v). Note that we do not assume independence between the idiosyncratic components and the rest of the primitives. Otherwise, we would need only 4th moments. \hyperref[A2]{A.2} is virtually the same as in the latter study, except for the lower moment requirement. \hyperref[A2]{A.2} vii) is the same as in \cite{bai2023approximate}, but it is required to hold uniformly in $t$. We formulate it in terms of $T$, because $t=\lfloor r T\rfloor$ for some $r\in (0,1)$ due to the recursive setup. It is primarily used to simplify proofs when the loading weakness is heterogeneous, but it also helps to improve convergence rates. \hyperref[A3]{A.3} i) - iii) ensure that the tests in \cite{pitarakis2023direct} and \cite{pitarakis2025novel} are asymptotically normal as required, but under lower-level conditions. For example, part iii) is similar to the one in \cite{clark2001tests} and \cite{stauskas2022tests}, where $\{ \mathrm{vec}(\*z_t\*z_t'-\+\Sigma_\*z)\}_{t=1}^T$ follows a mixing sequence of a specific size (see \citealp{hansen1992convergence}). In part iv), we treat the loadings as random and make them independent from $\*F_t$ in order to simplify some arguments. Alternatively, we can impose $\mathbb{E}(\*f_t\*f_t'|\+\Lambda)=\*I_r$. They can also be treated as fixed, similarly to \cite{gonccalves2017tests}. Parts v) - vii) ensure that the feasible variance estimator of the tests approximate the infeasible counterparts (observed factors) well. Furthermore, \hyperref[A4]{A.4} i) is needed to make convergence rates more transparent. Particularly, in the proofs we will need $\frac{N}{N^{\alpha}}\frac{1}{\sqrt{T}}=o(1)$, but at the same time $T^{1/4}\frac{N}{N^{\alpha}}\frac{1}{\sqrt{T}}=O(1)$. Thus, under \hyperref[A4]{A.4} i), $\frac{N}{N^{\alpha}}\frac{1}{\sqrt{T}}=O_p(T^{-1/4})$. Part ii) of the same assumption is identical to the one needed for inference in Lemma 4 of \cite{bai2023approximate}. Under $\alpha=1$ (strong loadings), it naturally coincides with the requirement of $\sqrt{T}N^{-1}=o(1)$ in \cite{gonccalves2017tests}.
 \section{Overview of the Tests}
 \begin{remark}
     In this supplementary material, we use the notation $g_j$ for $j=1,\ldots,4$ to denote our statistics for conciseness purposes only. The original notation in \cite{pitarakis2023direct} and \cite{pitarakis2025novel} uses $\varepsilon_f$, $z_f$ and $\overline{z}_f$, respectively (and the respective subscripts for the variance estimators). 
 \end{remark}
 \subsection{Tests for Forecast Encompassing and Accuracy}
We consider in total 4 tests in total. One is for the forecast encompassing and two are for equal forecast accuracy. The test for forecast encompassing is given by \cite{pitarakis2023direct} and it has the following form:
 \begin{align}\label{arxiv_test}
     \mathrm{(i.)}\quad g_{f,1}=\frac{1}{\widetilde{\omega}_1}\left(\frac{1}{\sqrt{n}}\sum_{t=k_0}^{T-1}\widetilde{u}^2_{1,t+1}-\frac{1}{2}\left[\frac{n}{m_0}\frac{1}{\sqrt{n}}\sum_{t=k_0}^{k_0+m_0-1}\widetilde{u}_{1,t+1}\widetilde{u}_{2,t+1} + \frac{n}{n-m_0}\frac{1}{\sqrt{n}}\sum_{t=k_0+m_0}^{T-1}\widetilde{u}_{1,t+1}\widetilde{u}_{2,t+1} \right] \right),
 \end{align}
 where $k_0=\lfloor T\pi_0 \rfloor$ for $\pi_0\in (0,1$) and $m_0=\lfloor n \mu_0\rfloor = \lfloor (T-k_0) \mu_0\rfloor$ is a cut-off point to split the average for $\mu_0\in (0,1)$. We must require $\mu_0\neq 1/2$, as otherwise the asymptotic variance of the statistic would degenerate. Also, $\widetilde{\omega}_1$ is the estimated standard deviation of the limiting distribution of the test statistic. 
 Two tests of forecasting accuracy come from \cite{pitarakis2025novel}: 
 \begin{align}\label{ET1}
      \mathrm{(ii.)}\quad g_{f,2}=\frac{1}{\widetilde{\omega}}_2\frac{n}{l^0_1}\left(\frac{1}{\sqrt{n}}\sum_{t=k_0}^{k_0+l^0_1-1}\widetilde{u}^2_{1,t+1}-\frac{l^0_1}{l^0_2}\frac{1}{\sqrt{n}}\sum_{t=k_0}^{k_0+l^0_2-1}\widetilde{u}^2_{2,t+1} \right)
 \end{align}
 and 
 \begin{align}\label{ET2}
   \mathrm{(iii.)}\quad g_{f,3}=\frac{1}{\widetilde{\omega}}_{3}\frac{1}{n(1-\tau_0)}\sum_{l_1=\lfloor n\tau_0 \rfloor+1}^n   g_{f,2}(l_1, \lfloor n\lambda^0_2 \rfloor ).
 \end{align}
 Here, $l^0_j=\lfloor n\lambda_j^0 \rfloor$ for $j=1,2$ and $\lambda _j^0\in (0,1)$, which represent two portions of the out-of-sample period of which the forecast errors of both models are compared. Note that we can have $l^0_1>l_2^0$ or vice-versa, which means that both portions are overlapping. However, we cannot have equality in order to avoid the asymptotic degeneracy of the variance. Note that (\ref{ET2}) is simply an average of (\ref{ET1}) over some chosen feasible set of $l_1$ for the fixed $l_2^0$ (or, fixed $\lambda_2^0$). The tuning parameter $\tau_0\in (0,1)$ helps to pick that set. Intuition behind the averaging is the following: if $l_2^0$ is fixed and $l_1$ changes, then the MSE of the restricted model accumulates. In effect, possible choices of $l_1$ are aggregated. Lastly, $\widetilde{\omega}_2$ and $\widetilde{\omega}_{3}$ represent estimators of the standard deviations of both test statistics. \\

 \noindent The next test is an alternative to (iii.). It is based on averaging over $l_2$ while the $l_1^0$ is fixed, such that the MSE of the unrestricted model accumulates. That is 
 \begin{align}\label{ET_new}
       \mathrm{(iv.)}\quad g_{f,4}(\tau_0,\lambda^0_1)=\frac{1}{\widetilde{\omega}}_{4}\frac{1}{n(1-\tau_0)}\sum_{l_2=\lfloor n\tau_0 \rfloor+1}^n   g_{f,2}(\lfloor n\lambda^0_1 \rfloor, l_2  ).
 \end{align}
 Note that this statistic is not present in \cite{pitarakis2025novel}, although it is suggested that alternative averaging choices can be made. In total, we can summarize both (\ref{ET2}) and (\ref{ET_new}) as 
 \begin{align}\label{ET_2_sum}
      g_{p,f}(\tau_0,\lambda^0_j)=\frac{1}{\widetilde{\omega}}_{p}\frac{1}{n(1-\tau_0)}\sum_{l_j=\lfloor n\tau_0 \rfloor+1}^n   g_{f,2}(\lfloor n\lambda^0_j \rfloor, l_k  ),
 \end{align}
 where $p=3,4$, $j,k=1,2$ with $j\neq k$. 
 
 \subsection{On the Infeasible Statistics}
The infeasible statistics (i.e., observed $\*f_t$) can be decomposed into two components, such that for $j=1,\ldots,4$ we have
\begin{align}
    g_{f,j}= \widetilde{\omega}_j^{-1}(g_{f,j,1}+g_{f,j,2}),
\end{align}
where $g_{f,j,1}$ generates the distribution under the null. The infeasible variance estimator is $\widetilde{\omega}_j^2=\widetilde{\mathbb{V}ar}(g_j)$. In particular, as $T\to \infty$,
\begin{align}
   \widetilde{\omega}_j^{-1}g_{f,j,1}\to_d \mathcal{N}(0,1).
\end{align}
At the same time, $g_{f,j,2}$ is negligible or generates statistical power under the null and the alternative, respectively. Firstly, we will explore $g_{f,j,1}$ and then show that $g_{f,j,2}\to_p f_j(\+\beta^0)=0$ if $\+\beta^0=\*0_r$. 
\subsubsection{The Null Distribution}
In this section, we will analyze the asymptotic null distributions of the infeasible statistics under our assumptions, whereas in \cite{pitarakis2023direct} and \cite{pitarakis2025novel} they are obtained under high-level conditions. The power terms will be explored in the upcoming section. In the proof, we will for simplicity let $\mathbb{E}(u_{t}^2|\mathcal{F}_{t-1})=\sigma^2$.  We will comment on what changes under conditional heteroskedasticity. The key is that $u^2_{t+1}-\sigma^2$ is no longer a martingale difference under conditional heteroskedasticity. Note how $\mathbb{E}[(u_{t+1}^2-\sigma^2)(u_{t}^2-\sigma^2)]=\mathbb{E}[\mathbb{E}(u_{t+1}^2-\sigma^2|\mathcal{F}_{t})(u_{t}^2-\sigma^2)]\neq 0$ unless $\mathbb{E}(u_{t+1}^2|\mathcal{F}_t)=\sigma^2$. 
\begin{proposition} \label{Prop1} Under Assumptions A.1 and A.3, as $T\to \infty$ one has
    \begin{align*}
        \widetilde{\omega}_j^{-1}g_{f,j,1}\to_d \mathcal{N}(0,1)
    \end{align*}
    for $j=1,\ldots,4$.
\end{proposition}
\noindent \textbf{Proof}. Let $j=1$. In particular, 
\begin{align}
     g_{f,1}&=\frac{1}{\widetilde{\omega}_1}\left(\frac{1}{\sqrt{n}}\sum_{t=k_0}^{T-1}\widetilde{u}^2_{1,t+1}-\frac{1}{2}\left[\frac{n}{m_0}\frac{1}{\sqrt{n}}\sum_{t=k_0}^{k_0+m_0-1}\widetilde{u}_{1,t+1}\widetilde{u}_{2,t+1} + \frac{n}{n-m_0}\frac{1}{\sqrt{n}}\sum_{t=k_0+m_0}^{T-1}\widetilde{u}_{1,t+1}\widetilde{u}_{2,t+1} \right] \right)\notag\\
     &= \frac{1}{\widetilde{\omega}_1}\left(\frac{1}{\sqrt{n}}\sum_{t=k_0}^{T-1}u_{t+1}^2-\frac{1}{2}\left[\frac{n}{m_0}\frac{1}{\sqrt{n}}\sum_{t=k_0}^{k_0+m_0-1}u_{t+1}^2 + \frac{n}{n-m_0}\frac{1}{\sqrt{n}}\sum_{t=k_0+m_0}^{T-1}u_{t+1}^2 \right] \right) \notag\\
     &+\frac{1}{\widetilde{\omega}_1}\Bigg(\frac{1}{\sqrt{n}}\sum_{t=k_0}^{T-1}(\widetilde{u}^2_{1,t+1}-u_{t+1}^2)-\frac{1}{2}\Bigg[\frac{n}{m_0}\frac{1}{\sqrt{n}}\sum_{t=k_0}^{k_0+m_0-1}(\widetilde{u}_{1,t+1}\widetilde{u}_{2,t+1}-u_{t+1}^2) \notag\\
     &+ \frac{n}{n-m_0}\frac{1}{\sqrt{n}}\sum_{t=k_0+m_0}^{T-1}(\widetilde{u}_{1,t+1}\widetilde{u}_{2,t+1}-u_{t+1}^2) \Bigg] \Bigg)\notag\\
     &= \frac{1}{\widetilde{\omega}_1}\left(\frac{1}{\sqrt{n}}\sum_{t=k_0}^{T-1}(u_{t+1}^2-\sigma^2)-\frac{1}{2}\left[\frac{n}{m_0}\frac{1}{\sqrt{n}}\sum_{t=k_0}^{k_0+m_0-1}(u_{t+1}^2-\sigma^2) + \frac{n}{n-m_0}\frac{1}{\sqrt{n}}\sum_{t=k_0+m_0}^{T-1}(u_{t+1}^2-\sigma^2) \right] \right) \notag\\
     &+\frac{1}{\widetilde{\omega}_1}\Bigg(\frac{1}{\sqrt{n}}\sum_{t=k_0}^{T-1}(\widetilde{u}^2_{1,t+1}-u_{t+1}^2)-\frac{1}{2}\Bigg[\frac{n}{m_0}\frac{1}{\sqrt{n}}\sum_{t=k_0}^{k_0+m_0-1}(\widetilde{u}_{1,t+1}\widetilde{u}_{2,t+1}-u_{t+1}^2) \notag\\
     &+ \frac{n}{n-m_0}\frac{1}{\sqrt{n}}\sum_{t=k_0+m_0}^{T-1}(\widetilde{u}_{1,t+1}\widetilde{u}_{2,t+1}-u_{t+1}^2) \Bigg] \Bigg)\notag\\
     &=\widetilde{\omega}_1^{-1}(g_{f,1,1} + g_{f,1,2}),
\end{align}
where $g_{f,1,1}$ generates the asymptotic distribution and $g_{f,1,2}$ gives the statistical power under the alternative, but it is negligible under the null. We start with $g_{f,1,1}$. In particular, notice how by splitting the first term appropriately, we obtain
\begin{align}
  g_{f,1,1}&= \frac{1}{\sqrt{n}}\sum_{t=k_0}^{T-1}(u_{t+1}^2-\sigma^2)-\frac{1}{2}\left[\frac{n}{m_0}\frac{1}{\sqrt{n}}\sum_{t=k_0}^{k_0+m_0-1}(u_{t+1}^2-\sigma^2) + \frac{n}{n-m_0}\frac{1}{\sqrt{n}}\sum_{t=k_0+m_0}^{T-1}(u_{t+1}^2-\sigma^2)\right]\notag\\
   &=  \frac{1}{\sqrt{n}}\sum_{t=k_0}^{k_0+m_0-1}(u_{t+1}^2-\sigma^2)+\frac{1}{\sqrt{n}}\sum_{t=k_0+m_0}^{T-1}(u_{t+1}^2-\sigma^2)\notag\\
   &-\frac{1}{2}\left[\frac{n}{m_0}\frac{1}{\sqrt{n}}\sum_{t=k_0}^{k_0+m_0-1}(u_{t+1}^2-\sigma^2) + \frac{n}{n-m_0}\frac{1}{\sqrt{n}}\sum_{t=k_0+m_0}^{T-1}(u_{t+1}^2-\sigma^2)\right]\notag\\
   &=\left(1-\frac{1}{2}\frac{n}{m_0} \right)\frac{1}{\sqrt{n}}\sum_{t=k_0}^{k_0+m_0-1}(u_{t+1}^2-\sigma^2)+\left(1-\frac{1}{2}\frac{n}{n-m_0} \right)\frac{1}{\sqrt{n}}\sum_{t=k_0+m_0}^{T-1}(u_{t+1}^2-\sigma^2)\notag\\
   &= \left(1-\frac{1}{2}\frac{n}{m_0} \right)\sqrt{\frac{T}{n}}\frac{1}{\sqrt{T}}\sum_{t=k_0}^{k_0+m_0-1}(u_{t+1}^2-\sigma^2)+\left(1-\frac{1}{2}\frac{n}{n-m_0} \right)\sqrt{\frac{T}{n}}\frac{1}{\sqrt{T}}\sum_{t=k_0+m_0}^{T-1}(u_{t+1}^2-\sigma^2),\notag\\
\end{align}
where under our martingale difference assumption and existence of the fourth moment, we have that both $\frac{1}{\sqrt{T}}\sum_{t=k_0}^{k_0+m_0-1}(u_{t+1}^2-\sigma^2)$ and $\frac{1}{\sqrt{T}}\sum_{t=k_0+m_0}^{T-1}(u_{t+1}^2-\sigma^2)$ are $O_p(1)$, plus they are uncorrelated. Therefore, by using the definitions of the tuning parameters and denoting $\phi^2=\mathbb{V}ar(u_{t+1}^2-\sigma^2)$, we obtain as $T\to \infty$
\begin{align}
    g_{f,1,1}&=\left(1-\frac{1}{2\mu_0} \right)\frac{1}{\sqrt{1-\pi_0}}\frac{1}{\sqrt{T}}\sum_{t=\lfloor T\pi_o \rfloor}^{\lfloor T\pi_o \rfloor+\lfloor (T-k_0)\mu_0 \rfloor-1}(u_{t+1}^2-\sigma^2)\notag\\
     &+\left(1-\frac{1}{2(1-\mu_0)} \right)\frac{1}{\sqrt{1-\pi_0}}\frac{1}{\sqrt{T}}\sum_{t=\lfloor T\pi_o \rfloor+\lfloor (T-k_0)\mu_0 \rfloor}^{T-1}(u_{t+1}^2-\sigma^2)+o_p(1)\notag\\
     &\Rightarrow \left(1-\frac{1}{2\mu_0} \right)\frac{1}{\sqrt{1-\pi_0}}\phi\int_{s=\pi_0}^{\pi_0+\mu_0-\pi_0\mu_0}dW_u(s)\notag\\
     &+\left(1-\frac{1}{2(1-\mu_0)} \right)\frac{1}{\sqrt{1-\pi_0}} \phi \int_{s=\pi_0+\mu_0-\pi_0\mu_0}^1dW_u(s)\notag\\
     &\overset{D}{=}\frac{2\mu_0-1}{2\mu_0}\frac{1}{\sqrt{1-\pi_0}}\phi \mathcal{N}\left(0,\mu_0(1-\pi_0) \right)+\frac{1-2\mu_0}{2(1-\mu_0)}\frac{1}{\sqrt{1-\pi_0}}\phi \mathcal{N}\left(0, (1-\pi_0)(1-\mu_0) \right)\notag\\
     &\overset{D}{=} \mathcal{N}\left(0, \phi^2\frac{(1-2\mu_0)^2}{4\mu_0} \right) + \mathcal{N}\left(0, \phi^2\frac{(1-2\mu_0)^2}{4(1-\mu_0)} \right)\notag\\
     &\overset{D}{=}\mathcal{N}\left(0, \phi^2\frac{(1-2\mu_0)^2(1-\mu_0)+(1-2\mu_0)^2\mu_0}{4\mu_0(1-\mu_0)}\right)\notag\\
     &\overset{D}{=} \mathcal{N}\left(0, \phi^2 \frac{(1-2\mu_0)^2}{4\mu_0(1-\mu_0)}\right),
\end{align}
where we used the fact that Brownian increments are normally distributed for the fixed tuning parameters. Also, we merged the normal variates because they are uncorrelated, and hence independent. This implies that as $T\to \infty$ 
\begin{align}
\widetilde{\omega}_1^{-1}g_{f,1,1}\to_d \mathcal{N}(0,1).
\end{align}
Note that if $\mathbb{E}(u_{t}^2|\mathcal{F}_{t-1})=\sigma_t^2$, then $\phi^2$ should be interpreted as a long-run variance, because $\frac{1}{\sqrt{T}}\sum_{s=1}^t(u_{s+1}^2-\sigma^2)$ still generates Brownian motion under ARCH/GARCH errors (see \citealp{pitarakis2025novel}). Also, $\widetilde{\omega}^2_1$ should have a HAC correction. The same applies for the remaining statistics. Further we go the the distribution part under the null of $ g_{f,2}$, that is $j=2$. Note how 
\begin{align}
     g_{f,2}&=\frac{1}{\widetilde{\omega}}_2\frac{n}{l^0_1}\left(\frac{1}{\sqrt{n}}\sum_{t=k_0}^{k_0+l^0_1-1}\widetilde{u}^2_{1,t+1}-\frac{l^0_1}{l^0_2}\frac{1}{\sqrt{n}}\sum_{t=k_0}^{k_0+l^0_2-1}\widetilde{u}^2_{2,t+1} \right)\notag\\
      &= \frac{1}{\widetilde{\omega}}_2\frac{n}{l^0_1}\left(\frac{1}{\sqrt{n}}\sum_{t=k_0}^{k_0+l^0_1-1}u_{t+1}^2-\frac{l^0_1}{l^0_2}\frac{1}{\sqrt{n}}\sum_{t=k_0}^{k_0+l^0_2-1}u_{t+1}^2 \right) \notag\\
      &+ \frac{1}{\widetilde{\omega}}_2\frac{n}{l^0_1}\left(\frac{1}{\sqrt{n}}\sum_{t=k_0}^{k_0+l^0_1-1}(\widetilde{u}^2_{1,t+1}-u_{t+1}^2)-\frac{l^0_1}{l^0_2}\frac{1}{\sqrt{n}}\sum_{t=k_0}^{k_0+l^0_2-1}(\widetilde{u}^2_{2,t+1}-u_{t+1}^2) \right)\notag\\
      &=\widetilde{\omega}_2^{-1}(g_{f,2,1}+g_{f,2,2}),
\end{align}
where $g_{f,2,1}$ is the component that is responsible for the distribution under the null. In particular, 
\begin{align}\label{b_1}
   g_{f,2,1}&= \frac{n}{l^0_1}\left(\frac{1}{\sqrt{n}}\sum_{t=k_0}^{k_0+l^0_1-1}u_{t+1}^2-\frac{l^0_1}{l^0_2}\frac{1}{\sqrt{n}}\sum_{t=k_0}^{k_0+l^0_2-1}u_{t+1}^2\right)\notag\\
    &= \frac{n}{l^0_1}\left(\frac{1}{\sqrt{n}}\sum_{t=k_0}^{k_0+l^0_1-1}(u_{t+1}^2-\sigma^2)-\frac{l^0_1}{l^0_2}\frac{1}{\sqrt{n}}\sum_{t=k_0}^{k_0+l^0_2-1}(u_{t+1}^2-\sigma^2)\right)\notag\\
    &+  \frac{n}{l^0_1}\left(\frac{1}{\sqrt{n}}\sum_{t=k_0}^{k_0+l^0_1-1}\sigma^2-\frac{l^0_1}{l^0_2}\frac{1}{\sqrt{n}}\sum_{t=k_0}^{k_0+l^0_2-1}\sigma^2\right)=g_{f,2,1,1}+g_{f,2,1,2},
\end{align}
where $g_{z,11}$ is precisely zero under homoskedasticity: 
\begin{align}\label{b_12}
   g_{f,2,1,2}&=\frac{n}{\lfloor n\lambda^0_1 \rfloor}\left(\frac{1}{\sqrt{n}}\lfloor n \lambda_1^0 \rfloor-\frac{\lfloor n \lambda_1^0\rfloor}{\lfloor n \lambda_2^0\rfloor} \frac{1}{\sqrt{n}}\lfloor n \lambda_2^0 \rfloor \right)\sigma^2= \frac{n}{\lfloor n\lambda^0_1 \rfloor} \sqrt{n}\left(\frac{\lfloor n \lambda_1^0 \rfloor}{n}-\frac{\lfloor n\lambda_1^0  \rfloor}{n} \right)\sigma^2=0.
\end{align}
Next, by using $\frac{n}{\lfloor n\lambda^0_1 \rfloor}=\frac{1}{\lambda_1^0}+O(n^{-1})$, $\frac{\lfloor n\lambda^0_1 \rfloor}{\lfloor n\lambda^0_2 \rfloor}=\frac{\lambda_1^0}{\lambda_2^0}+O(n^{-1})$ and $\sqrt{\frac{T}{n}}=\frac{1}{\sqrt{1-\pi_0}}+O(n^{-1})$ \footnote{For this and similar terms we use the fundamental relationship $\sup_{1\leq t\leq T}\sup_{c\in (0,1)}\left|\left(\frac{t}{T}\right)^k-c^k \right|=O(T^{-1})$ for all finite $k$ (see p. 512 in \citealp{moon2004gmm}).} and boundedness of the sums due to the martingale difference assumption, we get as $T\to \infty$
\begin{align}\label{b1_distribution}
    g_{f,2,1,1}&=\frac{n}{l^0_1}\left(\frac{1}{\sqrt{n}}\sum_{t=k_0}^{k_0+l^0_1-1}(u_{t+1}^2-\sigma^2)-\frac{l^0_1}{l^0_2}\frac{1}{\sqrt{n}}\sum_{t=k_0}^{k_0+l^0_2-1}(u_{t+1}^2-\sigma^2)\right) \notag\\
    &= \frac{1}{\lambda_1^0}\left(\frac{1}{\sqrt{1-\pi_0}}\frac{1}{\sqrt{T}}\sum_{t=\lfloor T \pi_0 \rfloor}^{\lfloor T \pi_0 \rfloor+\lfloor(T-k_0)\lambda_1^0 \rfloor-1}(u_{t+1}^2-\sigma^2) - \frac{\lambda_1^0}{\lambda_2^0}\frac{1}{\sqrt{1-\pi_0}}\frac{1}{\sqrt{T}} \sum_{t=\lfloor T \pi_0 \rfloor}^{\lfloor T \pi_0 \rfloor+\lfloor(T-k_0)\lambda_2^0 \rfloor-1}(u_{t+1}^2-\sigma^2)\right)\notag\\
    &+O(n^{-1})\notag\\
    &\Rightarrow  \frac{1}{\lambda_1^0} \left( \frac{\phi}{\sqrt{1-\pi_0}}\int_{s=\pi_0}^{\pi_0+\lambda_1^0(1-\pi_0)}dW_u(s) - \frac{\lambda_1^0}{\lambda_2^0}\frac{\phi}{\sqrt{1-\pi_0}} \int_{s=\pi_0}^{\pi_0+\lambda_2^0(1-\pi_0)}dW_u(s) \right)\notag\\
    &= \frac{\phi}{\lambda_1^0} \left(W_u(\lambda_1^0) - \frac{\lambda_1^0}{\lambda_2^0}W_u(\lambda_2^0) \right)=\phi\left(W_u\left(\frac{1}{\lambda_1^0}\right) -W_u\left(\frac{1}{\lambda_2^0}\right) \right)\notag\\
    &\overset{D}{=}\mathcal{N}\left(0,\phi^2\left|\frac{1}{\lambda_1^0} -\frac{1}{\lambda_2^0}\right| \right)\notag\\
    &\overset{D}{=}\mathcal{N}\left(0,\phi^2\frac{|\lambda_1^0-\lambda_2^0 |}{\lambda_1^0\lambda_2^0} \right)
\end{align} 
by the properties of the distribution of Brownian increments and the fact that $\lambda_1^0$ and $\lambda_2^0$ are positive. Again, this implies that 
\begin{align}
     \widetilde{\omega}_2^{-1}g_{f,2,1}\to_d \mathcal{N}(0,1).
\end{align}
Lastly, we work with $j=2,3 $. Respectively, by using the steps employed in (\ref{b_1}) and (\ref{b_12}), we obtain
\begin{align}\label{avg_stat1}
    g_{f,3}&=\frac{1}{\widetilde{\omega}}_{3}\frac{1}{n(1-\tau_0)}\sum_{l_1=\lfloor n\tau_0 \rfloor+1}^n   g_{f,2}(l_1, \lfloor n\lambda^0_2 \rfloor )\notag\\
     &= \frac{1}{\widetilde{\omega}}_{3}\Bigg[\frac{1}{n(1-\tau_0)}\sum_{l_1=\lfloor n\tau_0 \rfloor+1}^n\frac{n}{l_1}\left(\frac{1}{\sqrt{n}}\sum_{t=k_0}^{k_0+l_1-1}(u_{t+1}^2-\sigma^2)-\frac{l_1}{l^0_2}\frac{1}{\sqrt{n}}\sum_{t=k_0}^{k_0+l^0_2-1}(u_{t+1}^2-\sigma^2)\right) \notag\\
     &+\frac{1}{n(1-\tau_0)}\sum_{l_1=\lfloor n\tau_0 \rfloor+1}^n\frac{n}{l_1}\left(\frac{1}{\sqrt{n}}\sum_{t=k_0}^{k_0+l_1-1}(\widetilde{u}^2_{1,t+1}-u_{t+1}^2)-\frac{l_1}{l^0_2}\frac{1}{\sqrt{n}}\sum_{t=k_0}^{k_0+l^0_2-1}(\widetilde{u}^2_{2,t+1}-u_{t+1}^2) \right)\Bigg]\notag\\
     &= \widetilde{\omega}_{3}^{-1}(g_{f,3,1}+g_{f,3,2}),
\end{align}
and 
\begin{align}\label{avg_stat2}
     g_{f,4}&=\frac{1}{\widetilde{\omega}}_{4}\frac{1}{n(1-\tau_0)}\sum_{l_2=\lfloor n\tau_0 \rfloor+1}^n   g_{f,2}(\lfloor n\lambda^0_1 \rfloor, l_2)\notag\\
     &= \frac{1}{\widetilde{\omega}}_{4}\Bigg[\frac{1}{n(1-\tau_0)}\sum_{l_2=\lfloor n\tau_0 \rfloor+1}^n\frac{n}{l_1^0}\left(\frac{1}{\sqrt{n}}\sum_{t=k_0}^{k_0+l^0_1-1}(u_{t+1}^2-\sigma^2)-\frac{l^0_1}{l_2}\frac{1}{\sqrt{n}}\sum_{t=k_0}^{k_0+l_2-1}(u_{t+1}^2-\sigma^2)\right) \notag\\
     &+\frac{1}{n(1-\tau_0)}\sum_{l_2=\lfloor n\tau_0 \rfloor+1}^n\frac{n}{l_1^0}\left(\frac{1}{\sqrt{n}}\sum_{t=k_0}^{k_0+l^0_1-1}(\widetilde{u}^2_{1,t+1}-u_{t+1}^2)-\frac{l^0_1}{l_2}\frac{1}{\sqrt{n}}\sum_{t=k_0}^{k_0+l_2-1}(\widetilde{u}^2_{2,t+1}-u_{t+1}^2) \right)\Bigg]\notag\\
     &= \widetilde{\omega}_{4}^{-1}(g_{f,4,1}+g_{f,4,2})
\end{align}
where $l_1=\lfloor n \lambda_1 \rfloor$ and $l_2=\lfloor n \lambda_2 \rfloor$ with either $\lambda_1$ or $\lambda_2$ varying now depending on the averaging choice. Clearly, $g_{f,3,1}$ ($g_{f,4,1}$) is responsible for the distribution under the null, whereas $g_{f,3,2}$ ($g_{f,4,2}$) will stay negligible under the null as averaging over the out-of-sample (sub)set will not change the order of the remainder. Clearly, because $\lambda_2^0$ is fixed in (\ref{avg_stat1}), by using the results in (\ref{b1_distribution}), we have that as $T\to \infty$
\begin{align}
   g_{f,3,1}\Rightarrow \frac{\phi}{1-\tau_0}\int_{\lambda_1=\tau_0}^1\left(\frac{W_u(\lambda_1)}{\lambda_1}-\frac{W_u(\lambda_2^0)}{\lambda_2^0} \right) d\lambda_1,
\end{align}
whereas, because $\lambda_1^0$ is fixed in (\ref{avg_stat2}), we have 
\begin{align}
   g_{f,4,1}\Rightarrow \frac{\phi}{1-\tau_0}\int_{\lambda_2=\tau_0}^1\left( \frac{W_u(\lambda^0_1)}{\lambda^0_1}-\frac{W_u(\lambda_2)}{\lambda_2} \right) d\lambda_2,
\end{align}
which both are normal variables. By using $\mathrm{C}ov(W_u(s_1),W_u(s_2))=\mathrm{min}(s_1,s_2)$, we can obtain the variance of $g_{f,4,1}$ by the Fubini's Theorem:
\begin{align}
    \mathbb{V}ar(g_{f,4,1})&=\omega_{4}^2=\frac{\phi^2}{(1-\tau_0)^2}\notag\\
    &\times \int_{s_1=\tau_0}^1\int_{s_2=\tau_0}^1\mathbb{E}\left(\left[\frac{W_u(\lambda_1^0)}{\lambda_1^0}\right]^2-\frac{W_u(\lambda_1^0)}{\lambda_1^0}\frac{W_u(s_2)}{s_2}-\frac{W_u(s_1)}{s_1}\frac{W_u(\lambda_1^0)}{\lambda_1^0}+\frac{W_u(s_1)}{s_1}\frac{W_u(s_2)}{s_2} \right) ds_2ds_1\notag\\
    &=\frac{\phi^2}{(1-\tau_0)^2}\times \int_{s_1=\tau_0}^1\int_{s_2=\tau_0}^1 \left(\frac{1}{\lambda_1^0}-\frac{\mathrm{min}(\lambda_1^0,s_2)}{\lambda_1^0s_2}- \frac{\mathrm{min}(\lambda_1^0,s_1)}{\lambda_1^0s_1}+\frac{\mathrm{min}(s_1,s_2)}{s_1s_2}\right)ds_2ds_1\notag\\
    &= \frac{\phi^2}{(1-\tau_0)^2}\times \int_{s_1=\tau_0}^1\int_{s_2=\tau_0}^1\frac{1}{\lambda_1^0}ds_2ds_1\notag\\
    &+ \frac{\phi^2}{(1-\tau_0)^2}\times\int_{s_1=\tau_0}^1\int_{s_2=\tau_0}^1\left( -\frac{\mathrm{min}(\lambda_1^0,s_2)}{\lambda_1^0s_2}- \frac{\mathrm{min}(\lambda_1^0,s_1)}{\lambda_1^0s_1}+\frac{\mathrm{min}(s_1,s_2)}{s_1s_2}\right)ds_2ds_1\notag\\
    &= \frac{\phi^2}{\lambda_1^0} +  \frac{\phi^2}{(1-\tau_0)^2}(-I-II+III),
\end{align}
where the solution of $I - III$ will depend on whether $\lambda_1^0\leq \tau_0$ or vice versa. Let $\lambda_1^0\leq  \tau_0$, then $\mathrm{min}(\lambda_1^0,s_1)=\mathrm{min}(\lambda_1^0,s_2)=\lambda_1^0$, because both $s_1$ and $s_2$ belong to $[\tau_0,1]$. Then
\begin{align}
    I=\int_{s_1=\tau_0}^1\int_{s_2=\tau_0}^1\frac{\mathrm{min}(\lambda_1^0,s_2)}{\lambda_1^0s_2}ds_2ds_1=\int_{s_1=\tau_0}^1\int_{s_2=\tau_0}^1\frac{1}{s_2}ds_2ds_1=-(1-\tau_0)\ln (\tau_0),
\end{align}
which is positive. Next, 
\begin{align}
     II=\int_{s_1=\tau_0}^1\int_{s_2=\tau_0}^1\frac{\mathrm{min}(\lambda_1^0,s_1)}{\lambda_1^0s_1}ds_2ds_1=\int_{s_1=\tau_0}^1\int_{s_2=\tau_0}^1\frac{1}{s_1}ds_2ds_1=-(1-\tau_0)\ln (\tau_0)
\end{align}
by symmetry. Lastly, due to independence of coordinates:
\begin{align}
    III=\int_{s_1=\tau_0}^1\int_{s_2=\tau_0}^1\frac{\mathrm{min}(s_1,s_2)}{s_1s_2}ds_2ds_1&=\int_{s_2=\tau_0}^1\int_{s_1=\tau_0}^1\frac{\mathrm{min}(s_1,s_2)}{s_1s_2}ds_1ds_2\notag\\
    &=\int_{s_2=\tau_0}^1\left(\underbrace{\int_{s_1=\tau_0}^{s_2} \frac{s_1}{s_1s_2}ds_1}_{s_1\leq s_2}+ \underbrace{\int_{s_1=s_2}^{1} \frac{s_2}{s_1s_2}ds_1}_{s_2\leq s_1}\right)ds_2\notag\\
    &= \int_{s_2=\tau_0}^1 \left(\frac{s_2-\tau_0}{s_2}-\ln(s_2) \right)ds_2\notag\\
    &= (1-\tau_0)-\tau_0(\ln(1)-\ln(\tau_0))-\int_{s_2=\tau_0}^1\ln(s_2)ds_2\notag\\
    &= (1-\tau_0)-\tau_0(\ln(1)-\ln(\tau_0))-(-1-\tau_0\ln(\tau_0)+\tau_0)\notag\\
    &=2(1-\tau_0)+2\tau_0\ln(\tau_0).
\end{align}
Hence, if $\lambda_1^0\leq  \tau_0$, then 
\begin{align}
    \mathbb{V}ar(g_{f,4,1})=\phi^2\left(\frac{1}{\lambda_1^0}+\frac{2(1-\tau_0)+2\tau_0\ln(\tau_0)+2(1-\tau_0)\ln (\tau_0)}{(1-\tau_0)^2} \right)=\phi^2\frac{(1-\tau_0)^2+2\lambda_1^0(1-\tau_0+\ln(\tau_0))}{\lambda_1^0(1-\tau_0)^2}.
\end{align}
Now, we let $\lambda_1^0> \tau_0$. Clearly, we still have $III=2(1-\tau_0)+2\tau_0\ln(\tau_0)$ since it is not a function of $\lambda_1^0$. However, 
\begin{align}
    I=\int_{s_1=\tau_0}^1\int_{s_2=\tau_0}^1 \frac{\mathrm{min}(\lambda_1^0,s_2)}{\lambda_1^0s_2}ds_2ds_1&=\int_{s_1=\tau_0}^1\left(\int_{s_2=\tau_0}^{\lambda_1^0}\frac{s_2}{s_2\lambda_1^0}ds_2+ \int_{s_2=\lambda_1^0}^{1}\frac{\lambda_1^0}{s_2\lambda_1^0}ds_2\right)ds_1\notag\\
    &= \int_{s_1=\tau_0}^1 \left(\frac{1}{\lambda_1^0}(\lambda_1^0-\tau_0)-\ln (\lambda_1^0) \right)ds_1\notag\\
    &=\frac{(\lambda_1^0-\tau_0)(1-\tau_0)}{\lambda_1^0}-\ln(\lambda_1^0)(1-\tau_0),
\end{align}
while by changing integration order
\begin{align}
    II=\int_{s_1=\tau_0}^1\int_{s_2=\tau_0}^1 \frac{\mathrm{min}(\lambda_1^0,s_1)}{\lambda_1^0s_1}ds_2ds_1&=\int_{s_2=\tau_0}^1\int_{s_1=\tau_0}^1 \frac{\mathrm{min}(\lambda_1^0,s_1)}{\lambda_1^0s_1} ds_1 ds_2\notag\\
    &= \int_{s_2=\tau_0}^1 \left(\int_{s_1=\tau_0}^{\lambda_1^0}\frac{s_1}{s_1\lambda_1^0} ds_1+\int_{s_1=\lambda_1^0}^{1}\frac{\lambda_1^0}{s_1\lambda_1^0} ds_1\right) ds_2\notag\\
    &=\frac{(\lambda_1^0-\tau_0)(1-\tau_0)}{\lambda_1^0}-\ln(\lambda_1^0)(1-\tau_0)
\end{align}
by symmetry. Therefore, when $\lambda_1^0> \tau_0$, then 
\begin{align}
     \mathbb{V}ar(g_{f,4,1})&=\phi^2\left(\frac{1}{\lambda_1^0}+\frac{2(1-\tau_0)+2\tau_0\ln(\tau_0)}{(1-\tau_0)^2}+\frac{2\ln(\lambda_1^0)(1-\tau_0)}{(1-\tau_0)^2}-\frac{2(\lambda_1^0-\tau_0)(1-\tau_0)}{\lambda_1^0(1-\tau_0)^2} \right)\notag\\
     &= \phi^2\frac{1-\tau_0^2+2\lambda_1^0((1-\tau_0)\ln(\lambda_1^0)+\tau_0\ln(\tau_0))}{\lambda_1^0(1-\tau_0)^2}.
\end{align}
Note that $\mathbb{V}ar(g_{f,4,1})$ is symmetric to the variance of $g_{f,3,1}$ which was demonstrated in Proposition 2 in \cite{pitarakis2025novel}, because now it is the fixed $\lambda_1^0$ that enters the formulas. In summary, for $j=3,4$ and $p,q=1,2$, we have
\begin{align}
    \omega_{j,p}^2 = \begin{cases}
        \phi^2\frac{(1-\tau_0)^2+2\lambda_q^0(1-\tau_0+\ln(\tau_0))}{\lambda_q^0(1-\tau_0)^2} \hspace{2mm}\text{if} \hspace{2mm} \lambda_q^0\leq \tau_0,\\
        \phi^2\frac{1-\tau_0^2+2\lambda_q^0((1-\tau_0)\ln(\lambda_q^0)+\tau_0\ln(\tau_0))}{\lambda_q^0(1-\tau_0)^2} \hspace{2mm}\text{if} \hspace{2mm} \lambda_q^0> \tau_0
    \end{cases}
\end{align}
This implies that 
\begin{align}
 & \widetilde{\omega}_{3}^{-1}g_{f,3,1}\to_d \mathcal{N}(0,1), \\
    &\widetilde{\omega}_{4}^{-1}g_{f,4,1}\to_d \mathcal{N}(0,1)
\end{align}
as $T\to \infty$. 
\subsubsection{The Local Alternative}
\noindent In this section we will briefly explore the local power generating terms $g_{f,j,2}$ for $j=1,\ldots,4$.
\begin{proposition} \label{Prop2} Under Assumptions A.1 and A.3, as $T\to \infty$ one has 
    \begin{align*}
        g_{f,j,2}\to_p \sqrt{1-\pi_0}\+\beta^{0\prime}(\*I_r-\+\Sigma_{\*w\*f}'\+\Sigma^{-1}_\*w\+\Sigma_{\*w\*f})\+\beta^{0}
    \end{align*}
    for $j=1,\ldots,4$.
\end{proposition}
\noindent \textbf{Proof}. Let $j=1$. Note how 
\begin{align}
    g_{f,1,2}= \frac{1}{\sqrt{n}}\sum_{t=k_0}^{T-1}(\widetilde{u}^2_{1,t+1}-u_{t+1}^2)-&\frac{1}{2}\Bigg[\frac{n}{m_0}\frac{1}{\sqrt{n}}\sum_{t=k_0}^{k_0+m_0-1}(\widetilde{u}_{1,t+1}\widetilde{u}_{2,t+1}-u_{t+1}^2) \notag\\
     &+ \frac{n}{n-m_0}\frac{1}{\sqrt{n}}\sum_{t=k_0+m_0}^{T-1}(\widetilde{u}_{1,t+1}\widetilde{u}_{2,t+1}-u_{t+1}^2)\Bigg],
\end{align}
where 
\begin{align}\label{power_exp_uu}
    \frac{1}{\sqrt{n}}\sum_{t=k_0}^{T-1}(\widetilde{u}^2_{1,t+1}-u_{t+1}^2)&=\sqrt{\frac{n}{T}}\+\beta^{0\prime}\left(\frac{1}{n}\sum_{t=k_0}^{T-1}\*f_t\*f_t'\right) \+\beta^0 + \sqrt{\frac{n}{T}}\frac{1}{n}\sum_{t=k_0}^{T-1}T^{1/4}(\widetilde{\+\theta}_t-\+\theta)'\*w_t\*w_t'T^{1/4}(\widetilde{\+\theta}_t-\+\theta)\notag\\
    &-2\sqrt{\frac{n}{T}}\frac{1}{n}\sum_{t=k_0}^{T-1}T^{1/4}(\widetilde{\+\theta}_t-\+\theta)'\*w_t\*f_t'\+\beta^0-\frac{2}{T^{1/4}}\frac{1}{\sqrt{n}}\sum_{t=k_0}^{T-1}T^{1/4}(\widetilde{\+\theta}_t-\+\theta)'\*w_tu_{t+1}\notag\\
    &+\frac{2}{T^{1/4}}\+\beta^{0\prime}\frac{1}{\sqrt{n}}\sum_{t=k_0}^{T-1}\*f_tu_{t+1}.
\end{align}
Clearly, 
\begin{align}
  T^{1/4}(\widetilde{\+\theta}_t-\+\theta) &= \left(\frac{1}{t}\sum_{s=1}^{t-1}\*w_s\*w_s'\right)^{-1}\frac{1}{t}\sum_{s=1}^{t-1}\*w_s\*f_s'\+\beta^0+T^{-1/4}\frac{T}{t} \left(\frac{1}{t}\sum_{s=1}^{t-1}\*w_s\*w_s'\right)^{-1}\frac{1}{\sqrt{T}}\sum_{s=1}^{t-1}\*w_su_{s+1} \notag\\
&=\+\Sigma_\*w^{-1}\+\Sigma_{\*w\*f}\+\beta^0+O_p(T^{-1/4})
\end{align}
uniformly in $t$ since $\sup_{k_0\leq t \leq T-1}\frac{T}{t}=\sup_{s\in [\pi_0, 1]}s^{-1}+O(T^{-1})$. This and the fact that $\{ \*z_su_{s+1}\}_{s=1}^{t-1}$ is a heterogeneous MDS sequence immediately imply that 
\begin{align}
    &\left| \frac{2}{T^{1/4}}\frac{1}{\sqrt{n}}\sum_{t=k_0}^{T-1}T^{1/4}(\widetilde{\+\theta}_t-\+\theta)'\*w_tu_{t+1}\right|=O_p(T^{-1/4}),\label{(theta-theta)wu}\\
    & \left|\frac{2}{T^{1/4}}\+\beta^{0\prime}\frac{1}{\sqrt{n}}\sum_{t=k_0}^{T-1}\*f_tu_{t+1} \right|\leq \frac{2}{T^{1/4}}\left\|\+\beta^0 \right\| \left\| \frac{1}{\sqrt{n}}\sum_{t=k_0}^{T-1}\*f_tu_{t+1}\right\|=O_p(T^{-1/4}).\label{fu}
\end{align}
Note that (\ref{(theta-theta)wu}) - (\ref{fu}) hold under ARCH/GARCH effects since 
\begin{align}
    \mathbb{E}\left(\left\|\frac{1}{\sqrt{n}}\sum_{t=k_0}^{T-1}\*f_tu_{t+1} \right\|^2\right)&=\left(\frac{n}{T}\right)^{-1}\frac{1}{T}\sum_{t=k_0}^{T-1}\sum_{s=k_0}^{T-1}\mathbb{E}\left(\*f_t'\*f_su_{t+1}u_{s+1} \right)\notag\\
    &=\left(\frac{n}{T}\right)^{-1}\frac{1}{T}\sum_{t=k_0}^{T-1}\mathbb{E}\left(\*f_t'\*f_t\mathbb{E}(u_{t+1}^2|\mathcal{F}_t )\right)\notag\\
    &=\left(\frac{n}{T}\right)^{-1}\frac{1}{T}\sum_{t=k_0}^{T-1}\mathbb{E}\left(\sigma_{t+1}^2\left\|\*f_t\right\|^2\right)\notag\\
    &= \left(\frac{n}{T}\right)^{-1}\frac{1}{T}\sum_{t=k_0}^{T-1}\mathbb{E}(\sigma_{t+1}^2)\mathbb{E}\left(\left\|\*f_t\right\|^2\right)\notag\\
    &=\sigma^2\left(\frac{n}{T}\right)^{-1}\frac{1}{T}\sum_{t=k_0}^{T-1}\mathbb{E}\left(\left\|\*f_t\right\|^2\right)=O(1),
\end{align}
where we were able to split the expectation, because $\sigma^2_t$ is independent from other model primitives. Next, 
\begin{align}
    \sqrt{\frac{n}{T}}\frac{1}{n}\sum_{t=k_0}^{T-1}T^{1/4}(\widetilde{\+\theta}_t-\+\theta)'\*w_t\*w_t'&T^{1/4}(\widetilde{\+\theta}_t-\+\theta)= \sqrt{\frac{n}{T}}\frac{1}{n}\sum_{t=k_0}^{T-1}T^{1/4}(\widetilde{\+\theta}_t-\+\theta)'\+\Sigma_\*w T^{1/4}(\widetilde{\+\theta}_t-\+\theta)\notag\\
    &+ \underbrace{\sqrt{\frac{n}{T}}\frac{1}{\sqrt{n}}\sum_{t=k_0}^{T-1}\left[T^{1/4}(\widetilde{\+\theta}_t-\+\theta) \otimes T^{1/4}(\widetilde{\+\theta}_t-\+\theta) \right]'n^{-1/2}\mathrm{vec}(\*w_t\*w_t'-\+\Sigma_\*w)}_{O_p(n^{-1/2})}\notag\\
    &=\sqrt{\frac{n}{T}}\frac{1}{n}\sum_{t=k_0}^{T-1}\+\beta^{0\prime}\+\Sigma_{\*w\*f}'\+\Sigma^{-1}_\*w\+\Sigma_\*w \+\Sigma^{-1}_\*w\+\Sigma_{\*w\*f}\+\beta^0+o_p(1)\notag\\
    &= \sqrt{1-\pi_0}\+\beta^{0\prime}\+\Sigma_{\*w\*f}'\+\Sigma^{-1}_\*w\+\Sigma_{\*w\*f}\+\beta^0+o_p(1)
\end{align}
by the same argument as in Lemma A.2 of \cite{stauskas2022tests}. In particular,
\begin{align}
    \sum_{t=k_0}^{T-1}\left[T^{1/4}(\widetilde{\+\theta}_t-\+\theta) \otimes T^{1/4}(\widetilde{\+\theta}_t-\+\theta) \right]'&n^{-1/2}\mathrm{vec}(\*w_t\*w_t'-\+\Sigma_\*w)\notag\\
    &\Rightarrow \int_{s=\pi_0}^1 \left[\+\Sigma_\*w^{-1}\+\Sigma_{\*w\*f}\+\beta^0\otimes \+\Sigma_\*w^{-1}\+\Sigma_{\*w\*f}\+\beta^0 \right]'d\*B(s) + b,
\end{align}
where $\*B(s)$ is a Brownian motion and $b$ represents a bias term induced by the time dependence of $\*z_t$. Similarly, 
\begin{align}
    2\sqrt{\frac{n}{T}}\frac{1}{n}\sum_{t=k_0}^{T-1}T^{1/4}(\widetilde{\+\theta}_t-\+\theta)'\*w_t\*f_t'\+\beta^0&=2\sqrt{\frac{n}{T}}\frac{1}{n}\sum_{t=k_0}^{T-1}T^{1/4}(\widetilde{\+\theta}_t-\+\theta)'\+\Sigma_{\*w\*f}\+\beta^0\notag\\
    &+\underbrace{2 \sqrt{\frac{n}{T}}\frac{1}{\sqrt{n}}\sum_{t=k_0}^{T-1}\left[\+\beta^0 \otimes T^{1/4}(\widetilde{\+\theta}_t-\+\theta)\right]'n^{-1/2}\mathrm{vec}(\*w_t\*f_t'-\+\Sigma_{\*w\*f})}_{O_p(n^{-1/2})}\notag\\
    &= 2\sqrt{\frac{n}{T}}\frac{1}{n}\sum_{t=k_0}^{T-1}\+\beta^{0\prime}\+\Sigma_{\*w\*f}'\+\Sigma_\*w^{-1}\+\Sigma_{\*w\*f}\+\beta^0 + o_p(1)\notag\\
    &= 2\sqrt{1-\pi_0}\+\beta^{0\prime}\+\Sigma_{\*w\*f}'\+\Sigma^{-1}_\*w\+\Sigma_{\*w\*f}\+\beta^0+o_p(1).
\end{align}
Lastly, $\sqrt{\frac{n}{T}}\+\beta^{0\prime}\left(\frac{1}{n}\sum_{t=k_0}^{T-1}\*f_t\*f_t'\right) \+\beta^0=\sqrt{1-\pi_0}\+\beta^{0\prime}\+\beta^{0}+O_p(n^{-1/2})$. The component that remains is 
\begin{align}\label{negl_u_1u_2-u}
    \frac{1}{\sqrt{n}}\sum_{t=k_0}^{k_0+m_0-1}(\widetilde{u}_{1,t+1}\widetilde{u}_{2,t+1}&-u_{t+1}^2)= \frac{1}{T^{1/4}}\frac{1}{\sqrt{n}}\sum_{t=k_0}^{k_0+m_0-1} T^{1/4}(\widetilde{\+\theta}_t-\+\theta)'\*w_tu_{t+1}- \frac{1}{\sqrt{T}\sqrt{n}}\sum_{t=k_0}^{k_0+m_0-1} \sqrt{T}(\widetilde{\+\delta}_t-\delta)'\*z_tu_{t+1}\notag\\
    &- \frac{1}{T^{1/4}}\sqrt{\frac{n}{T}}\frac{1}{n}\sum_{t=k_0}^{k_0+m_0-1}\sqrt{T}(\widetilde{\+\delta}_t-\delta)'\mathbb{E}\left(\*z_t\*f_t'\right)\+\beta^0-\frac{1}{T^{1/4}}\frac{1}{\sqrt{n}}\sum_{t=k_0}^{k_0+m_0-1}\+\beta^{0\prime}\*f_tu_{t+1}\notag\\
    &- \frac{1}{T^{1/4}}\underbrace{\sqrt{\frac{n}{T}}\frac{1}{\sqrt{n}}\sum_{t=k_0}^{k_0+m_0-1}\left[\+\beta^0 \otimes\sqrt{T}(\widetilde{\+\delta}_t-\delta)  \right]'n^{-1/2}\mathrm{vec}(\*z_t\*f_t'-\mathbb{E}\left(\*z_t\*f_t'\right))}_{O_p(n^{-1/2})}\notag\\
    &+ \frac{1}{T^{1/4}}\sqrt{\frac{n}{T}}\frac{1}{n}\sum_{t=k_0}^{k_0+m_0-1}\sqrt{T}(\widetilde{\+\delta}_t-\delta)'\mathbb{E}(\*z_t\*w_t')T^{1/4}(\widetilde{\+\theta}_t-\+\theta)\notag\\
    &+ \frac{1}{T^{1/4}}\underbrace{\sqrt{\frac{n}{T}}\frac{1}{\sqrt{n}}\sum_{t=k_0}^{k_0+m_0-1}\left[T^{1/4}(\widetilde{\+\theta}_t-\+\theta)\otimes \sqrt{T}(\widetilde{\+\delta}_t-\delta)\right]'n^{-1/2}\mathrm{vec}(\*z_t\*w_t'-\mathbb{E}(\*z_t\*w_t'))}_{O_p(n^{-1/2})}\notag\\
    &=O_p(T^{-1/4})
\end{align}
by the same arguments as in (\ref{(theta-theta)wu}) and (\ref{fu}) since $\left\|\frac{1}{\sqrt{n}}\sum_{t=k_0}^{k_0+m_0-1} \sqrt{T}(\widetilde{\+\delta}_t-\delta)'\*z_tu_{t+1} \right\|=O_p(1)$ by the MDS sequence assumption, and 
\begin{align}
    \left|\frac{1}{T^{1/4}}\sqrt{\frac{n}{T}}\frac{1}{n}\sum_{t=k_0}^{k_0+m_0-1}\sqrt{T}(\widetilde{\+\delta}_t-\delta)'\mathbb{E}\left(\*z_t\*f_t'\right)\+\beta^0\right|&\leq \frac{1}{T^{1/4}}\sup_{k_0\leq t\leq T-1}\left\| \sqrt{T}(\widetilde{\+\delta}_t-\delta)\right\|\sqrt{\frac{n}{T}}\frac{1}{n}\sum_{t=k_0}^{k_0+m_0-1}\left\|\mathbb{E}\left(\*z_t\*f_t'\right)\right\|\left\| \+\beta^0\right\|\notag\\
    &=O_p(T^{-1/4}).
\end{align}
Also,
\begin{align}
   \sum_{t=k_0}^{k_0+m_0-1}\left[T^{1/4}(\widetilde{\+\theta}_t-\+\theta)\otimes \sqrt{T}(\widetilde{\+\delta}_t-\delta)\right]'&n^{-1/2}\mathrm{vec}(\*z_t\*w_t'-\mathbb{E}(\*z_t\*w_t'))\notag\\
    &\Rightarrow \int_{s=\pi_0}^{\pi_0+\mu_0}\left[\+\Sigma_\*w^{-1}\+\Sigma_{\*w\*f}\+\beta^0\otimes \*f(\*W(s)) \right]' d\*V(s) + b
\end{align}
for two uncorrelated Brownian motions $\*W(s)$ and $\*V(s)$, and so 
\begin{align}
\left|\frac{1}{T^{1/4}}\sqrt{\frac{n}{T}}\frac{1}{n}\sum_{t=k_0}^{k_0+m_0-1}\sqrt{T}(\widetilde{\+\delta}_t-\delta)'\mathbb{E}(\*z_t\*w_t')T^{1/4}(\widetilde{\+\theta}_t-\+\theta) \right|=O_p(T^{-1/4}),
\end{align}
 as well. By the same steps, by adjusting the summation limits, we can show that 
\begin{align}
   \left|\frac{1}{\sqrt{n}}\sum_{t=k_0+m_0}^{T-1}(\widetilde{u}_{1,t+1}\widetilde{u}_{2,t+1}-u_{t+1}^2)\right|=O_p(T^{-1/4})
\end{align}
 as well. In summary, we have that as $T\to \infty$,
\begin{align}
    g_{f,1,2}&=\sqrt{1-\pi_0}\+\beta^{0\prime}\+\beta^{0}+\sqrt{1-\pi_0}\+\beta^{0\prime}\+\Sigma_{\*w\*f}'\+\Sigma^{-1}_\*w\+\Sigma_{\*w\*f}\+\beta^0\notag\\
    &-2\sqrt{1-\pi_0}\+\beta^{0\prime}\+\Sigma_{\*w\*f}'\+\Sigma^{-1}_\*w\+\Sigma_{\*w\*f}\+\beta^0+o_p(1)\notag\\
    &\to_p \sqrt{1-\pi_0}\+\beta^{0\prime}(\*I_r-\+\Sigma_{\*w\*f}'\+\Sigma^{-1}_\*w\+\Sigma_{\*w\*f})\+\beta^{0}.
\end{align}
Based on these results, we immediately obtain as $T\to \infty$ for $j=2,3,4$: 
\begin{align}
    g_{f,2,2}&=\frac{n}{l^0_1}\left(\frac{1}{\sqrt{n}}\sum_{t=k_0}^{k_0+l^0_1-1}(\widetilde{u}^2_{1,t+1}-u_{t+1}^2)-\frac{l^0_1}{l^0_2}\frac{1}{\sqrt{n}}\sum_{t=k_0}^{k_0+l^0_2-1}(\widetilde{u}^2_{2,t+1}-u_{t+1}^2) \right)\notag\\
    &=\frac{n}{l^0_1}\frac{1}{\sqrt{n}}\sum_{t=k_0}^{k_0+l^0_1-1}(\widetilde{u}^2_{1,t+1}-u_{t+1}^2) + O_p(T^{-1/4})\notag\\
    &=\sqrt{\frac{n}{T}}\frac{1}{l_1^0}\sum_{t=k_0}^{k_0+l_1^0-1}\+\beta^{0\prime}(\*I_r-\+\Sigma_{\*w\*f}'\+\Sigma^{-1}_\*w\+\Sigma_{\*w\*f})\+\beta^0+O_p(T^{-1/4})\notag\\
    &\to_p \sqrt{1-\pi_0}\+\beta^{0\prime}(\*I_r-\+\Sigma_{\*w\*f}'\+\Sigma^{-1}_\*w\+\Sigma_{\*w\*f})\+\beta^{0},
\end{align}
because, similarly to (\ref{negl_u_1u_2-u}), we have
\begin{align}
    \left|\frac{1}{\sqrt{n}}\sum_{t=k_0}^{k_0+l^0_2-1}(\widetilde{u}^2_{2,t+1}-u_{t+1}^2)\right|&\leq \left|\frac{2}{\sqrt{T}\sqrt{n}}\sum_{t=k_0}^{k_0+l^0_2-1}\sqrt{T}(\widetilde{\+\delta}_t-\+\delta)'\*z_tu_{t+1}\right|\notag\\
    &+\left|\frac{1}{\sqrt{n}}\frac{1}{T}\sum_{t=k_0}^{k_0+l^0_2-1}\sqrt{T}(\widetilde{\+\delta}_t-\+\delta)'\*z_t\*z_t'\sqrt{T}(\widetilde{\+\delta}_t-\+\delta)\right|\notag\\
    &\leq \left|\frac{2}{\sqrt{T}\sqrt{n}}\sum_{t=k_0}^{k_0+l^0_2-1}\sqrt{T}(\widetilde{\+\delta}_t-\+\delta)'\*z_tu_{t+1}\right|\notag\\
    &+\left|\frac{1}{\sqrt{n}}\frac{1}{T}\sum_{t=k_0}^{k_0+l^0_2-1}\sqrt{T}(\widetilde{\+\delta}_t-\+\delta)'\+\Sigma_\*z\sqrt{T}(\widetilde{\+\delta}_t-\+\delta)\right|\notag\\
    &+\left|\frac{1}{T}\underbrace{\sum_{t=k_0}^{k_0+l^0_2-1}\left[\sqrt{T}(\widetilde{\+\delta}_t-\+\delta)\otimes \sqrt{T}(\widetilde{\+\delta}_t-\+\delta) \right]'n^{-1/2}\mathrm{vec}\left(\*z_t\*z_t'-\+\Sigma_\*z \right)}_{O_p(1)}\right|
    &=o_p(1).
\end{align}
Then, by inserting (\ref{power_exp_uu}), we have
\begin{align}
    g_{f,3,2}&=\frac{1}{n(1-\tau_0)}\sum_{l_1=\lfloor n\tau_0 \rfloor+1}^n\frac{n}{l_1}\left(\frac{1}{\sqrt{n}}\sum_{t=k_0}^{k_0+l_1-1}(\widetilde{u}^2_{1,t+1}-u_{t+1}^2)-\frac{l_1}{l^0_2}\frac{1}{\sqrt{n}}\sum_{t=k_0}^{k_0+l^0_2-1}(\widetilde{u}^2_{2,t+1}-u_{t+1}^2) \right)\notag\\
    &= \frac{1}{n(1-\tau_0)}\sum_{l_1=\lfloor n\tau_0 \rfloor+1}^n\frac{n}{l_1}\frac{1}{\sqrt{n}}\sum_{t=k_0}^{k_0+l_1-1}(\widetilde{u}^2_{1,t+1}-u_{t+1}^2) - \frac{1}{n(1-\tau_0)}\sum_{l_1=\lfloor n\tau_0 \rfloor+1}^n\frac{n}{l^0_2}\frac{1}{\sqrt{n}}\sum_{t=k_0}^{k_0+l^0_2-1}(\widetilde{u}^2_{2,t+1}-u_{t+1}^2)  \notag\\
    &= \frac{1}{n(1-\tau_0)}\sum_{l_1=\lfloor n\tau_0 \rfloor+1}^n\frac{n}{l_1}\frac{1}{\sqrt{n}}\sum_{t=k_0}^{k_0+l_1-1}(\widetilde{u}^2_{1,t+1}-u_{t+1}^2) + O_p(T^{-1/4})\notag\\
    &= \sqrt{\frac{n}{T}}\frac{1}{n(1-\tau_0)}\sum_{l_1=\lfloor n\tau_0 \rfloor+1}^n\frac{1}{l_1}\sum_{t=k_0}^{k_0+l_1-1}\+\beta^{0\prime}(\*I_r-\+\Sigma_{\*w\*f}'\+\Sigma^{-1}_\*w\+\Sigma_{\*w\*f})\+\beta^0+O_p(T^{-1/4})\notag\\
    &= \sqrt{\frac{n}{T}}\frac{1}{n(1-\tau_0)}\sum_{l_1=\lfloor n\tau_0 \rfloor+1}^n\+\beta^{0\prime}(\*I_r-\+\Sigma_{\*w\*f}'\+\Sigma^{-1}_\*w\+\Sigma_{\*w\*f})\+\beta^0+O_p(T^{-1/4})\notag\\
    &=\sqrt{\frac{n}{T}}\frac{n-\lfloor n\tau_0 \rfloor}{n}\frac{1}{(1-\tau_0)}\+\beta^{0\prime}(\*I_r-\+\Sigma_{\*w\*f}'\+\Sigma^{-1}_\*w\+\Sigma_{\*w\*f})\+\beta^0+O_p(T^{-1/4})\notag\\
    &\to_p\sqrt{1-\pi_0}\+\beta^{0\prime}(\*I_r-\+\Sigma_{\*w\*f}'\+\Sigma^{-1}_\*w\+\Sigma_{\*w\*f})\+\beta^{0},
\end{align}
because
\begin{align}
    \left|\frac{1}{n(1-\tau_0)}\sum_{l_1=\lfloor n\tau_0 \rfloor+1}^n\frac{n}{l^0_2}\frac{1}{\sqrt{n}}\sum_{t=k_0}^{k_0+l^0_2-1}(\widetilde{u}^2_{2,t+1}-u_{t+1}^2) \right|&\leq \frac{n-\lfloor n\tau_0 \rfloor}{n(1-\tau_0)}\frac{n}{l_2^0}\left| \frac{1}{\sqrt{n}}\sum_{t=k_0}^{k_0+l^0_2-1}(\widetilde{u}^2_{2,t+1}-u_{t+1}^2)\right|\notag\\
    &=O_p(T^{-1/4})
\end{align}
and by the same argument
\begin{align}\label{gf4_power}
     g_{f,4,2}&= \frac{1}{n(1-\tau_0)}\sum_{l_2=\lfloor n\tau_0 \rfloor+1}^n\frac{n}{l_1^0}\left(\frac{1}{\sqrt{n}}\sum_{t=k_0}^{k_0+l^0_1-1}(\widetilde{u}^2_{1,t+1}-u_{t+1}^2)-\frac{l^0_1}{l_2}\frac{1}{\sqrt{n}}\sum_{t=k_0}^{k_0+l_2-1}(\widetilde{u}^2_{2,t+1}-u_{t+1}^2) \right)\notag\\
     &= \frac{1}{n(1-\tau_0)}\sum_{l_2=\lfloor n\tau_0 \rfloor+1}^n\frac{n}{l_1^0}\frac{1}{\sqrt{n}}\sum_{t=k_0}^{k_0+l^0_1-1}(\widetilde{u}^2_{1,t+1}-u_{t+1}^2) - \frac{1}{n(1-\tau_0)}\sum_{l_2=\lfloor n\tau_0 \rfloor+1}^n\frac{n}{l_2}\frac{1}{\sqrt{n}}\sum_{t=k_0}^{k_0+l_2-1}(\widetilde{u}^2_{2,t+1}-u_{t+1}^2)\notag\\
     &=\sqrt{\frac{n}{T}}\frac{1}{n(1-\tau_0)}\sum_{l_2=\lfloor n\tau_0 \rfloor+1}^n\frac{1}{l_1^0}\sum_{t=k_0}^{k_0+l_1^0-1}\+\beta^{0\prime}(\*I_r-\+\Sigma_{\*w\*f}'\+\Sigma^{-1}_\*w\+\Sigma_{\*w\*f})\+\beta^0+O_p(T^{-1/4})\notag\\
    &= \sqrt{\frac{n}{T}}\frac{1}{n(1-\tau_0)}\sum_{l_2=\lfloor n\tau_0 \rfloor+1}^n\+\beta^{0\prime}(\*I_r-\+\Sigma_{\*w\*f}'\+\Sigma^{-1}_\*w\+\Sigma_{\*w\*f})\+\beta^0+O_p(T^{-1/4})\notag\\
    &=\sqrt{\frac{n}{T}}\frac{n-\lfloor n\tau_0 \rfloor}{n}\frac{1}{(1-\tau_0)}\+\beta^{0\prime}(\*I_r-\+\Sigma_{\*w\*f}'\+\Sigma^{-1}_\*w\+\Sigma_{\*w\*f})\+\beta^0+O_p(T^{-1/4})\notag\\
    &\to_p\sqrt{1-\pi_0}\+\beta^{0\prime}(\*I_r-\+\Sigma_{\*w\*f}'\+\Sigma^{-1}_\*w\+\Sigma_{\*w\*f})\+\beta^{0}
\end{align}
as $T\to \infty$, since
\begin{align}
    &\left|\frac{1}{n(1-\tau_0)}\sum_{l_2=\lfloor n\tau_0 \rfloor+1}^n\frac{n}{l_2}\frac{1}{\sqrt{n}}\sum_{t=k_0}^{k_0+l_2-1}(\widetilde{u}^2_{2,t+1}-u_{t+1}^2)\right|\notag\\
    &\leq \sup_{\lfloor n\tau_0 \rfloor+1\leq l_2 \leq n}\left(\frac{n}{l_2}\right)\frac{1}{n(1-\tau_0)}\sum_{l_2=\lfloor n\tau_0 \rfloor+1}^n \left|\frac{1}{\sqrt{n}}\sum_{t=k_0}^{k_0+l_2-1}(\widetilde{u}^2_{2,t+1}-u_{t+1}^2) \right|=O_p(T^{-1/4}),
\end{align}
as long as $\tau_0\neq 0$ and it is sufficiently far from it. The latter requirement is placed to make the supremum bounded. Alternatively, we have
\begin{align}
     &\left|\frac{1}{n(1-\tau_0)}\sum_{l_2=\lfloor n\tau_0 \rfloor+1}^n\frac{n}{l_2}\frac{1}{\sqrt{n}}\sum_{t=k_0}^{k_0+l_2-1}(\widetilde{u}^2_{2,t+1}-u_{t+1}^2)\right|\notag\\
    &\leq \sup_{\lfloor n\tau_0 \rfloor+1\leq l_2 \leq n}\left|\frac{1}{\sqrt{n}}\sum_{t=k_0}^{k_0+l_2-1}(\widetilde{u}^2_{2,t+1}-u_{t+1}^2) \right| \frac{1}{(1-\tau_0)}\int_{\lambda_2=\tau_0}^1\frac{1}{\lambda_2}d\lambda_2 + o_p(T^{-1/4})\notag\\
    &= -\sup_{\lfloor n\tau_0 \rfloor+1\leq l_2 \leq n}\left|\frac{1}{\sqrt{n}}\sum_{t=k_0}^{k_0+l_2-1}(\widetilde{u}^2_{2,t+1}-u_{t+1}^2) \right| \frac{1}{(1-\tau_0)}\ln(\tau_0)+o_p(T^{-1/4 })\notag\\
    &=O_p(T^{-1/4}).
\end{align}
More generally, they both follow directly based on the logic in \cite{pitarakis2025novel}, and in particular, the passage from (A.6) to (A.8), which demonstrates that the remainder still vanishes inside of the integral. 
\section{Theoretical Results I: PCA}
\subsection{Relevant Expansions}
\subsubsection{Homogeneous $\alpha$}
 Recall that since $\widehat{\*F}_t$ is the matrix of $r$ eigenvectors corresponding to the $r$ largest eigenvalues of $(Nt)^{-1} \*X_t\*X_t'$, we have \footnote{Let us note that the SVD (and the subsequent decomposition) is here conducted on the $t\times t$ covariance for $\*X'$ i.e., $\frac{1}{Nt}\*X\*X'$, following the work of \citet{bai2023approximate}. The same decomposition and asymptotic expansions below can be done in the (more traditional) case of taking eigenvectors of the $N\times N$ covariance $\frac{1}{Nt}\*X'\*X$  \citep[see][]{stock2002forecasting, barigozzi2023fnets}. Both SVDs return the same set of singular values, of course, and as long as the interest is in modeling static principal eigenvectors, the difference boils down to loadings and factors being exchanged in the various expressions.} 
\begin{align}\label{original_exp}
    \widehat{\*F}_t\*D_{Nt,r}^2=\frac{1}{Nt}\*X_t\*X_t'\widehat{\*F}_t&=\left(\frac{1}{Nt}\*F_t\+\Lambda'\+\Lambda \*F_t'+ \frac{1}{Nt}\*F_t\+\Lambda'\*E_t'+\frac{1}{Nt}\*E_t \+\Lambda \*F_t'+\frac{1}{Nt}\*E_t\*E_t'\right) \widehat{\*F}_t\notag\\
    &=\frac{1}{Nt}\*F_t\+\Lambda'\+\Lambda \*F_t'\widehat{\*F}_t+ \frac{1}{Nt}\*F_t\+\Lambda'\*E_t'\widehat{\*F}_t+\frac{1}{Nt}\*E_t \+\Lambda \*F_t'\widehat{\*F}_t+\frac{1}{Nt}\*E_t\*E_t'\widehat{\*F}_t,
\end{align}
which implies that by letting the rotation matrix $\*H_{Nt,r}=\*D_{Nt,r}^{-2}(t^{-1}\widehat{\*F}_t'\*F_t)(N^{-1}\+\Lambda'\+\Lambda )$
\begin{align}\label{Fact_Space_Stack}
    \widehat{\*F}_t-\*F_t\*H_{Nt,r}'&= \widehat{\*F}_t- \*F_t(N^{-1}\+\Lambda'\+\Lambda )(t^{-1}\*F_t'\widehat{\*F}_t)\*D_{Nt,r}^{-2}\notag\\
    &=  \left(\frac{1}{Nt}\*F_t\+\Lambda'\*E_t'\widehat{\*F}_t+\frac{1}{Nt}\*E_t \+\Lambda \*F_t'\widehat{\*F}_t+\frac{1}{Nt}\*E_t\*E_t'\widehat{\*F}_t \right)\*D_{Nt,r}^{-2}.
\end{align}
This is the same expansion as in \cite{bai2023approximate}. It is possible to obtain the uniform rate of $\frac{1}{t}\left\|  \widehat{\*F}_t-\*F_t\*H_{Nt,r}'\right\|^2=\frac{1}{t}\sum_{s=1}^t\left\|\widehat{\*f}_s-\*H_{Nt,r}\*f_s \right\|^2$ (Frobenius norm) by applying high-level conditions, especially on the divergence rate of the spectral norm of $\*E_t\*E_t'$. However, while the asymptotic behavior of the whole stack is important, we are more interested in the behavior of its $s$th row for $s=1,\ldots,t$, because we will need to understand the asymptotic behavior of their \textit{out-of-sample average}, where the rotation matrix will change for every $t=k_0, \ldots, T-1$. This will result in a generalization of Theorem 4.1 of \cite{gonccalves2017tests} to potentially weaker loadings. For this, we obtain an expansion similar to (\ref{Fact_Space_Stack}) by implementing further modifications from both \cite{bai2023approximate} and \cite{gonccalves2017tests}. In particular,
\begin{align}
     \widehat{\*F}_t-\*F_t\*H_{Nt,r}'&=\left(\frac{1}{N^{\alpha}t}\*F_t\+\Lambda'\*E_t'\widehat{\*F}_t+\frac{1}{N^{\alpha}t}\*E_t \+\Lambda \*F_t'\widehat{\*F}_t+\frac{1}{N^{\alpha}t}\*E_t\*E_t'\widehat{\*F}_t \right)\left(\frac{N}{N^\alpha}\*D_{Nt,r}^{2} \right)^{-1}\notag\\
     &= \left(\frac{1}{N^{\alpha}t}\*F_t\+\Lambda'\*E_t'\widehat{\*F}_t+\frac{1}{N^{\alpha}t}\*E_t \+\Lambda \*F_t'\widehat{\*F}_t+\frac{1}{N^{\alpha}t}\mathbb{E}(\*E_t\*E_t')\widehat{\*F}_t + \frac{1}{N^{\alpha}t}(\*E_t\*E_t'-\mathbb{E}(\*E_t\*E_t'))\widehat{\*F}_t \right)\left(\frac{N}{N^\alpha}\*D_{Nt,r}^{2} \right)^{-1},
\end{align}
which will not require imposing any rates of $\*E_t\*E_t'$ directly, but will shift focus to the covariance analysis of the panel idiosyncratic components, which will be needed in the analysis of rows. Also, $\alpha\in (0,1]$ is a device which accommodates potentially weaker factor loadings, where e.g. $\alpha=1$ gives the usual strong loading case, and $\alpha=0$ leads to absolutely uninformative loadings. For instance, let $r=1$, then if $\frac{1}{N^{\alpha}}\+\Lambda'\+\Lambda =\sum_{i=1}^N\lambda_i^2 <\infty$, we have square-summable loadings, which implies that individual loadings are practically zero for highly indexed individuals. Accommodation of weaker loadings will require specific assumptions, one of them being $\frac{1}{N^{\alpha}}\+\Lambda'\+\Lambda$ having a positive definite limit. There requirements are reflected in Assumption 3 iii). Note that now $\*H_{Nt,r}=\left(\frac{N}{N^\alpha}\*D_{Nt,r}^{2}\right)^{-1}(t^{-1}\widehat{\*F}_t'\*F_t)(N^{-\alpha}\+\Lambda'\+\Lambda )$. Further we define the following scalar quantities: 
\begin{align}
    &\gamma_{l,s,\alpha}=\frac{1}{N^\alpha}\sum_{i=1}^N\mathbb{E}(e_{i,l}e_{i,s})=\frac{N}{N^\alpha}\frac{1}{N}\sum_{i=1}^N\mathbb{E}(e_{i,l}e_{i,s})=\frac{N}{N^\alpha}\gamma_{l,s},\label{gamma_fact}\\
    & \xi_{l,s,\alpha}=\frac{N}{N^\alpha}\frac{1}{N}\sum_{i=1}^N( e_{i,l}e_{i,s}-\mathbb{E}(e_{i,l}e_{i,s}))=\frac{N}{N^\alpha}\xi_{l,s}\label{xi_fact}, \\
    &\eta_{l,s,\alpha}=\frac{1}{N^\alpha}\sum_{i=1}^N\*f_l'\+\lambda_ie_{i,s}\label{eta_fact},\\
    & \nu_{l,s,\alpha}=\frac{1}{N^\alpha}\sum_{i=1}^N\*f_s'\+\lambda_ie_{i,l}\label{nu_fact}.
\end{align}
Note that only the terms that depend on $\+\lambda_i$ are directly scaled by $N^{-\alpha}$. Because (\ref{gamma_fact}) and (\ref{xi_fact}) are functions of idiosyncratics only, $N/N^{\alpha}$ can be seen as a ''penalty'' term on the overall rate, because we estimate potentially weaker loadings with the usual PC method. Therefore, we have that 
\begin{align}\label{row_fact_space}
    \widehat{\*f}_s-\*H_{Nt,r}\*f_s=\left(\frac{N}{N^\alpha}\*D_{Nt,r}^{2}\right)^{-1}\left(\frac{1}{t}\sum_{l=1}^t\widehat{\*f}_l\gamma_{l,s,\alpha}+ \frac{1}{t}\sum_{l=1}^t\widehat{\*f}_l\xi_{l,s,\alpha}+\frac{1}{t}\sum_{l=1}^t\widehat{\*f}_l\eta_{l,s,\alpha}+\frac{1}{t}\sum_{l=1}^t\widehat{\*f}_l \nu_{l,s,\alpha}\right).
\end{align}
\subsubsection{Heterogeneous $\alpha$}
Note that above imposes that the loadings are weaker to the same degree. Alternatively, we  can allow for $\*B_N=\mathrm{diag}(N^{\alpha_1/2},\ldots, N^{\alpha_r/2})$, where $1\geq \alpha_1>\alpha_2>\ldots >\alpha_r>0$ (the weakest loading still cannot be absolutely uninformative). Note that $\|\*B_N \|\leq MN^{\alpha_1/2}$ and $\|\*B_N^{-1} \|\leq m N^{-\alpha_r/2}$ for some positive constants $M$ and $m$, which means that the order of (the inverse of) this normalization matrix is dominated by the (weakest) strongest factor loading. Again, recall that $ \widehat{\*F}_t\*D_{Nt,r}^2=\frac{1}{Nt}\*X_t\*X_t'\widehat{\*F}_t$ by the eigenvalue-eigenvector relationship. Then, by using the fact that both $\*B_N$ and $\*D_{Nt,r}^2$ are diagonal, we obtain
\begin{align}
    \widehat{\*F}_t\*D_{Nt,r}^2\*B_N^{-1}=\widehat{\*F}_t\*B_N^{-1}\*D_{Nt,r}^2 =\widehat{\*F}_t\*B_N(\*B_N^{-2}\*D_{Nt,r}^2)=\frac{1}{Nt}\*X_t\*X_t'\widehat{\*F}_t\*B_N^{-1}.
\end{align}
By inserting the expression for $\*X_t$, multiplying both sides by $N$ and multiplying $\widehat{\*F}_t$ in, we get 
\begin{align}\label{step1}
    \left(\*F_t\+\Lambda'\+\Lambda t^{-1}\*F_t'\widehat{\*F}_t+ \*F_t\+\Lambda't^{-1}\*E_t'\widehat{\*F}_t+t^{-1}\*E_t \+\Lambda \*F_t'\widehat{\*F}_t+t^{-1}\*E_t\*E_t'\widehat{\*F}_t \right)\*B_N^{-1}=\widehat{\*F}_t\*B_N(N\*B_N^{-2}\*D_{Nt,r}^2),
\end{align}
which can further be turned into 
\begin{align}\label{step2}
   \widehat{\*F}_t\*B_N(N\*B_N^{-2}\*D_{Nt,r}^2)- \*F_t \+\Lambda'\+\Lambda t^{-1}\*F_t'\widehat{\*F}_t\*B_N^{-1}=\*F_t\+\Lambda't^{-1}\*E_t'\widehat{\*F}_t\*B_N^{-1} +t^{-1}\*E_t \+\Lambda \*F_t\widehat{\*F}_t\*B_N^{-1}+t^{-1}\*E_t\*E_t'\widehat{\*F}_t\*B_N^{-1}.
\end{align}
By premultiplying both sides by $(N\*B_N^{-2}\*D_{Nt,r}^2)^{-1}$, we obtain 
\begin{align}\label{step3}
  \widehat{\*F}_t\*B_N&-   \*F_t \+\Lambda'\+\Lambda t^{-1}\*F_t'\widehat{\*F}_t\*B_N^{-1} (N\*B_N^{-2}\*D_{Nt,r}^2)^{-1}\notag\\
  &=  \widehat{\*F}_t\*B_N-   \*F_t\*B_N\left[\*B_N^{-1}\+\Lambda'\+\Lambda t^{-1}\*F_t'\widehat{\*F}_t\*B_N^{-1}(N\*B_N^{-2}\*D_{Nt,r}^2)^{-1}\right]\notag\\
  &=\left(\*F_t\+\Lambda't^{-1}\*E_t'\widehat{\*F}_t\*B_N^{-1}+t^{-1}\*E_t \+\Lambda \*F_t'\widehat{\*F}_t\*B_N^{-1}+t^{-1}\*E_t\*E_t'\widehat{\*F}_t\*B_N^{-1}\right) (N\*B_N^{-2}\*D_{Nt,r}^2)^{-1},
\end{align}
where we define $\overline{\*H}_{Nt,r}=(N\*B_N^{-2}\*D_{Nt,r}^2)^{-1}\*B_N^{-1}t^{-1}\widehat{\*F}_t'\*F_t\+\Lambda'\+\Lambda \*B_N^{-1} $. The component $\*B_N^{-1}t^{-1}\widehat{\*F}_t'\*F_t\+\Lambda'\+\Lambda \*B_N^{-1}$ is bounded in probability for $t=T$ as argued in \cite{bai2023approximate}, and we will show that it is bounded uniformly in the upcoming Lemma 1. At this stage, note that by using the fact that the product of $\*B_N$ and $\*D_{Nt,r}^2$ commutes, we have the property
\begin{align}
    \*B_N^{-1}\overline{\*H}_{Nt,r}\*B_N= \*B_N^{-1} N^{-1}\*B_N^{2} \*D_{Nt,r}^{-2} \*B_N^{-1}t^{-1}\widehat{\*F}_t'\*F_t\+\Lambda'\+\Lambda &=  \*B_N^{-1} N^{-1}\*B_N^{2}\*B_N^{-1} \*D_{Nt,r}^{-2} t^{-1}\widehat{\*F}_t'\*F_t\+\Lambda'\+\Lambda\notag\\
    &= \*D_{Nt,r}^{-2} (t^{-1}\widehat{\*F}_t'\*F_t) (N^{-1}\+\Lambda'\+\Lambda),
\end{align}
which is the original rotation defined below (\ref{original_exp}). Therefore, the operations performed in (\ref{step1}) - (\ref{step3}) are equivalent to the imposition of the $N/N^{\alpha}$ rate penalty in the homogeneously weak loadings case, where its heterogeneous equivalent is $N\*B_N^{-2}$. Clearly, $\lambda_{\mathrm{max}}(N\*B_N^{-2})=N/N^{\alpha_r}$, so out of $r$ penalties, the weakest loading weights the most. Overall,
\begin{align}
     &\widehat{\*F}_t-\*F_t\*H_{Nt,r}'= (\widehat{\*F}_t\*B_N-\*F_t\*B_N\overline{\*H}_{Nt,r}')\*B_N^{-1}\notag\\
     &= \left(\*F_t\+\Lambda't^{-1}\*E_t'\widehat{\*F}_t\*B_N^{-1}+t^{-1}\*E_t \+\Lambda \*F_t'\widehat{\*F}_t\*B_N^{-1}+t^{-1}\*E_t\*E_t'\widehat{\*F}_t\*B_N^{-1}\right) (N\*B_N^{-2}\*D_{Nt,r}^2)^{-1} \*B_N^{-1}\notag\\
     &= \left(\*F_t\+\Lambda't^{-1}\*E_t'\widehat{\*F}_t\*B_N^{-1}+t^{-1}\*E_t \+\Lambda \*F_t'\widehat{\*F}_t\*B_N^{-1}+t^{-1}\mathbb{E}(\*E_t\*E_t')\widehat{\*F}_t\*B_N^{-1} +t^{-1}(\*E_t\*E_t'-\mathbb{E}(\*E_t\*E_t'))\widehat{\*F}_t\*B_N^{-1} \right) (N\*B_N^{-2}\*D_{Nt,r}^2)^{-1} \*B_N^{-1}\notag\\
     &=N^{-\alpha_r}\left(\*F_t\+\Lambda't^{-1}\*E_t'\widehat{\*F}_t+t^{-1}\*E_t \+\Lambda \*F_t'\widehat{\*F}_t+t^{-1}\mathbb{E}(\*E_t\*E_t')\widehat{\*F}_t +t^{-1}(\*E_t\*E_t'-\mathbb{E}(\*E_t\*E_t'))\widehat{\*F}_t \right) (N\*B_N^{-2}\*D_{Nt,r}^2)^{-1} (N^{\alpha_r}\*B_N^{-2})
\end{align}
In what remains, we will use the same definitions of the scalar quantities in (\ref{gamma_fact}) - (\ref{nu_fact}). However, we replace $\alpha$ with $\alpha_r$ (in the interim, absence of a subscript $\alpha$ ($\alpha_r$) means that there is no scaling in terms of $N$). Next, to accommodate heterogeneity in loading weakness, we re-define 
\begin{align}
    &\eta_{l,s,\alpha_r}^D=\frac{\sqrt{N}}{N^{\alpha_r}}\sum_{i=1}^N(N^{-1/2}\*B_N\*f_l)'\*B_N^{-1}\+\lambda_ie_{i,s}=\frac{\sqrt{N}}{N^{\alpha_r}}\eta_{l,s}^D,\\
    &\nu_{l,s,\alpha_r}^D=\frac{\sqrt{N}}{N^{\alpha_r}}\sum_{i=1}^N(N^{-1/2}\*B_N\*f_s)'\*B_N^{-1}\+\lambda_ie_{i,l}=\frac{\sqrt{N}}{N^{\alpha_r}}\nu_{l,s}^D.
\end{align}
 Additionally, we introduce $\*Q_{N,\alpha_r}=\*B_N^{-1}\sqrt{N^{\alpha_r}}$, which is $O(1)$. Then, by again using the commutative product, we get 
\begin{align}\label{het_expand}
    & \widehat{\*f}_s-\*H_{Nt,r}\*f_s\notag\\
     &=\*B_N^{-1}(N\*B_N^{-2}\*D_{Nt,r}^2)^{-1}\left(\frac{1}{t}\sum_{l=1}^t\*B_N^{-1}\widehat{\*f}_l\gamma_{l,s}+ \frac{1}{t}\sum_{l=1}^t\*B_N^{-1}\widehat{\*f}_l\xi_{l,s}+\frac{1}{t}\sum_{l=1}^t\*B_N^{-1}\widehat{\*f}_l\eta_{l,s}+\frac{1}{t}\sum_{l=1}^t\*B_N^{-1}\widehat{\*f}_l \nu_{l,s}\right)\notag\\
     &=(N\*B_N^{-2}\*D_{Nt,r}^2)^{-1}\*Q^2_{N,\alpha_r}\left(\frac{1}{t}\sum_{l=1}^t\widehat{\*f}_l\gamma_{l,s,\alpha_r}+ \frac{1}{t}\sum_{l=1}^t\widehat{\*f}_l\xi_{l,s,\alpha_r}+\frac{1}{t}\sum_{l=1}^t\widehat{\*f}_l\eta_{l,s,\alpha_r}^D+\frac{1}{t}\sum_{l=1}^t\widehat{\*f}_l \nu_{l,s,\alpha_r}^D\right).
\end{align}
\subsection{Lemmas for PCA Convergence Rates}
\subsubsection{Boundedness of Eigenvalues and Rotation Matrices}
\begin{lemma}\label{Lemma1}
Under Assumptions A.1 - A.4, one has 
\begin{enumerate}[(a)]
    \item  if  $\alpha\in (0,1])$ (homogeneous)
    \begin{align*}
&\mathrm{(i.)}\quad\sup_{k_0\leq t\leq T-1}\left\|\left(\frac{N}{N^\alpha}\*D_{Nt,r}^{2}\right)^{-1} \right\|^2=O_p(1), \\
   &\mathrm{(ii.)}\quad\sup_{k_0\leq t\leq T-1} \norm{\* H_{Nt,r}}^q=\sup_{k_0\leq t\leq T-1} \left\|\left(\frac{N}{N^\alpha}\*D_{Nt,r}^{2}\right)^{-1}(t^{-1}\widehat{\*F}_t'\*F_t)(N^{-\alpha}\+\Lambda'\+\Lambda)\right\|^q=O_p(1),\quad \text{for any}\; q>0.
\end{align*}

\item Furthermore, under Assumptions A.1 - A.4 and heterogeneous loadings ($1\geq\alpha_1\geq\cdots\geq\alpha_r>1/2$), one has
\begin{align*}
   &\mathrm{(i.)}\quad\sup_{k_0\leq t\leq T-1}\left\|(N\*B_N^{-2}\*D_{Nt,r}^2)^{-1} \right\|^2=O_p(1), \\
   &\mathrm{(ii.)}\quad\sup_{k_0\leq t\leq T-1} \norm{\overline{\*H}_{Nt,r}}^q=\sup_{k_0\leq t\leq T-1} \left\|(N\*B_N^{-2}\*D_{Nt,r}^2)^{-1}\*B_N^{-1}t^{-1}\widehat{\*F}_t'\*F_t\+\Lambda'\+\Lambda \*B_N^{-1}\right\|^q=O_p(1),\quad \text{for any}\; q>0.
\end{align*}
\end{enumerate}
\end{lemma}
\noindent \textbf{Proof.} $\mathrm{(a)}$ To demonstrate $\mathrm{(i.)}$, we note that 
 $\*D_{Nt,r}^{2}$ is the squared matrix of the $r$ largest singular values --and therefore contains the $r$ largest eigenvalues-- of $(Nt)^{-1} \*X_t\*X_t'$, for $t=k_0,\ldots,T-1$. The term $\frac{N}{N^{\alpha}}$ is the rate penalty term, which takes into account the fact that we normalize by $N$ \textit{in practice}, while the weakness of the loadings requires $N^{\alpha}$, instead, which is unknown. Hence, $\frac{N}{N^{\alpha}}\*D_{Nt,r}^{2}$ is the matrix with the $r$ largest eigenvalues of $(N^\alpha t)^{-1} \*X_t\*X_t'$ (see Lemma 1 in \citealp{bai2023approximate}), which means that we can easily convert
\begin{align}
    \frac{N}{N^\alpha}\lambda_j\left( (Nt)^{-1} \*X_t\*X_t'\right)=\lambda_j\left((N^\alpha t)^{-1} \*X_t\*X_t' \right).
\end{align}
Without loss of generality we assume the \textit{original} eigenvalues in $\*D_{Nt,r}^{2}$ are arranged in non-decreasing order as: $d_{1,t}^2\geq\cdots\geq d_{r,t}^2$. It follows that $\left(\frac{N}{N^\alpha}\*D_{Nt,r}^{2}\right)^{-1}=\frac{N^\alpha}{N}\times \operatorname{diag}\left( d_{j,t}^{-2}\right)$ for $d_{1,t}^{-2}\leq\cdots\leq d_{r,t}^{-2}$, and 
\begin{align}
\left\|\left(\frac{N}{N^\alpha}\*D_{Nt,r}^{2}\right)^{-1} \right\|^2=\operatorname{tr}\left(N^{-2(1-\alpha)}\*D_{Nt,r}^{-4}\right)=N^{-2(1-\alpha)}\sum_{j=1}^rd_{j,t}^{-4}\leq N^{-2(1-\alpha)}rd_{r,t}^{-4}.    
\end{align}
In what follows, it is enough to show boundedness of power 1, because squaring will not change the outcome. By taking supremum over $t$ then 
\begin{align}\label{sup_eigen_bound}
\sup_{k_0\leq t\leq T-1} \left\|\left(\frac{N}{N^\alpha}\*D_{Nt,r}^{2}\right)^{-1} \right\|\leq r^{1/2} \sup_{k_0\leq t\leq T-1} |N^{(\alpha-1)}d_{r,t}^{-2}|=N^{(\alpha-1)}r^{1/2} \sup_{k_0\leq t\leq T-1} |d_{r,t}^{-2}|=r^{1/2} \frac{N^{(\alpha-1)}}{\inf_{k_0\leq t \leq T-1} |d_{r,t}^{2}|}. 
\end{align}
Therefore, to show ($i.$) it is sufficient to show that $\frac{N}{N^{\alpha}}\inf_{k_0\leq t \leq T-1} |d_{r,t}^{2}|=O_p(1)$, or, alternatively, that the leading component of $\inf_{k_0\leq t \leq T-1} |d_{r,t}^{2}|$ is $O_p\left(\frac{N^{\alpha}}{N} \right)$. For that, we use a similar strategy to Lemma A.5 in \citet{gonccalves2017tests}. For $t=k_0,\ldots,T-1$, $s=1,\ldots,t$ then we let $ \*x_s=\*\Lambda\*f_s+\*e_s$ as the horizontal slice of $\*X_t$, where $\*x_s, \*e_s\in\mathbb{R}^{N}$, $\*f_s\in \mathbb{R}^{r}$, $\*\Lambda\in \mathbb{R}^{N\times r}$. The scaled population covariance of $\*x_s$ is then achieved by the conditional law of iterated expectations \begin{align}
    \label{eq_scaledcov}
   N^{-1}\*\Sigma&=N^{(\alpha-1)}N^{-\alpha}\mathbb{E}(\*x_s\*x_s'|\+\Lambda)\notag\\
   &=N^{(\alpha-1)}\left[N^{-\alpha}\mathbb{E}(\*\Lambda \*f_s\*f_s'\*\Lambda'|\+\Lambda)+ N^{-\alpha}\mathbb{E}(\*\Lambda \*f_s\*e_s'|\+\Lambda)+ N^{-\alpha}\mathbb{E}(\*e_s\*f_s'\+\Lambda'|\+\Lambda)+ N^{-\alpha}\mathbb{E}(\*e_s\*e_s'|\+\Lambda)\right]\notag 
   \\
   &= N^{(\alpha-1)}\Big[N^{-\alpha}\*\Lambda\mathbb{E}( \*f_s\*f_s'|\+\Lambda)\*\Lambda' +N^{-\alpha}\mathbb{E}(\*\Lambda \*f_s\mathbb{E}(\*e_s'|\+\Lambda,\*F_t)|\+\Lambda)+N^{-\alpha}\mathbb{E}(\mathbb{E}(\*e_s|\+\Lambda,\*F_t)\*f_s'\+\Lambda'|\+\Lambda)\notag\\
   &+ N^{-\alpha}\mathbb{E}(\mathbb{E}(\*e_s\*e_s'|\+\Lambda,\*F_t)|\+\Lambda)\Big]\notag\\ 
   &=N^{(\alpha-1)}\left[N^{-\alpha}\*\Lambda \mathbb{E}(\*f_s\*f_s')\*\Lambda'+N^{-\alpha}\mathbb{E}(\*\Sigma_e|\+\Lambda)\right]\notag\\
   &=N^{(\alpha-1)}\left[N^{-\alpha}\*\Lambda\*\Lambda'+N^{-\alpha}\*\Sigma_e\right],
\end{align} 
where we used that the cross-product in expectation between $\*f_s$ and $\*e_s$ is zero by Assumption A.2,i) and the usual PC factors normalization condition $\mathbb{E}(\*f_s\*f_s'|\+\Lambda)=\mathbb{E}(\*f_s\*f_s')=\*I_r$ due to independence between the loadings and the factors, implying orthogonality of the factors. Now, by Assumption A.3,iii) we know that the $r\times r$ matrix $N^{-\alpha}\+\Lambda' \+\Lambda>0$ and diagonal, meaning its eigenvalues are all distinct and positive. The non-zero eigenvalues of $\*\Lambda\*\Lambda'$ and $\*\Lambda'\*\Lambda$ are necessarily the same\footnote{This is a standard result: by SVD of $\*\Lambda=\*U\*S \*V'$, then $\*\Lambda'\*\Lambda=\*V \*S'\*S\*V'$ and $\*\Lambda\*\Lambda'=\*U \*S\*S'\*U'$, where: $\*S\in \mathbb{R}^{N\times r}$ is diagonal with eigenvalues of $\*\Lambda$, $\*U\in \mathbb{R}^{N\times N}$ is an orthogonal matrix of left singular vectors, $\*V\in \mathbb{R}^{r\times r}$ is an orthogonal matrix of right singular vectors. It follows that the eigenvalues in $\*S\*S'$ and $\*S'\*S$ are the same, and the same of $\*\Lambda.$} and the remaining $N-r$ eigenvalues of $\*\Lambda\*\Lambda'$ are zero. Then, the idea --by Weil's theorem\footnote{For any two Hermitian matrices $\*A, \*B\in \mathbb{R}^n$, and their sum $\*C=\*A+\*B$, then for $j,i=1:n$ it holds that $\lambda_{j+i-1}(\*C)\leq \lambda_j(\*A)+\lambda_i(\*B)$, for $j+i\leq n+1$. In the present case $\*C:=N^{-\alpha}\*\Sigma,\; \*A:=N^{-\alpha}\*\Lambda' \*\Lambda,\; \*B:=N^{-\alpha}\*\Sigma_e$ and $i=\max=1.$ }-- is to bound the absolute difference between the $j$th eigenvalue of $N^{-\alpha}\*\Sigma$ and the $j$th eigenvalue of $N^{-\alpha}\*\Lambda\*\Lambda'$, $j=1,\ldots,r$, by the max eigenvalue of $N^{-\alpha}\*\Sigma_e$, which is bounded (away from infinity) by Assumption A.2,v). Namely, for any $\alpha\in(0,1]$
\begin{align}\label{Weil_1}
    \left|\lambda_j(N^{-\alpha}\*\Sigma)-\lambda_j(N^{-\alpha}\*\Lambda'\*\Lambda)\right|\leq N^{-\alpha}\lambda_{\max}\left(\*\Sigma_e\right)=O(N^{-\alpha})=o(1),
\end{align}
 where we used the notation $\lambda_j(\*A)$($\lambda_{\max}(\*A)$) the $j$th($\max$) eigenvalue of any matrix $\*A$, such that e.g., $d_{j,t}^2=\lambda_j((Nt)^{-1}\*X_t\*X_t')=N^{(\alpha-1)}\lambda_j((N^{\alpha}t)^{-1}\*X_t\*X_t')$ and $d_{r,t}^2=\lambda_{\max}((Nt)^{-1}\*X_t\*X_t')=N^{(\alpha-1)}\lambda_{\max}((N^{\alpha}t)^{-1}\*X_t\*X_t')$. Then, Weyl's theorem again (see footnote 3), and using as above the fact that $(Nt)^{-1}\*X_t\*X_t'$ and $(Nt)^{-1}\*X_t'\*X_t$ share the same non-zero eigenvalues and we can easily convert between the eigenvalues of $(Nt)^{-1}\*X_t\*X_t'$ and $(N^{\alpha}t)^{-1}\*X_t\*X_t'$ , yields 
\begin{align}\label{chain1}
\left|d_{j,t}^2-N^{(\alpha-1)}\lambda_j(N^{-\alpha}\*\Lambda'\*\Lambda)\right|&\leq \left\|(Nt)^{-1}\*X_t'\*X_t-N^{-1}\*\Sigma\right\|_{sp}    \notag\\
&= N^{(\alpha-1)}\left\|(N^{\alpha}t)^{-1}\*X_t'\*X_t-N^{-\alpha}\*\Sigma\right\|,
\end{align}\label{chain2}
 from which by Assumption A.3,iii) \begin{align}
    d_{r,t}^2&\geq N^{(\alpha-1)}\lambda_r(N^{-\alpha}\*\Lambda'\*\Lambda)-N^{(\alpha-1)}\left\|(N^{\alpha}t)^{-1}\*X_t'\*X_t-N^{-\alpha}\*\Sigma\right\|_{sp}\\
    &\geq N^{(\alpha-1)}\lambda_{\min}(\*\Sigma_{\+\Lambda})/2-N^{(\alpha-1)}\left\|(N^{\alpha}t)^{-1}\*X_t'\*X_t-N^{-\alpha}\*\Sigma\right\|_{sp}.
\end{align}
Then, taking infimum wrt $t$, we get 
\begin{align}\label{chain3}
    \inf_{k_0\leq t \leq T-1}|d_{r,t}^2|\geq N^{(\alpha-1)}\lambda_{\min}(\*\Sigma_{\+\Lambda})/2- N^{(\alpha-1)}\sup_{k_0\leq t\leq T-1}\left\|(N^{\alpha}t)^{-1}\*X_t'\*X_t-N^{-\alpha}\*\Sigma\right\|_{sp}
\end{align}
 As $\*\Sigma_{\+\Lambda}$ is positive definite as of Assumption A.3,iii), we can treat $N^{(\alpha-1)}\lambda_{\min}(\*\Sigma_{\+\Lambda})/2=O\left( \frac{N^{\alpha}}{N}\right)$ as the dominant component of $\inf_{k_0\leq t \leq T-1}|d_{r,t}^2|$, which is what we were supposed to achieve in the light of (\ref{sup_eigen_bound}). What remains to be shown to prove (i.) is that $\sup_{k_0\leq t\leq T-1}\left\|(N^{\alpha}t)^{-1}\*X_t'\*X_t-N^{-\alpha}\*\Sigma\right\|_{sp}=o_p(1).$ Expanding $N^{-\alpha}\*\Sigma$ using \eqref{eq_scaledcov}, and given that by the factor model decomposition one has 
 \begin{align}
 (N^{\alpha}t)^{-1}\*X_t'\*X_t=N^{-\alpha}\*\Lambda\left(t^{-1}\sum_{s=1}^t \*f_s\*f_s'\right)\*\Lambda'+N^{-\alpha}\left(t^{-1}\sum_{s=1}^t \*e_s\*e_s'\right)+N^{-\alpha}\*\Lambda t^{-1}\sum_{s=1}^t \*f_s\*e_s'+N^{-\alpha} t^{-1}\sum_{s=1}^t \*e_s\*f_s'\*\Lambda',    
 \end{align}
 thus
 \begin{align}
     \sup_{k_0\leq t\leq T-1}\left\|(N^{\alpha}t)^{-1}\*X_t'\*X_t-N^{-\alpha}\*\Sigma\right\|_{sp}\leq \sup_{k_0\leq t\leq T-1}\|G_{1,t}\|+\sup_{k_0\leq t\leq T-1}\|G_{2,t}\|+2\sup_{k_0\leq t\leq T-1}\|G_{3,t}\|
 \end{align}
for $G_{1,t}=N^{-\alpha}\*\Lambda\left(t^{-1}\sum_{s=1}^t \*f_s\*f_s'-\*I_r\right)\*\Lambda'$, using again the PC factors normalization condition (Assumption A3.2); $G_{2,t}=N^{-\alpha}\left(t^{-1}\sum_{s=1}^t \*e_s\*e_s'-\*\Sigma_e\right)$; $G_{3,t}=N^{-\alpha}\*\Lambda t^{-1}\sum_{s=1}^t \*f_s\*e_s',$ and using triangle inequality coupled with the fact that $\|\*A\|_{sp}\leq \|\*A\|$ for any matrix $\*A.$ We now proceed to show each of these $\sup$-terms is $o_p(1)$. Starting with $G_{1,t}$, by factoring out of the $\sup_{k_0\leq t\leq T-1}$ the scaled loadings, which do not depend on $t$, one has 
\begin{align*}
    \sup_{k_0\leq t\leq T-1}\left\|G_{1,t}\right\|&=\left\|\frac{\*\Lambda}{N^{\alpha/2
    }}\right\|^2 \sup_{k_0\leq t\leq T-1} \left\|\left(\frac{T}{t}\right) \left(\frac{1}{T}\sum_{s=1}^t \left(\*f_s\*f_s'-\*I_r\right)\right)\right\|\\
    &\leq C \frac{T}{k_0}  \sup_{k_0\leq t\leq T-1} \left\|\frac{1}{T}\sum_{s=1}^t \left(\*f_s\*f_s'-\*I_r\right)\right\|= O_p(1/\sqrt{T})=o_p(1),
\end{align*} which follows by the fact that $\left\|\frac{\*\Lambda}{N^{\alpha/2}}\right\|^2=\operatorname{tr}\left(\*\Lambda'\*\Lambda/N^{\alpha}\right)=O(1)$ by Assumption A3.3, the fact that $T/k_0=O(1)$ since $k_0=\lfloor T\pi_0 \rfloor$ for $\pi_0\in (0,1)$, and by Assumption A3,ii). Now onto $G_{2,t}$.
By union bound, for any $\varepsilon>0$: \begin{align*}
\mathbb{P}\left(\sup_{k_0\leq t\leq T-1}\|G_{2,t}\|>\varepsilon\right)=\mathbb{P}\left(\bigcup_{t=k_0}^{T-1}\{\|G_{2,t}\|>\varepsilon\}\right)\leq \sum_{t=k_0}^{T-1}  \mathbb{P}\left(\|G_{2,t}\|>\varepsilon\right)\leq \frac{1}{\varepsilon^2}\sum_{t=k_0}^{T-1}\mathbb{E}\left(\|G_{2,t}\|^2\right),  
\end{align*}
where the last inequality follows from squaring and applying Markov's inequality. So, we are onto showing that $\sum_{t=k_0}^{T-1}\mathbb{E}\|G_{2,t}\|^2=o(1).$ Now, using that for any square-$N$ matrix $\*A$, the Frobenius norm gives $\|\*A\|^2=\operatorname{tr}(\*A'\*A)=\sum_{i,j=1}^N a_{i,j}^2$, then for $t=k_0,\ldots,T-1$
\begin{align*}
\mathbb{E}\left(\|G_{2,t}\|^2\right)&=\frac{1}{N^{2\alpha}}\sum_{i=1}^N \sum_{j=1}^N\mathbb{E}\left(\frac{1}{t} \sum_{s=1}^t(e_{i,s}e_{j,s}-\mathbb{E}(e_{i,s}e_{j,s}))\right)^2\\
&=\frac{N}{N^{2\alpha}}\frac{1}{N}\sum_{i=1}^N \sum_{j=1}^N\frac{1}{t^2} \sum_{s_1=1}^t\sum_{s_2=1}^t \mathrm{C}ov\left( e_{i,s_1}e_{j,s_1}, e_{i,s_2}e_{j,s_2}\right)\\
&\leq \frac{T}{k_0^2} \frac{N}{N^{2\alpha}}\left(\frac{1}{N}\sum_{i=1}^N \sum_{j=1}^N\frac{1}{T} \sum_{s_1=1}^{T}\sum_{s_2=1}^{T}\left|\mathrm{C}ov\left( e_{i,s_1}e_{j,s_1}, e_{i,s_2}e_{j,s_2}\right)\right|\right),
\end{align*}
from which it follows that
\begin{align*}
    \sum_{t=k_0}^{T-1}\mathbb{E}\left(\|G_{2,t}\|^2\right)&\leq\sum_{t=1}^{T}\mathbb{E}\left(\|G_{2,t}\|^2\right)\notag\\
    &\leq  \left(\frac{T}{k_0}\right)^2 \frac{N}{N^{2\alpha}}\left(\frac{1}{N}\sum_{i=1}^N\sum_{j=1}^N\frac{1}{T} \sum_{s_1=1}^{T}\sum_{s_2=1}^{T}\left|\mathrm{C}ov\left( e_{i,s_1}e_{j,s_1}, e_{i,s_2}e_{j,s_2}\right)\right|\right)\\&=O\left( \frac{N}{N^{2\alpha}}\right)=o(1),
\end{align*}
by Assumption A2.iv) and for $\alpha>1/2$, which implies that overall
\begin{align}
    \sup_{k_0\leq t\leq T-1}\left\|G_{2,t} \right\|=O_p\left(\frac{\sqrt{N}}{N^{\alpha}} \right).
\end{align}
Finally, for $G_{3,t}$ we obtain
\begin{align*} \allowdisplaybreaks
    \sup_{k_0\leq t\leq T-1}\|G_{3,t}\|^2&\leq \left\|\frac{N^{\alpha/2}}{N^{\alpha}}\frac{1}{N^{\alpha/2}}\*\Lambda\right\|^2 \sup_{k_0\leq t\leq T-1}\left\|\frac{1}{t}\sum_{s=1}^t\*f_s\*e_s'\right\|^2\notag\\
    &= \frac{1}{N^{\alpha}} \left\|\frac{\*\Lambda}{N^{\alpha/2}}\right\|^2\sup_{k_0\leq t\leq T-1}\left\|\frac{1}{t}\sum_{s=1}^t\*f_s\*e_s'\right\|^2\\
    &\leq \frac{C}{N^{\alpha}} \sum_{i=1}^N \sup_{k_0\leq t\leq T-1}\left\|\frac{1}{t}\sum_{s=1}^t\*f_s e_{i,s}\right\|^2 \notag\\
    &\leq \frac{1}{k_0}\left(\frac{T}{k_0}\right)\frac{C}{N^{\alpha}}\sum_{i=1}^N \sup_{k_0\leq t\leq T-1}\left\|\frac{1}{\sqrt{T}}\sum_{s=1}^t\*f_s e_{i,s}\right\|^2 \\
    & = \frac{C}{N^{\alpha}} \sum_{i=1}^N \left\|\frac{1}{t}\sum_{s=1}^t\*f_s e_{i,s}\right\|^2 \notag\\
    &\leq \frac{1}{k_0}\left(\frac{T}{k_0}\right)C\frac{N}{N^\alpha}\frac{1}{N}\sum_{i=1}^N \sup_{k_0\leq t\leq T-1}\left\|\frac{1}{\sqrt{T}}\sum_{s=1}^t\*f_s e_{i,s}\right\|^2 \\
    &= O_p\left(\frac{N}{N^\alpha}\frac{1}{k_0}\right)=o_p(1),
\end{align*}
provided that the rate restriction is satisfied, where the latter inequalities follow from the fact that $\left\|\frac{\*\Lambda}{N^{\alpha/2}}\right\|^2=\operatorname{tr}\left(\*\Lambda'\*\Lambda/N^{\alpha}\right)=O(1)$ by Assumption A3.3, the fact that $T/k_0=O(1)$ since $k_0=\lfloor T\pi_0 \rfloor$ for $\pi_0\in (0,1)$, and by Assumption A.3,v). Hence, 
\begin{align}
    \sup_{k_0\leq t\leq T-1}\left\|G_{3,t} \right\|=O_p\left(\sqrt{\frac{N}{N^{\alpha}}}\frac{1}{\sqrt{k_0}} \right).
\end{align}
Finally, to put the results together,
\begin{align}
 \frac{N}{N^\alpha}\inf_{k_0\leq t \leq T-1}|d_{r,t}^2|&\geq\lambda_{\min}(\*\Sigma_{\+\Lambda})/2- \sup_{k_0\leq t\leq T-1}\left\|(N^{\alpha}t)^{-1}\*X_t'\*X_t-N^{-\alpha}\*\Sigma\right\|_{sp}\notag\\
 &=\lambda_{\min}(\*\Sigma_{\+\Lambda})/2 + O_p\left( \frac{1}{\sqrt{T}}\right) + O_p\left(\frac{\sqrt{N}}{N^{\alpha}} \right)+O_p\left(\sqrt{\frac{N}{N^{\alpha}}}\frac{1}{\sqrt{k_0}} \right)
\end{align}
and 
\begin{align}
    \sup_{k_0\leq t\leq T-1} \left\|\left(\frac{N}{N^\alpha}\*D_{Nt,r}^{2}\right)^{-1} \right\|\leq \frac{2r^{1/2}}{\lambda_{\min}(\*\Sigma_{\+\Lambda})}+o_p(1).
\end{align}
This proves (i.).\bigbreak

\noindent Now onto $\mathrm{(ii.)}$. As in \citet{gonccalves2017tests} we show the result for $q=2$, since $\sup_{k_0\leq t\leq T-1} \|\*H_{Nt,r}\|^q=\sup_{k_0\leq t\leq T-1} \left(\|\*H_{Nt,r}\|^2\right)^{q/2}=\left(\sup_{k_0\leq t\leq T-1} \|\*H_{Nt,r}\|^2\right)^{q/2}=O_p(1)$ if $\sup_{k_0\leq t\leq T-1} \|\*H_{Nt,r}\|^2=O_p(1)$, which holds for any $q>0.$ Hence, for $t=k_0,\ldots,T-1$ by submultiplicativity of the Frobenius norm we get
\begin{align*}
    \sup_{k_0\leq t\leq T-1}\|\*H_{Nt,r}\|^2\leq \underbrace{\sup_{k_0\leq t\leq T-1} \left\|\left(\frac{N}{N^\alpha}\*D_{Nt,r}^{2}\right)^{-1}\right\|^2}_{(I)} \underbrace{\sup_{k_0\leq t\leq T-1} \left\|\frac{1}{t}\widehat{\*F}_t'\*F_t \right\|^2}_{(II)}  \underbrace{\left\|\frac{1}{N^{\alpha}}\+\Lambda'\+\Lambda \right\|^2}_{(III)}.
\end{align*}
    The term $(I)$ is $O_p(1)$, as proved in (i.) of part (a), while $(III)=O(1)$ by Assumption A.3,iii). Therefore, it is sufficient to prove the term $(II)=O_p(1)$. Note how
    \begin{align}
        \sup_{k_0\leq t\leq T-1}\left\|\frac{1}{t}\widehat{\*F}_t'\*F_t \right\|^2\leq\sup_{k_0\leq t\leq T-1} t^{-1}\left\| \widehat{\*F}_t\right\|^2 \sup_{k_0\leq t\leq T-1} t^{-1}\left\| \*F_t\right\|^2&=\sup_{k_0\leq t\leq T-1} \mathrm{tr}\left(t^{-1}\widehat{\*F}_t'\widehat{\*F}_t \right)\sup_{k_0\leq t\leq T-1} \mathrm{tr}\left(t^{-1}\*F_t'\*F_t \right)\notag\\
        &=r\sup_{k_0\leq t\leq T-1} \mathrm{tr}\left(t^{-1}\*F_t'\*F_t \right)=O_p(1).
    \end{align}
     Now we turn to the heterogeneous loading cases in part $(\mathrm{b})$. To begin with $\mathrm{(i.)}$, we will follow a different approach since in the heterogeneous case we have different convergence rates for the loadings as given by $\*B_N$, which need to be taken into account in the eigenvalue analysis. The strategy in $\mathrm{(i.)}$ of part (a) crucially hinges on the homogeneous rate. Therefore, we will start with the slightly modified expansion as in \cite{bai2023approximate}:
     \begin{align}
         \frac{1}{Nt}\*X_t\*X_t'&=N^{-1}\*F_t\+\Lambda'\+\Lambda t^{-1}\*F_t'+ N^{-1}\*F_t\+\Lambda't^{-1}\*E_t'+N^{-1}t^{-1}\*E_t \+\Lambda \*F_t'+N^{-1}t^{-1}\*E_t\*E_t'\notag\\
         &=N^{-1}\*F_t\+\Lambda'\+\Lambda t^{-1}\*F_t'+ N^{-1}\*F_t\+\Lambda't^{-1}\*E_t'+N^{-1}t^{-1}\*E_t \+\Lambda \*F_t'+N^{-1}t^{-1}\mathbb{E}\left(\*E_t\*E_t'\right)\notag\\
         &+N^{-1}t^{-1}\left(\*E_t\*E_t'-\mathbb{E}\left(\*E_t\*E_t'\right) \right).
     \end{align}
By following the discussion in the proof of Lemma 1 in \cite{bai2023approximate}, if the four last terms on the right-hand side are negligible, then the eigenvalue matrix of $\frac{1}{Nt}\*X_t\*X_t$, represented by $\*D^2_{Nt,r}$, will be determined by the eigenvalues of $N^{-1}\*F_t\+\Lambda'\+\Lambda t^{-1}\*F_t'=N^{-1}\*F_t\*B_N\left(\*B_N^{-1}\+\Lambda'\+\Lambda\*B_N^{-1}\right) \*B_Nt^{-1}\*F_t'$. Note that the later asymtotically has the same eigenvalues as 
\begin{align}
    N^{-1}\left(\*B_N^{-1}\+\Lambda'\+\Lambda\*B_N^{-1}\right) \*B_Nt^{-1}\*F_t'\*F_t\*B_N&= \left(\*B_N^{-1}\+\Lambda'\+\Lambda\*B_N^{-1}\right) N^{-1}\*B^2_N\notag\\
    &+\left(\*B_N^{-1}\+\Lambda'\+\Lambda\*B_N^{-1}\right) N^{-1}\*B_N\left( t^{-1}\*F_t'\*F_t-\*I_r\right)\*B_N\notag\\
    &= N^{-1}\*B_N^2  \left(\*B_N^{-1}\+\Lambda'\+\Lambda\*B_N^{-1}\right) + O_p(T^{-1/2}), 
\end{align}
because $\*B_N$ and $\*B_N^{-1}\+\Lambda'\+\Lambda\*B_N^{-1}$ are diagonal, and 
\begin{align}\label{FF-I_het_rate_1}
    \sup_{k_0\leq t\leq T-1}\left\|\left(\*B_N^{-1}\+\Lambda'\+\Lambda\*B_N^{-1}\right) N^{-1}\*B_N\left( t^{-1}\*F_t'\*F_t-\*I_r\right)\*B_N \right\|&\leq \left\| \*B_N^{-1}\+\Lambda'\+\Lambda\*B_N^{-1}\right\| \left\|N^{-1/2}\*B_N \right\|^2\notag\\
    &\times \sup_{k_0\leq t\leq T-1}\left\|  t^{-1}\*F_t'\*F_t-\*I_r\right\|=O_p(T^{-1/2}). 
\end{align}
This implies that 
\begin{align}
  (N\*B_N^{-2})  N^{-1}\left(\*B_N^{-1}\+\Lambda'\+\Lambda\*B_N^{-1}\right) \*B_Nt^{-1}\*F_t'\*F_t\*B_N&=  (N\*B_N^{-2})\left(\*B_N^{-1}\+\Lambda'\+\Lambda\*B_N^{-1}\right) N^{-1}\*B^2_N\notag\\
    &+ (N\*B_N^{-2})\left(\*B_N^{-1}\+\Lambda'\+\Lambda\*B_N^{-1}\right) N^{-1}\*B_N\left( t^{-1}\*F_t'\*F_t-\*I_r\right)\*B_N\notag\\
    &=  \left(\*B_N^{-1}\+\Lambda'\+\Lambda\*B_N^{-1}\right) + O_p\left(\frac{N}{N^{\alpha_r}}\frac{1}{\sqrt{T}} \right), 
\end{align}
due to 
\begin{align}\label{FF-I_het_rate_2}
     \sup_{k_0\leq t\leq T-1}\left\|(N\*B_N^{-2})\left(\*B_N^{-1}\+\Lambda'\+\Lambda\*B_N^{-1}\right) N^{-1}\*B_N\left( t^{-1}\*F_t'\*F_t-\*I_r\right)\*B_N \right\|&\leq \left\| \*B_N^{-1}\+\Lambda'\+\Lambda\*B_N^{-1}\right\| \left\|N^{-1/2}\*B_N \right\|^2\notag\\
    &\times \frac{N}{N^{\alpha_r}}\left\| N^{\alpha_r}\*B_N^{-2}\right\|\sup_{k_0\leq t\leq T-1}\left\|  t^{-1}\*F_t'\*F_t-\*I_r\right\|\notag\\
    &=O_p\left(\frac{N}{N^{\alpha_r}}\frac{1}{\sqrt{T}} \right).
\end{align}
Recall that $\*B_N^{-1}\+\Lambda'\+\Lambda\*B_N^{-1}$ has positive and distinct eigenvalues by assumption, which implies that
\begin{align}
    \frac{N}{N^{\alpha_j}}\lambda_j\left(\frac{1}{Nt}\*X_t\*X_t' \right)=\frac{N}{N^{\alpha_j}}\*D_{Nt,r,jj}^2=\lambda_j\left(\*B_N^{-1}\+\Lambda'\+\Lambda\*B_N^{-1} \right)+o_p(1).
\end{align}
Lastly, since $\*B_N^{-1}\+\Lambda'\+\Lambda\*B_N^{-1}$ additionally is diagonal, we obtain that $ N\*B_N^{-2}\*D_{Nt,r}^2 =\*B_N^{-1}\+\Lambda'\+\Lambda\*B_N^{-1} +O_p\left(\frac{N}{N^{\alpha_r}}\frac{1}{\sqrt{T}} \right)$. Therefore, by the CMT
\begin{align}\label{NBD^2_approx}
  \left(N\*B_N^{-2}\*D_{Nt,r}^2 \right)^{-1}=\left(\*B_N^{-1}\+\Lambda'\+\Lambda\*B_N^{-1} \right)^{-1}+O_p\left(\frac{N}{N^{\alpha_r}}\frac{1}{\sqrt{T}} \right),
\end{align}
which exists by assumption. Hence, to demonstrate that $\sup_{k_0\leq t\leq T-1}\left\| \left(N\*B_N^{-2}\*D_{Nt,r}^2 \right)^{-1} \right\|=O_p(1)$, we need to check that the orders of the remainder terms hold in supremum sense, because $\left(\*B_N^{-1}\+\Lambda'\+\Lambda\*B_N^{-1} \right)^{-1}$ is independent of $t$. We already have (\ref{FF-I_het_rate_1}) and (\ref{FF-I_het_rate_2}), where the latter used the rate correction $\frac{N}{N^{\alpha_r}}$. Therefore, by using $\left\|\*A \right\|_{sp}\leq\left\|\*A \right\|$ and the rate correction, we continue with  
\begin{align}\label{spectral_rate_1}
   \frac{N}{N^{\alpha_r}}\sup_{k_0\leq t\leq T-1} \left\|  N^{-1}\*F_t\+\Lambda't^{-1}\*E_t'\right\|_{sp}&\leq \frac{N}{N^{\alpha_r}}N^{-1}\sup_{k_0\leq t\leq T-1} t^{-1/2}\left\|\*F_t \right\|_{sp}\left\|\+\Lambda \right\|_{sp}\sup_{k_0\leq t\leq T-1} t^{-1/2}\left\|\*E_t \right\|_{sp}\notag\\
   &\leq \frac{N}{N^{\alpha_r}}N^{-1}\sup_{k_0\leq t\leq T-1} t^{-1/2}\left\|\*F_t \right\|_{sp}\left\|\+\Lambda \right\|_{sp}\sup_{k_0\leq t\leq T-1} \sqrt{\frac{T}{t}}\sup_{k_0\leq t\leq T-1}T^{-1/2}\left\|\*E_t \right\|_{sp}\notag\\
   &\leq \frac{N}{N^{\alpha_r}}\sup_{k_0\leq t\leq T-1} t^{-1/2}\left\|\*F_t \right\|\left\|N^{-1/2}\+\Lambda \right\|N^{-1/2}\sqrt{\frac{T}{k_0}}T^{-1/2}\sup_{k_0\leq t\leq T-1} \left\|\*E_t \right\|_{sp}\notag\\
   &=  O_p\left(\frac{N^{1/2+\alpha_1/2}}{N^{\alpha_r}}\frac{1}{\sqrt{T}} \right) + O_p\left(N^{\alpha_1/2-\alpha_r} \right),
\end{align}
since $\sup_{k_0\leq t\leq T-1} \left\|\*E_t \right\|_{sp}=\max \{\sqrt{N},\sqrt{T} \}$. Also, because the trace gives a finite sum,
\begin{align}
    \left\| N^{-1/2}\+\Lambda\right\|=\sqrt{\mathrm{tr}\left(N^{-1}\+\Lambda'\+\Lambda \right)}=\sqrt{\sum_{j=1}^r\left(\frac{1}{N}\sum_{i=1}^N\lambda_{i,j}^2 \right)}&=\sqrt{\sum_{j=1}^r\frac{N^{\alpha_j}}{N}\left(\frac{1}{N^{\alpha_j}}\sum_{i=1}^N\lambda_{i,j}^2 \right)}\notag\\
    &\leq \sqrt{\sum_{j=1}^r\sup_{\alpha_r\leq j\leq \alpha_1}\left(\frac{N^{\alpha_j}}{N}\right)\left(\frac{1}{N^{\alpha_j}}\sum_{i=1}^N\lambda_{i,j}^2 \right)}\notag\\ 
    &=\sqrt{\frac{N^{\alpha_1}}{N}}\sqrt{\sum_{j=1}^r\left(\frac{1}{N^{\alpha_j}}\sum_{i=1}^N\lambda_{i,j}^2 \right)}\notag \\
    &=O_p\left( \sqrt{\frac{N^{\alpha_1}}{N}}\right)
\end{align}
even under heterogeneous $\alpha$, because $\alpha_1\leq 1$. While its limit is positive definite only under strong loadings, it is sufficient for boundedness.\footnote{Alternatively, by avoiding the high-level condition in \hyperref[A3]{A.3} vii),
\begin{align*}
    \frac{N}{N^{\alpha_r}}\sup_{k_0\leq t\leq T-1} \left\|  N^{-1}\*F_t\+\Lambda't^{-1}\*E_t'\right\|&=\frac{N}{N^{\alpha_r}}\sup_{k_0\leq t\leq T-1} \left\|  N^{-1}\*F_t\*B_N\*B_N^{-1}\+\Lambda't^{-1}\*E_t'\right\|\notag\\
    &\leq \frac{\sqrt{N}}{N^{\alpha_r}}\left\|N^{-1/2}\*B_N \right\|\sup_{k_0\leq t\leq T-1}\left\|t^{-1/2}\*F_t \right\|\sup_{k_0\leq t\leq T-1}\left\|t^{-1/2}\*E_t\+\Lambda\*B_N^{-1} \right\|=O_p\left(\frac{\sqrt{N}}{N^{\alpha_r}} \right)=o_p(1)
\end{align*}
if we additionally assume that $\sup_{k_0\leq t\leq T-1}\left\|t^{-1/2}\*E_t\+\Lambda\*B_N^{-1} \right\|=O_p(1)$ (see (9) in \citealp{bai2023approximate}), which is a slightly worse rate.} The upcoming term is just a transpose of the latter, and the analysis is identical. Next, 
\begin{align}\label{spectral_rate_2}
    \frac{N}{N^{\alpha_r}}\sup_{k_0\leq t\leq T-1}\left\|N^{-1}t^{-1}\mathbb{E}\left(\*E_t\*E_t'\right) \right\|_{sp}&\leq \frac{N}{N^{\alpha_r}}\frac{T}{k_0}\sup_{k_0\leq t\leq T-1}\left\|N^{-1}T^{-1}\mathbb{E}\left(\*E_t\*E_t'\right) \right\|\notag\\
    &=\frac{N}{N^{\alpha_r}}\frac{T}{k_0}\sup_{k_0\leq t\leq T-1}\sqrt{\mathrm{tr}\left(N^{-2}T^{-2} \mathbb{E}\left(\*E_t\*E_t'\right)\mathbb{E}\left(\*E_t\*E_t'\right)'\right)}\notag\\
    &=\frac{N}{N^{\alpha_r}}\frac{T}{k_0}\sup_{k_0\leq t\leq T-1}\sqrt{\frac{1}{N^{2}T^2}\sum_{s=1}^t\left(\sum_{i=1}^N\mathbb{E}(e_{i,s}^2) \right)^2}\notag\\
    &\leq \frac{N}{N^{\alpha_r}}\frac{T}{k_0} \frac{1}{\sqrt{T}}\sqrt{\frac{1}{T}\sum_{s=1}^T\left(\frac{1}{N}\sum_{i=1}^N\mathbb{E}(e_{i,s}^2) \right)^2}=O\left(\frac{N}{N^{\alpha_r}}\frac{1}{\sqrt{T}} \right)
\end{align}
and 
\begin{align}
    \frac{N}{N^{\alpha_r}}&\sup_{k_0\leq t\leq T-1}\left\|N^{-1}t^{-1}\left(\*E_t\*E_t'-\mathbb{E}\left(\*E_t\*E_t'\right) \right) \right\|_{sp}\leq\frac{T}{k_0} \sup_{k_0\leq t\leq T-1}\left\| N^{-1}T^{-1}\left(\*E_t\*E_t'-\mathbb{E}\left(\*E_t\*E_t'\right) \right)\right\|\notag\\
    &=  \frac{N}{N^{\alpha_r}}\frac{T}{k_0}\sup_{k_0\leq t\leq T-1}\sqrt{\mathrm{tr}\left[N^{-2}T^{-2} \left(\*E_t\*E_t'-\mathbb{E}\left(\*E_t\*E_t'\right) \right)\left(\*E_t\*E_t'-\mathbb{E}\left(\*E_t\*E_t'\right) \right)'\right]}\notag\\
    &=\frac{N}{N^{\alpha_r}}\frac{T}{k_0}\sup_{k_0\leq t\leq T-1}\sqrt{\frac{1}{N^{2}T^2}\sum_{s=1}^t\left(\sum_{i=1}^N(e_{i,s}^2-\mathbb{E}(e_{i,s}^2))\right)^2}\notag\\
    &\leq \frac{N}{N^{\alpha_r}}\frac{T}{k_0} \frac{1}{\sqrt{TN}}\sqrt{\frac{1}{T}\sum_{s=1}^T\left(\frac{1}{\sqrt{N}}\sum_{i=1}^N(e_{i,s}^2-\mathbb{E}(e_{i,s}^2)) \right)^2}\notag\\
    &=O_p\left(\frac{N}{N^{\alpha_r}}\frac{1}{\sqrt{TN}}\right).
\end{align}
This completes the proof. \\
\begin{remark}
    \textit{Note that the statement of Lemma \ref{Lemma1} is formulated in terms of Frobenius norm, but the remainder rate in (\ref{spectral_rate_1}) is based on the spectral norm. However, in this case they are equivalent, because we can use $\left\| \*A\*B\right\|\leq \left\|\*A \right\|_{sp}\left\|\*B \right\|$ (see Lemma A.1 in \citealp{bai2023approximate}). Particularly,
\begin{align*}
     \frac{N}{N^{\alpha_r}}\sup_{k_0\leq t\leq T-1} \left\|  N^{-1}\*F_t\+\Lambda't^{-1}\*E_t'\right\|&\leq \frac{N}{N^{\alpha_r}}\sup_{k_0\leq t\leq T-1}t^{-1/2}\left\|\*F_t\+\Lambda' \right \|\sup_{k_0\leq t \leq T-1}t^{-1/2}\left\|\*E_t \right\|_{sp}\notag\\
     &\leq \frac{N}{N^{\alpha_r}}\sup_{k_0\leq t\leq T-1} t^{-1/2}\left\|\*F_t \right\|\left\|N^{-1/2}\+\Lambda \right\|N^{-1/2}\sqrt{\frac{T}{k_0}}T^{-1/2}\sup_{k_0\leq t\leq T-1} \left\|\*E_t \right\|_{sp}\notag\\
   &=  O_p\left(\frac{N^{1/2+\alpha_1/2}}{N^{\alpha_r}}\frac{1}{\sqrt{T}} \right) + O_p\left(N^{\alpha_1/2-\alpha_r} \right).
\end{align*}
Therefore, we obtain the same result.}
\end{remark} 

\noindent Finally, to show $\mathrm{(ii.)}$ it is sufficient to observe how by submultiplicativity of the Frobenius norm \begin{align*}
    \sup_{k_0\leq t\leq T-1}\|\overline{\*H}_{Nt,r}\|^2&\leq \sup_{k_0\leq t\leq T-1} \left\|\left(N\*B_N^{-2}\*D_{Nt,r}^2\right)^{-1}\right\|^2\sup_{k_0\leq t\leq T-1} \left\|\*B_N^{-1}t^{-1}\widehat{\*F}_t'\*F_t\+\Lambda'\+\Lambda \*B_N^{-1} \right\|^2\notag\\
    &= \sup_{k_0\leq t\leq T-1} \left\|\left(N\*B_N^{-2}\*D_{Nt,r}^2\right)^{-1}\right\|^2 \sup_{k_0\leq t\leq T-1} \left\|\*B_N^{-1}t^{-1}\widehat{\*F}_t'\*F_t\*B_N\left(\*B_N^{-1}\+\Lambda'\+\Lambda \*B_N^{-1}\right) \right\|^2\notag\\
    &\leq \sup_{k_0\leq t\leq T-1} \left\|\left(N\*B_N^{-2}\*D_{Nt,r}^2\right)^{-1}\right\|^2\sup_{k_0\leq t\leq T-1} \left\|\*B_N^{-1}t^{-1}\widehat{\*F}_t'\*F_t\*B_N\right\|^2 \left\|\*B_N^{-1}\+\Lambda'\+\Lambda \*B_N^{-1} \right\|^2=O_p(1),
\end{align*}
because the component $\*B_N^{-1}t^{-1}\widehat{\*F}_t'\*F_t\*B_N$ is bounded in probability for $t=T$ as argued in \cite{bai2023approximate}, and it can be shown to be bounded in probability in supremum sense. In particular, note that by using eigenvector-eigenvalue relationship and the fact that $\widehat{\*F}_t'\widehat{\*F}_t=t\*I_r$, we get 
\begin{align}
   N\*B_N^{-2}\*D^2_{Nt,r}= \frac{1}{t^2}\*B_N^{-1}\widehat{\*F}_t'\*X_t\*X_t'\widehat{\*F}\*B_N^{-1}&=\frac{1}{t^2}\*B_N^{-1}\widehat{\*F}_t'\*F_t\*B_N\left(\*B_N^{-1}\+\Lambda'\+\Lambda\*B_N^{-1}\right)\*B_N\*F_t'\widehat{\*F}_t\*B_N^{-1}\notag\\
   &+ \frac{1}{t^2}\*B_N^{-1}\widehat{\*F}_t'\*F_t\+\Lambda'\*E_t'\widehat{\*F}_t\*B_N^{-1}+ \frac{1}{t^2}\*B_N^{-1}\widehat{\*F}_t'\*E_t\+\Lambda\*F_t'\widehat{\*F}_t\*B_N^{-1}\notag\\
   &+\frac{1}{t^{2}}\*B_N^{-1}\widehat{\*F}_t'\*E_t\*E_t'\widehat{\*F}_t\*B_N^{-1},
\end{align}
where 
\begin{align}
    \sup_{k_0\leq t\leq T-1}&\left\|\frac{1}{t^2}\*B_N^{-1}\widehat{\*F}_t'\*F_t\+\Lambda'\*E_t'\widehat{\*F}_t\*B_N^{-1} \right\|_{sp}\notag\\
    &\leq \frac{\sqrt{N}}{N^{\alpha_r}}\left\|N^{\alpha_r/2}\*B_N^{-1} \right\|^2\sup_{k_0\leq t\leq T-1} t^{-1}\left\| \widehat{\*F}_t\right\|_{sp}^2\sup_{k_0\leq t\leq T-1} t^{-1/2}\left\|\*F_t \right\|_{sp}\left\| N^{-1/2}\+\Lambda\right\|\sup_{k_0\leq t\leq T-1} t^{-1/2}\left\| \*E_t\right\|_{sp}\notag\\
    &\leq \frac{\sqrt{N}}{N^{\alpha_r}}\left\|N^{\alpha_r/2}\*B_N^{-1} \right\|^2\sup_{k_0\leq t\leq T-1} t^{-1}\left\| \widehat{\*F}_t\right\|^2\sup_{k_0\leq t\leq T-1} t^{-1/2}\left\|\*F_t \right\| \left\| N^{-1/2}\+\Lambda\right\|\notag\\
    &\times \sqrt{\frac{T}{k_0}}T^{-1/2}\sup_{k_0\leq t\leq T-1}\left\| \*E_t\right\|_{sp}\notag\\
    &=\frac{\sqrt{N}}{N^{\alpha_r}}\left\|N^{\alpha_r/2}\*B_N^{-1} \right\|^2 r \sup_{k_0\leq t\leq T-1} t^{-1/2}\left\|\*F_t \right\| \left\| N^{-1/2}\+\Lambda\right\|\times \sqrt{\frac{T}{k_0}}T^{-1/2}\sup_{k_0\leq t\leq T-1}\left\| \*E_t\right\|_{sp}\notag\\
    &= O_p\left(\sqrt{\frac{N^{\alpha_1}}{N} }\right)\times \left(O_p\left(\frac{N}{N^{\alpha_r}}\frac{1}{\sqrt{T}} \right) + O_p\left(N^{1/2-\alpha_r} \right)\right)\notag\\
    &= O_p\left(\frac{N^{1/2+\alpha_1/2}}{N^{\alpha_r}}\frac{1}{\sqrt{T}} \right) + O_p\left(N^{\alpha_1/2-\alpha_r} \right),
\end{align}
where the next term is just a transpose, so its order is the same. \footnote{Alternatively, by avoiding the high-level condition in \hyperref[A3]{A.3} vii),
\begin{align*}
     \sup_{k_0\leq t\leq T-1}\left\|\frac{1}{t^2}\*B_N^{-1}\widehat{\*F}_t'\*F_t\+\Lambda'\*E_t'\widehat{\*F}_t\*B_N^{-1} \right\|_{sp}&\leq \frac{\sqrt{N}}{N^{\alpha_r}}r\sup_{k_0\leq t\leq T-1}\left\|t^{-1/2} \*E_t\+\Lambda\*B_N^{-1}\right\|_{sp}\left\|N^{-1/2}\*B_N \right\|\left\| N^{\alpha_r/2}\*B_N^{-1}\right\|^2  \sup_{k_0\leq t\leq T-1}\left\| t^{-1/2}\*F_t\right\|_{sp}\notag\\
     &=O_p(N^{(1-2\alpha_r)/2}).
\end{align*}
} Next, 
\begin{align}
    \left\|\frac{1}{t^{2}}\*B_N^{-1}\widehat{\*F}_t'\*E_t\*E_t'\widehat{\*F}_t\*B_N^{-1} \right\|_{sp}&\leq N^{-\alpha_r} \left\|N^{\alpha_r/2}\*B_N^{-1} \right\|^2\sup_{k_0\leq t\leq T-1} t^{-1}\left\| \widehat{\*F}_t\right\|_{sp}^2 \sup_{k_0\leq t\leq T-1} t^{-1} \left\|\*E_t \right\|^2_{sp}\notag\\
    &\leq N^{-\alpha_r} \left\|N^{\alpha_r/2}\*B_N^{-1} \right\|^2\sup_{k_0\leq t\leq T-1} t^{-1}\left\| \widehat{\*F}_t\right\|^2 \frac{T}{k_0}T^{-1} \sup_{k_0\leq t\leq T-1} \left\|\*E_t \right\|^2_{sp}\notag\\
    & = N^{-\alpha_r} \left\|N^{\alpha_r/2}\*B_N^{-1} \right\|^2 r \frac{T}{k_0}T^{-1} \sup_{k_0\leq t\leq T-1} \left\|\*E_t \right\|^2_{sp}\notag\\
    &=O_p\left(\frac{N}{N^{\alpha_r}}\frac{1}{T} \right) + O_p(N^{-\alpha_r}). 
\end{align}
Therefore, by combining this analysis with (\ref{NBD^2_approx}), we have that 
\begin{align}\label{BFFB_implication}
N\*B_N^{-2}\*D^2_{Nt,r}=\*B_N^{-1}t^{-1}\widehat{\*F}_t'\*F_t\*B_N\left(\*B_N^{-1}\+\Lambda'\+\Lambda\*B_N^{-1}\right)\*B_Nt^{-1}\*F_t'\widehat{\*F}_t\*B_N^{-1}=\*B_N^{-1}\+\Lambda'\+\Lambda\*B_N^{-1}+o_p(1),
\end{align}
where the remainder is negligible uniformly in $t$. Because $\*B_N^{-1}\+\Lambda'\+\Lambda\*B_N^{-1}$ is bounded for all $t$, it must be that $\sup_{k_0\leq t\leq T-1}\left\| \*B_N^{-1}t^{-1}\widehat{\*F}_t'\*F_t\*B_N\right\|=O_p(1)$, as well. Note that (\ref{BFFB_implication}) implies that since $\*B_N^{-1}\+\Lambda'\+\Lambda\*B_N^{-1}$ is invertible asymptotically, $\*B_N^{-1}t^{-1}\widehat{\*F}_t'\*F_t\*B_N$ is invertble and has an invertible probability limit, as well (see also a discussion on p. 1903 in \citealp{bai2023approximate}). Consequently, $\overline{\*H}_{Nt,r}^{-1}$ exists.
\subsubsection{Approximation of the Factor Space}
\noindent For the further result, we observe that the statistics in \cite{pitarakis2025novel} and \cite{pitarakis2023direct} are comprised of the components of the form $\sum_{t=\lfloor f_1(T)\rfloor }^{\lfloor f_2(T)\rfloor}d_u(\widetilde{u}_{1,t+1}, \widetilde{u}_{2,t+1})$, where $\frac{\lfloor f_j(T)\rfloor}{T}\to q_j$ for $j=1,2$, such that $q_2>q_1$ and some loss-differential $d_u(.)$. We implicitly allow $\lfloor f_j(T)\rfloor$ to subsume additive constants as they will vanish in the limit when divided by $T$. Moreover, $\lfloor f_1(T)\rfloor\geq k_0$ and $\lfloor f_2(T)\rfloor\leq T-1$. For example, in case of $\varepsilon_f(m_0)$ in (\ref{arxiv_test}), the first component has $\lfloor f_1(T)\rfloor=\lfloor T\pi_0 \rfloor=k_0$ and $\lfloor f_2(T)\rfloor=T-1$ with $q_1=\pi_0$ and $q_2=1$. The second component also entails $\lfloor f_1(T)\rfloor=\lfloor T\pi_0 \rfloor=k_0$, $q_1=\pi_0$, but $\lfloor f_2(T)\rfloor=\lfloor f_1(T)\rfloor+\lfloor (T-k_0)\mu_0 \rfloor-1=k_0+m_0-1$ and $q_2=\pi_0+(1-\pi_0)\mu_0$. Similar summation limits apply to every component of every considered statistic. Therefore, it is convenient to formulate the following lemma with the bounds on $\frac{1}{\sqrt{d_T}}\sum_{t=\lfloor f_1(T)\rfloor }^{\lfloor f_2(T)\rfloor}d_u(.)$, where $d_T=(\lfloor f_2(T)\rfloor-\lfloor f_1(T)\rfloor+1)$ and $d_u(.)$ will obey expansions of the infeasible statistics.
\noindent \begin{lemma} \label{Lemma2} Under A.1 - A.4 as $(N,T)\to \infty$, 
\begin{enumerate}[(a)]
   \item  if  $\alpha\in (0,1])$ (homogeneous):
\begin{align*}
    &\mathrm{(i.)}\hspace{2mm}\frac{1}{d_T}\sum_{t=\lfloor f_1(T)\rfloor}^{\lfloor f_2(T)\rfloor}\left\|\widehat{\*f}_t - \*H_{Nt,r}\*f_t \right\|^2= O_p\left(\frac{N^2}{N^{2\alpha}}\frac{1}{T} \right) + O_p(N^{1-2\alpha}),\\
    & \mathrm{(ii.)}\hspace{2mm} \sup_{k_0\leq t\leq T-1} \frac{1}{t}\left\|\widehat{\*F}_t-\*F_t\*H'_{Nt,r} \right\|^2=\sup_{k_0\leq t\leq T-1} \frac{1}{t}\sum_{s=1}^t\left\| \widehat{\*f}_s - \*H_{Nt,r}\*f_s\right\|^2= O_p\left(\frac{N^2}{N^{2\alpha}}\frac{1}{k_0} \right) + O_p(N^{1-2\alpha});
\end{align*}
\item Furthermore, under Assumptions A.1 - A.4 and heterogeneous loadings ($1\geq\alpha_1\geq\cdots\geq\alpha_r>1/2$), one has
\begin{align*}
    &\mathrm{(i.)}\hspace{2mm}\frac{1}{d_T}\sum_{t=\lfloor f_1(T)\rfloor}^{\lfloor f_2(T)\rfloor}\left\|\widehat{\*f}_t - \*H_{Nt,r}\*f_t \right\|^2= O_p\left(\frac{N^2}{N^{2\alpha_r}}\frac{1}{T} \right) + O_p(N^{1-2\alpha_r}),\\
    & \mathrm{(ii.)}\hspace{2mm} \sup_{k_0\leq t\leq T-1} \frac{1}{t}\left\|\widehat{\*F}_t-\*F_t\*H'_{Nt,r} \right\|^2=\sup_{k_0\leq t\leq T-1} \frac{1}{t}\sum_{s=1}^t\left\| \widehat{\*f}_s - \*H_{Nt,r}\*f_s\right\|^2= O_p\left(\frac{N^2}{N^{2\alpha_r}}\frac{1}{k_0} \right) + O_p(N^{1-2\alpha_r}).
\end{align*}
\end{enumerate}
\end{lemma}

\noindent \textbf{Proof.} (a) To demonstrate $\mathrm{(i.)}$, we begin by following \cite{bai2002determining} and by using the expansion in (\ref{row_fact_space}) (with the time index shifted to the out-of-sample set) in connection to $(a+b+c+d)^2\leq 4(a^2+b^2+c^2+d^2)$: 
\begin{align}
     &\frac{1}{d_T}\sum_{t=\lfloor f_1(T)\rfloor}^{\lfloor f_2(T)\rfloor}\left\|\widehat{\*f}_t - \*H_{Nt,r}\*f_t \right\|^2\leq 4\sup_{k_0\leq t\leq T-1}\left\|\left(\frac{N}{N^\alpha}\*D_{Nt,r}^{2}\right)^{-1} \right\|^2\times \left( \*I + \mathbf{II} + \mathbf{III} + \mathbf{IV}  \right),
\end{align}
 where we will analyse $\*I - \mathbf{IV}$ separately. Indeed, by Cauchy-Schwarz (CS) inequality
\begin{align}
    \*I=\frac{1}{d_T}\sum_{t=\lfloor f_1(T)\rfloor}^{\lfloor f_2(T)\rfloor} \left\| \frac{1}{t}\sum_{l=1}^t\widehat{\*f}_l\gamma_{l,t,\alpha}\right\|^2&\leq \frac{1}{d_T}\sum_{t=\lfloor f_1(T)\rfloor}^{\lfloor f_2(T)\rfloor} \left(\frac{1}{t}\sum_{l=1}^t\left\|\widehat{\*f}_l \right\| |\gamma_{l,t,\alpha}| \right)^2\notag\\
    &\leq \frac{1}{d_T}\sum_{t=\lfloor f_1(T)\rfloor}^{\lfloor f_2(T)\rfloor} \left( \frac{1}{t}\sum_{l=1}^t\|\widehat{\*f}_l \|^2\right) \left(\frac{1}{t}\sum_{l=1}^t \gamma_{l,t,\alpha}^2\right)\notag\\
    &= r \frac{1}{d_T}\sum_{t=\lfloor f_1(T)\rfloor}^{\lfloor f_2(T)\rfloor} \frac{1}{t}\sum_{l=1}^t \gamma_{l,t,\alpha}^2\notag\\
    &\leq r \frac{1}{d_Tk_0}\sum_{t=1}^{T-1}\sum_{l=1}^{T-1}\gamma_{l,t,\alpha}^2\notag\\
    &= r \left(\frac{N}{N^\alpha} \right)^2\left(\frac{T^2}{d_Tk_0}\right)\frac{1}{T^2}\sum_{t=1}^{T-1}\sum_{l=1}^{T-1}\gamma_{l,t}^2\notag\\
    &= O\left(\frac{N^2}{N^{2\alpha}}\frac{1}{T} \right),
\end{align}
where we used the fact that 
\begin{align}
    \frac{1}{t}\sum_{l=1}^t\|\widehat{\*f}_l \|^2=\frac{1}{t}\sum_{l=1}^t\sum_{p=1}^r\widehat{f}_{p,l}^2= \sum_{p=1}^r\underbrace{\left(\frac{1}{t}\sum_{l=1}^t\widehat{f}_{p,l}^2 \right)}_{\text{1 by the restriction}}= r 
\end{align}
and 
\begin{align*}
    \left(\frac{T^2}{d_Tk_0}\right)=\left(\frac{d_Tk_0}{T^2} \right)^{-1} = \left((q_2-q_1)\pi_0\right)^{-1}+O(T^{-1}).
\end{align*}
We will use these two facts for the remainder. Note that this rate is worse than the one in \cite{bai2023approximate}, where they have an extra $T^{-1}$. This penalty comes from the fact that the rotation matrix changes for each iteration of the recursion in the current case. To continue, by using an identical chain of inequalities, we obtain 
\begin{align}
    \mathbf{II}=\frac{1}{d_T}\sum_{t=\lfloor f_1(T)\rfloor}^{\lfloor f_2(T)\rfloor} \left\| \frac{1}{t}\sum_{l=1}^t\widehat{\*f}_l\xi_{l,t,\alpha} \right\|^2&\leq  \frac{1}{d_T}\sum_{t=\lfloor f_1(T)\rfloor}^{\lfloor f_2(T)\rfloor} \left( \frac{1}{t}\sum_{l=1}^t\|\widehat{\*f}_l \|^2\right) \left(\frac{1}{t}\sum_{l=1}^t \xi_{l,t,\alpha}^2\right)\notag\\
    &\leq  r \frac{1}{d_Tk_0}\sum_{t=1}^{T-1}\sum_{l=1}^{T-1}\xi_{l,t,\alpha}^2\notag\\
    &=  r \left(\frac{N}{N^\alpha} \right)^2\left(\frac{T^2}{d_Tk_0}\right)\frac{1}{T^2}\sum_{t=1}^{T-1}\sum_{l=1}^{T-1}\xi_{l,t}^2\notag\\
    &= O_p\left(\frac{N}{N^{2\alpha}} \right) = O_p(N^{1-2\alpha}),
\end{align}
because $\frac{1}{T^2}\sum_{t=1}^{T-1}\sum_{l=1}^{T-1}\xi_{l,t}^2=O_p(N^{-1})$ by our assumptions, which implies that $\alpha >1/2$ in order to make this term negligible. Note that to obtain the mean-square out-of-sample factor estimation error we already cannot allow the loadings be too weak. In comparison, \cite{bai2023approximate} do not need this since they do not have rate punishment coming from the recursion. Further
\begin{align}
    \mathbf{III}= \frac{1}{d_T}\sum_{t=\lfloor f_1(T)\rfloor}^{\lfloor f_2(T)\rfloor} \left\| \frac{1}{t}\sum_{l=1}^t\widehat{\*f}_l\eta_{l,t,\alpha} \right\|^2&\leq  \frac{1}{d_T}\sum_{t=\lfloor f_1(T)\rfloor}^{\lfloor f_2(T)\rfloor} \left( \frac{1}{t}\sum_{l=1}^t\|\widehat{\*f}_l \|^2\right) \left(\frac{1}{t}\sum_{l=1}^t \eta_{l,t,\alpha}^2\right)\notag\\
    &\leq  r \frac{1}{d_Tk_0}\sum_{t=1}^{T-1}\sum_{l=1}^{T-1}\eta_{l,t,\alpha}^2\notag\\
    &=  r  \left(\frac{T^2}{d_Tk_0}\right)\frac{1}{T^2}\sum_{t=1}^{T-1}\sum_{l=1}^{T-1}\eta_{l,t,\alpha}^2\notag\\
    &= O_p(N^{-\alpha}),
\end{align}
and similarly 
\begin{align}
    \mathbf{IV}= \frac{1}{d_T}\sum_{t=\lfloor f_1(T)\rfloor}^{\lfloor f_2(T)\rfloor} \left\| \frac{1}{t}\sum_{l=1}^t\widehat{\*f}_l\nu_{l,t,\alpha} \right\|^2&\leq  \frac{1}{d_T}\sum_{t=\lfloor f_1(T)\rfloor}^{\lfloor f_2(T)\rfloor} \left( \frac{1}{t}\sum_{l=1}^t\|\widehat{\*f}_l \|^2\right) \left(\frac{1}{t}\sum_{l=1}^t \nu_{l,t,\alpha}^2\right)\notag\\
    &\leq  r \frac{1}{d_Tk_0}\sum_{t=1}^{T-1}\sum_{l=1}^{T-1}\nu_{l,t,\alpha}^2\notag\\
    &=  r  \left(\frac{T^2}{d_Tk_0}\right)\frac{1}{T^2}\sum_{t=1}^{T-1}\sum_{l=1}^{T-1}\nu_{l,t,\alpha}^2\notag\\
    &= O_p(N^{-\alpha})
\end{align}
by our assumptions. \\

\noindent For $\mathrm{(ii.)}$, we will apply a similar chain of arguments to create inequalities on the expansion in (\ref{row_fact_space}): 
\begin{align}
    \sup_{k_0\leq t\leq T-1} \frac{1}{t}\left\|\widehat{\*F}_t-\*F_t\*H'_{Nt,r} \right\|^2&=\sup_{k_0\leq t\leq T-1} \frac{1}{t}\sum_{s=1}^t\left\| \widehat{\*f}_s - \*H_{Nt,r}\*f_s\right\|^2\notag\\
    &\leq 4\sup_{k_0\leq t\leq T-1}\left\|\left(\frac{N}{N^\alpha}\*D_{Nt,r}^{2}\right)^{-1} \right\|^2\times \left( \*i + \mathbf{ii} + \mathbf{iii} + \mathbf{iv}  \right),
\end{align}
where we will investigate $\*i - \mathbf{iv}$ in turn. To begin with, 
\begin{align}
    \*i=\sup_{k_0\leq t\leq T-1} \frac{1}{t}\sum_{s=1}^t \left\| \frac{1}{t}\sum_{l=1}^t\widehat{\*f}_l\gamma_{l,s,\alpha}\right\|^2&\leq  \sup_{k_0\leq t\leq T-1} \frac{1}{t}\sum_{s=1}^t \left(\frac{1}{t}\sum_{l=1}^t\left\|\widehat{\*f}_l \right\| |\gamma_{l,s,\alpha}| \right)^2\notag\\
    &\leq  \sup_{k_0\leq t\leq T-1} \frac{1}{t}\sum_{s=1}^t \left( \frac{1}{t}\sum_{l=1}^t\|\widehat{\*f}_l \|^2\right) \left(\frac{1}{t}\sum_{l=1}^t \gamma_{l,s,\alpha}^2\right)\notag\\
    &= r  \sup_{k_0\leq t\leq T-1} \frac{1}{t}\sum_{s=1}^t \frac{1}{t}\sum_{l=1}^t \gamma_{l,s,\alpha}^2\notag\\
    &\leq r \frac{1}{k_0^2}\sum_{s=1}^{T-1}\sum_{l=1}^{T-1}\gamma_{l,s,\alpha}^2\notag\\
    &= r \frac{1}{k_0}\left(\frac{N}{N^\alpha} \right)^2\frac{T}{k_0}\frac{1}{T}\sum_{t=1}^{T-1}\sum_{l=1}^{T-1}\gamma_{l,s}^2\notag\\
    &= O\left(\frac{N^2}{N^{2\alpha}}\frac{1}{k_0} \right),
\end{align}
where we used the fact that $\frac{T}{k_0}=\left(\frac{k_0}{T} \right)^{-1}=\pi_0^{-1}+O(T^{-1})$. The rest of the terms follow a similar fashion. In particular, 
\begin{align}
    \mathbf{ii}=\sup_{k_0\leq t\leq T-1} \frac{1}{t}\sum_{s=1}^t \left\| \frac{1}{t}\sum_{l=1}^t\widehat{\*f}_l\xi_{l,s,\alpha} \right\|^2&\leq  \sup_{k_0\leq t\leq T-1} \frac{1}{t}\sum_{s=1}^t \left( \frac{1}{t}\sum_{l=1}^t\|\widehat{\*f}_l \|^2\right) \left(\frac{1}{t}\sum_{l=1}^t \xi_{l,s,\alpha}^2\right)\notag\\
    &\leq  r \frac{1}{k_0^2}\sum_{s=1}^{T-1}\sum_{l=1}^{T-1}\xi_{l,s,\alpha}^2\notag\\
    &=  r \left(\frac{N}{N^\alpha} \right)^2\left(\frac{T^2}{k_0^2}\right)\frac{1}{T^2}\sum_{s=1}^{T-1}\sum_{l=1}^{T-1}\xi_{l,s}^2\notag\\
    &= O_p\left(\frac{N}{N^{2\alpha}} \right) = O_p(N^{1-2\alpha}),
\end{align}
\begin{align}
     \mathbf{iii}=\sup_{k_0\leq t\leq T-1} \frac{1}{t}\sum_{s=1}^t \left\| \frac{1}{t}\sum_{l=1}^t\widehat{\*f}_l\eta_{l,s,\alpha} \right\|^2&\leq  \sup_{k_0\leq t\leq T-1} \frac{1}{t}\sum_{s=1}^t \left( \frac{1}{t}\sum_{l=1}^t\|\widehat{\*f}_l \|^2\right) \left(\frac{1}{t}\sum_{l=1}^t \eta_{l,s,\alpha}^2\right)\notag\\
    &\leq  r \frac{1}{k_0^2}\sum_{s=1}^{T-1}\sum_{l=1}^{T-1}\eta_{l,s,\alpha}^2\notag\\
    &=  r  \left(\frac{T^2}{k_0^2}\right)\frac{1}{T^2}\sum_{s=1}^{T-1}\sum_{l=1}^{T-1}\eta_{l,s,\alpha}^2\notag\\
    &= O_p(N^{-\alpha})
\end{align}
and 
\begin{align}
     \mathbf{iv}= \sup_{k_0\leq t\leq T-1} \frac{1}{t}\sum_{s=1}^t \left\| \frac{1}{t}\sum_{l=1}^t\widehat{\*f}_l\nu_{l,s,\alpha} \right\|^2&\leq \sup_{k_0\leq t\leq T-1} \frac{1}{t}\sum_{s=1}^t \left( \frac{1}{t}\sum_{l=1}^t\|\widehat{\*f}_l \|^2\right) \left(\frac{1}{t}\sum_{l=1}^t \nu_{l,s,\alpha}^2\right)\notag\\
    &\leq  r \frac{1}{k_0^2}\sum_{s=1}^{T-1}\sum_{l=1}^{T-1}\nu_{l,s,\alpha}^2\notag\\
    &=  r  \left(\frac{T^2}{k_0^2}\right)\frac{1}{T^2}\sum_{s=1}^{T-1}\sum_{l=1}^{T-1}\nu_{l,s,\alpha}^2\notag\\
    &= O_p(N^{-\alpha}),
\end{align}
which completes $(\mathrm{ii})$.
\\

\noindent (b) Both $(\mathrm{i.}) - (\mathrm{ii.})$ are very similar to part (a), as we only need the heterogeneity-adjusted expansion in (\ref{het_expand}) for both cases. Clearly,
\begin{align}
      &\frac{1}{d_T}\sum_{t=\lfloor f_1(T)\rfloor}^{\lfloor f_2(T)\rfloor}\left\|\widehat{\*f}_t - \*H_{Nt,r}\*f_t \right\|^2\leq 4\sup_{k_0\leq t\leq T-1}\left\|\left(N\*B_N^{-2}\*D_{Nt,r}^{2}\right)^{-1} \right\|^2 \left\| \*Q^2_{N,\alpha}\right\|^2\times \left( \*I + \mathbf{II} + \mathbf{III} + \mathbf{IV}  \right),
\end{align}
where each $\mathbf{I}$ - $\mathbf{IV}$ follows the same chain of arguments as the respective terms in part ($\mathrm{i.}$) of part (a). Instead of $N/N^\alpha$ we will have $N/N^{\alpha_r}$ penalizing the rates, which coincides with the largest eigenvalue of $N\*B_N^{-2}$, meaning that the weakest loading is the dominant one in determining the rates. The rate is slower for $\mathbf{III}$ and $\mathbf{IV},$, however. For instance, 
\begin{align}
    \mathbf{III}=\frac{1}{d_T}\sum_{t=\lfloor f_1(T)\rfloor}^{\lfloor f_2(T)\rfloor} \left\| \frac{1}{t}\sum_{l=1}^t\widehat{\*f}_l\eta_{l,t,\alpha_r}^D \right\|^2&\leq  \frac{1}{d_T}\sum_{t=\lfloor f_1(T)\rfloor}^{\lfloor f_2(T)\rfloor} \left( \frac{1}{t}\sum_{l=1}^t\|\widehat{\*f}_l \|^2\right) \left(\frac{1}{t}\sum_{l=1}^t (\eta_{l,t,\alpha_r}^D)^2\right)\notag\\
    &\leq  r \frac{N}{N^{2\alpha_r}}\frac{1}{d_Tk_0}\sum_{t=1}^{T-1}\sum_{l=1}^{T-1}(\eta_{l,t}^D)^2\notag\\
    &=  r \frac{N}{N^{2\alpha_r}} \left(\frac{T^2}{d_Tk_0}\right)\frac{1}{T^2}\sum_{t=1}^{T-1}\sum_{l=1}^{T-1}(\eta_{l,t}^D)^2\notag\\
    &=O_p\left(\frac{N}{N^{2\alpha_r}}\right)
\end{align}
and 
\begin{align}
    \mathbf{iii}=\sup_{k_0\leq t\leq T-1} \frac{1}{t}\sum_{s=1}^t \left\| \frac{1}{t}\sum_{l=1}^t\widehat{\*f}_l\eta_{l,s,\alpha_r}^D \right\|^2&\leq  \sup_{k_0\leq t\leq T-1} \frac{1}{t}\sum_{s=1}^t \left( \frac{1}{t}\sum_{l=1}^t\|\widehat{\*f}_l \|^2\right) \left(\frac{1}{t}\sum_{l=1}^t (\eta_{l,s,\alpha_r}^D)^2\right)\notag\\
    &\leq  r \frac{1}{k_0^2}\sum_{s=1}^{T-1}\sum_{l=1}^{T-1}(\eta_{l,s,\alpha_r}^D)^2\notag\\
    &=  r  \frac{N}{N^{2\alpha_r}}\left(\frac{T^2}{k_0^2}\right)\frac{1}{T^2}\sum_{s=1}^{T-1}\sum_{l=1}^{T-1}(\eta_{l,s}^D)^2\notag\\
    &=O_p\left(\frac{N}{N^{2\alpha_r}}\right)
\end{align}
and the same for $\mathbf{IV}$ ($\mathbf{iv}$). For the rest, we use A.3 iv) for the heterogeneous case. Part ($\mathrm{ii.}$) follows the same discussion. Note that the change from $\alpha$ to $\alpha_r$ is the same as in \cite{bai2023approximate}, while other changes occur due to the recursive setup. 
\subsubsection{Limits of Eigenvalues and Rotation Matrices}
\begin{lemma} \label{Lemma3} Under Assumptions A.1 - A.4 as $(N,T)\to \infty$
\begin{enumerate}[(a)]
\item If  $\alpha\in (0,1])$ (homogeneous):
    \begin{align*}
    &\mathrm{(i.)} \quad \sup_{k_0\leq t\leq T-1}\left\|\*H_{Nt,r}^{-1}-\*H_{r}^{-1} \right\|=O_p\left(\frac{N}{N^\alpha}\frac{1}{\sqrt{k_0}} \right)+O_p(N^{(1-2\alpha)/2}),\\
    &\mathrm{(ii.)} \quad \sup_{k_0\leq t\leq T-1}\left\|\left(\frac{N}{N^\alpha}\*D_{Nt,r}^{2}\right)^{-1}-\+\Sigma_{\+\Lambda}^{-1} \right\|= O_p\left(\frac{N}{N^\alpha}\frac{1}{\sqrt{k_0}} \right)+ O_p(N^{(1-2\alpha)/2}).
\end{align*}
\item Furthermore, under Assumptions A.1 - A.4 and heterogeneous loadings ($1\geq\alpha_1\geq\cdots\geq\alpha_r>1/2$), one has
\begin{align*}
    &\mathrm{(i.)} \quad \sup_{k_0\leq t\leq T-1}\left\|\*H_{Nt,r}^{-1}-\*H_{r}^{-1} \right\|=O\left(\frac{\sqrt{N^{\alpha_1+\alpha_r}}}{N} \right)+O_p\left(N^{[(1+\alpha_1)-3\alpha_r]/2}\right),\\
    &\mathrm{(ii.)} \quad \sup_{k_0\leq t\leq T-1}\left\|\left(N\*B_N^{-2}\*D_{Nt,r}^{2}\right)^{-1}-\+\Sigma_{\+\Lambda}^{-1} \right\|= O_p\left(\frac{N}{N^{\alpha_r}}\frac{1}{\sqrt{k_0}} \right)+ O_p(N^{(1-2\alpha_r)/2}).
    \end{align*}
    \end{enumerate}
\end{lemma}
\bigskip

\noindent \textbf{Proof}. $\mathrm{(a)}$ To see $\mathrm{(i.)}$, we base the approach on Lemma 2 in \cite{bai2002determining}, and decomposition $t^{-1}\*H_{Nt,r}\*F'\*F_t\*H_{Nt,r}'=\*H_{Nt,r}\*H_{Nt,r}'+\*H_{Nt,r}\left(t^{-1}\*F'\*F-\*I_r \right)\*H_{Nt,r}'$. We then obtain
\begin{align}\label{HH'_approx}
   \*H_{Nt,r}\*H_{Nt,r}'&=-\*H_{Nt,r}\left(t^{-1}\*F_t'\*F_t-\*I_r \right)\*H_{Nt,r}' - t^{-1}\*H_{Nt,r}\*F'\*F_t\*H_{Nt,r}'\notag\\
   &= \*I_r-\*H_{Nt,r}\left(t^{-1}\*F_t'\*F_t-\*I_r \right)\*H_{Nt,r}'-t^{-1}\widehat{\*F}_t'(\widehat{\*F}_t-\*F_t\*H_{Nt,r}')\notag\\
   &-(\widehat{\*F}_t-\*F_t\*H_{Nt,r}')'t^{-1}\*F_t\*H_{Nt,r}'\notag\\
   &=\*I_r+O_p\left(\frac{N}{N^\alpha}\frac{1}{\sqrt{k_0}} \right)+O_p(N^{(1-2\alpha)/2})+O_p(T^{-1/2})\notag\\
   &= \*I_r+O_p\left(\frac{N}{N^\alpha}\frac{1}{\sqrt{k_0}} \right)+O_p(N^{(1-2\alpha)/2}),
\end{align}
uniformly in $t$, since 
\begin{align}
   \sup_{k_0\leq t\leq T-1} \left\| \*H_{Nt,r}\left(t^{-1}\*F_t'\*F_t-\*I_r \right)\*H_{Nt,r}'\right\|\leq \sup_{k_0\leq t\leq T-1}\left\|\*H_{Nt,r} \right\|^2\sup_{k_0\leq t\leq T-1}\left\|\frac{1}{t}\sum_{s=1}^t\*f_s\*f_s'-\*I_r \right\|=O_p(T^{-1/2}).
   \end{align}
   Next,
   \begin{align}
       \sup_{k_0\leq t\leq T-1}\left\|\*H_{Nt,r} \right\|^2&=\sup_{k_0\leq t\leq T-1}\left\|\|\*D_{Nt,r}^{-2}(t^{-1}\widehat{\*F}_t'\*F_t)(N^{-1}\+\Lambda'\+\Lambda )\right\|^2\notag\\
       &=\sup_{k_0\leq t\leq T-1}\left\|\left(\frac{N}{N^\alpha}\*D_{Nt,r}^{2}\right)^{-1}(t^{-1}\widehat{\*F}_t'\*F_t)(N^{-\alpha}\+\Lambda'\+\Lambda ) \right\|^2=O_p(1)
   \end{align}
  as was demonstrated in Lemma 1 $\mathrm{(a)}$ part $\mathrm{(ii.)}$. Moreover, 
   \begin{align}
       \sup_{k_0\leq t\leq T-1} \left\|t^{-1}\widehat{\*F}_t'(\widehat{\*F}_t-\*F_t\*H_{Nt,r}') \right\|&\leq \sup_{k_0\leq t\leq T-1} t^{-1/2}\left\|\widehat{\*F}_t \right\|\sup_{k_0\leq t\leq T-1} t^{-1/2}\left\| \widehat{\*F}_t-\*F_t\*H_{Nt,r}'\right\|\notag\\
  &=\sqrt{r}\sup_{k_0\leq t\leq T-1} t^{-1/2}\left\| \widehat{\*F}_t-\*F_t\*H_{Nt,r}'\right\|= O_p\left(\frac{N}{N^\alpha}\frac{1}{\sqrt{k_0}} \right)+O_p(N^{(1-2\alpha)/2}),
   \end{align}
and 
\begin{align}
    \sup_{k_0\leq t\leq T-1}\left\|(\widehat{\*F}_t-\*F_t\*H_{Nt,r}')'t^{-1}\*F_t\*H_{Nt,r}'\right\|&\leq \sup_{k_0\leq t\leq T-1} t^{-1/2}\left\| \widehat{\*F}_t-\*F_t\*H_{Nt,r}'\right\|\sup_{k_0\leq t\leq T-1} t^{-1/2}\left\| \*F_t \right\|\sup_{k_0\leq t\leq T-1} \left\| \*H_{Nt,r}\right\|\notag\\
  &=O_p\left(\frac{N}{N^\alpha}\frac{1}{\sqrt{k_0}} \right)+O_p(N^{(1-2\alpha)/2}).
\end{align}
This implies that $\*H_{Nt,r}=\*H_r+O_p\left(\frac{N}{N^\alpha}\frac{1}{\sqrt{k_0}} \right)+O_p(N^{(1-2\alpha)/2})$ uniformly in $t$, where $\*H_r$ is an orthogonal matrix with only $\pm 1$ on the main diagonal. Then
\begin{align}
    \sup_{k_0\leq t\leq T-1}\left\|\*H_{Nt,r}^{-1}-\*H_{r}^{-1} \right\|&\leq \sup_{k_0\leq t\leq T-1}\left\|\*H_{Nt,r}^{-1} \right\|\left\| \*H_{r}^{-1}\right\|\sup_{k_0\leq t\leq T-1}\left\| \*H_{Nt,r}-\*H_{r}\right\|\notag\\
    &=O_p\left(\frac{N}{N^\alpha}\frac{1}{\sqrt{k_0}} \right)+O_p(N^{(1-2\alpha)/2}).
\end{align}

\noindent For $\mathrm{(ii.)}$, we will now use the argument based on the equations (20) - (24) of \cite{bai2013principal} to firstly demonstrate that $\*H_r$ is also diagonal only. To begin with, we know that 
\begin{align}\label{FhatF}
    t^{-1}\widehat{\*F}_t'\*F_t&=t^{-1}(\widehat{\*F}_t-\*F_t\*H_{Nt,r}')'\*F_t+\*H_{Nt,r}t^{-1}\*F_t'\*F_t\notag\\
&=\*H_{Nt,r}t^{-1}\*F_t'\*F_t+O_p\left(\frac{N}{N^\alpha}\frac{1}{\sqrt{k_0}} \right)+O_p(N^{(1-2\alpha)/2})\notag\\
    &=\*H_{Nt,r} +O_p\left(\frac{N}{N^\alpha}\frac{1}{\sqrt{k_0}} \right)+O_p(N^{(1-2\alpha)/2}) 
\end{align}
uniformly in $t$, based on the same bounds used above. Now, by the definition of $\*H_{Nt,r}$ and by using (\ref{FhatF}) we get 
\begin{align}
   \*H_{Nt,r}= \left(\frac{N}{N^\alpha}\*D_{Nt,r}^{2}\right)^{-1}(t^{-1}\widehat{\*F}_t'\*F_t)(N^{-\alpha}\+\Lambda'\+\Lambda)&=\left(\frac{N}{N^\alpha}\*D_{Nt,r}^{2}\right)^{-1}\*H_{Nt,r}(N^{-\alpha}\+\Lambda'\+\Lambda)\notag\\
   &+O_p\left(\frac{N}{N^\alpha}\frac{1}{\sqrt{k_0}} \right)+O_p(N^{(1-2\alpha)/2}),
\end{align}
and hence uniformly in $t$,
\begin{align}
    \left(\frac{N}{N^\alpha}\*D_{Nt,r}^{2}\right)\*H_{Nt,r}=\*H_{Nt,r}(N^{-\alpha}\+\Lambda'\+\Lambda)+O_p\left(\frac{N}{N^\alpha}\frac{1}{\sqrt{k_0}} \right)+O_p(N^{(1-2\alpha)/2}),
\end{align}
and by using $\*H_{Nt,r}=\*H_r+O_p\left(\frac{N}{N^\alpha}\frac{1}{\sqrt{k_0}} \right)+O_p(N^{(1-2\alpha)/2})$,
\begin{align}\label{DH_HLambda}
    \left(\frac{N}{N^\alpha}\*D_{Nt,r}^{2}\right)\*H_{r}=\*H_{r}(N^{-\alpha}\+\Lambda'\+\Lambda)+O_p\left(\frac{N}{N^\alpha}\frac{1}{\sqrt{k_0}} \right)+O_p(N^{(1-2\alpha)/2}).
\end{align}
This chain of arguments tells us that up to a negligible term, $\*H_{r}$ is a matrix consisting of eigenvectors of $N^{-\alpha}\+\Lambda'\+\Lambda$, where the latter is diagonal and by assumption has all distinct eigenvalues. This implies that each eigenvalue is associated with a unique unitary eigenvector up to the sign and each of the eigenvectors has a single non-zero element (see the discussion on p. 27 in \citealp{bai2013principal}). Therefore, $\*H_{r}$ is a diagonal matrix with eigenvalues of $\pm 1$ up to the remainder. This again verifies that $\*H_r^{-1}$ exists. 
Therefore, $\mathrm{(ii.)}$ comes as an important implication of (\ref{DH_HLambda}), because it leads to
\begin{align}
      \left(\frac{N}{N^\alpha}\*D_{Nt,r}^{2}\right)&=(N^{-\alpha}\+\Lambda'\+\Lambda)+O_p\left(\frac{N}{N^\alpha}\frac{1}{\sqrt{k_0}} \right) +O_p(N^{(1-2\alpha)/2})\notag\\
      &= \+\Sigma_{\+\Lambda} + O_p\left(\frac{N}{N^\alpha}\frac{1}{\sqrt{k_0}} \right)+ O_p(N^{(1-2\alpha)/2}),
\end{align}
uniformly in $t$, because $ O_p(N^{(1-2\alpha)/2})$ dominates $O_p(N^{-\alpha/2})$, and so 
\begin{align}
    \sup_{k_0\leq t\leq T-1}\left\|\left(\frac{N}{N^\alpha}\*D_{Nt,r}^{2}\right)^{-1}-\+\Sigma_{\+\Lambda}^{-1} \right\|&\leq \sup_{k_0\leq t\leq T-1}\left\|\left(\frac{N}{N^\alpha}\*D_{Nt,r}^{2}\right)^{-1} \right\|\left\|\+\Sigma_{\+\Lambda}^{-1} \right\|\sup_{k_0\leq t\leq T-1}\left\|\left(\frac{N}{N^\alpha}\*D_{Nt,r}^{2}\right)-\+\Sigma_{\+\Lambda} \right\|\notag\\
    &= O_p\left(\frac{N}{N^\alpha}\frac{1}{\sqrt{k_0}} \right)+ O_p(N^{(1-2\alpha)/2}).
\end{align}
\noindent $\mathrm{(b)}$ To demonstrate $\mathrm{(i.)}$, we can invoke exactly the same decomposition as in $(\mathrm{i.})$ of part (a) with an important modification, to obtain   
\begin{align}
 \*H_{Nt,r}\*H_{Nt,r}'&= \*I_r-\*H_{Nt,r}\left(t^{-1}\*F_t'\*F_t-\*I_r \right)\*H_{Nt,r}'-t^{-1}\widehat{\*F}_t'(\widehat{\*F}_t-\*F_t\*H_{Nt,r}')\notag\\
   &-(\widehat{\*F}_t-\*F_t\*H_{Nt,r}')'t^{-1}\*F_t\*H_{Nt,r}'\notag\\
   &= \*I_r-(\*B_N^{-1}\overline{\*H}_{Nt,r}\*B_N)\left(t^{-1}\*F_t'\*F_t-\*I_r \right)(\*B_N^{-1}\overline{\*H}_{Nt,r}\*B_N)'-t^{-1}\widehat{\*F}_t'(\widehat{\*F}_t-\*F_t\*H_{Nt,r}')\notag\\
   &-(\widehat{\*F}_t-\*F_t\*H_{Nt,r}')'t^{-1}\*F_t(\*B_N^{-1}\overline{\*H}_{Nt,r}\*B_N)'\notag\\
   &=O\left(\frac{\sqrt{N^{\alpha_1+\alpha_r}}}{N} \right)+O_p\left(N^{[(1+\alpha_1)-3\alpha_r]/2}\right),
   \end{align}
   because 
   \begin{align}
       \sup_{k_0\leq t\leq T-1}\left\| (\*B_N^{-1}\overline{\*H}_{Nt,r}\*B_N)\left(t^{-1}\*F_t'\*F_t-\*I_r \right)(\*B_N^{-1}\overline{\*H}_{Nt,r}\*B_N)'\right\|&\leq \left\|\*B_N^{-1}\overline{\*H}_{Nt,r}\*B_N \right\|^2 \sup_{k_0\leq t\leq T-1}\left\| t^{-1}\*F_t'\*F_t-\*I_r\right\|\notag\\
       &= O_p\left(\frac{N^{\alpha_1}}{N^{\alpha_r}}\frac{1}{\sqrt{T}}\right),
   \end{align}
   which is clearly dominated by $ O_p\left(\frac{N}{N^{\alpha_r}}\frac{1}{\sqrt{T}}\right)$.
   Moreover, \begin{align}
      \sup_{k_0\leq t\leq T-1} \left\|(\widehat{\*F}_t-\*F_t\*H_{Nt,r}')'t^{-1}\*F_t(\*B_N^{-1}\overline{\*H}_{Nt,r}\*B_N)\right\|&\leq  \sup_{k_0\leq t\leq T-1} t^{-1/2} \left\|(\widehat{\*F}_t-\*F_t\*H_{Nt,r}')\right\|\notag\\
      &\times  \sup_{k_0\leq t\leq T-1} t^{-1/2}\left\|\*F_t \right\|\left\| \*B_N^{-1}\overline{\*H}_{Nt,r}\*B_N\right\| \notag\\
      &= \left(O_p\left(\frac{N}{N^{\alpha_r}}\frac{1}{\sqrt{k_0}} \right)+O_p(N^{(1-2\alpha_r)/2}) \right)\times O_p\left(\sqrt{\frac{N^{\alpha_1}}{N^{\alpha_r}}} \right)\notag\\
      &= O\left(\frac{\sqrt{N^{\alpha_1+\alpha_r}}}{N} \right)+O_p\left(N^{[(1+\alpha_1)-3\alpha_r]/2}\right),
   \end{align}
which clearly dominates the rate of $\sup_{k_0\leq t\leq T-1}\left\|t^{-1}\widehat{\*F}_t'(\widehat{\*F}_t-\*F_t\*H_{Nt,r}') \right\|$. The order comes from
\begin{align}
    \frac{N}{N^{\alpha_r}}\frac{1}{\sqrt{k_0}}\times \sqrt{\frac{N^{\alpha_1}}{N^{\alpha_r}}}=\frac{N}{N^{\alpha_r}}\frac{1}{k_0^{1/4}}\times \sqrt{\frac{N^{\alpha_1}}{N^{\alpha_r}}}\frac{1}{k_0^{1/4}}&=O(1)\times\frac{N}{N^{\alpha_r}}\frac{1}{k_0^{1/4}} \times \frac{N^{\alpha_1}}{N}\times \sqrt{\frac{N^{\alpha_r}}{N^{\alpha_1}}}\notag\\
    &=O(1)\times \frac{N^{\alpha_1}}{N}\times \sqrt{\frac{N^{\alpha_r}}{N^{\alpha_1}}}=O\left(\frac{\sqrt{N^{\alpha_1+\alpha_r}}}{N} \right)\notag\\
    &=o(1)
\end{align}
under \hyperref[A4]{A.4} and because $\alpha_1\leq 1$. Additionally,  $3\alpha_r>1+\alpha_1\Longleftrightarrow \alpha_1<3\alpha_r-1$ for $\alpha_r>0.5$. Then
    \begin{align}
    \sup_{k_0\leq t\leq T-1}\left\|\*H_{Nt,r}^{-1}-\*H_{r}^{-1} \right\|=O\left(\frac{\sqrt{N^{\alpha_1+\alpha_r}}}{N} \right)+O_p\left(N^{[(1+\alpha_1)-3\alpha_r]/2}\right).
\end{align}
Note that the rate can be improved by employing high-level conditions, as discussed in (\ref{boost_1}) and (\ref{boost_2}). 
\bigskip

\noindent For $\mathrm{(ii.)}$, we can use different strategies, but it is convenient to employ the conclusion in (\ref{NBD^2_approx}) directly in combination with $\left(\*B_N^{-1}\+\Lambda'\+\Lambda\*B_N^{-1} \right)^{-1}=\+\Sigma_{\+\Lambda}^{-1}+O_p(N^{-\alpha_r/2})$ by the CMT.
\subsubsection{Extra Results: Better Rates for In-Sample Factor Approximation and Eigenvalues  }
\noindent Note that in order to prove part $\mathrm{(i.)}$ of $\mathrm{(a)}$ and $\mathrm{(b)}$ of Lemma 2, we need to rely on the strategy of \cite{bai2002determining}. In particular, we average over the out-of-sample expansions in (\ref{row_fact_space}) and (\ref{het_expand}), which allows to track the behavior of the rotation matrix, which changes for every $t=k_0,\ldots,T-1$. For the in-sample analysis (parts $\mathrm{(ii.)}$), we can utilize a high-level condition in \hyperref[A3]{A.3} viii) similarly to \cite{bai2023approximate} and improve the rate of convergence in both directions of $N$ and $T$.
\begin{lemma} \label{Lemma4} Under A.1 - A.4 as $(N,T)\to \infty$, 
\begin{enumerate}[(a)]
   \item  if  $\alpha\in (0,1])$ (homogeneous):
\begin{align*}
   \sup_{k_0\leq t\leq T-1} \frac{1}{t}\left\|\widehat{\*F}_t-\*F_t\*H'_{Nt,r} \right\|^2=\sup_{k_0\leq t\leq T-1} \frac{1}{t}\sum_{s=1}^t\left\| \widehat{\*f}_s - \*H_{Nt,r}\*f_s\right\|^2= O_p\left(\frac{N^2}{N^{2\alpha}}\frac{1}{T^2} \right) + O_p(N^{-\alpha});
\end{align*}
\item Furthermore, under Assumptions A.1 - A.4 and heterogeneous loadings ($1\geq\alpha_1\geq\cdots\geq\alpha_r>1/2$), one has
\begin{align*}
\sup_{k_0\leq t\leq T-1} \frac{1}{t}\left\|\widehat{\*F}_t-\*F_t\*H'_{Nt,r} \right\|^2=\sup_{k_0\leq t\leq T-1} \frac{1}{t}\sum_{s=1}^t\left\| \widehat{\*f}_s - \*H_{Nt,r}\*f_s\right\|^2= O_p\left(\frac{N^2}{N^{2\alpha_r}}\frac{1}{T^2} \right) + O_p(N^{-\alpha_r}).
\end{align*}
\end{enumerate}
\end{lemma}

\noindent \textbf{Proof.} $\mathrm{(a)}$. In homogeneous case, note that by using $\|\*A \|_{sp}\leq \left\| \*A\right\|$, we have
\begin{align}\label{boost_1}
   \sup_{k_0\leq t\leq T-1} t^{-1/2}\left\|\widehat{\*F}_t-\*F_t\*H_{Nt,r}' \right\|_{sp}&\leq \sup_{k_0\leq t\leq T-1}\left\|\left(\frac{1}{N^{\alpha}t^{3/2}}\*F_t\+\Lambda'\*E_t'\widehat{\*F}_t+\frac{1}{N^{\alpha}t^{3/2}}\*E_t \+\Lambda \*F_t'\widehat{\*F}_t+\frac{1}{N^{\alpha}t^{3/2}}\*E_t\*E_t'\widehat{\*F}_t \right)\right\|_{sp} \notag\\
   &\times \sup_{k_0\leq t\leq T-1}\left\|\left(\frac{N}{N^\alpha}\*D_{Nt,r}^{2} \right)^{-1}\right\|_{sp}\notag\\
   &\leq \left(2\sup_{k_0\leq t\leq T-1} \frac{1}{N^{\alpha}t^{3/2}}\left\|\*E_t \+\Lambda \*F_t'\widehat{\*F}_t \right\| _{sp}+\sup_{k_0\leq t\leq T-1}\frac{1}{N^{\alpha}t^{3/2}}\left\|\*E_t\*E_t'\widehat{\*F}_t\right\|_{sp}\right)\notag\\
   &\times \sup_{k_0\leq t\leq T-1}\left\| \left(\frac{N}{N^\alpha}\*D_{Nt,r}^{2} \right)^{-1}\right\|_{sp}\notag\\
   &\leq O_p(N^{-\alpha/2})\times 2\sup_{k_0\leq t\leq T-1}\left\|t^{-1/2}N^{-\alpha/2}\*E_t\+\Lambda \right\| \sup_{k_0\leq t\leq T-1}\left\|t^{-1/2}\*F_t \right\| \sqrt{r}\notag\\
   &+ \sqrt{r}\frac{T}{k_0}N^{-\alpha}T^{-1}\sup_{k_0\leq t\leq T-1}\left\|\*E_t \right\|^2\notag\\
   &= O_p\left(\frac{N}{N^\alpha}\frac{1}{T} \right)+O_p(N^{-\alpha/2}),
\end{align}
since $\sup_{k_0\leq t\leq T-1}\left\|t^{-1/2}N^{-\alpha/2}\*E_t\+\Lambda \right\|=O_p(1)$ as argued in equation (9) in \cite{bai2023approximate}. \\

\noindent $\mathrm{(b)}$. In the heterogeneous case, we can follow the exact same logic as in Section 5.1 in \cite{bai2023approximate} and redefine $\*H_{Nt,r}$ and $\overline{\*H}_{Nt,r}$, such that still $\*B_N^{-1}\overline{\*H}_{Nt,r}\*B_N=\*H_{Nt,r}$, and it is possible to obtain faster convergence rate. In particular now, 
\begin{align}
    \overline{\*H}_{Nt,r}=(N\*B_N^{-2}\*D_{Nt,r}^2)^{-1}\left(\*B_N^{-1}t^{-1}\widehat{\*F}_t'\*F_t\+\Lambda'\+\Lambda\*B_N^{-1}+\*B_N^{-1}\widehat{\*F}_t'\*E_t\+\Lambda\*B_N^{-1} \right)
\end{align}
and so 
\begin{align}\label{boost_2}
    \sup_{k_0\leq t\leq T-1} t^{-1/2}\left\|\widehat{\*F}_t-\*F_t\*H_{Nt,r}' \right\|_{sp}&\leq \sup_{k_0\leq t\leq T-1} t^{-1/2}\left\|\widehat{\*F}_t\*B_N-\*F_t\*B_N\overline{\*H}_{Nt,r}' \right\|_{sp}\left\|\*B_N^{-1} \right\|\notag\\
    &\leq \left(\sup_{k_0\leq t\leq T-1} \frac{1}{t^{3/2}}\left\|\*E_t \+\Lambda\*B_N^{-1}\*B_N \*F_t'\widehat{\*F}_t\*B_N^{-1} \right\| _{sp}+\sup_{k_0\leq t\leq T-1}\frac{1}{t^{3/2}}\left\|\*E_t\*E_t'\widehat{\*F}_t\*B_N^{-1}\right\|_{sp}\right)\notag\\
   &\times \sup_{k_0\leq t\leq T-1}\left\| \left(N\*B_N^{-2}\*D_{Nt,r}^{2} \right)^{-1}\right\|_{sp}\left\|\*B_N^{-1} \right\|\notag\\
   &=O_p(N^{-\alpha_r/2})\times \sup_{k_0\leq t\leq T-1}\left\|t^{-1/2}\*E_t \+\Lambda\*B_N^{-1} \right\|\sup_{k_0\leq t\leq T-1}\left\|\*B_N t^{-1}\*F_t'\widehat{\*F}_t\*B_N^{-1} \right\|\notag\\
   &+O_p(N^{-\alpha_r})\times \sqrt{r}\frac{T}{k_0} T^{-1}\sup_{k_0\leq t\leq T-1}\left\|\*E_t \right\|_{sp}^2\notag\\
   &=O_p\left(\frac{N}{N^{\alpha_r}}\frac{1}{T} \right)+O_p(N^{-\alpha_r/2}),
\end{align}
if $\sup_{k_0\leq t\leq T-1}\left\|t^{-1/2}\*E_t \+\Lambda\*B_N^{-1} \right\|=O_p(1)$, where a similar argument is used in the proof of the statement (25) in \cite{bai2023approximate}.\\

\noindent Similarly, the rates of the eigenvalues and the rotation matrices can be improved upon. 
\begin{lemma} \label{Lemma5} Under A.1 - A.4 as $(N,T)\to \infty$
\begin{enumerate}[(a)]
\item If  $\alpha\in (0,1])$ (homogeneous):
    \begin{align*}
    &\mathrm{(i.)} \quad \sup_{k_0\leq t\leq T-1}\left\|\*H_{Nt,r}^{-1}-\*H_{r}^{-1} \right\|=O_p\left(\frac{N}{N^\alpha}\frac{1}{T} \right)+O_p(N^{-\alpha/2}),\\
    &\mathrm{(ii.)} \quad \sup_{k_0\leq t\leq T-1}\left\|\left(\frac{N}{N^\alpha}\*D_{Nt,r}^{2}\right)^{-1}-\+\Sigma_{\+\Lambda}^{-1} \right\|= O_p\left(\frac{N}{N^\alpha}\frac{1}{T} \right)+ O_p(N^{-\alpha/2}).
\end{align*}
\item Furthermore, under Assumptions A.1 - A.4 and heterogeneous loadings ($1\geq\alpha_1\geq\cdots\geq\alpha_r>1/2$), one has
\begin{align*}
    &\mathrm{(i.)} \quad \sup_{k_0\leq t\leq T-1}\left\|\*H_{Nt,r}^{-1}-\*H_{r}^{-1} \right\|=O_p\left(\frac{N^{\alpha_1}}{N^{\alpha_r}}\frac{1}{\sqrt{T}} \right)+O_p \left(\frac{\sqrt{N^{\alpha_1}}}{N^{\alpha_r}} \right),\\
    &\mathrm{(ii.)} \quad \sup_{k_0\leq t\leq T-1}\left\|\left(N\*B_N^{-2}\*D_{Nt,r}^{2}\right)^{-1}-\+\Sigma_{\+\Lambda}^{-1} \right\|= O_p\left(\frac{N}{N^{\alpha_r}}\frac{1}{\sqrt{T}} \right)+ O_p(N^{(1-2\alpha_r)/2}).
    \end{align*}
    \end{enumerate}
\end{lemma}
\bigskip

\noindent \textbf{Proof}. $\mathrm{(a)}$ In (i.) We will use the result (i.) from Lemma \ref{Lemma4}, where a faster convergence rate for the in-sample factor approximation is obtained. Specifically,
\begin{align}\label{HH'_approx_better}
   \*H_{Nt,r}\*H_{Nt,r}'&=-\*H_{Nt,r}\left(t^{-1}\*F_t'\*F_t-\*I_r \right)\*H_{Nt,r}' - t^{-1}\*H_{Nt,r}\*F'\*F_t\*H_{Nt,r}'\notag\\
   &= \*I_r-\*H_{Nt,r}\left(t^{-1}\*F_t'\*F_t-\*I_r \right)\*H_{Nt,r}'-t^{-1}\widehat{\*F}_t'(\widehat{\*F}_t-\*F_t\*H_{Nt,r}')\notag\\
   &-(\widehat{\*F}_t-\*F_t\*H_{Nt,r}')'t^{-1}\*F_t\*H_{Nt,r}'\notag\\
   &=\*I_r+O_p\left(\frac{N}{N^\alpha}\frac{1}{T} \right)+ O_p(N^{-\alpha/2})+O_p(T^{-1/2})\notag\\
   &= \*I_r+O_p\left(\frac{N}{N^\alpha}\frac{1}{T} \right)+ O_p(N^{-\alpha/2}).
   \end{align}
   Therefore,  $\*H_{Nt,r}=\*H_r+O_p\left(\frac{N}{N^\alpha}\frac{1}{T} \right)+ O_p(N^{-\alpha/2})$ uniformly in $t$, where $\*H_r$ is again an orthogonal matrix with only $\pm 1$ on the main diagonal. Ultimately,
\begin{align}
    \sup_{k_0\leq t\leq T-1}\left\|\*H_{Nt,r}^{-1}-\*H_{r}^{-1} \right\|&\leq \sup_{k_0\leq t\leq T-1}\left\|\*H_{Nt,r}^{-1} \right\|\left\| \*H_{r}^{-1}\right\|\sup_{k_0\leq t\leq T-1}\left\| \*H_{Nt,r}-\*H_{r}\right\|\notag\\
    &=O_p\left(\frac{N}{N^\alpha}\frac{1}{T} \right)+ O_p(N^{-\alpha/2}).
\end{align}
\noindent For (ii.), we can again use exactly the same chain of arguments, which now lead to 
\begin{align}
     \left(\frac{N}{N^\alpha}\*D_{Nt,r}^{2}\right)\*H_{Nt,r}=\*H_{Nt,r}(N^{-\alpha}\+\Lambda'\+\Lambda)+O_p\left(\frac{N}{N^\alpha}\frac{1}{T} \right)+ O_p(N^{-\alpha/2}),
\end{align}
and hence $ \left(\frac{N}{N^\alpha}\*D_{Nt,r}^{2}\right)\*H_{r}=\*H_{r}(N^{-\alpha}\+\Lambda'\+\Lambda)+O_p\left(\frac{N}{N^\alpha}\frac{1}{T} \right)+ O_p(N^{-\alpha/2})$, leading to 
\begin{align}
     \sup_{k_0\leq t\leq T-1}\left\|\left(\frac{N}{N^\alpha}\*D_{Nt,r}^{2}\right)^{-1}-\+\Sigma_{\+\Lambda}^{-1} \right\|&\leq \sup_{k_0\leq t\leq T-1}\left\|\left(\frac{N}{N^\alpha}\*D_{Nt,r}^{2}\right)^{-1} \right\|\left\|\+\Sigma_{\+\Lambda}^{-1} \right\|\sup_{k_0\leq t\leq T-1}\left\|\left(\frac{N}{N^\alpha}\*D_{Nt,r}^{2}\right)-\+\Sigma_{\+\Lambda} \right\|\notag\\
     &=O_p\left(\frac{N}{N^\alpha}\frac{1}{T} \right)+ O_p(N^{-\alpha/2}).
\end{align}
\noindent (b) In (i.), we use use the same approach
\begin{align}
     \*H_{Nt,r}\*H_{Nt,r}'&= \*I_r-\*H_{Nt,r}\left(t^{-1}\*F_t'\*F_t-\*I_r \right)\*H_{Nt,r}'-t^{-1}\widehat{\*F}_t'(\widehat{\*F}_t-\*F_t\*H_{Nt,r}')\notag\\
   &-(\widehat{\*F}_t-\*F_t\*H_{Nt,r}')'t^{-1}\*F_t\*H_{Nt,r}'\notag\\
   &= \*I_r-(\*B_N^{-1}\overline{\*H}_{Nt,r}\*B_N)\left(t^{-1}\*F_t'\*F_t-\*I_r \right)(\*B_N^{-1}\overline{\*H}_{Nt,r}\*B_N)'-t^{-1}\widehat{\*F}_t'(\widehat{\*F}_t-\*F_t\*H_{Nt,r}')\notag\\
   &-(\widehat{\*F}_t-\*F_t\*H_{Nt,r}')'t^{-1}\*F_t(\*B_N^{-1}\overline{\*H}_{Nt,r}\*B_N)'\notag\\
   &=O_p\left(\frac{N^{\alpha_1}}{N^{\alpha_r}}\frac{1}{\sqrt{T}} \right)+O_p \left(\frac{\sqrt{N^{\alpha_1}}}{N^{\alpha_r}} \right),
\end{align}
because 
 \begin{align}
       \sup_{k_0\leq t\leq T-1}\left\| (\*B_N^{-1}\overline{\*H}_{Nt,r}\*B_N)\left(t^{-1}\*F_t'\*F_t-\*I_r \right)(\*B_N^{-1}\overline{\*H}_{Nt,r}\*B_N)'\right\|&\leq \left\|\*B_N^{-1}\overline{\*H}_{Nt,r}\*B_N \right\|^2 \sup_{k_0\leq t\leq T-1}\left\| t^{-1}\*F_t'\*F_t-\*I_r\right\|\notag\\
       &= O_p\left(\frac{N^{\alpha_1}}{N^{\alpha_r}}\frac{1}{\sqrt{T}}\right),
   \end{align}
For the final rate, it is sufficient to demonstrate the rate of the second component. In particular, 
\begin{align}
     \sup_{k_0\leq t\leq T-1} \left\|(\widehat{\*F}_t-\*F_t\*H_{Nt,r}')'t^{-1}\*F_t(\*B_N^{-1}\overline{\*H}_{Nt,r}\*B_N)\right\|&\leq  \sup_{k_0\leq t\leq T-1} t^{-1/2} \left\|(\widehat{\*F}_t-\*F_t\*H_{Nt,r}')\right\|\notag\\
      &\times  \sup_{k_0\leq t\leq T-1} t^{-1/2}\left\|\*F_t \right\|\left\| \*B_N^{-1}\overline{\*H}_{Nt,r}\*B_N\right\| \notag\\
      &=  \left(O_p\left(\frac{N}{N^{\alpha_r}}\frac{1}{T} \right)+O_p(N^{-\alpha_r/2}) \right)\times O_p\left(\sqrt{\frac{N^{\alpha_1}}{N^{\alpha_r}}} \right)\notag\\
      &= O\left(\sqrt{\frac{N^{\alpha_r}}{N^{\alpha_1}}}\frac{1}{\sqrt{T}} \right)\times \underbrace{O_p\left(\frac{N}{N^{\alpha_r}}\frac{1}{T^{1/4}} \right)}_{O_p(1)}\times O_p\left(\frac{N^{\alpha_1}}{N^{\alpha_r}}\frac{1}{T^{1/4}} \right) \notag\\
      &+ O_p \left(\frac{\sqrt{N^{\alpha_1}}}{N^{\alpha_r}} \right)\notag\\
      &=O_p \left(\frac{\sqrt{N^{\alpha_1}}}{N^{\alpha_r}} \right)+ o_p\left(\sqrt{\frac{N^{\alpha_r}}{N^{\alpha_1}}}\frac{1}{\sqrt{T}} \right)=o_p(1)
\end{align}
under $\alpha_r>1/2$ and the fact that under \hyperref[A4]{A.4} and $\alpha_1\leq 1$, we have $ O_p\left(\frac{N^{\alpha_1}}{N^{\alpha_r}}\frac{1}{T^{1/4}} \right)= \frac{N^{\alpha_1}}{N}O_p\left(\frac{N}{N^{\alpha_r}}\frac{1}{T^{1/4}} \right)=O_p\left(\frac{N^{\alpha_1}}{N} \right)$. Hence, $\*H_{Nt,r}=\*H_r+ O_p\left(\frac{N^{\alpha_1}}{N^{\alpha_r}}\frac{1}{\sqrt{T}} \right)+O_p \left(\frac{\sqrt{N^{\alpha_1}}}{N^{\alpha_r}} \right)$ and so 
\begin{align}
   \sup_{k_0\leq t \leq T-1}\left\|\*H_{Nt,r}^{-1} -\*H_r^{-1}\right\| = O_p\left(\frac{N^{\alpha_1}}{N^{\alpha_r}}\frac{1}{\sqrt{T}} \right)+O_p \left(\frac{\sqrt{N^{\alpha_1}}}{N^{\alpha_r}} \right).
\end{align}
\\
\noindent For (ii.), we leave the same convergence rate. We conjecture that it can be improved, via derivation of the exact limit of $\*B_Nt^{-1}\widehat{\*F}_t'\*F_t\*B_{N}^{-1}$.
\section{Theoretical Results II: Asymptotic Equivalence}
\subsection{Expansions of the Feasible Statistics}
 Subscript ``$f$'' (``$\widehat{f}$'') indicates that the test is \textit{infeasible} (\textit{feasible}) since we need to use the estimated factors. The \emph{feasible} statistic of forecast encompassing is 
 \begin{align}
       g_{\widehat{f},1}=\frac{1}{\widehat{\omega}_1}\left(\frac{1}{\sqrt{n}}\sum_{t=k_0}^{T-1}\widetilde{u}^2_{1,t+1}-\frac{1}{2}\left[\frac{n}{m_0}\frac{1}{\sqrt{n}}\sum_{t=k_0}^{k_0+m_0-1}\widetilde{u}_{1,t+1}\widehat{u}_{2,t+1} + \frac{n}{n-m_0}\frac{1}{\sqrt{n}}\sum_{t=k_0+m_0}^{T-1}\widetilde{u}_{1,t+1}\widehat{u}_{2,t+1} \right] \right).
 \end{align}
 By following \cite{gonccalves2017tests}, the latter can be conveniently decomposed into 
 \begin{align}\label{arxiv_exp}
         g_{\widehat{f},1}&=  g_{f,1}+  g_{f,1}\left(\frac{\widetilde{\omega}_1}{\widehat{\omega}_1}-1 \right)\notag\\
          &+\frac{1}{2\widehat{\omega}_1}\left(\frac{n}{m_0}\frac{1}{\sqrt{n}}\sum_{t=k_0}^{k_0+m_0-1}\widetilde{u}_{1,t+1}(\widetilde{u}_{2,t+1}- \widehat{u}_{2,t+1}) + \frac{n}{n-m_0}\frac{1}{\sqrt{n}}\sum_{t=k_0+m_0}^{T-1}\widetilde{u}_{1,t+1}(\widetilde{u}_{2,t+1}- \widehat{u}_{2,t+1}) \right).
 \end{align}
 Similarly, the test for forecasting accuracy in (\ref{ET1}) admits the following expansion: 
 \begin{align}\label{ET_1_exp}
       g_{\widehat{f},2}&=\frac{1}{\widehat{\omega}}_2\frac{n}{l^0_1}\left(\frac{1}{\sqrt{n}}\sum_{t=k_0}^{k_0+l^0_1-1}\widetilde{u}^2_{1,t+1}-\frac{l^0_1}{l^0_2}\frac{1}{\sqrt{n}}\sum_{t=k_0}^{k_0+l^0_2-1}\widehat{u}^2_{2,t+1} \right)\notag\\
       &=\frac{1}{\widehat{\omega}}_2\frac{n}{l^0_1}\left(\frac{1}{\sqrt{n}}\sum_{t=k_0}^{k_0+l^0_1-1}\widetilde{u}^2_{1,t+1}-\frac{l^0_1}{l^0_2}\frac{1}{\sqrt{n}}\sum_{t=k_0}^{k_0+l^0_2-1}\widetilde{u}^2_{2,t+1} \right)+\frac{1}{\widehat{\omega}}_z\frac{n}{l^0_1} \frac{l_1^0}{l_2^0}\frac{1}{\sqrt{n}}\sum_{t=k_0}^{k_0+l_2^0-1}(\widetilde{u}^2_{2,t+1}-\widehat{u}^2_{2,t+1})\notag\\
       &= g_{f,2}+g_{f,2}\left(\frac{\widetilde{\omega}_2}{\widehat{\omega}_2}-1 \right)+\frac{1}{\widehat{\omega}}_2 \frac{n}{l_2^0}\frac{1}{\sqrt{n}}\sum_{t=k_0}^{k_0+l_2^0-1}(\widetilde{u}^2_{2,t+1}-\widehat{u}^2_{2,t+1})\notag\\
       &=g_{f,2}+g_{f,2}\left(\frac{\widetilde{\omega}_2}{\widehat{\omega}_2}-1 \right) + \frac{2}{\widehat{\omega}}_2 \frac{n}{l_2^0}\frac{1}{\sqrt{n}}\sum_{t=k_0}^{k_0+l_2^0-1}\widetilde{u}_{2,t+1}(\widetilde{u}_{2,t+1}-\widehat{u}_{2,t+1})\notag\\
       &-\frac{1}{\widehat{\omega}}_2 \frac{n}{l_2^0}\frac{1}{\sqrt{n}}\sum_{t=k_0}^{k_0+l_2^0-1}(\widetilde{u}_{2,t+1}-\widehat{u}_{2,t+1})^2.
 \end{align}
\noindent Generally, the expansion of (\ref{ET_2_sum}) for some $j=3,4$, and $k,p=1,2$ with $k\neq p$ can be represented with 
\begin{align}
     g_{\widehat{f},j}(\lambda_k^0)&=\frac{1}{\widehat{\omega}}_{j}\frac{1}{n(1-\tau_0)}\sum_{l_p=\lfloor n\tau_0 \rfloor+1}^n  g_{\widehat{f},2}(\lfloor n\lambda^0_k \rfloor, l_p  )\notag\\
     &= g_{f,j}+g_{f,j}\left(\frac{\widetilde{\omega}_{j}}{\widehat{\omega}_{j}} -1\right)+ \frac{1}{n(1-\tau_0)}\frac{2}{\widehat{\omega}}_{j}\frac{1}{\sqrt{n}}\sum_{l_p=\lfloor n\tau_0 \rfloor+1}^n \frac{n}{l_2}\sum_{t=k_0}^{k_0+l_2-1}\widetilde{u}_{2,t+1}(\widetilde{u}_{2,t+1}-\widehat{u}_{2,t+1})\notag\\
       &-\frac{1}{n(1-\tau_0)}\frac{1}{\widehat{\omega}}_{j}\frac{1}{\sqrt{n}}\sum_{l_p=\lfloor n\tau_0 \rfloor+1}^n \frac{n}{l_2}\sum_{t=k_0}^{k_0+l_2-1}(\widetilde{u}_{2,t+1}-\widehat{u}_{2,t+1})^2,
\end{align}
where we use the fact that the estimation error enters only the second model component. Now, if we set $j=3$ and $k=2$, which means that we fix $l^0=\lfloor n \lambda_2^0\rfloor$ and are left with $p=1$, we obtain 
 \begin{align}\label{ET_2_exp}
       g_{\widehat{f},3}&=\frac{1}{\widehat{\omega}}_{3}\frac{1}{n(1-\tau_0)}\sum_{l_1=\lfloor n\tau_0 \rfloor+1}^n  g_{\widehat{f},2}(\lfloor n\lambda^0_2 \rfloor, l_1  )\notag\\
       &= g_{f,3}+g_{f,3}\left(\frac{\widetilde{\omega}_{3}}{\widehat{\omega}_{3}} -1\right)+ \frac{1}{n(1-\tau_0)}\frac{2}{\widehat{\omega}}_{3} \frac{n}{l_2^0}\frac{1}{\sqrt{n}}\sum_{l_1=\lfloor n\tau_0 \rfloor+1}^n\sum_{t=k_0}^{k_0+l_2^0-1}\widetilde{u}_{2,t+1}(\widetilde{u}_{2,t+1}-\widehat{u}_{2,t+1})\notag\\
       &-\frac{1}{n(1-\tau_0)}\frac{1}{\widehat{\omega}}_{3} \frac{n}{l_2^0}\frac{1}{\sqrt{n}}\sum_{l=\lfloor n\tau_0 \rfloor+1}^n\sum_{t=k_0}^{k_0+l_2^0-1}(\widetilde{u}_{2,t+1}-\widehat{u}_{2,t+1})^2\notag\\
       &= g_{f,3}+g_{f,3}\left(\frac{\widetilde{\omega}_{3}}{\widehat{\omega}_{3}} -1\right)+ \left(\frac{n-\lfloor n\tau_0 \rfloor}{n(1-\tau_0)} \right)\frac{2}{\widehat{\omega}}_{3} \frac{n}{l_2^0}\frac{1}{\sqrt{n}}\sum_{t=k_0}^{k_0+l_2^0-1}\widetilde{u}_{2,t+1}(\widetilde{u}_{2,t+1}-\widehat{u}_{2,t+1})\notag\\
       &-\left(\frac{n-\lfloor n\tau_0 \rfloor}{n(1-\tau_0)} \right)\frac{1}{\widehat{\omega}}_{3} \frac{n}{l_2^0}\frac{1}{\sqrt{n}}\sum_{t=k_0}^{k_0+l_2^0-1}(\widetilde{u}_{2,t+1}-\widehat{u}_{2,t+1})^2\notag\\
       &= g_{f,3}+g_{f,3}\left(\frac{\widetilde{\omega}_{3}}{\widehat{\omega}_{3}} -1\right)\notag\\
       &+\left(\frac{n-\lfloor n\tau_0 \rfloor}{n(1-\tau_0)} \right)\left(\frac{2}{\widehat{\omega}}_{3} \frac{n}{l_2^0}\frac{1}{\sqrt{n}}\sum_{t=k_0}^{k_0+l_2^0-1}\widetilde{u}_{2,t+1}(\widetilde{u}_{2,t+1}-\widehat{u}_{2,t+1})- \frac{1}{\widehat{\omega}}_{3} \frac{n}{l_2^0}\frac{1}{\sqrt{n}}\sum_{t=k_0}^{k_0+l_2^0-1}(\widetilde{u}_{2,t+1}-\widehat{u}_{2,t+1})^2\right).
 \end{align}
 Note that because $\frac{n-\lfloor n\tau_0 \rfloor}{n}=1-\tau_0+O(n^{-1})$, we have that $\frac{n-\lfloor n\tau_0 \rfloor}{n(1-\tau_0)}=1+O(n^{-1})$, which means that analysis of (\ref{ET_2_exp}) is almost identical to (\ref{ET_1_exp}). Intuition of this is that the portion of the expansion involving the factor estimation error does not depend on $l_1$. Naturally, expansion of (\ref{ET_new}) is based on the fact that we fix $l_1^0=\lfloor n\lambda_1^0 \rfloor$:
 \begin{align}\label{ET_new_exp}
      g_{\widehat{f},4}&=\frac{1}{\widehat{\omega}}_{4}\frac{1}{n(1-\tau_0)}\sum_{l_2=\lfloor n\tau_0 \rfloor+1}^n   g_{\widehat{f},2}(\lfloor n\lambda^0_1 \rfloor ,l_2 )\notag\\
      &= g_{f,4}+g_{f,4}\left(\frac{\widetilde{\omega}_{4}}{\widehat{\omega}_{4}} -1\right)\notag\\
       &+\frac{1}{n(1-\tau_0)}\sum_{l_2=\lfloor n\tau_0 \rfloor}^n\frac{n}{l_2}\left(\frac{2}{\widehat{\omega}}_{4} \frac{1}{\sqrt{n}}\sum_{t=k_0}^{k_0+l_2-1}\widetilde{u}_{2,t+1}(\widetilde{u}_{2,t+1}-\widehat{u}_{2,t+1})- \frac{1}{\widehat{\omega}}_{4} \frac{1}{\sqrt{n}}\sum_{t=k_0}^{k_0+l_2-1}(\widetilde{u}_{2,t+1}-\widehat{u}_{2,t+1})^2\right).
 \end{align}
 \noindent We are left to expand $|\widetilde{\omega}^2_j-\widehat{\omega}^2_j|$ for $j=1,\ldots,4$ to demonstrate that such terms are negligible. A crucial step here is to expand all the components of the estimated variances in terms of the quantities that constitute (\ref{ET_1_exp}) - (\ref{ET_2_exp}), so that the rates would be known. Let $\+\zeta_j$ for $j=1,\ldots,4$ be a vector of tuning parameters associated with the respective statistic. For instance $\+\zeta_{1}=(\mu_0,\pi_0)'$. It can be shown that for a \textit{known} positive function $f_j$ the difference can be written as
 \begin{align}
     |\widetilde{\omega}^2_j-\widehat{\omega}^2_j|=|f_j(\+\zeta_j)\widetilde{\phi}^2- f_j(\+\zeta_j)\widehat{\phi}^2|=f_j(\+\zeta_j)|\widetilde{\phi}^2-\widehat{\phi}^2|,
 \end{align}
 where $\phi^2=\lim_{T\to \infty}\mathbb{V}ar\left[\frac{1}{\sqrt{n}}\sum_{t=k_0}^{T-1}(u_{2,t+1}^2-\mathbb{E}(u_{2,t+1}^2))\right]$. Finding the estimator of the later requires imposing some dependence structure. For instance, \cite{gonccalves2017tests} and \cite{stauskas2022tests} use a martingale difference assumption directly (their estimand is $\mathbb{V}ar(u_{2,t+1})$). Then
 \begin{align}
 \phi^2=\lim_{T\to \infty}\frac{1}{n}\sum_{t=k_0}^{T-1}\mathbb{E}(u_{2,t+1}^2-\mathbb{E}(u_{2,t+1}^2))^2=\mathbb{V}ar(u_{2,t+1}^2-\sigma^2), 
 \end{align}
 under conditional homoskedasticity, whose natural feasible counterpart is 
 \begin{align}
     \widehat{\phi}^2&=\frac{1}{n}\sum_{t=k_0}^{T-1}\left(\widehat{u}_{2,t+1}^2-\frac{1}{n}\sum_{t=k_0}^{T-1}\widehat{u}_{2,t+1}^2\right)^2\notag\\
     &=\frac{1}{n}\sum_{t=k_0}^{T-1}\left(\widetilde{u}^2_{2,t+1} -\frac{1}{n}\sum_{t=k_0}^{T-1}\widetilde{u}_{2,t+1}^2 - (\widetilde{u}^2_{2,t+1}-\widehat{u}^2_{2,t+1}) + \left[\frac{1}{n}\sum_{t=k_0}^{T-1}(\widetilde{u}_{2,t+1}^2-\widehat{u}_{2,t+1}^2) \right] \right)^2\notag\\
     &=\frac{1}{n}\sum_{t=k_0}^{T-1}\left(\widetilde{u}^2_{2,t+1} -\frac{1}{n}\sum_{t=k_0}^{T-1}\widetilde{u}_{2,t+1}^2\right)^2 + q(\widetilde{u}^2_{2,t+1},\widehat{u}^2_{2,t+1})\notag\\
     &= \widetilde{\phi}^2 + q(\widetilde{u}^2_{2,t+1},\widehat{u}^2_{2,t+1}),
 \end{align}
 where
 \begin{align}
     q(\widetilde{u}^2_{2,t+1},\widehat{u}^2_{2,t+1})&=-\frac{2}{n}\sum_{t=k_0}^{T-1}\left(\widetilde{u}^2_{2,t+1} -\frac{1}{n}\sum_{t=k_0}^{T-1}\widetilde{u}_{2,t+1}^2 \right)\left( (\widetilde{u}^2_{2,t+1}-\widehat{u}^2_{2,t+1}) + \left[\frac{1}{n}\sum_{t=k_0}^{T-1}(\widetilde{u}_{2,t+1}^2-\widehat{u}_{2,t+1}^2) \right] \right)\notag\\
     &+\frac{1}{n}\sum_{t=k_0}^{T-1}\left( (\widetilde{u}^2_{2,t+1}-\widehat{u}^2_{2,t+1}) + \left[\frac{1}{n}\sum_{t=k_0}^{T-1}(\widetilde{u}_{2,t+1}^2-\widehat{u}_{2,t+1}^2) \right] \right)^2\notag\\
     &=q_1(\widetilde{u}^2_{2,t+1},\widehat{u}^2_{2,t+1})+q_2(\widetilde{u}^2_{2,t+1},\widehat{u}^2_{2,t+1}),
 \end{align}
 where we can analyze both components separately. For instance, by Cauchy-Schwarz inequality and the definition of $ \widetilde{\phi}^2$ we obtain 
 \begin{align}
     |q_1(\widetilde{u}^2_{2,t+1},\widehat{u}^2_{2,t+1})|&\leq 2 \left(\frac{1}{n}\sum_{t=k_0}^{T-1}\left(\widetilde{u}^2_{2,t+1} -\frac{1}{n}\sum_{t=k_0}^{T-1}\widetilde{u}_{2,t+1}^2 \right)^2\right)^{1/2}\notag\\
     &\times \left( \frac{1}{n}\sum_{t=k_0}^{T-1}\left((\widetilde{u}^2_{2,t+1}-\widehat{u}^2_{2,t+1}) + \left[\frac{1}{n}\sum_{t=k_0}^{T-1}(\widetilde{u}_{2,t+1}^2-\widehat{u}_{2,t+1}^2) \right]\right)^2\right)^{1/2}\notag\\
     &= 2 \widetilde{\phi}q_2(\widetilde{u}^2_{2,t+1},\widehat{u}^2_{2,t+1})^{1/2},
 \end{align}
 where $\widetilde{\phi}$ is clearly consistent and hence bounded. This can be seen by using the definition of $\widetilde{u}_{2,t+1}^2$:
 \begin{align}
   \sup_{k_0\leq t \leq T-1}\left| \widetilde{u}_{2,t+1}^2-u_{t+1}^2\right|& \leq 2T^{-1/2}\sup_{k_0\leq t \leq T-1}\left\| \sqrt{T}(\widetilde{\+\delta}_t-\+\delta)\right\|\sup_{k_0\leq t \leq T-1}\left\| \*z_t\right\|\sup_{k_0\leq t \leq T-1} |u_{t+1}|\notag\\
   &+T^{-1}\sup_{k_0\leq t \leq T-1}\left\| \sqrt{T}(\widetilde{\+\delta}_t-\+\delta)\right\|^2\sup_{k_0\leq t \leq T-1}\left\| \*z_t\right\|^2\notag\\
   &=O_p(T^{-1/2}).
 \end{align}
 therefore, 
 \begin{align}
     \widetilde{\phi}=\frac{1}{n}\sum_{t=k_0}^{T-1}\left(u_{t+1}^2-\frac{1}{n}\sum_{t=k_0}^{T-1}u_{t+1}^2\right)^{2}+o_p(1)\to_p \phi.
 \end{align}
 Therefore, the key component to analyze is $q_2(\widetilde{u}^2_{2,t+1},\widehat{u}^2_{2,t+1})$. For this, the following expansion is the key:
 \begin{align}\label{u-u_exp}
     \widetilde{u}^2_{2,t+1}-\widehat{u}^2_{2,t+1}= 2\widetilde{u}_{2,t+1}( \widetilde{u}_{2,t+1}- \widehat{u}_{2,t+1})-(\widetilde{u}_{2,t+1}- \widehat{u}_{2,t+1})^2
 \end{align}
 as in the expansion (\ref{ET_1_exp}). In combination, 
 \begin{align}\label{var_exp_final}
      |\widetilde{\omega}^2_j-\widehat{\omega}^2_j|=f_j(\+\zeta_j)|q(\widetilde{u}^2_{2,t+1},\widehat{u}^2_{2,t+1}) | \leq f_j(\+\zeta_j) [2\widetilde{\phi}q_2(\widetilde{u}^2_{2,t+1},\widehat{u}^2_{2,t+1})^{1/2}+ q_2(\widetilde{u}^2_{2,t+1},\widehat{u}^2_{2,t+1})].
 \end{align}
\subsection{Interim Results}
We notice that the expansions of the feasible statistics in (\ref{arxiv_exp}) - (\ref{ET_new_exp}) are essentially driven by 
\begin{align*}
&A=\frac{1}{\sqrt{d_T}}\sum_{t=\lfloor f_1(T) \rfloor}^{\lfloor f_2(T)\rfloor}(\widetilde{u}_{2,t+1}- \widehat{u}_{2,t+1})^2,\\
  &B=\frac{1}{\sqrt{d_T}}\sum_{t=\lfloor f_1(T)\rfloor }^{\lfloor f_2(T) \rfloor}\widetilde{u}_{1,t+1}(\widetilde{u}_{2,t+1}- \widehat{u}_{2,t+1}),\\
  & C=\frac{1}{\sqrt{d_T}}\sum_{t=\lfloor f_1(T)\rfloor}^{\lfloor f_2(T) \rfloor}\widetilde{u}_{2,t+1}(\widetilde{u}_{2,t+1}- \widehat{u}_{2,t+1}),\\
  &D=|\widetilde{\omega}^2_j-\widehat{\omega}^2_j | \quad \text{for} \quad j=1,\ldots,4,
\end{align*}
which we write in the general notation. We will formulate our results in terms of $A, B, C$, which will immediately give $D$. Similarly to \cite{gonccalves2017tests} and \cite{stauskas2022tests} (in CCE setup), $\widetilde{u}_{2,t+1}- \widehat{u}_{2,t+1}$ is the central component that we will use. Introduce $\+\Phi_{Nt,r}=\mathrm{diag}(\*I_k, \*H_{Nt,r})\in \mathbb{R}^{(k+r)\times (k+r)}$, where $\*H_{Nt,r}$ is a rotation matrix for either homogeneous or heterogeneous cases defined in Section 3. Then
\begin{align}\label{u_diff_exp}
    \widetilde{u}_{2,t+1}- \widehat{u}_{2,t+1}&=y_{t+1}-\widetilde{\+\delta}_t'\*z_t - (y_{t+1}-\widehat{\+\delta}_t'\widehat{\*z}_t)=\widehat{\+\delta}_t'\widehat{\*z}_t-\widetilde{\+\delta}_t'\*z_t\notag\\
    & =(\widehat{\*z}_t-\+\Phi_{Nt,r} \*z_t)'(\+\Phi_{Nt,r}^{-1})'(\widetilde{\+\delta}_t-\+\delta)+\widehat{\*z}_t'(\widehat{\+\delta}_t-(\+\Phi_{Nt,r}^{-1})'\widetilde{\+\delta}_t)+(\widehat{\*z}_t-\+\Phi_{Nt,r} \*z_t)'(\+\Phi_{Nt,r}^{-1})'\+\delta \notag\\
    &=(\widehat{\*z}_t-\+\Phi_{Nt,r} \*z_t)'(\+\Phi_{Nt,r}^{-1})'(\widetilde{\+\delta}_t-\+\delta)+\widehat{\*z}_t'(\widehat{\+\delta}_t-(\+\Phi_{Nt,r}^{-1})'\widetilde{\+\delta}_t)+(\widehat{\*f}_t-\*H_{Nt,r} \*f_t)'(\*H_{Nt,r}^{-1})'\+\beta^0 T^{-1/4},
\end{align}
where we used the local-to-alternative parameterization. The following lemma is the key in deriving the orders of $A, B$ and $C$.
    \begin{lemma}\label{Lemma6}
Under Assumptions A.1 - A.4 with homogeneous $\alpha$ we have \begin{align*}
   &\mathrm{(i.)}\quad \sup_{k_0\leq t\leq T-1}\left\|T^{1/4}(\widehat{\+\delta}_t-(\+\Phi_{Nt,r}^{-1})'\widetilde{\+\delta}_t)\right\|^2=O_p\left(\frac{N^2}{N^{2\alpha}}\frac{1}{k_0} \right) + O_p(N^{1-2\alpha})+O_p(T^{-1/2}).
\end{align*}
Under heterogeneous $\alpha$, we have
 \begin{align*}
   &\mathrm{(ii.)}\quad  \sup_{k_0\leq t\leq T-1}\left\|T^{1/4}(\widehat{\+\delta}_t-(\+\Phi_{Nt,r}^{-1})'\widetilde{\+\delta}_t)\right\|^2=O_p\left(\frac{N^2}{N^{2\alpha_r}}\frac{1}{k_0} \right) + O_p(N^{1-2\alpha_r})+O_p(T^{-1/2}).
   \end{align*}
\end{lemma}
\bigskip 

\noindent \textbf{Proof.} $\mathrm{(i.)}$ 
It is convenient to utilize the stacked notation as used before. We additionally introduce $\*y_t=(y_2,\ldots, y_t)'$ and $\*u_t=(u_2\ldots, u_t)'$ both in $\mathbb{R}^{t-1}$. Moreover, to keep the notation compact, we now let $\*F_t=(\*f_1',\ldots,\*f_{t-1}')$, and $\widehat{\*F}_t=(\widehat{\*f}_1',\ldots, \widehat{\*f}_{t-1}')$ both in $\mathbb{R}^{t-1}$. We normalize $t^{-1}\widehat{\*F}_t'\widehat{\*F}_t=\*I_r$ for compactness, again. Implementation-wise, we can employ the restriction $(t-1)^{-1}\widehat{\*F}_t'\widehat{\*F}_t=\*I_r$, but then $t^{-1}\widehat{\*F}_t'\widehat{\*F}_t=\frac{t-1}{t}\frac{1}{t-1}\widehat{\*F}_t'\widehat{\*F}_t=\frac{(t-1)/T}{t/T}\*I_r=\*I_r+O(T^{-1})$, and the conclusions would stay the same. Hence, we have 
\begin{align}\label{y_exp}
    \*y_t=\*Z_t\+\delta + \*u_t&= \widehat{\*Z}_t(\+\Phi^{-1}_{Nt,r})'\+\delta -(\widehat{\*Z}_t- \*Z_t\+\Phi_{Nt,r}')(\+\Phi^{-1}_{Nt,r})'\+\delta+\*u_t\notag\\
    &=\widehat{\*Z}_t(\+\Phi^{-1}_{Nt,r})'\+\delta-(\widehat{\*F}_t-\*F_t\*H_{Nt,r}')(\*H_{Nt,r}^{-1})'\+\beta^0T^{-1/4} + \*u_t
\end{align}
Further, by the FWL Theorem and the fact that for any conformable and invertible $\*A$ and $\*B$ matrices we have $\*B^{-1}=\*A'(\*A\*B\*A')^{-1}\*A$, we obtain 
\begin{align}
  (\+\Phi^{-1}_{Nt,r})'\widetilde{\+\delta}_t&= \begin{bmatrix}\*I_k & \*0_{k\times r}\\ \*0_{r\times k} & (\*H_{Nt,r}^{-1})' \end{bmatrix}\begin{bmatrix} \left(\*W_t'\*M_{\*F}\*W_t\right)^{-1}\*W_t'\*M_{\*F}\*y_t\\ 
    \left(\*F_t'\*M_{\*W}\*F_t \right)^{-1}\*F_t'\*M_{\*W}\*y_t\end{bmatrix}\notag\\
    &=\begin{bmatrix}\*I_k & \*0_{k\times r}\\ \*0_{r\times k} & (\*H_{Nt,r}^{-1})' \end{bmatrix}\begin{bmatrix} \left(\*W_t'\*M_{\*F}\*W_t\right)^{-1}\*W_t'\*M_{\*F}\*y_t\\ 
    \*H_{Nt,r}'\left(\*H_{Nt,r}\*F_t'\*M_{\*W}\*F_t\*H_{Nt,r}' \right)^{-1}\*H_{Nt,r}\*F_t'\*M_{\*W}\*y_t \end{bmatrix}\notag\\
    &= \begin{bmatrix} \left(\*W_t'\*M_{\*F}\*W_t\right)^{-1}\*W_t'\*M_{\*F}\*y_t\\ 
   \left(\*H_{Nt,r}\*F_t'\*M_{\*W}\*F_t\*H_{Nt,r}' \right)^{-1}\*H_{Nt,r}\*F_t'\*M_{\*W}\*y_t
    \end{bmatrix},
\end{align}
which gives 
\begin{align}
   T^{-1/4} \sqrt{T}\left(\widehat{\+\delta}_t-(\+\Phi_{Nt,r}^{-1})'\widetilde{\+\delta}_t\right)&= T^{-1/4}\begin{bmatrix}
\sqrt{T}\left(\*W'\*M_{\widehat{\*F}}\*W_t\right)^{-1}\*W_t'\*M_{\widehat{\*F}}\*y_t - \sqrt{T}\left(\*W_t'\*M_{\*F}\*W_t\right)^{-1}\*W_t'\*M_{\*F}\*y_t\\
      \sqrt{T}(\widehat{\*F}_t'\*M_{\*W}\widehat{\*F}_t )^{-1}\widehat{\*F}_t'\*M_{\*W}\*y_t -\sqrt{T}\left(\*H_{Nt,r}\*F_t'\*M_{\*W}\*F_t\*H_{Nt,r}' \right)^{-1}\*H_{Nt,r}\*F_t'\*M_{\*W}\*y_t
    \end{bmatrix}\notag\\
    &=\begin{bmatrix}
          \*a\\
          \*b
      \end{bmatrix}.
\end{align}
Clearly, since we have a vector,
   $ \left\| T^{-1/4}\sqrt{T}\left(\widehat{\+\delta}_t-(\+\Phi_{Nt,r}^{-1})'\widetilde{\+\delta}_t\right) \right\|^2=\left\|\*a \right\|^2+\left\|\*b \right\|^2$, therefore, we will analyse the sub-vectors. By using (\ref{y_exp}), we can further decompose: 
   \begin{align}\label{core_decomp}
        & T^{-1/4} \sqrt{T}\left(\widehat{\+\delta}_t-(\+\Phi_{Nt,r}^{-1})'\widetilde{\+\delta}_t\right)\notag\\
         &= T^{-1/4}\begin{bmatrix}
\left(\sqrt{T}\left(\*W'\*M_{\widehat{\*F}}\*W_t\right)^{-1}\*W_t'\*M_{\widehat{\*F}} - \sqrt{T}\left(\*W_t'\*M_{\*F}\*W_t\right)^{-1}\*W_t'\*M_{\*F}\right)\widehat{\*Z}_t(\+\Phi^{-1}_{Nt,r})'\+\delta\\
      \left(\sqrt{T}(\widehat{\*F}_t'\*M_{\*W}\widehat{\*F}_t )^{-1}\widehat{\*F}_t'\*M_{\*W} -\sqrt{T}\left(\*H_{Nt,r}\*F_t'\*M_{\*W}\*F_t\*H_{Nt,r}' \right)^{-1}\*H_{Nt,r}\*F_t'\*M_{\*W}\right)\widehat{\*Z}_t(\+\Phi^{-1}_{Nt,r})'\+\delta
    \end{bmatrix}\notag\\
    &- T^{-1/4}\begin{bmatrix}
\left(\sqrt{T}\left(\*W'\*M_{\widehat{\*F}}\*W_t\right)^{-1}\*W_t'\*M_{\widehat{\*F}} - \sqrt{T}\left(\*W_t'\*M_{\*F}\*W_t\right)^{-1}\*W_t'\*M_{\*F}\right)(\widehat{\*F}_t-\*F_t\*H_{Nt,r}')(\*H_{Nt,r}^{-1})'\+\beta^0T^{-1/4}\\
      \left(\sqrt{T}(\widehat{\*F}_t'\*M_{\*W}\widehat{\*F}_t )^{-1}\widehat{\*F}_t'\*M_{\*W} -\sqrt{T}\left(\*H_{Nt,r}\*F_t'\*M_{\*W}\*F_t\*H_{Nt,r}' \right)^{-1}\*H_{Nt,r}\*F_t'\*M_{\*W}\right)(\widehat{\*F}_t-\*F_t\*H_{Nt,r}')(\*H_{Nt,r}^{-1})'\+\beta^0T^{-1/4}
    \end{bmatrix}\notag\\
    &+ T^{-1/4}\begin{bmatrix}
\left(\sqrt{T}\left(\*W'\*M_{\widehat{\*F}}\*W_t\right)^{-1}\*W_t'\*M_{\widehat{\*F}} - \sqrt{T}\left(\*W_t'\*M_{\*F}\*W_t\right)^{-1}\*W_t'\*M_{\*F}\right)\*u_t\\
      \left(\sqrt{T}(\widehat{\*F}_t'\*M_{\*W}\widehat{\*F}_t )^{-1}\widehat{\*F}_t'\*M_{\*W} -\sqrt{T}\left(\*H_{Nt,r}\*F_t'\*M_{\*W}\*F_t\*H_{Nt,r}' \right)^{-1}\*H_{Nt,r}\*F_t'\*M_{\*W}\right)\*u_t
      \end{bmatrix}\notag\\
      &= \begin{bmatrix}
          \*I \\
          \mathbf{II}
      \end{bmatrix} - \begin{bmatrix}
          \mathbf{III} \\
          \mathbf{IV}
      \end{bmatrix} +\begin{bmatrix}
          \mathbf{V} \\
          \mathbf{VI}
      \end{bmatrix},
   \end{align}
   where each of the terms can be split into two again. Starting from $\*I$, we can simply it with 
   \begin{align*}
\*I_1&= T^{-1/4}\sqrt{T}\left(\*W'\*M_{\widehat{\*F}}\*W_t\right)^{-1}\*W_t'\*M_{\widehat{\*F}}\widehat{\*Z}_t(\+\Phi^{-1}_{Nt,r})'\+\delta\notag\\
&=  T^{-1/4}\sqrt{T}\left(\*W'\*M_{\widehat{\*F}}\*W_t\right)^{-1}\*W_t'\*M_{\widehat{\*F}} \*W_t\+\theta + T^{-1/4}\sqrt{T}\left(\*W'\*M_{\widehat{\*F}}\*W_t\right)^{-1}\*W_t'\*M_{\widehat{\*F}} \widehat{\*F}_t(\*H_{Nt,r}^{-1})'\+\beta \notag\\
&= T^{1/4}\+\theta 
   \end{align*}
 while
   \begin{align}
\*I_2&= T^{-1/4}\sqrt{T}\left(\*W_t'\*M_{\*F}\*W_t\right)^{-1}\*W_t'\*M_{\*F}\widehat{\*Z}_t(\+\Phi^{-1}_{Nt,r})'\+\delta \notag\\
&= T^{-1/4} \sqrt{T}\left(\*W_t'\*M_{\*F}\*W_t\right)^{-1}\*W_t'\*M_{\*F}\*W_t\+\theta + \sqrt{T}\left(\*W_t'\*M_{\*F}\*W_t\right)^{-1}\*W_t'\*M_{\*F} \widehat{\*F}_t(\*H_{Nt,r}^{-1})'\+\beta\notag\\
&=T^{1/4}\+\theta + T^{-1/4}\sqrt{T}\left(\*W_t'\*M_{\*F}\*W_t\right)^{-1}\*W_t'\*M_{\*F} \widehat{\*F}_t(\*H_{Nt,r}^{-1})'\+\beta^0T^{-1/4}\notag\\
&= T^{1/4}\+\theta+  T^{-1/4}\sqrt{T}\left(\*W_t'\*M_{\*F}\*W_t\right)^{-1}\*W_t'\*M_{\*F}\*F_t\*H_{Nt,r}'(\*H_{Nt,r}^{-1})'\+\beta^0T^{-1/4} \notag\\
&- T^{-1/4}\sqrt{T}\left(\*W_t'\*M_{\*F}\*W_t\right)^{-1}\*W_t'\*M_{\*F}(\*F_t\*H_{Nt,r}'-\widehat{\*F}_t)(\*H_{Nt,r}^{-1})'\+\beta^0T^{-1/4}\notag\\
&=  T^{1/4}\+\theta-\left(\*W_t'\*M_{\*F}\*W_t\right)^{-1}\*W_t'\*M_{\*F}(\*F_t\*H_{Nt,r}'-\widehat{\*F}_t)(\*H_{Nt,r}^{-1})'\+\beta^0,
   \end{align}
   which implies that 
   \begin{align}
       \*I=\*I_1-\*I_2=\left(\*W_t'\*M_{\*F}\*W_t\right)^{-1}\*W_t'\*M_{\*F}(\*F_t\*H_{Nt,r}'-\widehat{\*F}_t)(\*H_{Nt,r}^{-1})'\+\beta^0
   \end{align}
   and so 
   \begin{align}\label{rate_I}
       \sup_{k_0\leq t\leq T-1}\left\| \*I\right\| &=\sup_{k_0\leq t\leq T-1}\left\|\left(\*W_t'\*M_{\*F}\*W_t\right)^{-1}\*W_t'\*M_{\*F}(\widehat{\*F}_t-\*F_t\*H_{Nt,r}')(\*H_{Nt,r}^{-1})'\+\beta^0\right\|\notag\\
       &\leq \sup_{k_0\leq t\leq T-1}\left\| \left(t^{-1}\*W_t'\*M_{\*F}\*W_t\right)^{-1}\right\|\sup_{k_0\leq t\leq T-1}\left\|t^{-1}\*W_t'(\widehat{\*F}_t-\*F_t\*H_{Nt,r}')\right\|\sup_{k_0\leq t\leq T-1}\left\|\*H_{Nt,r}^{-1} \right\| \|\+\beta^0 \|\notag\\
       &+ \sup_{k_0\leq t\leq T-1}\left\| \left(t^{-1}\*W_t'\*M_{\*F}\*W_t\right)^{-1}\right\|\sup_{k_0\leq t\leq T-1}\left\|t^{-1}\*W_t'\*F_t \right\|\sup_{k_0\leq t\leq T-1}\left\| (t^{-1}\*F_t'\*F_t)^{-1}\right\|\notag\\
       &\times \sup_{k_0\leq t\leq T-1}\left\|t^{-1}\*F_t'(\widehat{\*F}_t-\*F_t\*H_{Nt,r}') \right\|\sup_{k_0\leq t\leq T-1}\left\|\*H_{Nt,r}^{-1} \right\| \|\+\beta^0 \|\notag\\
       &= O_p\left(\frac{N}{N^{\alpha}}\frac{1}{\sqrt{k_0}} \right) + O_p(N^{1/2-\alpha})
   \end{align}
   by \hyperref[Lemma2]{Lemma 2} (ii). In particular, we used
   \begin{align}
       \sup_{k_0\leq t\leq T-1}\left\|t^{-1}\*Z_t'(\widehat{\*F}_t-\*F_t\*H_{Nt,r}') \right\|&\leq\sup_{k_0\leq t\leq T-1} t^{-1/2}\left\| \*Z_t\right\| \sup_{k_0\leq t\leq T-1} \left\|t^{-1/2}(\widehat{\*F}_t-\*F_t\*H_{Nt,r}') \right\|\notag\\
       &=\sup_{k_0\leq t\leq T-1} \sqrt{\mathrm{tr}\left(\frac{1}{t}\*Z_t'\*Z_t\right)} \sup_{k_0\leq t\leq T-1} \left\|t^{-1/2}(\widehat{\*F}_t-\*F_t\*H_{Nt,r}') \right\|\notag\\
       &= O_p(1)\left( O_p\left(\frac{N}{N^{\alpha}}\frac{1}{\sqrt{k_0}} \right) + O_p(N^{1/2-\alpha}) \right).
   \end{align}
Moving to the first component of $\mathbf{II}$, we obtain 
\begin{align}
    \mathbf{II}_1&= T^{-1/4}\sqrt{T}(\widehat{\*F}_t'\*M_{\*W}\widehat{\*F}_t )^{-1}\widehat{\*F}_t'\*M_{\*W}\widehat{\*Z}_t(\+\Phi^{-1}_{Nt,r})'\+\delta\notag\\
    &= T^{-1/4}\sqrt{T}(\widehat{\*F}_t'\*M_{\*W}\widehat{\*F}_t )^{-1}\widehat{\*F}_t'\*M_{\*W}\*W_t\+\theta +T^{-1/4}\sqrt{T}(\widehat{\*F}_t'\*M_{\*W}\widehat{\*F}_t )^{-1}\widehat{\*F}_t'\*M_{\*W}\widehat{\*F}_t(\*H_{Nt,r}^{-1})'\+\beta \notag\\
    &= T^{1/4}(\*H_{Nt,r}^{-1})'\+\beta=(\*H_{Nt,r}^{-1})'\+\beta^0.
\end{align}
Next,
\begin{align}
    \mathbf{II}_2&= T^{-1/4}\sqrt{T}\left(\*H_{Nt,r}\*F_t'\*M_{\*W}\*F_t\*H_{Nt,r}' \right)^{-1}\*H_{Nt,r}\*F_t'\*M_{\*W}\widehat{\*Z}_t(\+\Phi^{-1}_{Nt,r})'\+\delta \notag\\
    &= T^{-1/4}\sqrt{T}\left(\*H_{Nt,r}\*F_t'\*M_{\*W}\*F_t\*H_{Nt,r}' \right)^{-1}\*H_{Nt,r}\*F_t'\*M_{\*W}\*W_t\+\theta \notag\\
    &+T^{-1/4}\sqrt{T}\left(\*H_{Nt,r}\*F_t'\*M_{\*W}\*F_t\*H_{Nt,r}' \right)^{-1}\*H_{Nt,r}\*F_t'\*M_{\*W}\*F_t\*H_{Nt,r}'(\*H_{Nt,r}^{-1})'\+\beta \notag\\
    &- T^{-1/4}\sqrt{T}\left(\*H_{Nt,r}\*F_t'\*M_{\*W}\*F_t\*H_{Nt,r}^{-1} \right)'\*H_{Nt,r}\*F_t'\*M_{\*W}(\*F_t\*H_{Nt,r}'-\widehat{\*F}_t)(\*H_{Nt,r}^{-1})'\+\beta\notag\\
    &= (\*H_{Nt,r}')^{-1}\+\beta^0-\left(\*H_{Nt,r}\*F_t'\*M_{\*W}\*F_t\*H_{Nt,r}' \right)^{-1}\*H_{Nt,r}\*F_t'\*M_{\*W}(\*F_t\*H_{Nt,r}'-\widehat{\*F}_t)(\*H_{Nt,r}^{-1})'\+\beta^0,
\end{align}
which follows the analysis in (\ref{rate_I}):
\begin{align}\label{rate_II}
    &\sup_{k_0\leq t\leq T-1}\left\| \mathbf{II}\right\|=\sup_{k_0\leq t\leq T-1}\left\|\left(\*H_{Nt,r}\*F_t'\*M_{\*W}\*F_t\*H_{Nt,r}' \right)^{-1}\*H_{Nt,r}\*F_t'\*M_{\*W}(\widehat{\*F}_t-\*F_t\*H_{Nt,r}')(\*H_{Nt,r}^{-1})'\+\beta^0 \right\|\notag\\
    &\leq \sup_{k_0\leq t\leq T-1}\left\|\left(\*H_{Nt,r}t^{-1}\*F_t'\*M_{\*W}\*F_t\*H_{Nt,r}' \right)^{-1} \right\|\sup_{k_0\leq t\leq T-1}\left\| \*H_{Nt,r}\right\|\sup_{k_0\leq t\leq T-1}\left\| \*H_{Nt,r}^{-1}\right\||\| \+\beta^0\|\notag\\
    &\times \sup_{k_0\leq t\leq T-1}\left\|t^{-1/2}\*F_t \right\|\sup_{k_0\leq t\leq T-1}\left\|t^{-1/2}(\widehat{\*F}_t-\*F_t\*H_{Nt,r}' )\right\|\notag\\
    &+ \sup_{k_0\leq t\leq T-1}\left\|\left(\*H_{Nt,r}t^{-1}\*F_t'\*M_{\*W}\*F_t\*H_{Nt,r}' \right)^{-1} \right\|\sup_{k_0\leq t\leq T-1}\left\| \*H_{Nt,r}\right\|\sup_{k_0\leq t\leq T-1}\left\| \*H_{Nt,r}^{-1}\right\|\notag\\
    &\times \sup_{k_0\leq t\leq T-1}\left\|t^{-1}\*F_t'\*W_t \right\|\sup_{k_0\leq t\leq T-1}\left\|(t^{-1}\*W_t'\*W_t)^{-1} \right\|\sup_{k_0\leq t\leq T-1}\left\|t^{-1/2}\*W_t \right\|\sup_{k_0\leq t\leq T-1}\left\|t^{-1/2}(\widehat{\*F}_t-\*F_t\*H_{Nt,r}' )\right\|\| \+\beta^0\|\notag\\
    &=O_p\left(\frac{N}{N^{\alpha}}\frac{1}{\sqrt{k_0}} \right) + O_p(N^{1/2-\alpha}).
\end{align}
For $\mathbf{III}_1$, we note that by the PC estimation restriction we get 
\begin{align}
    \mathbf{III}_1 &=T^{-1/4}\sqrt{T}\left(\*W'\*M_{\widehat{\*F}}\*W_t\right)^{-1}\*W_t'\*M_{\widehat{\*F}}(\widehat{\*F}_t-\*F_t\*H_{Nt,r}')(\*H_{Nt,r}^{-1})'\+\beta^0 T^{-1/4} \notag\\
    &= \left(\*W_t'\*M_{\widehat{\*F}}\*W_t\right)^{-1}\*W_t'\*M_{\widehat{\*F}}(\widehat{\*F}_t-\*F_t\*H_{Nt,r}')(\*H_{Nt,r}^{-1})'\+\beta^0\notag\\
    &=\left(t^{-1}\*W_t'\*M_{\widehat{\*F}}\*W_t\right)^{-1}t^{-1}\*W_t'(\widehat{\*F}_t-\*F_t\*H_{Nt,r}')(\*H_{Nt,r}^{-1})'\+\beta^0 \notag\\
   &+ \left(t^{-1}\*W_t'\*M_{\widehat{\*F}}\*W_t\right)^{-1}t^{-1}\*W_t'\widehat{\*F}_t t^{-1}\widehat{\*F}_t'(\widehat{\*F}_t-\*F_t\*H_{Nt,r}')(\*H_{Nt,r}^{-1})'\+\beta^0
\end{align}
and so, 
\begin{align}
    &\sup_{k_0\leq t\leq T-1}\left\| \mathbf{III}_1\right\|\notag\\
    &\leq \sup_{k_0\leq t\leq T-1} \left\|\left(t^{-1}\*W_t'\*M_{\widehat{\*F}}\*W_t\right)^{-1} \right\| \sup_{k_0\leq t\leq T-1} t^{-1/2}\left\|\*W_t \right\|\sup_{k_0\leq t\leq T-1} \left\|t^{-1/2}(\widehat{\*F}_t-\*F_t\*H_{Nt,r}' )\right\| \sup_{k_0\leq t\leq T-1} \left\|\*H_{Nt,r}^{-1} \right\| \|\+\beta^0 \| \notag\\
    &+\sup_{k_0\leq t\leq T-1} \left\|\left(t^{-1}\*W_t'\*M_{\widehat{\*F}}\*W_t\right)^{-1} \right\| \sup_{k_0\leq t\leq T-1}\left\|t^{-1}\*W_t'\widehat{\*F}_t \right\|\sup_{k_0\leq t\leq T-1} t^{-1/2}\left\|\widehat{\*F}_t \right\|\notag\\
    &\times \sup_{k_0\leq t\leq T-1} \left\|t^{-1/2}(\widehat{\*F}_t-\*F_t\*H_{Nt,r}' )\right\| \sup_{k_0\leq t\leq T-1} \left\|\*H_{Nt,r}^{-1} \right\| \|\+\beta^0 \| \notag\\
    &= O_p\left(\frac{N}{N^{\alpha}}\frac{1}{\sqrt{k_0}} \right) + O_p(N^{1/2-\alpha}),
\end{align}
where we used the following facts. Firstly, $\sup_{k_0\leq t\leq T-1} t^{-1/2}\left\|\widehat{\*F}_t \right\|=\sup_{k_0\leq t\leq T-1} \sqrt{\mathrm{tr}\left(t^{-1}\widehat{\*F}_t'\widehat{\*F}_t \right)}=\sqrt{\mathrm{tr}(\*I_r)}=\sqrt{r}$. Next, 
\begin{align}
    \sup_{k_0\leq t\leq T-1}\left\|t^{-1}\*W_t'\widehat{\*F}_t \right\|= \sup_{k_0\leq t\leq T-1}\left\| \frac{1}{t}\sum_{s=1}^{t-1}\*w_s\widehat{\*f}_s'\right\|&\leq \sup_{k_0\leq t\leq T-1}\frac{1}{t}\sum_{s=1}^{t-1}\left\| \*w_s\right\|\left\| \widehat{\*f}_s\right\|\notag\\
    &\leq \left(\sup_{k_0\leq t\leq T-1}\frac{1}{t}\sum_{s=1}^{t-1}\left\|\*w_s \right\|^2 \right)^{1/2} \left(\sup_{k_0\leq t\leq T-1}\frac{1}{t}\sum_{s=1}^{t-1}\left\|\widehat{\*f}_s \right\|^2 \right)^{1/2}\notag\\
    &=\sqrt{r} \left(\sup_{k_0\leq t\leq T-1}\frac{1}{t}\sum_{s=1}^{t-1}\left\|\*w_s \right\|^2 \right)^{1/2}=O_p(1).
\end{align}
Finally, we need to verify that $\sup_{k_0\leq t\leq T-1} \left\|\left(t^{-1}\*W_t'\*M_{\widehat{\*F}}\*W_t\right)^{-1} \right\| =O_p(1)$. For this, we can use two decompositions similar to the ones in Lemma 2 of \cite{bai2002determining}: 
\begin{align}\label{sup_check_W}
    t^{-1}\*W_t'(\*M_\*F-\*M_{\widehat{\*F}})\*W_t&= t^{-1}\*W_t'(\*M_{\*F\*H}-\*M_{\widehat{\*F}})\*W_t\notag\\
    &= t^{-2}\*W_t'(\widehat{\*F}_t-\*F_t\*H_{Nt,r}')(t^{-1}\widehat{\*F}_t'\widehat{\*F}_t)^{-1}(\widehat{\*F}_t-\*F_t\*H_{Nt,r}')'\*W_t\notag\\
    &+t^{-2}\*W_t'(\widehat{\*F}_t-\*F_t\*H_{Nt,r}')(t^{-1}\widehat{\*F}_t'\widehat{\*F}_t)^{-1}\*H_{Nt,r}\*F_t'\*W_t\notag\\
    &+t^{-2}\*W_t'\*F_t\*H_{Nt,r}'(t^{-1}\widehat{\*F}_t'\widehat{\*F}_t)^{-1}(\widehat{\*F}_t-\*F_t\*H_{Nt,r}')'\*W_t\notag\\
    &+t^{-2}\*W_t'\*F_t\*H_{Nt,r}'\left[(t^{-1}\widehat{\*F}_t'\widehat{\*F}_t)^{-1}-(\*H_{Nt,r}t^{-1}\*F_t'\*F_t\*H_{Nt,r}') ^{-1}\right]\*H_{Nt,r}\*F_t'\*W_t\notag\\
    &=\mathbf{i+ii+iii+iv},
\end{align}
where $\*i-\mathbf{iii}$ are negligible based on Lemma \hyperref[Lemma2]{2}. For instance, by noting that $(t^{-1}\widehat{\*F}_t'\widehat{\*F}_t)^{-1}=\*I_r$, 
\begin{align}
    \sup_{k_0\leq t\leq T-1}\|\*i \|\leq \sup_{k_0\leq t\leq T-1} t^{-1}\left\|\*W_t \right\|^2 \sup_{k_0\leq t\leq T-1} t^{-1}\left\|\widehat{\*F}_t-\*F_t\*H_{Nt,r}' \right\|^2 = O_p\left(\frac{N^2}{N^{2\alpha}}\frac{1}{k_0} \right) + O_p(N^{1-2\alpha}),
\end{align}
whereas 
\begin{align}
    \sup_{k_0\leq t\leq T-1}\|\mathbf{ii} \|&\leq \sup_{k_0\leq t\leq T-1} t^{-1/2}\left\| \*W_t\right\| \sup_{k_0\leq t\leq T-1}\left\|t^{-1}\*F_t'\*W_t \right\|\sup_{k_0\leq t\leq T-1}\|\*H_{Nt,r} \| \sup_{k_0\leq t\leq T-1} t^{-1/2}\left\|\widehat{\*F}_t-\*F_t\*H_{Nt,r}' \right\|\notag\\
    &=O_p\left(\frac{N}{N^{\alpha}}\frac{1}{\sqrt{k_0}} \right) + O_p(N^{1/2-\alpha})
\end{align}
and $\|\mathbf{iii} \|$ is of the same order. We are left with $\mathbf{iv}$, whose central component can be decomposed in a similar fashion: 
\begin{align}\label{hatFF-FF}
    \sup_{k_0\leq t\leq T-1} \left\| t^{-1}\widehat{\*F}_t'\widehat{\*F}_t-\*H_{Nt,r}t^{-1}\*F_t'\*F_t\*H_{Nt,r}' \right\| &\leq \sup_{k_0\leq t\leq T-1} t^{-1}\left\|\widehat{\*F}_t-\*F_t\*H_{Nt,r}' \right\|^2\notag\\
    &+2\sup_{k_0\leq t\leq T-1}t^{-1/2}\left\| \widehat{\*F}_t-\*F_t\*H_{Nt,r}'\right\|\sup_{k_0\leq t\leq T-1} t^{-1/2}\| \*F_t\|\sup_{k_0\leq t\leq T-1}\|\*H_{Nt,r} \|\notag\\
    &=O_p\left(\frac{N}{N^{\alpha}}\frac{1}{\sqrt{k_0}} \right) + O_p(N^{1/2-\alpha})
\end{align}
with the same holding for the inverses, which in connection to $\sup_{k_0\leq t\leq T-1} t^{-2}\|\*W_t'\*F_t \|^2=O_p(1)$ means that $\|\mathbf{iv} \|$ is of the same order. This discussion implies that $ t^{-1}\*W_t'\*M_{\widehat{\*F}}\*W_t=t^{-1}\*W_t'\*M_{\*F}\*W_t + o_p(1)$ uniformly in $t$ and by the Continuous Mapping Theorem (CMT)
\begin{align}
   (t^{-1}\*W_t'\*M_{\widehat{\*F}}\*W_t)^{-1}=(t^{-1}\*W_t'\*M_{\*F}\*W_t)^{-1} + o_p(1)
\end{align}
uniformly in $t$. Clearly, 
    $\sup_{k_0\leq t\leq T-1}\|\mathbf{III}_2 \|= O_p\left(\frac{N}{N^{\alpha}}\frac{1}{\sqrt{k_0}} \right) + O_p(N^{1/2-\alpha})$,
because it has exactly the same form as (\ref{rate_I}):
\begin{align}
   \sup_{k_0\leq t\leq T-1} \left\| \mathbf{III}_2\right\|&=\sup_{k_0\leq t\leq T-1}\left\|\left(\*W_t'\*M_{\*F}\*W_t\right)^{-1}\*W_t'\*M_{\*F}(\widehat{\*F}_t-\*F_t\*H_{Nt,r}')(\*H_{Nt,r}^{-1})'\+\beta^0\right\|\notag\\
       &\leq \sup_{k_0\leq t\leq T-1}\left\| \left(t^{-1}\*W_t'\*M_{\*F}\*W_t\right)^{-1}\right\|\sup_{k_0\leq t\leq T-1}\left\|t^{-1}\*W_t'(\widehat{\*F}_t-\*F_t\*H_{Nt,r}')\right\|\sup_{k_0\leq t\leq T-1}\left\|\*H_{Nt,r}^{-1} \right\| \|\+\beta^0 \|\notag\\
       &+ \sup_{k_0\leq t\leq T-1}\left\| \left(t^{-1}\*W_t'\*M_{\*F}\*W_t\right)^{-1}\right\|\sup_{k_0\leq t\leq T-1}\left\|t^{-1}\*W_t'\*F_t \right\|\sup_{k_0\leq t\leq T-1}\left\| (t^{-1}\*F_t'\*F_t)^{-1}\right\|\notag\\
       &\times \sup_{k_0\leq t\leq T-1}\left\|t^{-1}\*F_t'(\widehat{\*F}_t-\*F_t\*H_{Nt,r}') \right\|\sup_{k_0\leq t\leq T-1}\left\|\*H_{Nt,r}^{-1} \right\| \|\+\beta^0 \|\notag\\
       &= O_p\left(\frac{N}{N^{\alpha}}\frac{1}{\sqrt{k_0}} \right) + O_p(N^{1/2-\alpha}),
\end{align}
and therefore, in total 
\begin{align}
    \sup_{k_0\leq t\leq T-1}\|\mathbf{III} \|= O_p\left(\frac{N}{N^{\alpha}}\frac{1}{\sqrt{k_0}} \right) + O_p(N^{1/2-\alpha}).
\end{align}
Moving on to $\mathbf{IV}$, we get 
\begin{align}
    \sup_{k_0\leq t\leq T-1}& \| \mathbf{IV}_1\|\notag\\
    &\leq \sup_{k_0\leq t\leq T-1} \left\| (\widehat{\*F}_t'\*M_{\*W}\widehat{\*F}_t )^{-1}\widehat{\*F}_t'\*M_{\*W}(\widehat{\*F}_t-\*F_t\*H_{Nt,r}')(\*H_{Nt,r}^{-1})'\+\beta^0\right\|\notag\\
    & \leq  \sup_{k_0\leq t\leq T-1}\left\|(t^{-1}\widehat{\*F}_t'\*M_{\*W}\widehat{\*F}_t )^{-1} \right\| \sup_{k_0\leq t\leq T-1} t^{-1/2}\left\|\widehat{\*F}_t \right\|\sup_{k_0\leq t\leq T-1} t^{-1/2}\left\| \widehat{\*F}_t-\*F_t\*H_{Nt,r}'\right\| \sup_{k_0\leq t\leq T-1}\|\*H_{Nt,r}^{-1} \|\|\+\beta^0 \|\notag \\
    &+ \sup_{k_0\leq t\leq T-1}\left\|(t^{-1}\widehat{\*F}_t'\*M_{\*W}\widehat{\*F}_t )^{-1} \right\| \sup_{k_0\leq t\leq T-1}\left\|t^{-1}\widehat{\*F}_t'\*W_t \right\|\sup_{k_0\leq t\leq T-1}\left\|(t^{-1}\*W_t'\*W_t)^{-1} \right\|\sup_{k_0\leq t\leq T-1} t^{-1/2}\left\|\*W_t \right\|\notag\\
    &\times \sup_{k_0\leq t\leq T-1} t^{-1/2}\left\| \widehat{\*F}_t-\*F_t\*H_{Nt,r}'\right\| \sup_{k_0\leq t\leq T-1}\|\*H_{Nt,r}^{-1} \|\|\+\beta^0 \|\notag\\
    &= O_p\left(\frac{N}{N^{\alpha}}\frac{1}{\sqrt{k_0}} \right) + O_p(N^{1/2-\alpha})
\end{align}
based on Lemma \hyperref[Lemma2]{2} and the results we used previously. Similarly to (\ref{sup_check_W}), we need to check whether $\sup_{k_0\leq t\leq T-1}\left\|(t^{-1}\widehat{\*F}_t'\*M_{\*W}\widehat{\*F}_t )^{-1} \right\|=O_p(1)$. Note how similarly to (\ref{hatFF-FF}) we obtain 
\begin{align}\label{hatFWF-FWF}
    &\sup_{k_0\leq t\leq T-1}\left\|t^{-1}\widehat{\*F}_t'\*M_{\*W}\widehat{\*F}_t-t^{-1}\*H_{Nt,r}\*F_t'\*M_{\*W}\*F_t\*H_{Nt,r}'\right\| \notag \\
    &\leq \sup_{k_0\leq t\leq T-1} t^{-1}\left\| \widehat{\*F}_t-\*F_t\*H_{Nt,r}'\right\|^2+\sup_{k_0\leq t\leq T-1} t^{-1}\left\| \widehat{\*F}_t-\*F_t\*H_{Nt,r}'\right\|^2 \sup_{k_0\leq t\leq T-1}t^{-1}\| \*W_t\|^2\sup_{k_0\leq t\leq T-1} \|(t^{-1}\*W_t'\*W_t)^{-1} \|\notag\\
    &+2  \sup_{k_0\leq t\leq T-1} t^{-1/2}\left\| \widehat{\*F}_t-\*F_t\*H_{Nt,r}'\right\|\sup_{k_0\leq t\leq T-1} t^{-1/2}\|\*F_t \|\sup_{k_0\leq t\leq T-1}\|\*H_{Nt,r} \|\notag\\
    &+2  \sup_{k_0\leq t\leq T-1} t^{-1/2}\left\| \widehat{\*F}_t-\*F_t\*H_{Nt,r}'\right\|\sup_{k_0\leq t\leq T-1}t^{-1/2}\| \*W_t\|\sup_{k_0\leq t\leq T-1} \|(t^{-1}\*W_t'\*W_t)^{-1} \|\notag\\
    &\times \sup_{k_0\leq t\leq T-1}\left\|t^{-1}\*W_t'\*F_t \right\|\sup_{k_0\leq t\leq T-1}\| \*H_{Nt,r}\| \notag\\
    &= O_p\left(\frac{N}{N^{\alpha}}\frac{1}{\sqrt{k_0}} \right) + O_p(N^{1/2-\alpha})
\end{align}
and the same rate holds for the inverses. This discussion implies that 
\begin{align}
    (t^{-1}\widehat{\*F}_t'\*M_{\*W}\widehat{\*F}_t)^{-1}=(t^{-1}\*H_{Nt,r}\*F_t'\*M_{\*W}\*F_t\*H_{Nt,r}')^{-1}+o_p(1)
\end{align}
uniformly in $t$. By noting that $\mathbf{IV}_2$ has exactly the same form as (\ref{rate_II}), we can deduct that overall 
\begin{align}
    \sup_{k_0\leq t\leq T-1}\|\mathbf{IV} \|=  O_p\left(\frac{N}{N^{\alpha}}\frac{1}{\sqrt{k_0}} \right) + O_p(N^{1/2-\alpha}).
\end{align}
For the remaining terms, we continue with $\mathbf{V}$ and focus on $\mathbf{V}_2$ first, to illustrate the approach.  Because $\sup_{k_0\leq t\leq T-1} \left\|\left(t^{-1}\*W_t'\*M_\*F\*W_t\right)^{-1} \right\| =O_p(1)$ by assumption, we can focus directly on its numerator. Particularly, 
\begin{align}
    \sup_{k_0\leq t\leq T-1} &\left\|T^{-1/4}\sqrt{T}t^{-1}\*W_t'\*M_\*F\*u_t \right\|\leq   \sup_{k_0\leq t\leq T-1}\left\|T^{-1/4}\sqrt{T}t^{-1}\*W_t'\*u_t \right\| +   \sup_{k_0\leq t\leq T-1}\left\| T^{-1/4}\sqrt{T}t^{-1}\*W_t'\*P_\*F\*u_t\right\|\notag\\
    &\leq T^{-1/4}\sup_{k_0\leq t\leq T-1}\frac{T}{t}\sup_{k_0\leq t\leq T-1}\left\| \frac{1}{\sqrt{T}}\sum_{s=1}^{t-1}\*w_su_{s+1} \right\|\notag\\
     &+T^{-1/4}\sup_{k_0\leq t\leq T-1}\frac{T}{t}\sup_{k_0\leq t\leq T-1}\left\|t^{-1}\*W_t'\*F_t \right\|\sup_{k_0\leq t\leq T-1}\left\|(t^{-1}\*F_t'\*F_t)^{-1}\right\|\sup_{k_0\leq t\leq T-1}\left\| \frac{1}{\sqrt{T}}\sum_{s=1}^{t-1}\*f_su_{s+1} \right\|\notag\\
     &= T^{-1/4}\frac{T}{k_0}\sup_{k_0\leq t\leq T-1}\left\| \frac{1}{\sqrt{T}}\sum_{s=1}^{t-1}\*w_su_{s+1} \right\|\notag\\
    &+T^{-1/4}\frac{T}{k_0}\sup_{k_0\leq t\leq T-1}\left\|t^{-1}\*W_t'\*F_t \right\|\sup_{k_0\leq t\leq T-1}\left\|(t^{-1}\*F_t'\*F_t)^{-1}\right\|\sup_{k_0\leq t\leq T-1}\left\| \frac{1}{\sqrt{T}}\sum_{s=1}^{t-1}\*f_su_{s+1} \right\|\notag\\
    &=O_p(T^{-1/4}),
\end{align}
by Lemma 2.1 of \cite{corradi2001predictive} and $Tk_0^{-1}=\frac{1}{\sqrt{1-\pi_0}}+O(T^{-1})$. The analysis of $\mathbf{V}_1$ is similar. Because $\sup_{k_0\leq t\leq T-1} \left\|\left(t^{-1}\*W_t'\*M_{\widehat{\*F}}\*W_t\right)^{-1} \right\| =O_p(1)$ by (\ref{sup_check_W}) and $(t^{-1}\widehat{\*F}_t'\widehat{\*F}_t)^{-1}=\*I_r$, we have
    \begin{align}
    \sup_{k_0\leq t\leq T-1} \left\|T^{-1/4}\sqrt{T}t^{-1}\*W_t'\*M_{\widehat{\*F}}\*u_t \right\|&\leq \sup_{k_0\leq t\leq T-1}\left\|T^{-1/4}\sqrt{T}t^{-1}\*W_t'\*u_t \right\| + \sup_{k_0\leq t\leq T-1}\left\|T^{-1/4}\sqrt{T}t^{-1}\*W_t'\*P_{\widehat{\*F}}\*u_t \right\| \notag\\
    &\leq T^{-1/4} \sup_{k_0\leq t\leq T-1}\frac{T}{t}\sup_{k_0\leq t\leq T-1}\left\| \frac{1}{\sqrt{T}}\sum_{s=1}^{t-1}\*w_su_{s+1} \right\|\notag\\
    &+T^{-1/4} \sup_{k_0\leq t\leq T-1}\frac{T}{t}\sup_{k_0\leq t\leq T-1}\left\|t^{-1}\*W_t'\widehat{\*F}_t \right\|\sup_{k_0\leq t\leq T-1}\left\| \frac{1}{\sqrt{T}}\sum_{s=1}^{t-1}\widehat{\*f}_su_{s+1} \right\|\notag\\
    &\leq T^{-1/4}\frac{T}{k_0}\sup_{k_0\leq t\leq T-1}\left\| \frac{1}{\sqrt{T}}\sum_{s=1}^{t-1}\*w_su_{s+1} \right\|\notag\\
    &+T^{-1/4}\frac{T}{k_0}\sup_{k_0\leq t\leq T-1}\left\|t^{-1}\*W_t'\widehat{\*F}_t \right\|\sup_{k_0\leq t\leq T-1}\left\| \frac{1}{\sqrt{T}}\sum_{s=1}^{t-1}\widehat{\*f}_su_{s+1} \right\|\notag\\
    &=O_p(T^{-1/4}),
\end{align}
because $\{\widehat{\*f}_su_{s+1} \}_{s=1}^{t-1}$ is still an MDS process, because $\*u_t$ is mean-independent from $(Nt)^{-1}\*X_t\*X_t'$, and hence its eigenvectors. In particular, 
\begin{align}
    \mathbb{E}\left(\left\|\frac{1}{\sqrt{T}}\sum_{s=1}^{t-1}\widehat{\*f}_su_{s+1} \right\|^2 \right)&=\frac{1}{T}\sum_{s=1}^{t-1}\sum_{l=1}^{t-1}\mathbb{E}(u_{s+1}u_{l+1}\widehat{\*f}_s'\widehat{\*f}_l)\notag\\
    &=\frac{1}{T}\sum_{s=1}^{t-1}\mathbb{E}\left(u_{s+1}^2\widehat{\*f}_s'\widehat{\*f}_s \right)=\frac{t}{T}\mathbb{E}\left(\frac{1}{t}\sum_{s=1}^{t-1}\sigma^2_{t+1}\left\|\widehat{\*f}_s \right\|^2 \right)\notag\\
    &=\frac{t}{T}\mathbb{E}(\sigma^2_{t+1})\mathbb{E}\left(\frac{1}{t}\sum_{s=1}^{t-1}\left\|\widehat{\*f}_s \right\|^2 \right)\notag\\
    &=\sigma^2 \frac{t}{T}\mathbb{E}\left(\frac{1}{t}\sum_{s=1}^{t-1}\left\|\widehat{\*f}_s \right\|^2 \right)\notag\\
    &=tT^{-1} \sigma^2 r = s\sigma^2 r+O(T^{-1})
\end{align}
for $s\in (\pi_0, 1]$, and this results holds uniformly in $t$. Again, we were able to split the expectation, because $\sigma_t^2$ is independent of other model primitives, and hence of eigenvectors thereof. Therefore
\begin{align}
     \mathbb{E}\left(\sup_{k_0\leq t\leq T-1}\left\|\frac{1}{\sqrt{T}}\sum_{s=1}^{t-1}\widehat{\*f}_su_{s+1} \right\|^2 \right)=O(1).
\end{align}
Thus, 
\begin{align}
    \sup_{k_0\leq t\leq T-1}\|\*V \|=O_p(T^{-1/4}).
\end{align}
Ultimately, due to (\ref{hatFWF-FWF}), the numerator of $\mathbf{VI}_2$ follows 
\begin{align}
    &\sup_{k_0\leq t\leq T-1}\left\| T^{-1/4}\sqrt{T}\*H_{Nt,r}t^{-1}\*F_t'\*M_{\*W}\*u_t\right\|\notag\\
    &\leq    \sup_{k_0\leq t\leq T-1}\left\| T^{-1/4}\sqrt{T}\*H_{Nt,r}t^{-1}\*F_t'\*u_t \right\| +    \sup_{k_0\leq t\leq T-1}\left\| T^{-1/4}\sqrt{T}\*H_{Nt,r}t^{-1}\*F_t'\*P_\*W \*u_t\right\| \notag\\
    &\leq \sup_{k_0\leq t\leq T-1}\|\*H_{Nt,r} \|T^{-1/4}\sup_{k_0\leq t\leq T-1}\frac{T}{t}\sup_{k_0\leq t\leq T-1}\left\|\frac{1}{\sqrt{T}}\sum_{s=1}^{t-1}\*f_su_{s+1} \right\|\notag\\
    &+\sup_{k_0\leq t\leq T-1}\|\*H_{Nt,r} \|T^{-1/4}\sup_{k_0\leq t\leq T-1}\frac{T}{t}\sup_{k_0\leq t\leq T-1}\left\|t^{-1}\*W_t'\*F_t \right\|\sup_{k_0\leq t\leq T-1}\left\|(t^{-1}\*W_t'\*W_t)^{-1}\right\|\sup_{k_0\leq t\leq T-1}\left\| \frac{1}{\sqrt{T}}\sum_{s=1}^{t-1}\*w_su_{s+1} \right\|\notag\\
    &\leq \sup_{k_0\leq t\leq T-1}\|\*H_{Nt,r} \|T^{-1/4}\frac{T}{k_0}\sup_{k_0\leq t\leq T-1}\left\|\frac{1}{\sqrt{T}}\sum_{s=1}^{t-1}\*f_su_{s+1} \right\|\notag\\
    &+\sup_{k_0\leq t\leq T-1}\|\*H_{Nt,r} \|T^{-1/4}\frac{T}{k_0}\sup_{k_0\leq t\leq T-1}\left\|t^{-1}\*W_t'\*F_t \right\|\sup_{k_0\leq t\leq T-1}\left\|(t^{-1}\*W_t'\*W_t)^{-1}\right\|\sup_{k_0\leq t\leq T-1}\left\| \frac{1}{\sqrt{T}}\sum_{s=1}^{t-1}\*w_su_{s+1} \right\|\notag\\
    &= O_p(T^{-1/4}),
\end{align}
while the one of $\mathbf{VI}_1$ follows 
\begin{align}
    \sup_{k_0\leq t\leq T-1}&\left\| T^{-1/4}\sqrt{T}t^{-1}\widehat{\*F}_t'\*M_{\*W}\*u_t\right\|\notag\\
    &\leq   \sup_{k_0\leq t\leq T-1}\left\| T^{-1/4}\sqrt{T}t^{-1}\widehat{\*F}_t'\*u_t \right\| +   \sup_{k_0\leq t\leq T-1}\left\| T^{-1/4}\sqrt{T}t^{-1}\widehat{\*F}_t'\*P_\*W\*u_t\right\|\notag\\
    &\leq T^{-1/4} \sup_{k_0\leq t\leq T-1}\frac{T}{t}\sup_{k_0\leq t\leq T-1}\left\|\frac{1}{\sqrt{T}}\sum_{s=1}^{t-1}\widehat{\*f}_su_{s+1} \right\|\notag\\
    &+T^{-1/4} \sup_{k_0\leq t\leq T-1}\frac{T}{t}\sup_{k_0\leq t\leq T-1}\left\|t^{-1}\*W_t'\widehat{\*F}_t \right\|\sup_{k_0\leq t\leq T-1}\left\|(t^{-1}\*W_t'\*W_t)^{-1}\right\|\sup_{k_0\leq t\leq T-1}\left\| \frac{1}{\sqrt{T}}\sum_{s=1}^{t-1}\*w_su_{s+1} \right\|\notag\\
    &\leq T^{-1/4}\frac{T}{k_0}\sup_{k_0\leq t\leq T-1}\left\|\frac{1}{\sqrt{T}}\sum_{s=1}^{t-1}\widehat{\*f}_su_{s+1} \right\|\notag\\
    &+T^{-1/4}\frac{T}{k_0}\sup_{k_0\leq t\leq T-1}\left\|t^{-1}\*W_t'\widehat{\*F}_t \right\|\sup_{k_0\leq t\leq T-1}\left\|(t^{-1}\*W_t'\*W_t)^{-1}\right\|\sup_{k_0\leq t\leq T-1}\left\| \frac{1}{\sqrt{T}}\sum_{s=1}^{t-1}\*w_su_{s+1} \right\|\notag\\
    &= O_p(T^{-1/4}),
\end{align}
meaning that overall 
\begin{align}
    \sup_{k_0\leq t\leq T-1}\|\mathbf{VI} \|=O_p(T^{-1/4}).
\end{align}
Combining the orders of $\*I$ - $\mathbf{VI}$, implies that
\begin{align}
    \sup_{k_0\leq t\leq T-1}\left\|(\widehat{\+\delta}_t-(\+\Phi_{Nt,r}^{-1})'\widetilde{\+\delta}_t) \right\|=o_p(T^{-1/4}).
\end{align}
Because we will apply this result in sequences of Cauchy-Schwarz inequalities, we can combine the same terms $\*I$ - $\mathbf{VI}$ again. Then by the CMT and $(\sum_{j=1}^qa_j)^2\leq q \sum_{j=1}^qa_j^2$, we have that 
\begin{align}
    \sup_{k_0\leq t\leq T-1}\left\|T^{-1/4}\sqrt{T}(\widehat{\+\delta}_t-(\+\Phi_{Nt,r}^{-1})'\widetilde{\+\delta}_t) \right\|^2 =O_p\left(\frac{N^2}{N^{2\alpha}}\frac{1}{k_0} \right) + O_p(N^{1-2\alpha})+O_p(T^{-1/2}),
\end{align}
which completes the proof. \\

\noindent $\mathrm{(ii.)}$ The proof is nearly identical to part $\mathrm{(i.)}$, because the only change comes from applying the rates from part (b) of Lemma \ref{Lemma1}, Lemma \ref{Lemma2} and Lemma \ref{Lemma3} responsible for the heterogeneous weakness of factor loadings. 
\noindent \begin{lemma}\label{Lemma7}
Under Assumption A.1-A.4 with either homogeneous or heterogeneous loading weakness we have \begin{align*}
   &\mathrm{(i.)} \quad A= \frac{1}{\sqrt{d_T}}\sum_{t=\lfloor f_1(T)\rfloor }^{\lfloor f_2(T) \rfloor}(\widetilde{u}_{2,t+1}- \widehat{u}_{2,t+1})^2 =o_p(1),\\
   & \mathrm{(ii.)} \quad B = \frac{1}{\sqrt{d_T}}\sum_{t=\lfloor f_1(T)\rfloor }^{\lfloor f_2(T) \rfloor}\widetilde{u}_{1,t+1}(\widetilde{u}_{2,t+1}- \widehat{u}_{2,t+1}) =o_p(1),\\
   & \mathrm{(iii.)} \quad C =\frac{1}{\sqrt{d_T}}\sum_{t=\lfloor f_1(T)\rfloor }^{\lfloor f_2(T) \rfloor}\widetilde{u}_{2,t+1}(\widetilde{u}_{2,t+1}- \widehat{u}_{2,t+1})=o_p(1),\notag\\
   & \mathrm{(iv.)} \quad D=|\widetilde{\omega}^2_j-\widehat{\omega}^2_j | =o_p(1)\quad \text{for} \quad j=1,\ldots,4.
 \end{align*}
\end{lemma}
We will conduct the proof by assuming homogeneous $\alpha\in (0,1]$. The heterogeneous case is almost identical, where we apply the appropriate rate lemmas. \\

\noindent \textbf{Proof.} $\mathrm{(i.)}$  By using (\ref{u_diff_exp}) and $(a+b+c)^2\leq 3(a^2+b^2+c^2)$ we get 
\begin{align}
   A= \frac{1}{\sqrt{d_T}}\sum_{t=\lfloor f_1(T)\rfloor }^{\lfloor f_2(T) \rfloor}( \widetilde{u}_{2,t+1}- \widehat{u}_{2,t+1})^2&\leq  \frac{3}{\sqrt{d_T}}\sum_{t=\lfloor f_1(T)\rfloor }^{\lfloor f_2(T) \rfloor}\left[(\widehat{\*z}_t-\+\Phi_{Nt,r} \*z_t)'(\+\Phi_{Nt,r}^{-1})'(\widetilde{\+\delta}_t-\+\delta)\right]^2\notag\\
    &+ \frac{3}{\sqrt{d_T}}\sum_{t=\lfloor f_1(T)\rfloor }^{\lfloor f_2(T) \rfloor}\left[\widehat{\*z}_t'(\widehat{\+\delta}_t-(\+\Phi_{Nt,r}^{-1})'\widetilde{\+\delta}_t)\right]^2\notag\\
    & +\frac{3}{\sqrt{d_T}}\sum_{t=\lfloor f_1(T)\rfloor }^{\lfloor f_2(T) \rfloor}\left[(\widehat{\*f}_t-\*H_{Nt,r} \*f_t)'(\*H_{Nt,r}^{-1})'\+\beta^0 T^{-1/4}\right]^2\notag\\
    &=3(A1+A2+A3),
\end{align}
where $A_3$ is straightforward: 
\begin{align}
    A_3&\leq \sup_{k_0\leq t\leq T-1}\left\| \*H_{Nt,r}^{-1}\right\|^2 \left\| \+\beta^0\right\|^2\frac{1}{\sqrt{Td_T}}\sum_{t=\lfloor f_1(T)\rfloor }^{\lfloor f_2(T) \rfloor}\left\|\widehat{\*f}_t-\*H_{Nt,r} \*f_t \right\|^2 \notag\\
    &= \sup_{k_0\leq t\leq T-1}\left\| \*H_{Nt,r}^{-1}\right\|^2 \left\| \+\beta^0\right\|^2\sqrt{\frac{d_T}{T}}\frac{1}{d_T}\sum_{t=\lfloor f_1(T)\rfloor }^{\lfloor f_2(T) \rfloor}\left\|\widehat{\*f}_t-\*H_{Nt,r} \*f_t \right\|^2\notag\\
    &= O_p\left(\frac{N^2}{N^{2\alpha}}\frac{1}{T} \right) + O_p(N^{1-2\alpha}).
\end{align}
Next,
\begin{align}
    A_1&\leq  \sup_{k_0\leq t\leq T-1}\left\|\+\Phi_{Nt,r}^{-1} \right\|^2 \sup_{k_0\leq t\leq T-1}\left\|\widetilde{\+\delta}_t-\+\delta \right\|^2 \frac{1}{\sqrt{d_T}}\sum_{t=\lfloor f_1(T)\rfloor }^{\lfloor f_2(T) \rfloor}\left\| \widehat{\*z}_t-\+\Phi_{Nt,r} \*z_t\right\|^2\notag\\
    &\leq  \sup_{k_0\leq t\leq T-1}\left\|\+\Phi_{Nt,r}^{-1} \right\|^2 \sup_{k_0\leq t\leq T-1}\left\|\sqrt{T}(\widetilde{\+\delta}_t-\+\delta) \right\|^2 T^{-1/2}\sqrt{\frac{d_T}{T}}\frac{1}{d_T}\sum_{t=\lfloor f_1(T)\rfloor }^{\lfloor f_2(T) \rfloor}\left\|\widehat{\*f}_t-\*H_{Nt,r} \*f_t \right\|^2\notag\\
    &= O_p\left(\frac{N^2}{N^{2\alpha}}\frac{1}{T^{3/2}} \right) + O_p(N^{1-2\alpha}T^{-1/2}),
\end{align}
because 
\begin{align}
    \sup_{k_0\leq t\leq T-1}\left\|\sqrt{T}(\widetilde{\+\delta}_t-\+\delta) \right\|\leq \sup_{k_0\leq t\leq T-1}\left\| \left(\frac{1}{t}\sum_{s=1}^{t-1}\*z_s\*z_s' \right)^{-1}\right\| \frac{T}{k_0}\sup_{k_0\leq t\leq T-1}\left\|\frac{1}{\sqrt{T}}\sum_{s=1}^{t-1}\*z_su_{s+1} \right\|=O_p(1),
\end{align}
because by the MDS Functional Central Limit Theorem (FCLT) as $T\to \infty$
\begin{align}\label{FCLT_corradi}
    \sup_{k_0\leq t\leq T-1}\left\|\frac{1}{\sqrt{T}}\sum_{s=1}^{t-1}\*z_su_{s+1} \right\|\Rightarrow \sigma\sup_{s\in (\pi_0, 1)}\left\|\+\Sigma_\*z^{1/2}\*W(s) \right\|,
\end{align}
where $\*B(s)$ is a standard Brownian Motion (see a similar argument in Lemma 2.1 of \citealp{corradi2001predictive}). Lastly, 
\begin{align}
    A_2&\leq \sup_{k_0\leq t\leq T-1}\left\|\sqrt{T}(\widehat{\+\delta}_t-(\+\Phi_{Nt,r}^{-1})'\widetilde{\+\delta}_t) \right\|^2\frac{1}{T\sqrt{d_T}}\sum_{t=\lfloor f_1(T)\rfloor }^{\lfloor f_2(T) \rfloor}\left\| \widehat{\*z}_t \right\|^2\notag\\
    &\leq \sup_{k_0\leq t\leq T-1}\left\|\sqrt{T}(\widehat{\+\delta}_t-(\+\Phi_{Nt,r}^{-1})'\widetilde{\+\delta}_t) \right\|^2\frac{1}{\sqrt{d_T}}\frac{1}{T}\sum_{t=1 }^{T}\left\| \widehat{\*z}_t \right\|^2\notag\\
    &= \sqrt{\frac{T}{d_T}}\sup_{k_0\leq t\leq T-1}\left\|T^{-1/4}\sqrt{T}(\widehat{\+\delta}_t-(\+\Phi_{Nt,r}^{-1})'\widetilde{\+\delta}_t) \right\|^2 \frac{1}{T}\sum_{t=1 }^{T}\left\| \widehat{\*z}_t \right\|^2\notag\\
    &=O_p\left(\frac{N^2}{N^{2\alpha}}\frac{1}{k_0} \right) + O_p(N^{1-2\alpha})+O_p(T^{-1/2}),
\end{align}
which is driven by the supremum component. 
Because $k_0<T$ in finite samples, we have that overall 
\begin{align}
    |A|=O_p\left(\frac{N^2}{N^{2\alpha}}\frac{1}{k_0} \right) + O_p(N^{1-2\alpha})+O_p(T^{-1/2}),
\end{align}
which completes $\mathrm{(i.)}$. \\

\noindent We move to $\mathrm{(ii.)}$. We have 
\begin{align}
    B&=\frac{1}{\sqrt{d_T}}\sum_{t=\lfloor f_1(T)\rfloor }^{\lfloor f_2(T) \rfloor}\widetilde{u}_{1,t+1}(\widetilde{u}_{2,t+1}- \widehat{u}_{2,t+1}) \notag\\
    &= \frac{1}{\sqrt{d}_T}\sum_{t=\lfloor f_1(T)\rfloor }^{\lfloor f_2(T) \rfloor} (u_{t+1}+\+\beta'\*f_t-(\widetilde{\+\theta}_t-\+\theta)'\*w_t)(\widetilde{u}_{2,t+1}- \widehat{u}_{2,t+1})\notag\\
    &= \frac{1}{\sqrt{d}_T}\sum_{t=\lfloor f_1(T)\rfloor }^{\lfloor f_2(T) \rfloor} u_{t+1}(\widetilde{u}_{2,t+1}- \widehat{u}_{2,t+1}) +  \+\beta'\frac{1}{\sqrt{d}_T}\sum_{t=\lfloor f_1(T)\rfloor }^{\lfloor f_2(T) \rfloor}\*f_t(\widetilde{u}_{2,t+1}- \widehat{u}_{2,t+1})\notag\\
    &- \frac{1}{\sqrt{d}_T}\sum_{t=\lfloor f_1(T)\rfloor }^{\lfloor f_2(T) \rfloor}(\widetilde{\+\theta}_t-\+\theta)'\*w_t(\widetilde{u}_{2,t+1}- \widehat{u}_{2,t+1})=B1+B2-B3,
\end{align}
where we begin with $B2$ and $B3$. In particular,
\begin{align}
    \left|B2 \right|&=\left|T^{-1/4}\+\beta^0\frac{1}{\sqrt{d}_T} \sum_{t=\lfloor f_1(T)\rfloor }^{\lfloor f_2(T) \rfloor}\*f_t(\widetilde{u}_{2,t+1}- \widehat{u}_{2,t+1})\right|\notag\\
    &\leq \left\| \+\beta^0\right\|T^{-1/4}\frac{1}{\sqrt{d}_T}\sum_{t=\lfloor f_1(T)\rfloor }^{\lfloor f_2(T) \rfloor}\left\|\*f_t \right\| |\widetilde{u}_{2,t+1}- \widehat{u}_{2,t+1}| \notag\\
    &\leq \left\| \+\beta^0\right\|\left(\frac{1}{d_T}\sum_{t=\lfloor f_1(T)\rfloor }^{\lfloor f_2(T) \rfloor}\left\|\*f_t \right\|^2 \right)^{1/2} \left(\frac{1}{\sqrt{T}}\sum_{t=\lfloor f_1(T)\rfloor }^{\lfloor f_2(T) \rfloor}(\widetilde{u}_{2,t+1}- \widehat{u}_{2,t+1})^2 \right)^{1/2}\notag\\
    &\leq  \left\| \+\beta^0\right\|\left(\frac{T}{d_T}\frac{1}{T}\sum_{t=1 }^{T-1 }\left\|\*f_t \right\|^2 \right)^{1/2} \left(\sqrt{\frac{d_T}{T}}\frac{1}{\sqrt{d}_T}\sum_{t=\lfloor f_1(T)\rfloor }^{\lfloor f_2(T) \rfloor}(\widetilde{u}_{2,t+1}- \widehat{u}_{2,t+1})^2 \right)^{1/2}\notag\\
    &= \left(\frac{d_T}{T} \right)^{-1/4}\left\| \+\beta^0\right\| \left(\frac{1}{T}\sum_{t=1 }^{T-1 }\left\|\*f_t \right\|^2 \right)^{1/2} \left(\underbrace{\frac{1}{\sqrt{d}_T}\sum_{t=\lfloor f_1(T)\rfloor }^{\lfloor f_2(T) \rfloor}(\widetilde{u}_{2,t+1}- \widehat{u}_{2,t+1})^2}_{A} \right)^{1/2}\\
    &=O_p\left(\frac{N}{N^{\alpha}}\frac{1}{\sqrt{k_0}} \right) + O_p(N^{(1-2\alpha)/2})+O_p(T^{-1/4}),
\end{align} 
where $\frac{d_T}{T}=q_2-q_1+O(T^{-1})$, and a similar bound holds for $B3$: 
\begin{align}
    &|B3 | \leq \sup_{k_0\leq t\leq T-1}\left\|\sqrt{T}(\widetilde{\+\theta}_t-\+\theta) \right\|\frac{1}{\sqrt{T}}\frac{1}{\sqrt{d}_T}\sum_{t=\lfloor f_1(T)\rfloor }^{\lfloor f_2(T) \rfloor}\left\|\*w_t \right\| | \widetilde{u}_{2,t+1}- \widehat{u}_{2,t+1}| \notag\\
    &\leq  \sup_{k_0\leq t\leq T-1}\left\|T^{-1/4}\sqrt{T}(\widetilde{\+\theta}_t-\+\theta) \right\| \frac{1}{T^{1/4}}\frac{1}{\sqrt{d_T}} \left( \sum_{t=\lfloor f_1(T)\rfloor }^{\lfloor f_2(T) \rfloor}\left\| \*w_t\right\|^2\right)^{1/2}\left(\sum_{t=\lfloor f_1(T)\rfloor }^{\lfloor f_2(T) \rfloor}( \widetilde{u}_{2,t+1}- \widehat{u}_{2,t+1})^2 \right)^{1/2}\notag\\
    &\leq \sup_{k_0\leq t\leq T-1}\left\|T^{-1/4}\sqrt{T}(\widetilde{\+\theta}_t-\+\theta) \right\| \left(\frac{d_T}{T} \right)^{1/4}\left(\frac{T}{d_T} \right)^{1/2}\left( \frac{1}{T}\sum_{t=1 }^{T-1}\left\| \*w_t\right\|^2\right)^{1/2}\left(\frac{1}{\sqrt{d_T}}\sum_{t=\lfloor f_1(T)\rfloor }^{\lfloor f_2(T) \rfloor}( \widetilde{u}_{2,t+1}- \widehat{u}_{2,t+1})^2 \right)^{1/2}\notag\\
    &= \left(\frac{d_T}{T} \right)^{-1/4}\sup_{k_0\leq t\leq T-1}\left\|T^{-1/4}\sqrt{T}(\widetilde{\+\theta}_t-\+\theta) \right\|\left( \frac{1}{T}\sum_{t=1 }^{T-1}\left\| \*w_t\right\|^2\right)^{1/2}\left(\underbrace{\frac{1}{\sqrt{d_T}}\sum_{t=\lfloor f_1(T)\rfloor }^{\lfloor f_2(T) \rfloor}( \widetilde{u}_{2,t+1}- \widehat{u}_{2,t+1})^2}_{A} \right)^{1/2}\notag\\
    &=O_p\left(\frac{N}{N^{\alpha}}\frac{1}{\sqrt{k_0}} \right) + O_p(N^{(1-2\alpha)/2})+O_p(T^{-1/4}),
\end{align}
where 
\begin{align}
    \sup_{k_0\leq t\leq T-1}\left\|T^{-1/4}\sqrt{T}(\widetilde{\+\theta}_t-\+\theta) \right\|&\leq\sup_{k_0\leq t\leq T-1}\left\| \left(\frac{1}{t}\sum_{s=1}^{t-1}\*w_s\*w_s'\right)^{-1}\frac{1}{t}\sum_{s=1}^{t-1}\*w_s\*f_s'\+\beta^0\right\|\notag\\
    &+T^{-1/4}\sup_{k_0\leq t\leq T-1}\left\|\frac{T}{t} \left(\frac{1}{t}\sum_{s=1}^{t-1}\*w_s\*w_s'\right)^{-1}\frac{1}{\sqrt{T}}\sum_{s=1}^{t-1}\*w_su_{s+1} \right\|\notag\\
    &\leq \sup_{k_0\leq t\leq T-1} \left\|\left(\frac{1}{t}\sum_{s=1}^{t-1}\*w_s\*w_s'\right)^{-1} \right\|\sup_{k_0\leq t\leq T-1}\left\|\frac{1}{t}\sum_{s=1}^{t-1}\*w_s\*f_s' \right\|\left\|\+\beta^0 \right\|\notag\\
    &+T^{-1/4} \frac{T}{k_0}\sup_{k_0\leq t\leq T-1} \left\|\left(\frac{1}{t}\sum_{s=1}^{t-1}\*w_s\*w_s'\right)^{-1} \right\| \sup_{k_0\leq t\leq T-1}\left\|\frac{1}{\sqrt{T}}\sum_{s=1}^{t-1}\*w_su_{s+1} \right\|\notag\\
    &=O_p(1),
\end{align}
because by the MDS Functional Central Limit Theorem (FCLT) as $T\to \infty$
\begin{align}
    \sup_{k_0\leq t\leq T-1}\left\|\frac{1}{\sqrt{T}}\sum_{s=1}^{t-1}\*w_su_{s+1} \right\|\Rightarrow \sigma\sup_{s\in (\pi_0, 1)}\left\|\+\Sigma_\*w^{1/2}\*B(s) \right\|,
\end{align}
where $\*B(s)$ is a standard Brownian Motion (see a similar argument in Lemma 2.1 of \citealp{corradi2001predictive}). Next is $B_1$, which follows a martingale structure and will require a more involved analysis. Particularly, 
\begin{align}
    B_1&= \frac{1}{\sqrt{d}_T}\sum_{t=\lfloor f_1(T)\rfloor }^{\lfloor f_2(T) \rfloor} u_{t+1}(\widetilde{u}_{2,t+1}- \widehat{u}_{2,t+1})\notag\\
    &= \frac{1}{\sqrt{d}_T}\sum_{t=\lfloor f_1(T)\rfloor }^{\lfloor f_2(T) \rfloor}u_{t+1}(\widehat{\*z}_t-\+\Phi_{Nt,r} \*z_t)'(\+\Phi_{Nt,r}^{-1})'(\widetilde{\+\delta}_t-\+\delta)+\frac{1}{\sqrt{d}_T}\sum_{t=\lfloor f_1(T)\rfloor }^{\lfloor f_2(T) \rfloor}u_{t+1}\widehat{\*z}_t'(\widehat{\+\delta}_t-(\+\Phi_{Nt,r}^{-1})'\widetilde{\+\delta}_t)\notag\\
    &+ \frac{1}{\sqrt{d}_T}\sum_{t=\lfloor f_1(T)\rfloor }^{\lfloor f_2(T) \rfloor}u_{t+1}(\widehat{\*f}_t-\*H_{Nt,r} \*f_t)'(\*H_{Nt,r}^{-1})'\+\beta^0 T^{-1/4}\notag\\
    &=B_{11}+B_{12}+B_{13},
\end{align}
where
\begin{align}
    |B_{11}|&\leq \frac{1}{\sqrt{d}_T}\sum_{t=\lfloor f_1(T)\rfloor }^{\lfloor f_2(T) \rfloor}|u_{t+1}|\left\|(\widehat{\*z}_t-\+\Phi_{Nt,r} \*z_t)\right\|\left\|(\+\Phi_{Nt,r}^{-1})\right\|\left\|(\widetilde{\+\delta}_t-\+\delta)\right\|\notag\\
    &\leq \left(\frac{d_T}{T} \right)^{-1/2}\sup_{k_0\leq t\leq T-1}\left\|\sqrt{T}(\widetilde{\+\delta}_t-\+\delta) \right\|\sup_{k_0\leq t\leq T-1}\left\|(\+\Phi_{Nt,r}^{-1}) \right\|\frac{1}{T}\sum_{t=\lfloor f_1(T)\rfloor }^{\lfloor f_2(T) \rfloor}|u_{t+1}|\left\|(\widehat{\*z}_t-\+\Phi_{Nt,r} \*z_t)\right\|\notag\\
    &= \left(\frac{d_T}{T} \right)^{-1/2}\sup_{k_0\leq t\leq T-1}\left\|\sqrt{T}(\widetilde{\+\delta}_t-\+\delta) \right\|\sup_{k_0\leq t\leq T-1}\left\|(\+\Phi_{Nt,r}^{-1}) \right\|\left(\frac{d_T}{T}\right)\frac{1}{d_T}\sum_{t=\lfloor f_1(T)\rfloor }^{\lfloor f_2(T) \rfloor}|u_{t+1}|\left\|(\widehat{\*z}_t-\+\Phi_{Nt,r} \*z_t)\right\|\notag\\
    &\leq  \left(\frac{d_T}{T} \right)^{1/2} \sup_{k_0\leq t\leq T-1}\left\|\sqrt{T}(\widetilde{\+\delta}_t-\+\delta) \right\|\sup_{k_0\leq t\leq T-1}\left\|(\+\Phi_{Nt,r}^{-1}) \right\|\left(\frac{1}{d_T}\sum_{t=\lfloor f_1(T)\rfloor }^{\lfloor f_2(T) \rfloor}u^2_{t+1} \right)^{1/2}\notag\\
    &\times \left(\frac{1}{d_T}\sum_{t=\lfloor f_1(T)\rfloor }^{\lfloor f_2(T) \rfloor}\left\|(\widehat{\*z}_t-\+\Phi_{Nt,r} \*z_t)\right\|^2\right)^{1/2}\notag\\
    &=  O_p\left(\frac{N}{N^{\alpha}}\frac{1}{\sqrt{k_0}} \right) + O_p(N^{1/2-\alpha}),
\end{align}
because $\left\|(\widehat{\*z}_t-\+\Phi_{Nt,r} \*z_t)\right\|^2=\left\|(\widehat{\*f}_t-\*H_{Nt,r} \*f_t)\right\|^2$. Next, we move to $A_{13}$. Let $\*H_r$ be the probability limit of $\*H_{Nt,r}$, and so 
\begin{align}\label{B13}
    |B_{13}|&\leq \left|\frac{1}{\sqrt{d}_T}\sum_{t=\lfloor f_1(T)\rfloor }^{\lfloor f_2(T) \rfloor}u_{t+1}(\widehat{\*f}_t-\*H_{Nt,r} \*f_t)'(\*H_{r}^{-1})'\+\beta^0 T^{-1/4}\right|\notag\\
    &+\left|\frac{1}{\sqrt{d}_T}\sum_{t=\lfloor f_1(T)\rfloor }^{\lfloor f_2(T) \rfloor}u_{t+1}(\widehat{\*f}_t-\*H_{Nt,r} \*f_t)'(\*H_{Nt,r}^{-1}-\*H_{r}^{-1})'\+\beta^0 T^{-1/4} \right|\notag\\
    &=|B_{13a}| + |B_{13b}|,
\end{align}
where by using the definition of $\widehat{\*f}_t-\*H_{Nt,r} \*f_t$, we obtain 
\begin{align}
    |B_{13a}|&\leq\left|\frac{1}{T^{1/4}}\frac{1}{\sqrt{d}_T}\sum_{t=\lfloor f_1(T)\rfloor }^{\lfloor f_2(T) \rfloor}\left(\frac{1}{t}\sum_{l=1}^t\widehat{\*f}_l\gamma_{l,t,\alpha} \right)'\left(\frac{N}{N^\alpha}\*D_{Nt,r}^{2}\right)^{-1}(\*H_{r}^{-1})'\+\beta^0u_{t+1}\right|\notag\\
    &+ \left|\frac{1}{T^{1/4}}\frac{1}{\sqrt{d}_T}\sum_{t=\lfloor f_1(T)\rfloor }^{\lfloor f_2(T) \rfloor}\left( \frac{1}{t}\sum_{l=1}^t\widehat{\*f}_l\xi_{l,t,\alpha} \right)'\left(\frac{N}{N^\alpha}\*D_{Nt,r}^{2}\right)^{-1}(\*H_{r}^{-1})'\+\beta^0u_{t+1}\right|\notag\\
    &+  \left|\frac{1}{T^{1/4}}\frac{1}{\sqrt{d}_T}\sum_{t=\lfloor f_1(T)\rfloor }^{\lfloor f_2(T) \rfloor}\left( \frac{1}{t}\sum_{l=1}^t\widehat{\*f}_l\eta_{l,t,\alpha} \right)'\left(\frac{N}{N^\alpha}\*D_{Nt,r}^{2}\right)^{-1}(\*H_{r}^{-1})'\+\beta^0u_{t+1}\right|\notag\\
    &+ \left|\frac{1}{T^{1/4}}\frac{1}{\sqrt{d}_T}\sum_{t=\lfloor f_1(T)\rfloor }^{\lfloor f_2(T) \rfloor}\left( \frac{1}{t}\sum_{l=1}^t\widehat{\*f}_l \nu_{l,t,\alpha} \right)'\left(\frac{N}{N^\alpha}\*D_{Nt,r}^{2}\right)^{-1}(\*H_{r}^{-1})'\+\beta^0u_{t+1}\right|\notag\\
    &=|B_{13aa}|+|B_{13ab}|+|B_{13ac}|+|B_{13ad}|.
\end{align}

\noindent Subsequently, we will invoke \hyperref[A4]{A.4} in the further derivations. Recall that we impose $\frac{N}{N^\alpha}\frac{1}{T^{1/4}}\to c>0$ as $(N,T)\to \infty$. By using this, we get that $\frac{N}{N^\alpha }\frac{1}{\sqrt{T}}=O(T^{-1/4})$. In addition, we will require that $N^{(1-2\alpha)}k_0^{1/4}=o(1)$. Clearly, $N^{(1-2\alpha)}k_0^{1/4}=\left(\frac{N}{N^\alpha}\frac{1}{k_0^{1/4}}\right)\frac{\sqrt{k_0}}{N^{\alpha}}=o(1)$ if $\frac{\sqrt{k_0}}{N^\alpha}=o(1)$ under our assumptions. Note that the latter restriction resembles $\sqrt{T}N^{-1}=o(1)$ in \cite{gonccalves2017tests}. It is stronger because of possibly weaker loadings, but both are identical under $\alpha=1$.\\

\noindent We will obtain the rates one-by-one. To begin with, 
\begin{align}
    |B_{13aa}|&=\left|\frac{1}{T^{1/4}}\frac{1}{\sqrt{d}_T}\sum_{t=\lfloor f_1(T)\rfloor }^{\lfloor f_2(T) \rfloor}\left(\frac{1}{t}\sum_{l=1}^t\widehat{\*f}_l\gamma_{l,t,\alpha} \right)'\left(\frac{N}{N^\alpha}\*D_{Nt,r}^{2}\right)^{-1}(\*H_{r}^{-1})'\+\beta^0u_{t+1}\right|\notag\\
    &\leq \left|\frac{1}{T^{1/4}}\frac{1}{\sqrt{d}_T}\sum_{t=\lfloor f_1(T)\rfloor }^{\lfloor f_2(T) \rfloor}\left(\frac{1}{t}\sum_{l=1}^t\*H_r\*f_l\gamma_{l,t,\alpha} \right)'\left(\frac{N}{N^\alpha}\*D_{Nt,r}^{2}\right)^{-1}(\*H_{r}^{-1})'\+\beta^0u_{t+1}\right|\notag\\
    &+\left|\frac{1}{T^{1/4}}\frac{1}{\sqrt{d}_T}\sum_{t=\lfloor f_1(T)\rfloor }^{\lfloor f_2(T) \rfloor}\left(\frac{1}{t}\sum_{l=1}^t(\widehat{\*f}_l-\*H_r\*f_l)\gamma_{l,t,\alpha} \right)'\left(\frac{N}{N^\alpha}\*D_{Nt,r}^{2}\right)^{-1}(\*H_{r}^{-1})'\+\beta^0u_{t+1}\right|\notag\\
    &= |B_{13aai}|+ |B_{13aaii}|,
\end{align}
where, by using 
\begin{align*}
    \frac{1}{d_T}&\sum_{t=\lfloor f_1(T)\rfloor }^{\lfloor f_2(T) \rfloor}\left\| \frac{1}{t}\sum_{l=1}^t(\widehat{\*f}_l-\*H_{r}\*f_l)\gamma_{l,t,\alpha} \right\|^2\leq \frac{1}{d_T}\sum_{t=\lfloor f_1(T)\rfloor }^{\lfloor f_2(T) \rfloor} \left(\frac{1}{t}\sum_{l=1}^t\left\|(\widehat{\*f}_l-\*H_{r}\*f_l)\right\||\gamma_{l,t,\alpha}| \right)^2\\
    &\leq \frac{1}{d_T}\sum_{t=\lfloor f_1(T)\rfloor }^{\lfloor f_2(T) \rfloor} \left(\frac{1}{t}\sum_{l=1}^t\left\|(\widehat{\*f}_l-\*H_{r}\*f_l)\right\|^2 \right)\left(\frac{1}{t}\sum_{l=1}^t\gamma_{l,t,\alpha}^2 \right)\leq \sup_{k_0\leq t\leq T-1}\frac{1}{t}\sum_{l=1}^t\left\|(\widehat{\*f}_l-\*H_{r}\*f_l)\right\|^2 \times \frac{1}{d_Tk_0}\sum_{t=\lfloor f_1(T)\rfloor }^{\lfloor f_2(T) \rfloor}\sum_{l=1}^T\gamma_{l,t,\alpha}^2
\end{align*}
we get 
\begin{align}\label{B13aaii}
    |B_{13aaii}|&\leq d_T^{1/4}\left(\frac{d_T}{T} \right)^{1/4}\left\|\*H_{r}^{-1}\right\|\left\|\+\beta^0 \right\|\sup_{k_0\leq t\leq T-1}\left\|\left(\frac{N}{N^\alpha}\*D_{Nt,r}^{2}\right)^{-1} \right\|\frac{1}{d_T}\sum_{t=\lfloor f_1(T)\rfloor }^{\lfloor f_2(T) \rfloor}\left\| \frac{1}{t}\sum_{l=1}^t(\widehat{\*f}_l-\*H_{r}\*f_l)\gamma_{l,t,\alpha} \right\||u_{t+1}|\notag\\
    &\leq d_T^{1/4}\left(\frac{d_T}{T} \right)^{1/4}\left\|\*H_{r}^{-1}\right\|\left\|\+\beta^0 \right\|\sup_{k_0\leq t\leq T-1}\left\|\left(\frac{N}{N^\alpha}\*D_{Nt,r}^{2}\right)^{-1} \right\|\left(\frac{1}{d_T}\sum_{t=\lfloor f_1(T)\rfloor }^{\lfloor f_2(T) \rfloor}\left\| \frac{1}{t}\sum_{l=1}^t(\widehat{\*f}_l-\*H_{r}\*f_l)\gamma_{l,t,\alpha} \right\|^2 \right)^{1/2}\notag\\
    &\times \left(\frac{1}{d_T}\sum_{t=\lfloor f_1(T)\rfloor }^{\lfloor f_2(T) \rfloor}u_{t+1}^2 \right)^{1/2}\notag\\
    &\leq O_p(d_T^{1/4})\times \left(\frac{1}{d_T}\sum_{t=\lfloor f_1(T)\rfloor }^{\lfloor f_2(T) \rfloor}u_{t+1}^2 \right)^{1/2} \left(\sup_{k_0\leq t\leq T-1}\frac{1}{t}\sum_{l=1}^t\left\|\widehat{\*f}_l-\*H_{r}\*f_l \right\|^2 \right)^{1/2}\notag\\
    &\times \left(\frac{N^2}{N^{2\alpha}}\frac{1}{d_Tk_0} \sum_{t=\lfloor f_1(T)\rfloor }^{\lfloor f_2(T) \rfloor}\sum_{l=1}^T\gamma_{l,t}^2\right)^{1/2}\notag\\
    &= O_p(d_T^{1/4})\times \left(O_p\left(\frac{N}{N^\alpha}\frac{1}{\sqrt{k_0}} \right)+O_p(N^{(1-2\alpha)/2})\right)\times  O_p(N^{(1-\alpha)}T^{-1/2})\notag\\
    &=  \left(O_p\left(\frac{N}{N^\alpha}\frac{1}{\sqrt{k_0}} \right)+O_p(N^{(1-2\alpha)/2})\right)\times  O_p(N^{(1-\alpha)}T^{-1/4})=o_p(1),
\end{align}
since $N^{(1-\alpha)}T^{-1/4}=O(1)$ by assumption. Note that the rate also comes from 
\begin{align}
    \frac{1}{d_Tk_0} \sum_{t=\lfloor f_1(T)\rfloor }^{\lfloor f_2(T) \rfloor}\sum_{l=1}^T\gamma_{l,t}^2\leq \frac{T^2}{d_Tk_0}\frac{
1}{T^2}\sum_{t=1 }^{T}\sum_{l=1}^T\gamma_{l,t}^2=O(T^{-1})
\end{align}
and, based on the results in Lemma \ref{Lemma2} and Lemma \ref{Lemma3}
\begin{align}
    \sup_{k_0\leq t\leq T-1}\frac{1}{t}\sum_{l=1}^t\left\|\widehat{\*f}_l-\*H_{r}\*f_l \right\|^2 &=\sup_{k_0\leq t\leq T-1} \frac{1}{t}\sum_{l=1}^t\left\|\widehat{\*f}_l-\*H_{Nt,r}\*f_l+(\*H_{Nt,r}-\*H_{r})\*f_l \right\|^2\notag\\
    &\leq 2\sup_{k_0\leq t\leq T-1} \frac{1}{t}\sum_{l=1}^t\left\|\widehat{\*f}_l-\*H_{Nt,r}\*f_l\right\|^2 +2\sup_{k_0\leq t\leq T-1}\left\|\*H_{Nt,r}-\*H_{r} \right\|^2\frac{T}{k_0}\frac{1}{T}\sum_{l=1}^t\left\|\*f_l \right\|^2\notag\\
    &= O_p\left(\frac{N^2}{N^{2\alpha}}\frac{1}{k_0} \right) + O_p(N^{1-2\alpha}).
\end{align}
Up next, 
\begin{align}
    B_{13aai}&=\frac{1}{T^{1/4}}\frac{1}{\sqrt{d}_T}\sum_{t=\lfloor f_1(T)\rfloor }^{\lfloor f_2(T) \rfloor}\left(\frac{1}{t}\sum_{l=1}^t\*H_r\*f_l\gamma_{l,t,\alpha} \right)'\left(\frac{N}{N^\alpha}\*D_{Nt,r}^{2}\right)^{-1}(\*H_{r}^{-1})'\+\beta^0u_{t+1}\notag\\
    &=\frac{1}{T^{1/4}}\frac{1}{\sqrt{d}_T}\sum_{t=\lfloor f_1(T)\rfloor }^{\lfloor f_2(T) \rfloor}\left(\frac{1}{t}\sum_{l=1}^t\*H_r\*f_l\gamma_{l,t,\alpha} \right)'\+\Sigma_{\+\Lambda}^{-1}(\*H_{r}^{-1})'\+\beta^0u_{t+1}\notag\\
    &+\frac{1}{T^{1/4}}\frac{1}{\sqrt{d}_T}\sum_{t=\lfloor f_1(T)\rfloor }^{\lfloor f_2(T) \rfloor}\left(\frac{1}{t}\sum_{l=1}^t\*H_r\*f_l\gamma_{l,t,\alpha} \right)'\left[\left(\frac{N}{N^\alpha}\*D_{Nt,r}^{2}\right)^{-1}-\+\Sigma_{\+\Lambda}^{-1}\right](\*H_{r}^{-1})'\+\beta^0u_{t+1}\notag\\
    &=  B_{13aai,\Lambda}+ B_{13aai,D-\Lambda},
\end{align}
such that, by letting $\+\Sigma_{\+\Lambda}^{-1}(\*H_{r}^{-1})'\+\beta^0=\*q$ and using the fact that $\+\gamma_{l,t,\alpha}$ is non-random, we obtain
\begin{align}
\mathbb{V}ar(B_{13aai,\Lambda})&=\mathbb{E}\left(\left\|\frac{1}{T^{1/4}}\frac{1}{\sqrt{d}_T}\sum_{t=\lfloor f_1(T)\rfloor }^{\lfloor f_2(T) \rfloor}\left(\frac{1}{t}\sum_{l=1}^t\*H_r\*f_l\gamma_{l,t,\alpha} \right)'\*qu_{t+1} \right\|^2 \right)\notag\\
&=\mathbb{E}\left(\frac{1}{\sqrt{T}}\frac{1}{d_T}\sum_{t=\lfloor f_1(T)\rfloor }^{\lfloor f_2(T) \rfloor}\sum_{s=\lfloor f_1(T)\rfloor }^{\lfloor f_2(T) \rfloor}\left(\frac{1}{t}\sum_{l=1}^t\*H_r\*f_l\gamma_{l,t,\alpha} \right)'\*qu_{t+1}u_{s+1}\*q'\left(\frac{1}{s}\sum_{j=1}^s\*H_r\*f_j\gamma_{j,s,\alpha} \right) \right)\notag\\
     &= \frac{1}{\sqrt{T}}\frac{1}{d_T}\sum_{t=\lfloor f_1(T)\rfloor }^{\lfloor f_2(T) \rfloor}\mathbb{E}\left(\sigma^2_{t+1}\left[\left(\frac{1}{t}\sum_{l=1}^t\*H_{r}\*f_l\gamma_{l,t,\alpha}\right)' \*q \right]^2\right)\notag\\
    &\leq  \left\|\*q \right\|^2\frac{1}{\sqrt{T}}\frac{1}{d_T}\sum_{t=\lfloor f_1(T)\rfloor }^{\lfloor f_2(T) \rfloor}\mathbb{E}(\sigma^2_{t+1})\mathbb{E}\left(\left\| \frac{1}{t}\sum_{l=1}^t\*H_{r}\*f_l\gamma_{l,t,\alpha}\right\|^2 \right)\notag\\
    &\leq \left\|\*q \right\|^2\left\| \*H_r\right\|^2\frac{1}{\sqrt{T}}\frac{1}{d_T}\sum_{t=\lfloor f_1(T)\rfloor }^{\lfloor f_2(T) \rfloor}\sigma^2\mathbb{E}\left(\frac{1}{t}\sum_{l=1}^t\left\| \*f_l\right\|^2\left| \gamma_{l,t,\alpha}\right|^2 \right)\notag\\
    &=  \left\|\*q \right\|^2\left\| \*H_r\right\|^2\frac{1}{\sqrt{T}}\frac{1}{d_T}\sum_{t=\lfloor f_1(T)\rfloor }^{\lfloor f_2(T) \rfloor}\sigma^2\frac{1}{t}\sum_{l=1}^t\mathbb{E}(\left\| \*f_l\right\|^2) \gamma_{l,t,\alpha}^2 \notag\\
    &\leq\sigma^2\left\|\*q \right\|^2\left\| \*H_r\right\|^2 \sup_{k_0\leq l \leq T-1}\mathbb{E}(\left\| \*f_l\right\|^2) \frac{N^2}{N^{2\alpha}}\frac{1}{\sqrt{T}}\frac{T^2}{d_Tk_0}\frac{1}{T^2}\sum_{t=1 }^{T}\sum_{l=1}^T\gamma_{l,t}^2\notag\\
    &= O\left(\frac{N^2}{N^{2\alpha}}\frac{1}{\sqrt{T}}\frac{1}{T} \right)=O(T^{-1}),
\end{align}
meaning that $|B_{13aai,\Lambda}|=O_p(T^{-1/2})$. Lastly, by using the same chain of inequalities as in (\ref{B13aaii}),
\begin{align}
    |B_{13aai,D-\Lambda}|&\leq T^{1/4} \left(\frac{d_T}{T}\right)^{1/2}\left\| \*H_r^{-1}\right\|\left\| \*H_{r}\right\|\left\|\+\beta_0 \right\| \sup_{k_0\leq t\leq T-1} \left\|\left(\frac{N}{N^\alpha}\*D_{Nt,r}^{2}\right)^{-1}-\+\Sigma_{\+\Lambda}^{-1} \right\|\left(\sup_{k_0\leq t\leq T-1}\frac{1}{t}\sum_{l=1}^t\left\|\*f_l \right\|^2 \right)^{1/2}\notag\\
    &\times \left(\frac{N^2}{N^{2\alpha}}\frac{1}{d_Tk_0} \sum_{t=\lfloor f_1(T)\rfloor }^{\lfloor f_2(T) \rfloor}\sum_{l=1}^T\gamma^2_{l,t}\right)^{1/2} \left( \frac{1}{d_T}\sum_{t=\lfloor f_1(T)\rfloor }^{\lfloor f_2(T) \rfloor}u_{t+1}^2\right)^{1/2}\notag\\
    &= T^{1/4}O_p\left(\frac{N}{N^{\alpha}}\frac{1}{\sqrt{T}} \right)\times \left( O_p\left(\frac{N}{N^\alpha}\frac{1}{\sqrt{k_0}} \right)+ O_p(N^{(1-2\alpha)/2})\right)\notag\\
    &= O_p\left(\frac{N}{N^\alpha}\frac{1}{\sqrt{k_0}} \right)+ O_p(N^{(1-2\alpha)/2})
\end{align}
when $\frac{N}{N^{\alpha}}\frac{1}{T^{1/4}}\to c>0$. We move on to 
\begin{align}
   |B_{13ab}|&= \left|\frac{1}{T^{1/4}}\frac{1}{\sqrt{d}_T}\sum_{t=\lfloor f_1(T)\rfloor }^{\lfloor f_2(T) \rfloor}\left( \frac{1}{t}\sum_{l=1}^t\widehat{\*f}_l\xi_{l,t,\alpha} \right)'\left(\frac{N}{N^\alpha}\*D_{Nt,r}^{2}\right)^{-1}(\*H_{r}^{-1})'\+\beta^0u_{t+1}\right|\notag\\
   &\leq \left|\frac{1}{T^{1/4}}\frac{1}{\sqrt{d}_T}\sum_{t=\lfloor f_1(T)\rfloor }^{\lfloor f_2(T) \rfloor} \left(\frac{1}{t}\sum_{l=1}^t\*H_{r}\*f_l\xi_{l,t,\alpha}\right)' \left(\frac{N}{N^\alpha}\*D_{Nt,r}^{2}\right)^{-1}(\*H_{r}^{-1})'\+\beta^0u_{t+1}\right|\notag\\
   &+\left|\frac{1}{T^{1/4}}\frac{1}{\sqrt{d}_T}\sum_{t=\lfloor f_1(T)\rfloor }^{\lfloor f_2(T) \rfloor}\left( \frac{1}{t}\sum_{l=1}^t(\widehat{\*f}_l-\*H_{r}\*f_l)\xi_{l,t,\alpha} \right)'\left(\frac{N}{N^\alpha}\*D_{Nt,r}^{2}\right)^{-1}(\*H_{r}^{-1})'\+\beta^0u_{t+1}\right|\notag\\
   &= |B_{13abi}|+ |B_{13abii}|, 
\end{align}
where by two iterations of Cauchy-Schwarz inequality we get 
\begin{align}\label{a13abii}
    |B_{13abii}| &= d_T^{1/4}\left(\frac{d_T}{T} \right)^{1/4}\left\|\*H_{r}^{-1}\right\|\left\|\+\beta^0 \right\|\sup_{k_0\leq t\leq T-1}\left\|\left(\frac{N}{N^\alpha}\*D_{Nt,r}^{2}\right)^{-1} \right\|\frac{1}{d_T}\sum_{t=\lfloor f_1(T)\rfloor }^{\lfloor f_2(T) \rfloor}\left\| \frac{1}{t}\sum_{l=1}^t(\widehat{\*f}_l-\*H_{r}\*f_l)\xi_{l,t,\alpha}\right\||u_{t+1}|\notag\\
    &\leq d_T^{1/4}\left(\frac{d_T}{T} \right)^{1/4}\left\|\*H_{r}^{-1}\right\|\left\|\+\beta^0 \right\|\sup_{k_0\leq t\leq T-1}\left\|\left(\frac{N}{N^\alpha}\*D_{r}^{2}\right)^{-1} \right\|\left(\frac{1}{d_T}\sum_{t=\lfloor f_1(T)\rfloor }^{\lfloor f_2(T) \rfloor}\left\| \frac{1}{t}\sum_{l=1}^t(\widehat{\*f}_l-\*H_{r}\*f_l)\xi_{l,t,\alpha}\right\|^2 \right)^{1/2}\notag\\
    &\times \left(\frac{1}{d_T}\sum_{t=\lfloor f_1(T)\rfloor }^{\lfloor f_2(T) \rfloor}u_{t+1}^2 \right)^{1/2}\notag\\
    &\leq O_p(d_T^{1/4})\times \left(\frac{1}{d_T}\sum_{t=\lfloor f_1(T)\rfloor }^{\lfloor f_2(T) \rfloor}u_{t+1}^2 \right)^{1/2} \left(\sup_{k_0\leq t\leq T-1}\frac{1}{t}\sum_{l=1}^t\left\|\widehat{\*f}_l-\*H_{r}\*f_l \right\|^2 \right)^{1/2}\notag\\
    &\times \left(\frac{N^2}{N^{2\alpha}}\frac{1}{d_Tk_0} \sum_{t=\lfloor f_1(T)\rfloor }^{\lfloor f_2(T) \rfloor}\sum_{l=1}^T\xi^2_{l,t}\right)^{1/2}\notag\\
    &= O_p(d_T^{1/4})\times \left(O_p\left(\frac{N}{N^\alpha}\frac{1}{\sqrt{k_0}} \right)+O_p(N^{(1-2\alpha)/2})\right)\times  O_p(N^{(1-2\alpha)/2})\notag\\
    &=\left(\frac{d_T}{k_0}\right)^{1/4}O_p(k_0^{1/4})\times \left(O_p\left(N^{(3/2-2\alpha)}k_0^{-1/2} \right)+O_p(N^{(1-2\alpha)}) \right)\notag\\
    &= O_p\left(N^{(3/2-2\alpha)}k_0^{-1/4} \right)+O_p(N^{(1-2\alpha)}k_0^{1/4}),
\end{align}
which comes from
\begin{align}
    \frac{1}{d_Tk_0} \sum_{t=\lfloor f_1(T)\rfloor }^{\lfloor f_2(T) \rfloor}\sum_{l=1}^T\xi^2_{l,t}\leq \frac{T^2}{d_Tk_0}\frac{1}{T^2}\sum_{t=1}^T\sum_{l=1}^T\xi_{l,t}^2=O_p(N^{-1}).
\end{align}
 Hence, under $\frac{N}{N^\alpha}\frac{1}{T^{1/4}}\to c$ and $\frac{\sqrt{T}}{N^\alpha}=o(1)$ for $\alpha>0.5$, we have that 
\begin{align}
      |B_{13abii}|&=O_p\left(N^{(3/2-2\alpha)}k_0^{-1/4} \right)+O_p(N^{(1-2\alpha)}k_0^{1/4})\notag\\
      &= O_p(N^{1/2-\alpha})+O_p(\sqrt{k_0}N^{-\alpha})=o_p(1).
\end{align}
Note that with Lemma \ref{Lemma4}, we can make the latter less restrictive as we would require $\frac{T^{1/4}}{N^{(1-3\alpha)/2}}\to 0$. We further go to $B_{13abi}$, where, by letting $\+\Sigma_{\+\Lambda}^{-1}(\*H_{r}^{-1})'\+\beta^0=\*q$, we have 
\begin{align}
    B_{13abi}&=\frac{1}{T^{1/4}}\frac{1}{\sqrt{d}_T}\sum_{t=\lfloor f_1(T)\rfloor }^{\lfloor f_2(T) \rfloor} \left(\frac{1}{t}\sum_{l=1}^t\*H_{r}\*f_l\xi_{l,t,\alpha}\right)' \left(\frac{N}{N^\alpha}\*D_{Nt,r}^{2}\right)^{-1}(\*H_{r}^{-1})'\+\beta^0u_{t+1}\notag\\
    &= \frac{1}{T^{1/4}}\frac{1}{\sqrt{d}_T}\sum_{t=\lfloor f_1(T)\rfloor }^{\lfloor f_2(T) \rfloor} \left(\frac{1}{t}\sum_{l=1}^t\*H_{r}\*f_l\xi_{l,t,\alpha}\right)' \*q u_{t+1}\notag\\
    &+\frac{1}{T^{1/4}}\frac{1}{\sqrt{d}_T}\sum_{t=\lfloor f_1(T)\rfloor }^{\lfloor f_2(T) \rfloor} \left(\frac{1}{t}\sum_{l=1}^t\*H_{r}\*f_l\xi_{l,t,\alpha}\right)' \left[\left(\frac{N}{N^\alpha}\*D_{Nt,r}^{2}\right)^{-1}-\+\Sigma_{\+\Lambda}^{-1}\right](\*H_{r}^{-1})'\+\beta^0u_{t+1}\notag\\
    &=B_{13abi,\Lambda}+B_{13abi,D-\Lambda}
\end{align}
where 
\begin{align}
    &\mathbb{V}ar\left(B_{13abi,\Lambda} \right)=\mathbb{E}\left(\left\| \frac{1}{T^{1/4}}\frac{1}{\sqrt{d}_T}\sum_{t=\lfloor f_1(T)\rfloor }^{\lfloor f_2(T) \rfloor} \left(\frac{1}{t}\sum_{l=1}^t\*H_{r}\*f_l\xi_{l,t,\alpha}\right)' \*q u_{t+1} \right\|^2 \right) \notag\\
    &=\mathbb{E}\left(\frac{1}{\sqrt{T}}\frac{1}{d_T}\sum_{t=\lfloor f_1(T)\rfloor }^{\lfloor f_2(T) \rfloor}\sum_{s=\lfloor f_1(T)\rfloor }^{\lfloor f_2(T) \rfloor} \left(\frac{1}{t}\sum_{l=1}^t\*H_{r}\*f_l\xi_{l,t,\alpha}\right)' \*q u_{t+1}u_{s+1}\*q'\left(\frac{1}{s}\sum_{j=1}^s\*H_{r}\*f_j\xi_{j,s,\alpha}\right) \right)\notag\\
    &= \frac{1}{\sqrt{T}}\frac{1}{d_T}\sum_{t=\lfloor f_1(T)\rfloor }^{\lfloor f_2(T) \rfloor}\mathbb{E}\left(\sigma_{t+1}^2\left[\left(\frac{1}{t}\sum_{l=1}^t\*H_{r}\*f_l\xi_{l,t,\alpha}\right)' \*q \right]^2\right)\notag\\
    &\leq  \left\|\*q \right\|^2\frac{1}{\sqrt{T}}\frac{1}{d_T}\sum_{t=\lfloor f_1(T)\rfloor }^{\lfloor f_2(T) \rfloor}\mathbb{E}\left(\sigma_{t+1}^2\left\| \frac{1}{t}\sum_{l=1}^t\*H_{r}\*f_l\xi_{l,t,\alpha}\right\|^2 \right)\notag\\
    &\leq \left\|\*q \right\|^2\left\| \*H_r\right\|^2\frac{1}{\sqrt{T}}\frac{1}{d_T}\sum_{t=\lfloor f_1(T)\rfloor }^{\lfloor f_2(T) \rfloor}\mathbb{E}\left(\sigma_{t+1}^2\frac{1}{t}\sum_{l=1}^t\left\| \*f_l\right\|^2\left| \xi_{l,t,\alpha}\right|^2 \right)\notag\\
    &\leq \left\|\*q \right\|^2\left\| \*H_r\right\|^2\frac{1}{\sqrt{T}}\frac{1}{d_T}\sum_{t=\lfloor f_1(T)\rfloor }^{\lfloor f_2(T) \rfloor}\mathbb{E}(\sigma_{t+1}^2)\frac{1}{t}\sum_{l=1}^t\left(\mathbb{E}\left(\left\|\*f_t \right\|^4\right) \right)^{1/2}\left(\frac{N^4}{N^{4\alpha}}\mathbb{E}\left[\left(\frac{1}{N} \sum_{i=1}^N(e_{i,l}e_{i,t}-\mathbb{E}(e_{i,l}e_{i,t}))\right)^4\right] \right)^{1/2}\notag\\
    &\leq \left\|\*q \right\|^2\left\| \*H_r\right\|^2\sigma^2\left(\mathbb{E}\left(\left\|\*f_t \right\|^4\right) \right)^{1/2}\frac{1}{\sqrt{T}}\frac{T}{d_Tk_0}\frac{1}{T^2}\sum_{t=1 }^{T}\sum_{l=1}^T\left(\frac{N^4}{N^{4\alpha}}\mathbb{E}\left[\left(\frac{1}{N} \sum_{i=1}^N(e_{i,l}e_{i,t}-\mathbb{E}(e_{i,l}e_{i,t}))\right)^4\right] \right)^{1/2}\notag\\
    &=O(T^{-1/2}N^{(1-2\alpha)}),
\end{align}
which means that 
\begin{align}
    \left|B_{13abi,\Lambda} \right|=\left| \frac{1}{T^{1/4}}\frac{1}{\sqrt{d}_T}\sum_{t=\lfloor f_1(T)\rfloor }^{\lfloor f_2(T) \rfloor} \left(\frac{1}{t}\sum_{l=1}^t\*H_{r}\*f_l\xi_{l,t,\alpha}\right)' \*q u_{t+1}\right|=O_p(T^{-1/4}N^{(1-2\alpha)/2})=o_p(1)
\end{align}
provided that $\alpha>0.5$. Also, 
\begin{align}
    |B_{13abi,D-\Lambda} |&=\left| \frac{1}{T^{1/4}}\frac{1}{\sqrt{d}_T}\sum_{t=\lfloor f_1(T)\rfloor }^{\lfloor f_2(T) \rfloor} \left(\frac{1}{t}\sum_{l=1}^t\*H_{r}\*f_l\xi_{l,t,\alpha}\right)' \left[\left(\frac{N}{N^\alpha}\*D_{Nt,r}^{2}\right)^{-1}-\+\Sigma_{\+\Lambda}^{-1}\right](\*H_{r}^{-1})'\+\beta^0u_{t+1}\right|\notag\\
    &\leq \left(\frac{d_T}{T}\right)^{1/2}\left\| \*H_r^{-1}\right\|\left\|\+\beta_0 \right\|\frac{1}{d_T}\sum_{t=\lfloor f_1(T)\rfloor }^{\lfloor f_2(T) \rfloor} \left\|\frac{1}{t}\sum_{l=1}^t\*H_{r}\*f_l\xi_{l,t,\alpha}\right\|T^{1/4}\left\|\left(\frac{N}{N^\alpha}\*D_{Nt,r}^{2}\right)^{-1}-\+\Sigma_{\+\Lambda}^{-1} \right\|| u_{t+1}|\notag\\
    &\leq \left(\frac{d_T}{T}\right)^{1/2}\left\| \*H_r^{-1}\right\|\left\|\+\beta_0 \right\| \sup_{k_0\leq t\leq T-1} T^{1/4}\left\|\left(\frac{N}{N^\alpha}\*D_{Nt,r}^{2}\right)^{-1}-\+\Sigma_{\+\Lambda}^{-1} \right\|\left( \frac{1}{d_T}\sum_{t=\lfloor f_1(T)\rfloor }^{\lfloor f_2(T) \rfloor} \left\|\frac{1}{t}\sum_{l=1}^t\*H_{r}\*f_l\xi_{l,t,\alpha}\right\|^2\right)^{1/2}\notag\\
    &\times \left( \frac{1}{d_T}\sum_{t=\lfloor f_1(T)\rfloor }^{\lfloor f_2(T) \rfloor}u_{t+1}^2\right)^{1/2}\notag\\
    &\leq \left(\frac{d_T}{T}\right)^{1/2}\left\| \*H_r^{-1}\right\|\left\| \*H_{r}\right\|\left\|\+\beta_0 \right\| \sup_{k_0\leq t\leq T-1} T^{1/4}\left\|\left(\frac{N}{N^\alpha}\*D_{Nt,r}^{2}\right)^{-1}-\+\Sigma_{\+\Lambda}^{-1} \right\|\left(\sup_{k_0\leq t\leq T-1}\frac{1}{t}\sum_{l=1}^t\left\|\*f_l \right\|^2 \right)^{1/2}\notag\\
    &\times \left(\frac{N^2}{N^{2\alpha}}\frac{1}{d_Tk_0} \sum_{t=\lfloor f_1(T)\rfloor }^{\lfloor f_2(T) \rfloor}\sum_{l=1}^T\xi^2_{l,t}\right)^{1/2} \left( \frac{1}{d_T}\sum_{t=\lfloor f_1(T)\rfloor }^{\lfloor f_2(T) \rfloor}u_{t+1}^2\right)^{1/2}\notag\\
    &= O_p(1) \times T^{1/4}\times \left( O_p\left(\frac{N}{N^\alpha}\frac{1}{\sqrt{k_0}} \right)+ O_p(N^{(1-2\alpha)/2}) \right)\times O_p(N^{(1-2\alpha)/2})\notag\\
    &=  O_p(N^{(1-2\alpha)/2}) + O_p(\sqrt{k_0}N^{-\alpha})
\end{align}
under $\alpha>0.5$, $\frac{N}{N^\alpha}\frac{1}{T^{1/4}}\to c>0$ and $\sqrt{T}N^{-\alpha}=o(1)$. Hence, overall,
\begin{align}
    |B_{13ab}|=\left|\frac{1}{T^{1/4}}\frac{1}{\sqrt{d}_T}\sum_{t=\lfloor f_1(T)\rfloor }^{\lfloor f_2(T) \rfloor}\left( \frac{1}{t}\sum_{l=1}^t\widehat{\*f}_l\xi_{l,t,\alpha} \right)'\left(\frac{N}{N^\alpha}\*D_{Nt,r}^{2}\right)^{-1}(\*H_{r}^{-1})'\+\beta^0u_{t+1}\right|=o_p(1).
\end{align}
Next, we move to 
\begin{align}
    |B_{13ac}|&=\left|\frac{1}{T^{1/4}}\frac{1}{\sqrt{d}_T}\sum_{t=\lfloor f_1(T)\rfloor }^{\lfloor f_2(T) \rfloor}\left( \frac{1}{t}\sum_{l=1}^t\widehat{\*f}_l\eta_{l,t,\alpha} \right)'\left(\frac{N}{N^\alpha}\*D_{Nt,r}^{2}\right)^{-1}(\*H_{r}^{-1})'\+\beta^0u_{t+1}\right|\notag\\
    &\leq \left| \frac{1}{T^{1/4}}\frac{1}{\sqrt{d}_T}\sum_{t=\lfloor f_1(T)\rfloor }^{\lfloor f_2(T) \rfloor}\left( \frac{1}{t}\sum_{l=1}^t\*H_r\*f_l\eta_{l,t,\alpha} \right)'\left(\frac{N}{N^\alpha}\*D_{Nt,r}^{2}\right)^{-1}(\*H_{r}^{-1})'\+\beta^0u_{t+1}\right|\notag\\
    &+\left|\frac{1}{T^{1/4}}\frac{1}{\sqrt{d}_T}\sum_{t=\lfloor f_1(T)\rfloor }^{\lfloor f_2(T) \rfloor}\left( \frac{1}{t}\sum_{l=1}^t(\widehat{\*f}_l-\*H_r\*f_l)\eta_{l,t,\alpha} \right)'\left(\frac{N}{N^\alpha}\*D_{Nt,r}^{2}\right)^{-1}(\*H_{r}^{-1})'\+\beta^0u_{t+1}\right|\notag\\
    &=|B_{13aci}| + |B_{13acii}|,
\end{align}
where similarly to $B_{13abii}$:
\begin{align}\label{B13acii}
     |B_{13acii}|&\leq  d_T^{1/4}\left(\frac{d_T}{T} \right)^{1/4}\left\|\*H_{r}^{-1}\right\|\left\|\+\beta^0 \right\|\sup_{k_0\leq t\leq T-1}\left\|\left(\frac{N}{N^\alpha}\*D_{Nt,r}^{2}\right)^{-1} \right\|\frac{1}{d_T}\sum_{t=\lfloor f_1(T)\rfloor }^{\lfloor f_2(T) \rfloor}\left\| \frac{1}{t}\sum_{l=1}^t(\widehat{\*f}_l-\*H_{r}\*f_l)\eta_{l,t,\alpha}\right\||u_{t+1}|\notag\\
     &\leq  O_p(d_T^{1/4})\times \left(\frac{1}{d_T}\sum_{t=\lfloor f_1(T)\rfloor }^{\lfloor f_2(T) \rfloor}u_{t+1}^2 \right)^{1/2} \left(\sup_{k_0\leq t\leq T-1}\frac{1}{t}\sum_{l=1}^t\left\|\widehat{\*f}_l-\*H_{r}\*f_l \right\|^2 \right)^{1/2}\notag\\
    &\times \left(\frac{1}{d_Tk_0} \sum_{t=\lfloor f_1(T)\rfloor }^{\lfloor f_2(T) \rfloor}\sum_{l=1}^T\eta^2_{l,t,\alpha}\right)^{1/2}\notag\\
    &= O_p(d_T^{1/4})\times \left(O_p\left(\frac{N}{N^\alpha}\frac{1}{\sqrt{k_0}} \right)+O_p(N^{(1-2\alpha)/2}) \right)\times O_p(N^{-\alpha/2})\notag\\
    &=o_p(1)
\end{align}
under $\alpha>0.5$, $\frac{N}{N^\alpha}\frac{1}{T^{1/4}}\to c>0$ and $T^{1/4}N^{-\alpha/2}=\sqrt{T^{1/2}N^{-\alpha}}=o(1)$. Next, to obtain the sharpest possible rate,
\begin{align}
    |B_{13aci}|&\leq \left| \frac{1}{T^{1/4}}\frac{1}{\sqrt{d}_T}\sum_{t=\lfloor f_1(T)\rfloor }^{\lfloor f_2(T) \rfloor}\left( \frac{1}{t}\sum_{l=1}^t\*H_r\*f_l\eta_{l,t,\alpha} \right)'\+\Sigma_{\+\Lambda}^{-1}(\*H_{r}^{-1})'\+\beta^0u_{t+1}\right|\notag\\
    &+ \left| \frac{1}{T^{1/4}}\frac{1}{\sqrt{d}_T}\sum_{t=\lfloor f_1(T)\rfloor }^{\lfloor f_2(T) \rfloor}\left( \frac{1}{t}\sum_{l=1}^t\*H_r\*f_l\eta_{l,t,\alpha} \right)'\left[\left(\frac{N}{N^\alpha}\*D_{Nt,r}^{2}\right)^{-1}-\+\Sigma_{\+\Lambda}^{-1}\right](\*H_{r}^{-1})'\+\beta^0u_{t+1}\right|\notag\\
    &=|B_{13aci,\Lambda}|+|B_{13aci,D-\Lambda}|
\end{align}
where 
\begin{align}\label{B13aci_D-Lambda}
  |B_{13aci,D-\Lambda}|= &\left| \frac{1}{T^{1/4}}\frac{1}{\sqrt{d}_T}\sum_{t=\lfloor f_1(T)\rfloor }^{\lfloor f_2(T) \rfloor} \left(\frac{1}{t}\sum_{l=1}^t\*H_{r}\*f_l\eta_{l,t,\alpha}\right)' \left[\left(\frac{N}{N^\alpha}\*D_{Nt,r}^{2}\right)^{-1}-\+\Sigma_{\+\Lambda}^{-1}\right](\*H_{r}^{-1})'\+\beta^0u_{t+1}\right|\notag\\
      &\leq  \left(\frac{d_T}{T}\right)^{1/2}T^{1/4}\left\| \*H_r^{-1}\right\|\left\| \*H_{r}\right\|\left\|\+\beta_0 \right\| \sup_{k_0\leq t\leq T-1} \left\|\left(\frac{N}{N^\alpha}\*D_{Nt,r}^{2}\right)^{-1}-\+\Sigma_{\+\Lambda}^{-1} \right\|\left(\sup_{k_0\leq t\leq T-1}\frac{1}{t}\sum_{l=1}^t\left\|\*f_l \right\|^2 \right)^{1/2}\notag\\
    &\times \left(\frac{1}{d_Tk_0} \sum_{t=\lfloor f_1(T)\rfloor }^{\lfloor f_2(T) \rfloor}\sum_{l=1}^T\eta^2_{l,t,\alpha}\right)^{1/2} \left( \frac{1}{d_T}\sum_{t=\lfloor f_1(T)\rfloor }^{\lfloor f_2(T) \rfloor}u_{t+1}^2\right)^{1/2}\notag\\
    &=O_p(1)\times T^{1/4}\times O_p(N^{-\alpha/2})\times  \left(O_p\left(\frac{N}{N^\alpha}\frac{1}{\sqrt{k_0}} \right)+O_p(N^{(1-2\alpha)/2}) \right)\notag\\
    &= o_p(T^{1/4}N^{-\alpha/2})=o_p(1)
\end{align}
under the needed restrictions. Furthermore, 
\begin{align}
     B_{13aci,\Lambda}=&\frac{1}{T^{1/4}}\frac{1}{\sqrt{d}_T}\sum_{t=\lfloor f_1(T)\rfloor }^{\lfloor f_2(T) \rfloor}\left( \frac{1}{t}\sum_{l=1}^t\*H_r\*f_l\eta_{l,t,\alpha} \right)'\+\Sigma_{\+\Lambda}^{-1}(\*H_{r}^{-1})'\+\beta^0u_{t+1}\notag\\
     &=  \frac{1}{T^{1/4}}\frac{1}{\sqrt{d}_T}\sum_{t=\lfloor f_1(T)\rfloor }^{\lfloor f_2(T) \rfloor}\left( \*H_r\frac{1}{t}\sum_{l=1}^t\*f_l\*f_l'\frac{1}{N^{\alpha}} \sum_{i=1}^N\+\lambda_ie_{i,t}\right)'\+\Sigma_{\+\Lambda}^{-1}(\*H_{r}^{-1})'\+\beta^0u_{t+1}\notag\\
     &= \frac{1}{T^{1/4}}\frac{1}{\sqrt{d}_T}\sum_{t=\lfloor f_1(T)\rfloor }^{\lfloor f_2(T) \rfloor}\left( \*H_r\left[\frac{1}{t}\sum_{l=1}^t\*f_l\*f_l'-\*I_r\right]\frac{1}{N^{\alpha}} \sum_{i=1}^N\+\lambda_ie_{i,t}\right)'\+\Sigma_{\+\Lambda}^{-1}(\*H_{r}^{-1})'\+\beta^0u_{t+1}\notag\\
     &+\frac{1}{T^{1/4}}\frac{1}{\sqrt{d}_T}\sum_{t=\lfloor f_1(T)\rfloor }^{\lfloor f_2(T) \rfloor}\left( \*H_r\frac{1}{N^{\alpha}} \sum_{i=1}^N\+\lambda_ie_{i,t}\right)'\+\Sigma_{\+\Lambda}^{-1}(\*H_{r}^{-1})'\+\beta^0u_{t+1}\notag\\
     &= B_{13aci,\Lambda,FF-I}+B_{13aci,\Lambda,I}.
\end{align}
This, by letting again $\*q=\+\Sigma_{\+\Lambda}^{-1}(\*H_{r}^{-1})'\+\beta^0$, results in 
\begin{align}
    |B_{13aci,\Lambda,FF-I}|&\leq d_T^{1/4}\left(\frac{d_T}{T} \right)^{1/4}\left\| \*H_r\right\|\left\|\*q \right\|\sup_{k_0\leq t\leq T-1}\left\| \frac{1}{t}\sum_{l=1}^t\*f_l\*f_l'-\*I_r\right\|\frac{1}{d_T}\sum_{t=\lfloor f_1(T)\rfloor }^{\lfloor f_2(T) \rfloor}\left\|\frac{1}{N^{\alpha}} \sum_{i=1}^N\+\lambda_ie_{i,t} \right\||u_{t+1}|\notag\\
    &\leq \left(\frac{d_T}{T} \right)^{1/4}\left\| \*H_r\right\|\left\|\*q \right\|d_T^{1/4}\sup_{k_0\leq t\leq T-1}\left\| \frac{1}{t}\sum_{l=1}^t\*f_l\*f_l'-\*I_r\right\|\left(\frac{1}{d_T}\sum_{t=\lfloor f_1(T)\rfloor }^{\lfloor f_2(T) \rfloor}\left\|\frac{1}{N^{\alpha}} \sum_{i=1}^N\+\lambda_ie_{i,t} \right\|^2 \right)^{1/2}\notag\\
    &\times \left(\frac{1}{d_T}\sum_{t=\lfloor f_1(T)\rfloor }^{\lfloor f_2(T) \rfloor} u_{t+1}^2\right)^{1/2}\notag\\
    &=O_p(T^{-1/4}N^{-\alpha/2})
\end{align}
and 
\begin{align}
\mathbb{V}ar(B_{13aci,\Lambda,I})&=\mathbb{E}\left(\left\| \frac{1}{T^{1/4}}\frac{1}{\sqrt{d}_T}\sum_{t=\lfloor f_1(T)\rfloor }^{\lfloor f_2(T) \rfloor}\left( \*H_r\frac{1}{N^{\alpha}} \sum_{i=1}^N\+\lambda_ie_{i,t}\right)'\+\Sigma_{\+\Lambda}^{-1}(\*H_{r}^{-1})'\+\beta^0u_{t+1}\right\|^2 \right)\notag\\
&\leq\sigma^2\left\| \*H_r\right\|^2\left\| \*q\right\|^2\frac{1}{\sqrt{T}}\frac{1}{d_T}\sum_{t=\lfloor f_1(T)\rfloor }^{\lfloor f_2(T) \rfloor}\mathbb{E}\left(\left\| \frac{1}{N^{\alpha}} \sum_{i=1}^N\+\lambda_ie_{i,t}\right\|^2 \right)\notag\\
&= O(T^{-1/2}N^{-\alpha}),
\end{align}
which implies that $|B_{13aci,\Lambda,I}|=O_p(T^{-1/4}N^{-\alpha/2})$. Therefore, overall,
\begin{align}
   |B_{13ac}|= \left|\frac{1}{T^{1/4}}\frac{1}{\sqrt{d}_T}\sum_{t=\lfloor f_1(T)\rfloor }^{\lfloor f_2(T) \rfloor}\left( \frac{1}{t}\sum_{l=1}^t\widehat{\*f}_l\eta_{l,t,\alpha} \right)'\left(\frac{N}{N^\alpha}\*D_{Nt,r}^{2}\right)^{-1}(\*H_{r}^{-1})'\+\beta^0u_{t+1}\right|=o_p(1).
\end{align}
Eventually, we will adopt a similar strategy in case of
\begin{align}
     |B_{13ad}|&=\left|\frac{1}{T^{1/4}}\frac{1}{\sqrt{d}_T}\sum_{t=\lfloor f_1(T)\rfloor }^{\lfloor f_2(T) \rfloor}\left( \frac{1}{t}\sum_{l=1}^t\widehat{\*f}_l \nu_{l,t,\alpha} \right)'\left(\frac{N}{N^\alpha}\*D_{Nt,r}^{2}\right)^{-1}(\*H_{r}^{-1})'\+\beta^0u_{t+1}\right|\notag\\
     &\leq \left|\frac{1}{T^{1/4}}\frac{1}{\sqrt{d}_T}\sum_{t=\lfloor f_1(T)\rfloor }^{\lfloor f_2(T) \rfloor}\left( \frac{1}{t}\sum_{l=1}^t\*H_r\*f_l \nu_{l,t,\alpha} \right)'\left(\frac{N}{N^\alpha}\*D_{Nt,r}^{2}\right)^{-1}(\*H_{r}^{-1})'\+\beta^0u_{t+1}\right|\notag\\
     &+\left|\frac{1}{T^{1/4}}\frac{1}{\sqrt{d}_T}\sum_{t=\lfloor f_1(T)\rfloor }^{\lfloor f_2(T) \rfloor}\left( \frac{1}{t}\sum_{l=1}^t(\widehat{\*f}_l-\*H_r\*f_l) \nu_{l,t,\alpha} \right)'\left(\frac{N}{N^\alpha}\*D_{Nt,r}^{2}\right)^{-1}(\*H_{r}^{-1})'\+\beta^0u_{t+1}\right|\notag\\
     &=|B_{13adi}|+|B_{13adii}|,
\end{align}
where 
\begin{align}
   |B_{13adii}|& \leq  O_p(d_T^{1/4})\times \left(\frac{1}{d_T}\sum_{t=\lfloor f_1(T)\rfloor }^{\lfloor f_2(T) \rfloor}u_{t+1}^2 \right)^{1/2} \left(\sup_{k_0\leq t\leq T-1}\frac{1}{t}\sum_{l=1}^t\left\|\widehat{\*f}_l-\*H_{r}\*f_l \right\|^2 \right)^{1/2}\notag\\
    &\times \left(\frac{1}{d_Tk_0} \sum_{t=\lfloor f_1(T)\rfloor }^{\lfloor f_2(T) \rfloor}\sum_{l=1}^T\nu^2_{l,t,\alpha}\right)^{1/2}\notag\\
    &= O_p(d_T^{1/4})\times \left(O_p\left(\frac{N}{N^\alpha}\frac{1}{\sqrt{k_0}} \right)+O_p(N^{(1-2\alpha)/2}) \right)\times O_p(N^{-\alpha/2})\notag\\
    &=o_p(1),
\end{align}
as $\nu_{l,t,\alpha}$ has the same behavior as $\eta_{l,t,\alpha}$ in (\ref{B13acii}). Next, 
\begin{align}
    |B_{13adi}|&\leq \left|\frac{1}{T^{1/4}}\frac{1}{\sqrt{d}_T}\sum_{t=\lfloor f_1(T)\rfloor }^{\lfloor f_2(T) \rfloor}\left( \frac{1}{t}\sum_{l=1}^t\*H_r\*f_l \nu_{l,t,\alpha} \right)'\+\Sigma_{\+\Lambda}^{-1}(\*H_{r}^{-1})'\+\beta^0u_{t+1} \right|\notag\\
    &+\left|\frac{1}{T^{1/4}}\frac{1}{\sqrt{d}_T}\sum_{t=\lfloor f_1(T)\rfloor }^{\lfloor f_2(T) \rfloor}\left( \frac{1}{t}\sum_{l=1}^t\*H_r\*f_l \nu_{l,t,\alpha} \right)'\left[\left(\frac{N}{N^\alpha}\*D_{Nt,r}^{2}\right)^{-1}-\+\Sigma_{\+\Lambda}^{-1}\right](\*H_{r}^{-1})'\+\beta^0u_{t+1}\right|\notag\\
    &=|B_{13adi,\Lambda}|+|B_{13adi,D-\Lambda}|,
\end{align}
where by following the exact same steps in (\ref{B13aci_D-Lambda}), we have
\begin{align}
    |B_{13adi,D-\Lambda}|&\leq O_p(1) \times  T^{1/4} \sup_{k_0\leq t\leq T-1} \left\|\left(\frac{N}{N^\alpha}\*D_{Nt,r}^{2}\right)^{-1}-\+\Sigma_{\+\Lambda}^{-1} \right\|\times \left(\frac{1}{d_Tk_0} \sum_{t=\lfloor f_1(T)\rfloor }^{\lfloor f_2(T) \rfloor}\sum_{l=1}^T\eta^2_{l,t,\alpha}\right)^{1/2} \notag\\
    &=O_p(1)\times T^{1/4}\times O_p(N^{-\alpha/2})\times  \left(O_p\left(\frac{N}{N^\alpha}\frac{1}{\sqrt{k_0}} \right)+O_p(N^{(1-2\alpha)/2}) \right)\notag\\
    &= o_p(T^{1/4}N^{-\alpha/2})=o_p(1).
\end{align}
To obtain the sharp rate, notice how by definition of $\nu_{l,t,\alpha}$
\begin{align}
  B_{13adi,\Lambda}&=  \frac{1}{T^{1/4}}\frac{1}{\sqrt{d}_T}\sum_{t=\lfloor f_1(T)\rfloor }^{\lfloor f_2(T) \rfloor}\left( \frac{1}{t}\sum_{l=1}^t\*H_r\*f_l \nu_{l,t,\alpha} \right)'\+\Sigma_{\+\Lambda}^{-1}(\*H_{r}^{-1})'\+\beta^0u_{t+1} \notag\\
  &= \frac{1}{T^{1/4}}\frac{1}{\sqrt{d}_T}\sum_{t=\lfloor f_1(T)\rfloor }^{\lfloor f_2(T) \rfloor}\left(\*H_r \frac{1}{t}\sum_{l=1}^t\*f_l \frac{1}{N^\alpha}\sum_{i=1}^N\*f_t'\+\lambda_ie_{i,l} \right)'\+\Sigma_{\+\Lambda}^{-1}(\*H_{r}^{-1})'\+\beta^0u_{t+1}\notag\\
  &= \frac{1}{T^{1/4}}\frac{1}{\sqrt{d}_T}\sum_{t=\lfloor f_1(T)\rfloor }^{\lfloor f_2(T) \rfloor}\*f_t'\left(\*H_r \frac{1}{t}\sum_{l=1}^t\*f_l \frac{1}{N^\alpha}\sum_{i=1}^N\+\lambda_i'e_{i,l} \right)'\*qu_{t+1},
\end{align}
and so, by using the same chain of inequalities as in \cite{gonccalves2017tests}, we get
\begin{align}
    \mathbb{V}ar( B_{13adi,\Lambda})&= \mathbb{E}\left(\left\|\frac{1}{T^{1/4}}\frac{1}{\sqrt{d}_T}\sum_{t=\lfloor f_1(T)\rfloor }^{\lfloor f_2(T) \rfloor}\*f_t'\left(\*H_r \frac{1}{t}\sum_{l=1}^t\*f_l \frac{1}{N^\alpha}\sum_{i=1}^N\+\lambda_i'e_{i,l} \right)'\*qu_{t+1} \right\|^2 \right)\notag\\
    &\leq \left\|\*H_r \right\|^2\left\|\*q \right\|^2\frac{1}{\sqrt{T}}\frac{1}{d_T}\sum_{t=\lfloor f_1(T)\rfloor }^{\lfloor f_2(T) \rfloor}\mathbb{E}\left(\sigma_{t+1}^2\left\|\*f_t \right\|^2\left\| \frac{1}{t}\sum_{l=1}^t\*f_l \frac{1}{N^\alpha}\sum_{i=1}^N\+\lambda_i'e_{i,l}\right\|^2 \right)\notag\\
    &\leq \left\|\*H_r \right\|^2\left\|\*q \right\|^2\frac{1}{\sqrt{T}}\frac{1}{d_T}\sum_{t=\lfloor f_1(T)\rfloor }^{\lfloor f_2(T) \rfloor}\mathbb{E}(\sigma_{t+1}^2)\mathbb{E}\left(\sigma_t^2\left\|\*f_t \right\|^2\left\| \frac{1}{t}\sum_{l=1}^t\*f_l \frac{1}{N^\alpha}\sum_{i=1}^N\+\lambda_i'e_{i,l}\right\|^2 \right)\notag\\
    &\leq \sigma^2 \left\|\*H_r \right\|^2\left\|\*q \right\|^2\sup_{k_0\leq t\leq T-1}\left(\mathbb{E}\left(\left\|\*f_t \right\|^4 \right) \right)^{1/2} \frac{1}{\sqrt{T}}\left[\mathbb{E}\left(\left\|\frac{1}{t}\sum_{l=1}^t\*f_l \frac{1}{N^\alpha}\sum_{i=1}^N\+\lambda_i'e_{i,l} \right\|^4 \right) \right]^{1/2}\notag\\
    &\leq O(1) \frac{1}{\sqrt{T}}\left[\mathbb{E}\left(\frac{1}{t}\sum_{l=1}^t\left\|\*f_l\right\|^4 \left\|\frac{1}{N^\alpha}\sum_{i=1}^N\+\lambda_i'e_{i,l} \right\|^4 \right) \right]^{1/2}\notag\\
    &\leq O(1)  \frac{1}{\sqrt{T}}\frac{T}{k_0}\frac{1}{T}\sum_{l=1}^T\left(\mathbb{E}\left(\left\|\*f_l \right\|^8 \right)\right)^{1/4}\left(\mathbb{E}\left(\left\|\frac{1}{N^\alpha}\sum_{i=1}^N\+\lambda_i'e_{i,l} \right\|^8 \right)\right)^{1/4}\notag\\
    &= O(T^{-1/2}N^{-\alpha}),
\end{align}
implying that 
\begin{align}
    |B_{13adi,\Lambda}|=O_p(T^{-1/4}N^{-\alpha/2}). 
\end{align}
Note that under independence of $\*f_t$ and the idiosynratics, we would need only the 4th moment. This analysis brings us back to (\ref{B13}), and therefore
\begin{align}
     |B_{13}|&\leq \left|\frac{1}{\sqrt{d}_T}\sum_{t=\lfloor f_1(T)\rfloor }^{\lfloor f_2(T) \rfloor}u_{t+1}(\widehat{\*f}_t-\*H_{Nt,r} \*f_t)'(\*H_{r}^{-1})'\+\beta^0 T^{-1/4}\right|\notag\\
    &+\left|\frac{1}{\sqrt{d}_T}\sum_{t=\lfloor f_1(T)\rfloor }^{\lfloor f_2(T) \rfloor}u_{t+1}(\widehat{\*f}_t-\*H_{Nt,r} \*f_t)'(\*H_{Nt,r}^{-1}-\*H_{r}^{-1})'\+\beta^0 T^{-1/4} \right|\notag\\
    &=|B_{13a}| + |B_{13b}|=o_p(1),
\end{align}
because by Lemma \ref{Lemma3}, we have 
\begin{align}
    |B_{13b}|&\leq O_p(1) T^{1/4}\sup_{k_0\leq t\leq T-1}\left\|\*H_{Nt,r}^{-1}-\*H_{r}^{-1} \right\|\left(\frac{1}{d_T}\sum_{t=\lfloor f_1(T)\rfloor }^{\lfloor f_2(T) \rfloor}u_{t+1}^2 \right)^{1/2} \left(\frac{1}{d_T}\sum_{t=\lfloor f_1(T)\rfloor }^{\lfloor f_2(T) \rfloor}\left\|\widehat{\*f}_t-\*H_{Nt,r} \*f_t \right\|^2 \right)^{1/2}\notag\\
    &\leq O_p(1) T^{1/4} \left(O_p\left(\frac{N^2}{N^{2\alpha}}\frac{1}{k_0} \right) + O_p(N^{1-2\alpha}) \right)=o_p(1)
\end{align}
under $\frac{N}{N^\alpha}\frac{1}{k_0^{1/4}}\to c>0$ and $\sqrt{k_0}N^{-\alpha}=o(1)$. \\

\noindent We are left to analyze $B_{12}$, which will now follow based on the interim results developed before, Let $\+\Phi_r=\mathrm{diag}(\*I_k,\*H_r)$, then 
\begin{align}
    |B_{12} |&=\left|\frac{1}{\sqrt{d}_T}\sum_{t=\lfloor f_1(T)\rfloor }^{\lfloor f_2(T) \rfloor}u_{t+1}\widehat{\*z}_t'(\widehat{\+\delta}_t-(\+\Phi_{Nt,r}^{-1})'\widetilde{\+\delta}_t)\right|\notag\\
    &\leq \left|\frac{1}{\sqrt{d}_T}\sum_{t=\lfloor f_1(T)\rfloor }^{\lfloor f_2(T) \rfloor}u_{t+1}\*z_t'\+\Phi_r'(\widehat{\+\delta}_t-(\+\Phi_{Nt,r}^{-1})'\widetilde{\+\delta}_t)\right| \notag\\
    &+ \left|\frac{1}{\sqrt{d}_T}\sum_{t=\lfloor f_1(T)\rfloor }^{\lfloor f_2(T) \rfloor}u_{t+1}(\widehat{\*z}_t-\+\Phi_r\*z_t)'(\widehat{\+\delta}_t-(\+\Phi_{Nt,r}^{-1})'\widetilde{\+\delta}_t)\right|\notag\\
    &=|B_{12,a}|+|B_{12,b}|,
\end{align}
where 
\begin{align}
|B_{12,b}|&\leq \frac{1}{T^{1/4}}\frac{1}{\sqrt{d_T}}    \sum_{t=\lfloor f_1(T)\rfloor }^{\lfloor f_2(T) \rfloor}|u_{t+1}|\left\|\widehat{\*f}_t-\*H_r\*f_t \right\|\left\|T^{1/4}(\widehat{\+\delta}_t-(\+\Phi_{Nt,r}^{-1})'\widetilde{\+\delta}_t) \right\|\notag\\
&= d_{T}^{1/4}\left(\frac{d_T}{T} \right)^{1/4}\frac{1}{d_T}\sum_{t=\lfloor f_1(T)\rfloor }^{\lfloor f_2(T) \rfloor}|u_{t+1}|\left\|\widehat{\*f}_t-\*H_r\*f_t \right\|\left\|T^{1/4}(\widehat{\+\delta}_t-(\+\Phi_{Nt,r}^{-1})'\widetilde{\+\delta}_t) \right\|\notag\\
&\leq d_{T}^{1/4}\left(\frac{d_T}{T} \right)^{1/4}\sup_{k_0\leq t\leq T-1}\left\|T^{1/4}(\widehat{\+\delta}_t-(\+\Phi_{Nt,r}^{-1})'\widetilde{\+\delta}_t) \right\| \left(\frac{1}{d_t}\sum_{t=\lfloor f_1(T)\rfloor }^{\lfloor f_2(T) \rfloor}\left\|\widehat{\*f}_t-\*H_r\*f_t \right\| \right)^{1/2}\left(\frac{1}{d_T}\sum_{t=\lfloor f_1(T)\rfloor }^{\lfloor f_2(T) \rfloor}u_{t+1}^2 \right)^{1/2}\notag\\
&=O(1)\times d_T^{1/4}\left(O_p\left(\frac{N}{N^{\alpha}}\frac{1}{\sqrt{k_0}} \right) + O_p(N^{(1-2\alpha)/2})+O_p(T^{-1/4})\right)\left(O_p\left(\frac{N}{N^{\alpha}}\frac{1}{\sqrt{k_0}} \right) + O_p(N^{(1-2\alpha)/2}) \right)\notag\\
&= o_p(1)
\end{align}
under $\frac{N}{N^{\alpha}}\frac{1}{T^{1/4}}\to c>0$ and $\sqrt{k_0}N^{-\alpha}=o(1)$. Also, by the existence of moments, 
\begin{align}
\mathbb{V}ar(B_{12,a})&=\mathbb{E}\left(\left\|\frac{1}{T^{1/4}}\frac{1}{\sqrt{d}_T}\sum_{t=\lfloor f_1(T)\rfloor }^{\lfloor f_2(T) \rfloor}u_{t+1}\+\Phi_r'\*z_t'T^{1/4}(\widehat{\+\delta}_t-(\+\Phi_{Nt,r}^{-1})'\widetilde{\+\delta}_t) \right\|^2 \right)\notag\\
&=\mathbb{E}\left(\frac{1}{\sqrt{T}}\frac{1}{d_T}\sum_{t=\lfloor f_1(T)\rfloor }^{\lfloor f_2(T) \rfloor}\sum_{s=\lfloor f_1(T)\rfloor }^{\lfloor f_2(T) \rfloor}u_{t+1}u_{s+1} \*z_t'\+\Phi_r'T^{1/4}(\widehat{\+\delta}_t-(\+\Phi_{Nt,r}^{-1})'\widetilde{\+\delta}_t)T^{1/4}(\widehat{\+\delta}_s-(\+\Phi_{Ns,r}^{-1})'\widetilde{\+\delta}_s)'\+\Phi_r\*z_s\right)\notag\\
    &\leq \left\| \+\Phi_r\right\|^2\frac{1}{\sqrt{T}}\frac{1}{d_T}\sum_{t=\lfloor f_1(T)\rfloor }^{\lfloor f_2(T) \rfloor}\mathbb{E}\left(\sigma_{t+1}^2\left\| \*z_t\right\|^2 \left\| T^{1/4}(\widehat{\+\delta}_t-(\+\Phi_{Nt,r}^{-1})'\widetilde{\+\delta}_t)\right\|^2\right)\notag\\
    &\leq  \left\| \+\Phi_r\right\|^2\frac{1}{\sqrt{T}}\frac{1}{d_T}\sum_{t=\lfloor f_1(T)\rfloor }^{\lfloor f_2(T) \rfloor}\left( \mathbb{E}\left(\sigma_{t+1}^4\left\|\*z_t \right\|^4 \right)\right)^{1/2}\left(\mathbb{E}\left(\left\|T^{1/4}(\widehat{\+\delta}_t-(\+\Phi_{Nt,r}^{-1})'\widetilde{\+\delta}_t) \right\|^4 \right)\right)^{1/2}\notag\\
    &=(\mathbb{E}(\sigma_{t+1}^4))^{1/2}\left\| \+\Phi_r\right\|^2\frac{1}{\sqrt{T}}\frac{1}{d_T}\sum_{t=\lfloor f_1(T)\rfloor }^{\lfloor f_2(T) \rfloor}\left( \mathbb{E}\left(\left\|\*z_t \right\|^4 \right)\right)^{1/2}\left(\mathbb{E}\left(\left\|T^{1/4}(\widehat{\+\delta}_t-(\+\Phi_{Nt,r}^{-1})'\widetilde{\+\delta}_t) \right\|^4 \right)\right)^{1/2}\notag\\
    &=o(T^{-1/2})
\end{align}
under the appropriate restrictions, which implies that $|B_{12,a}|=o_p(1)$. Note that here we needed to implement Cauchy-Schwarz inequality on $\sigma_{t+1}^2$, as well, because due to GARCH effects, it includes lags of $u_{t}^2$, which is not independent of $T^{1/4}(\widehat{\+\delta}_t-(\+\Phi_{Nt,r}^{-1})'\widetilde{\+\delta}_t)$ since it is a function of $u_{t-1}$ and deeper lags. This completes the analysis of $B_1$, which means that overall 
\begin{align}
    |B|\leq \left|\frac{1}{\sqrt{d}_T}\sum_{t=\lfloor f_1(T)\rfloor }^{\lfloor f_2(T) \rfloor} u_{t+1}(\widetilde{u}_{2,t+1}- \widehat{u}_{2,t+1})\right| &+  \left|\+\beta'\frac{1}{\sqrt{d}_T}\sum_{t=\lfloor f_1(T)\rfloor }^{\lfloor f_2(T) \rfloor}\*f_t(\widetilde{u}_{2,t+1}- \widehat{u}_{2,t+1})\right|\notag\\
    &+ \left|\frac{1}{\sqrt{d}_T}\sum_{t=\lfloor f_1(T)\rfloor }^{\lfloor f_2(T) \rfloor}(\widetilde{\+\theta}_t-\+\theta)'\*w_t(\widetilde{u}_{2,t+1}- \widehat{u}_{2,t+1})\right|=o_p(1).
\end{align}
\\
\noindent Next, we go to $\mathrm{(iii.)}$, where
\begin{align}
    C=\frac{1}{\sqrt{d_T}}\sum_{t=\lfloor f_1(T)\rfloor}^{\lfloor f_2(T) \rfloor}\widetilde{u}_{2,t+1}(\widetilde{u}_{2,t+1}- \widehat{u}_{2,t+1})&=\underbrace{\frac{1}{\sqrt{d_T}}\sum_{t=\lfloor f_1(T)\rfloor}^{\lfloor f_2(T) \rfloor}u_{t+1}(\widetilde{u}_{2,t+1}
    - \widehat{u}_{2,t+1})}_{\text{$B_1=o_p(1)$}}\notag\\
    &-\frac{1}{\sqrt{d_T}}\sum_{t=\lfloor f_1(T)\rfloor}^{\lfloor f_2(T) \rfloor}(\widetilde{\+\delta}_t-\+\delta)'\*z_t(\widetilde{u}_{2,t+1}- \widehat{u}_{2,t+1})\notag\\
    &=C_1-C_2,
\end{align}
 and 
\begin{align}
    \left|C_2 \right|&=\left|\frac{1}{\sqrt{d_T}}\sum_{t=\lfloor f_1(T)\rfloor}^{\lfloor f_2(T) \rfloor}(\widetilde{\+\delta}_t-\+\delta)'\*z_t(\widetilde{u}_{2,t+1}- \widehat{u}_{2,t+1}) \right|\notag\\
    &\leq \sup_{k_0\leq t\leq T-1}\left\|\sqrt{T}(\widetilde{\+\delta}_t-\+\delta) \right\| \left(\frac{d_T}{T}\right)^{1/2}\frac{1}{d_T}\sum_{t=\lfloor f_1(T)\rfloor}^{\lfloor f_2(T) \rfloor}\left\|\*z_t \right\||\widetilde{u}_{2,t+1}- \widehat{u}_{2,t+1}|\notag\\
    &\leq  \sup_{k_0\leq t\leq T-1}\left\|\sqrt{T}(\widetilde{\+\delta}_t-\+\delta) \right\| \left(\frac{d_T}{T}\right)^{1/2} \frac{1}{d_T^{1/4}}\left(\frac{1}{d_T} \sum_{t=\lfloor f_1(T)\rfloor}^{\lfloor f_2(T) \rfloor}\left\|\*z_t \right\|^2\right)^{1/2} \left(\frac{1}{\sqrt{d}_T}\sum_{t=\lfloor f_1(T)\rfloor}^{\lfloor f_2(T) \rfloor}(\widetilde{u}_{2,t+1}- \widehat{u}_{2,t+1})^2 \right)^{1/2}\notag\\
    &= \sup_{k_0\leq t\leq T-1}\left\|\sqrt{T}(\widetilde{\+\delta}_t-\+\delta) \right\| \left(\frac{d_T}{T}\right)^{1/4} \frac{1}{T^{1/4}}\left(\frac{1}{d_T} \sum_{t=\lfloor f_1(T)\rfloor}^{\lfloor f_2(T) \rfloor}\left\|\*z_t \right\|^2\right)^{1/2} \left(\underbrace{\frac{1}{\sqrt{d_T}}\sum_{t=\lfloor f_1(T)\rfloor }^{\lfloor f_2(T) \rfloor}( \widetilde{u}_{2,t+1}- \widehat{u}_{2,t+1})^2}_{A} \right)^{1/2}\notag\\
    &= O_p\left(\frac{N}{N^{\alpha}}\frac{1}{T^{1/4}\sqrt{k_0}} \right) + O_p(T^{-1/4}N^{(1-2\alpha)/2})+O_p(T^{-1/2}).
\end{align}
 Lastly, $C_1=B_1$, which was shown to be negligible in $\mathrm{(ii.)}$. Therefore, overall
 \begin{align}
     |C|&\leq \left|\frac{1}{\sqrt{d_T}}\sum_{t=\lfloor f_1(T)\rfloor}^{\lfloor f_2(T) \rfloor}u_{t+1}(\widetilde{u}_{2,t+1}
    - \widehat{u}_{2,t+1})\right|\notag\\
    &+\left|\frac{1}{\sqrt{d_T}}\sum_{t=\lfloor f_1(T)\rfloor}^{\lfloor f_2(T) \rfloor}(\widetilde{\+\delta}_t-\+\delta)'\*z_t(\widetilde{u}_{2,t+1}- \widehat{u}_{2,t+1})\right|=o_p(1)
 \end{align}
 as expected.\\

 \noindent Ultimately, we analyze $\mathrm{(iv.)}$. By following the result in (\ref{var_exp_final}), we obtain 
 \begin{align}
      |\widetilde{\omega}^2_j-\widehat{\omega}^2_j|=f_j(\+\zeta_j)|q(\widetilde{u}^2_{2,t+1},\widehat{u}^2_{2,t+1}) | &\leq O(1)\left(q_2(\widetilde{u}^2_{2,t+1},\widehat{u}^2_{2,t+1})^{1/2}+ q_2(\widetilde{u}^2_{2,t+1},\widehat{u}^2_{2,t+1})\right) \notag\\
      &=o_p(1),
 \end{align}
where we only need to analyze $q_2(\widetilde{u}^2_{2,t+1},\widehat{u}^2_{2,t+1})$. By using $(a+b)^2\leq 2(a^2+b^2)$, we get
 \begin{align}
q_2(\widetilde{u}^2_{2,t+1},\widehat{u}^2_{2,t+1})&= \frac{1}{n}\sum_{t=k_0}^{T-1}\left( (\widetilde{u}^2_{2,t+1}-\widehat{u}^2_{2,t+1}) + \left[\frac{1}{n}\sum_{t=k_0}^{T-1}(\widetilde{u}_{2,t+1}^2-\widehat{u}_{2,t+1}^2) \right] \right)^2\notag\\
&\leq 2\left(\frac{1}{n}\sum_{t=k_0}^{T-1} (\widetilde{u}^2_{2,t+1}-\widehat{u}^2_{2,t+1})^2 + \left[\frac{1}{n}\sum_{t=k_0}^{T-1}(\widetilde{u}_{2,t+1}^2-\widehat{u}_{2,t+1}^2) \right]^2\right)\notag\\
&= 2(q_{21}(\widetilde{u}^2_{2,t+1},\widehat{u}^2_{2,t+1})+q_{22}(\widetilde{u}^2_{2,t+1},\widehat{u}^2_{2,t+1}))=o_p(1),
 \end{align}
where, by inserting the expansion in (\ref{u-u_exp}), we get
\begin{align}
   q_{22}(\widetilde{u}^2_{2,t+1},\widehat{u}^2_{2,t+1}) = \left(\frac{2}{n}\sum_{t=k_0}^{T-1}\widetilde{u}_{2,t+1}( \widetilde{u}_{2,t+1}- \widehat{u}_{2,t+1})- \frac{1}{n}\sum_{t=k_0}^{T-1}(\widetilde{u}_{2,t+1}- \widehat{u}_{2,t+1})^2\right)^2=o_p(n^{-1/2})
\end{align}
by the parts (i.) and (iii.) of this Lemma, while again by $(a+b)^2\leq 2(a^2+b^2)$ and (\ref{u-u_exp})
\begin{align}
    q_{21}(\widetilde{u}^2_{2,t+1},\widehat{u}^2_{2,t+1})&=\frac{1}{n}\sum_{t=k_0}^{T-1}\left[2\widetilde{u}_{2,t+1}( \widetilde{u}_{2,t+1}- \widehat{u}_{2,t+1})-(\widetilde{u}_{2,t+1}- \widehat{u}_{2,t+1})^2\right]^2\notag\\
    &\leq 4\frac{1}{n}\sum_{t=k_0}^{T-1}\widetilde{u}_{2,t+1}^2( \widetilde{u}_{2,t+1}- \widehat{u}_{2,t+1})^2+2 \frac{1}{n}\sum_{t=k_0}^{T-1}( \widetilde{u}_{2,t+1}- \widehat{u}_{2,t+1})^4 \notag\\
    &\leq 4\sup_{k_0\leq t \leq T-1}\widetilde{u}_{2,t+1}^2 \underbrace{\frac{1}{n}\sum_{t=k_0}^{T-1}( \widetilde{u}_{2,t+1}- \widehat{u}_{2,t+1})^2}_{o_p(1)} + 2 \frac{1}{n}\sum_{t=k_0}^{T-1}( \widetilde{u}_{2,t+1}- \widehat{u}_{2,t+1})^4\notag\\
    &=  4 \frac{1}{n}\sum_{t=k_0}^{T-1}( \widetilde{u}_{2,t+1}- \widehat{u}_{2,t+1})^4+o_p(n^{-1/2})=o_p(1).
\end{align}
The latter comes from the fact that $\widetilde{u}_{2,t+1}^2$ is consistent for $u_{t+1}^2$ uniformly in $t$:
\begin{align}
    \sup_{k_0\leq t \leq T-1}\widetilde{u}_{2,t+1}^2= \sup_{k_0\leq t \leq T-1}(y_{t+1}-\widetilde{\+\delta_t}'\*z_t)^2&= \sup_{k_0\leq t \leq T-1}(u_{t+1}-(\widetilde{\+\delta}_t-\+\delta)'\*z_t)^2\notag\\
    &\leq 2 \sup_{k_0\leq t \leq T-1} u_{t+1}^2+2T^{-1} \sup_{k_0\leq t \leq T-1}\left\|\sqrt{T}(\widetilde{\+\delta}_t-\+\delta) \right\|^2 \sup_{k_0\leq t \leq T-1}\left\|\*z_t \right\|^2\notag\\
    &=2\sup_{k_0\leq t \leq T-1} u_{t+1}^2+O_p(T^{-1})=O_p(1),
    \end{align}
and also, by using $(a+b+c)^4\leq 4(a^4+b^4+c^4)$, we obtain
\begin{align}\label{variance_expansion}
    \frac{1}{n}\sum_{t=k_0}^{T-1}( \widetilde{u}_{2,t+1}- \widehat{u}_{2,t+1})^4&\leq  4\frac{1}{T^2n}\sum_{t=k_0}^{T-1}\left\|(\widehat{\*z}_t-\+\Phi_{4,Nt,r} \*z_t)\right\|^4\left\|(\+\Phi_{4,Nt,r}^{-1})\right\|^4\left\|\sqrt{T}(\widetilde{\+\delta}_t-\+\delta)\right\|^4\notag\\
    &+ 4\frac{1}{Tn}\sum_{t=k_0}^{T-1}\left\|\widehat{\*z}_t\right\|^4\left\|T^{1/4}(\widehat{\+\delta}_t-(\+\Phi_{4,Nt,r}^{-1})'\widetilde{\+\delta}_t)\right\|^4 \notag\\
    &+4\frac{1}{Tn}\sum_{t=k_0}^{T-1} \left\|(\widehat{\*f}_t-\*H_{4,Nt,r} \*f_t)\right\|^4\left\|(\*H_{4,Nt,r}^{-1})\right\|^4\left\|\+\beta^0 \right\|^4=o_p(1),
\end{align}
where subscript $4$ indicates that we chose the fourth asymptotically equivalent rotation matrix described in Lemma 3 in \cite{bai2023approximate}, such that $\*H_{4,Nt,r}=(\widetilde{\+\Lambda}_t'\widetilde{\+\Lambda}_t)^{-1}\+\Lambda'\widetilde{\+\Lambda}_t$. Therefore, the first three terms in (\ref{variance_expansion}) are negligible based on the previous results, while the equivalent rotation matrix allows to write 
\begin{align}
    \sup_{k_0\leq t \leq T-1}\left\|(\widehat{\*f}_t-\*H_{4,Nt,r} \*f_t)\right\|^4&\leq 4  \sup_{k_0\leq t \leq T-1}\left\|\left( \frac{1}{N^\alpha}\widehat{\+\Lambda}_t'\widehat{\+\Lambda}_t\right)^{-1}(\*H_{4,Nt,r}^{-1})'\frac{1}{N^\alpha}\sum_{i=1}^N\+\lambda_ie_{i,t}\right\|^4 \notag\\
    &+ 4 \sup_{k_0\leq t \leq T-1}\left\|\left( \frac{1}{N^\alpha}\widehat{\+\Lambda}_t'\widehat{\+\Lambda}_t\right)^{-1}\frac{1}{N^{\alpha}}\left(\widehat{\+\Lambda}_t-\+\Lambda \*H_{4,Nt,r}^{-1}\right)'\*e_t  \right\|^4\notag\\
    &\leq 4 \sup_{k_0\leq t \leq T-1}\left\|\left( \frac{1}{N^\alpha}\widehat{\+\Lambda}_t'\widehat{\+\Lambda}_t\right)^{-1}\right\|  \sup_{k_0\leq t \leq T-1}\left\|\frac{1}{N^{\alpha}}\left(\widehat{\+\Lambda}_t-\+\Lambda \*H_{4,Nt,r}^{-1}\right)'\*e_t  \right\|^4+O_p(N^{-2\alpha})\notag\\
    &\leq \left( O_p\left(\frac{N}{N^\alpha}\frac{1}{T} \right)+O_p\left(\frac{\sqrt{N}}{N^{3\alpha/2}} \right)+O_p\left( \frac{N}{N^{2\alpha}}\frac{1}{\sqrt{T}}\right) \right)^4=o_p(1)
\end{align}
based on (18) on p. 1901 in \cite{bai2023approximate} and our assumptions, where \hyperref[A3]{A.3} plays the central role. Similarly, under heterogeneous weakness, by using the usual rotation matrix,
\begin{align}
     \sup_{k_0\leq t \leq T-1}\left\|(\widehat{\*f}_t-\*H_{Nt,r} \*f_t)\right\|^4&\leq 4\sup_{k_0\leq t \leq T-1}\left\|\*B_{N}^{-1}(N\*B_{N}^{-2}\*D^2_{Nt,r})^{-1}(\*B_{N}^{-1}\widehat{\*F}_t'\*F_t\*B_N)\*B_{N}^{-1}\+\Lambda'\*e_t \right\|^4\notag\\
     &+4 \sup_{k_0\leq t \leq T-1}\left\| \*B_{N}^{-1}(N\*B_{N}^{-2}\*D^2_{Nt,r})^{-1}\*B_{N}^{-1}\widehat{\*F}_t'\*E_t\*e_t\right\|^4\notag\\
     &\leq 4 \left\| \*B_{N}^{-1}\right\|^4 \sup_{k_0\leq t \leq T-1}\left\|(N\*B_{N}^{-2}\*D^2_{Nt,r})^{-1} \right\|^4\sup_{k_0\leq t \leq T-1}\left\| \*B_{N}^{-1}\widehat{\*F}_t'\*E_t\*e_t\right\|^4+O_p(N^{-2\alpha_r})\notag\\
     &\leq \left(O_p\left(\sqrt{\frac{N}{N^{\alpha_r}}}\frac{1}{\sqrt{T}} \right)\times O_p\left(\frac{N}{N^{\alpha_r}}\frac{1}{\sqrt{T}}\right)+O_p(N^{1/2-3\alpha_r/2}) \right)^4\notag\\
     &=o_p(1)
\end{align}
based on (27) on p. 1905 in \cite{bai2023approximate} and our assumptions. 
\begin{remark}
    It is important to comment on the way the proof proceeds in case $\{u_t\}$ is arbitrarily serially correlated. To illustrate the logic, we turn to the term $B_{12,a}=\frac{1}{\sqrt{d}_T}\sum_{t=\lfloor f_1(T)\rfloor }^{\lfloor f_2(T) \rfloor}u_{t+1}\*z_t'\+\Phi_r'(\widehat{\+\delta}_t-(\+\Phi_{Nt,r}^{-1})'\widetilde{\+\delta}_t)$, which in turn can be decomposed into several parts along the decomposition in (\ref{core_decomp}). In particular, we focus on $\*V$ and $\*V_2$ specifically, as the argument for the rest of the terms will be identical. Once applied to $\*V_2$, we get 
    \begin{align}
       B_{12,a}(\*V_2)&= \frac{1}{\sqrt{d_T}}\sum_{t=\lfloor f_1(T)\rfloor}^{\lfloor f_2(T)\rfloor} \*w_t'\left(\*W_t'\*M_{\*F}\*W_t\right)^{-1}\*W_t'\*M_{\*F}\*u_tu_{t+1}\notag\\
       &=  \frac{1}{\sqrt{d_T}}\sum_{t=\lfloor f_1(T)\rfloor}^{\lfloor f_2(T)\rfloor} \+\mu_\*w'\left(\*W_t'\*M_{\*F}\*W_t\right)^{-1}\*W_t'\*M_{\*F}\*u_tu_{t+1}\notag\\
       &+ \frac{1}{\sqrt{d_T}}\sum_{t=\lfloor f_1(T)\rfloor}^{\lfloor f_2(T)\rfloor} \left[\*w_t-\+\mu_\*w\right]'\left(\*W_t'\*M_{\*F}\*W_t\right)^{-1}\*W_t'\*M_{\*F}\*u_tu_{t+1}\notag\\
       &=\underbrace{\frac{1}{\sqrt{d_T}}\frac{1}{\sqrt{T}}\sum_{t=\lfloor f_1(T)\rfloor}^{\lfloor f_2(T)\rfloor} \+\mu_\*w'\frac{T}{t}\left(t^{-1}\*W_t'\*M_{\*F}\*W_t\right)^{-1}\frac{1}{\sqrt{T}}\*W_t'\*M_{\*F}\*u_tu_{t+1}}_{I}\notag\\
       &+ \underbrace{\frac{1}{\sqrt{d_T}}\frac{1}{\sqrt{T}}\sum_{t=\lfloor f_1(T)\rfloor}^{\lfloor f_2(T)\rfloor} \left[\*w_t-\+\mu_\*w\right]'\frac{T}{t}\left(t^{-1}\*W_t'\*M_{\*F}\*W_t\right)^{-1}\frac{1}{\sqrt{T}}\*W_t'\*M_{\*F}\*u_tu_{t+1}}_{II}\notag.
    \end{align}
    Let $t^{-1}\*W_t'\*M_{\*F}\*W_t=\+\Sigma_{\*w.\*f}+o_p(1)$. The second term then 
    \begin{align}
       \sqrt{d_T} II&= \sum_{t=\lfloor f_1(T)\rfloor}^{\lfloor f_2(T)\rfloor} \frac{T}{t}\left[\left(t^{-1}\*W_t'\*M_{\*F}\*W_t\right)^{-1}\frac{1}{\sqrt{T}}\*W_t'\*M_{\*F}\*u_t\right]'T^{-1/2}\left[\*w_t-\+\mu_\*w\right]u_{t+1}\notag\\
       &\Rightarrow \int_{s=q_1}^{q_2}s^{-1}\left(\+\Sigma_{\*w.\*f}^{-1/2}\*B(s)\right)'d\*W(s) + b,
    \end{align}
    where $\+\mu_\*w=\mathbb{E}\left(\*w_t\right)$, $b$ is a bias term, and $\*B(s)$ and $\*W_s$ represent Brownian motions. This convergence will be satisfied if $\{ \*z_t, u_{t+1}\}_{t=1}^T$ is a mixing process with a certain size (see \citealp{hansen1992convergence}). Therefore, $|II|=O_p(d_T^{-1/2})$. The similar argument applies to $I$:
    \begin{align}
        \sqrt{d_T}I&=\sum_{t=\lfloor f_1(T)\rfloor}^{\lfloor f_2(T)\rfloor} \+\mu_\*w'\frac{T}{t}\left(t^{-1}\*W_t'\*M_{\*F}\*W_t\right)^{-1}\frac{1}{\sqrt{T}}\*W_t'\*M_{\*F}\*u_tT^{-1/2}u_{t+1}\notag\\
        &\Rightarrow \+\mu_\*w'\int_{s=q_1}^{q_2}s^{-1}\+\Sigma_{\*w.\*f}^{-1/2}\*B(s)dv(s) + c
    \end{align},
   as $T\to \infty$, where $v(s)$ is another (scalar) Brownian motion. In total, we then have that
    \begin{align}
        \left| B_{12,a}(\*V_2)\right|= \left|\frac{1}{\sqrt{d_T}}\sum_{t=\lfloor f_1(T)\rfloor}^{\lfloor f_2(T)\rfloor} \*w_t'\left(\*W_t'\*M_{\*F}\*W_t\right)^{-1}\*W_t'\*M_{\*F}\*u_tu_{t+1}\right|=O_p(d_T^{-1/2}).
    \end{align}
    Similar arguments, where and MDS assumption is altered to a high-level condition, can be applied to other terms in Lemma 7.
\end{remark}
\subsection{Main Results}
\begin{theorem} \label{Theorem1} Under Assumption A.1 - A.4 with either homogeneous or heterogeneous loading weakness we have \begin{align*}
&\mathrm{(i.)} \quad  g_{\widehat{f},1}=  g_{f,1}+o_p(1),\\
&\mathrm{(ii.)} \quad  g_{\widehat{f},2}= g_{f,2}+o_p(1),\\
&\mathrm{(iii.)}\quad g_{\widehat{f},3}=g_{f,3}+o_p(1),\\
&\mathrm{(iv.)}\quad g_{\widehat{f},4}=g_{f,4}+o_p(1).
\end{align*}
\end{theorem}
\textbf{Proof}. $\mathrm{(i.)}$. We apply Lemma \ref{Lemma7} (iv.) and define $d_T=(k_0+m_0-1-k_0+1)=m_0$, such that $\sqrt{\frac{d_T}{n}}=\sqrt{\frac{\mu_0(1-\pi_0)}{1-\pi_0}}+O(T^{-1/2})=\sqrt{\mu_0}+O(T^{-1/2})$. Then
\begin{align}
    \frac{1}{\sqrt{n}}\sum_{t=k_0}^{k_0+m_0-1}\widetilde{u}_{1,t+1}(\widetilde{u}_{2,t+1}- \widehat{u}_{2,t+1})=\sqrt{\frac{d_T}{n}}\frac{1}{\sqrt{d_T}}\sum_{t=\lfloor f_1(T)\rfloor }^{\lfloor f_2(T) \rfloor}\widetilde{u}_{1,t+1}(\widetilde{u}_{2,t+1}- \widehat{u}_{2,t+1})=o_p(1).
\end{align}
Next, we let $d_T=(T-1-k_0-m_0+1)=T-k_0-m_0=n-m_0$, such that $\sqrt{\frac{d_T}{n}}=\sqrt{1-\frac{m_0}{n}}=\sqrt{1-\mu_0}+O(T^{-1/2})$. Then
\begin{align}
    \frac{1}{\sqrt{n}}\sum_{t=k_0+m_0}^{T-1}\widetilde{u}_{1,t+1}(\widetilde{u}_{2,t+1}- \widehat{u}_{2,t+1})=\sqrt{\frac{d_T}{n}}\frac{1}{\sqrt{d_T}}\sum_{t=\lfloor f_1(T)\rfloor }^{\lfloor f_2(T) \rfloor}\widetilde{u}_{1,t+1}(\widetilde{u}_{2,t+1}- \widehat{u}_{2,t+1})=o_p(1).
\end{align}
$\mathrm{(ii.)}$ By application of  Lemma \ref{Lemma7} (iv.) and by defining $d_T=k_0+l_2^0-1-k_0+1=l_2^0$, such that $\sqrt{\frac{l_2^0}{n}}=\sqrt{\lambda_2^0}+O(n^{-1/2})$. Then
\begin{align}
   & \frac{1}{\sqrt{n}}\sum_{t=k_0}^{k_0+l_2^0-1}\widetilde{u}_{2,t+1}(\widetilde{u}_{2,t+1}-\widehat{u}_{2,t+1})=\sqrt{\frac{d_T}{n}}\frac{1}{\sqrt{d_T}}\sum_{t=\lfloor f_1(T)\rfloor }^{\lfloor f_2(T) \rfloor}\widetilde{u}_{2,t+1}(\widetilde{u}_{2,t+1}-\widehat{u}_{2,t+1})=o_p(1),\\
   &\frac{1}{\sqrt{n}}\sum_{t=k_0}^{k_0+l_2^0-1}(\widetilde{u}_{2,t+1}-\widehat{u}_{2,t+1})^2=\sqrt{\frac{d_T}{n}}\frac{1}{\sqrt{d_T}}\sum_{t=\lfloor f_1(T)\rfloor }^{\lfloor f_2(T) \rfloor}(\widetilde{u}_{2,t+1}-\widehat{u}_{2,t+1})^2=o_p(1).
\end{align}
\noindent $(\mathrm{iii.})$ Follows by the exact same steps as $\mathrm{(ii.)}$, because averaging occurs only over $l_1$ while holding $l_2^0$ fixed. \\

\noindent $\mathrm{(iv.)}$ Again, we apply  Lemma \ref{Lemma7} $\mathrm{(iv.)}$ and define  $d_T=k_0+l_2-1-k_0+1=l_2$, such that $\sqrt{\frac{l_2}{n}}=\sqrt{\lambda_2}+O(n^{-1/2})$. Then, given that $ \sup_{\lfloor n\tau_0 \rfloor +1\leq l_2\leq n} \frac{n}{l_2}=\frac{n}{\lfloor n\tau_0 \rfloor}=\tau_0^{-1}+O(n^{-1})$ provided that $\tau_0\neq 0$ and making use of the fact that $f_2(T)$ also depends on $l_2$, we have
\begin{align}
    &\left|\frac{1}{n(1-\tau_0)}\sum_{l_2=\lfloor n\tau_0 \rfloor+1}^n\frac{n}{l_2}\left(\frac{2}{\widehat{\omega}}_{4} \frac{1}{\sqrt{n}}\sum_{t=k_0}^{k_0+l_2-1}\widetilde{u}_{2,t+1}(\widetilde{u}_{2,t+1}-\widehat{u}_{2,t+1})- \frac{1}{\widehat{\omega}}_{4} \frac{1}{\sqrt{n}}\sum_{t=k_0}^{k_0+l_2-1}(\widetilde{u}_{2,t+1}-\widehat{u}_{2,t+1})^2\right) \right|\notag\\
    &\leq \sup_{\lfloor n\tau_0 \rfloor+1\leq l_2\leq n} \left|\sqrt{\frac{d_T}{n}} \frac{1}{\sqrt{d_T}}\sum_{t=\lfloor f_1(T)\rfloor }^{\lfloor f_2(T,l_2) \rfloor}\widetilde{u}_{2,t+1}(\widetilde{u}_{2,t+1}-\widehat{u}_{2,t+1})\right| \frac{2}{\widehat{\omega}_{4}(1-\tau_0)}\frac{1}{n}\sum_{l_2=\lfloor n\tau_0 \rfloor+1}^n\frac{l_2}{n}\notag\\
    &+\sup_{\lfloor n\tau_0 \rfloor+1\leq l_2\leq n} \left|\sqrt{\frac{d_T}{n}} \frac{1}{\sqrt{d_T}}\sum_{t=\lfloor f_1(T)\rfloor }^{\lfloor f_2(T,l_2) \rfloor}(\widetilde{u}_{2,t+1}-\widehat{u}_{2,t+1})^2\right| \frac{2}{\widehat{\omega}_{4}(1-\tau_0)}\frac{1}{n}\sum_{l_2=\lfloor n\tau_0 \rfloor+1}^n\frac{n}{l_2}\notag\\
    &=o_p(1),
\end{align}
because 
\begin{align}\label{negligible_integral}
     \sup_{\lfloor n\tau_0 \rfloor+1\leq l_2\leq n} \left|\sqrt{\frac{d_T}{n}} \frac{1}{\sqrt{d_T}}\sum_{t=\lfloor f_1(T)\rfloor }^{\lfloor f_2(T,l_2) \rfloor}(\widetilde{u}_{2,t+1}-\widehat{u}_{2,t+1})^2\right|&\leq O(1)\times \frac{1}{\sqrt{k_0}}\sum_{t=k_0}^{T-1}(\widetilde{u}_{2,t+1}-\widehat{u}_{2,t+1})^2\notag\\
     &=o_p(1),
\end{align}
while the supremum in the first term is negligible, since in the $g_{\widehat{f},2}$ we demonstrated that such terms are negligible uniformly in $l_2$. More generally, they both follow directly based on the logic in \cite{pitarakis2025novel}, and in particular, the passage from (A.6) to (A.8), which demonstrates that the remainder still vanishes inside of the integral. Lastly,
\begin{align}
    \frac{1}{n}\sum_{l_2=\lfloor n\tau_0 \rfloor+1}^n\frac{l_2}{n}=\int_{s=\tau_0}^1s^{-1}ds + o(1)=O(1).
\end{align}
\subsection{Equivalence of the Power Enhancement}
The statistics $g_{f,2}$, 
$g_{f,3}$ and 
$g_{f,4}$ can be power-enhanced, as argued in \cite{pitarakis2025novel}. In particular, the choice of the infeasible adjustment terms are created by defining $\breve{u}_{2,t+1}^2=\widetilde{u}_{2,t+1}^2-(\widetilde{u}_{1,t+1}-\widetilde{u}_{2,t+1})^2$ (see (33) in \citealp{pitarakis2025novel}). It is equivalent to adding the following terms to the respective statistics:
\begin{align}
&\widetilde{\zeta}(\lambda_1^0,\lambda_2^0)=\frac{1}{\widetilde{\omega}_2}\frac{1}{\lambda_2^0}\frac{1}{\sqrt{n}}\sum_{t=k_0}^{k_0+l_2^0-1}(\widetilde{u}_{1,t+1}-\widetilde{u}_{2,t+1})^2,\\
&\widetilde{\psi}(\tau_0,\lambda_1^0,\lambda_2^0)=\frac{1}{\widetilde{\omega}}_{3}\frac{1}{\lambda_2^0}\frac{1}{\sqrt{n}}\sum_{t=k_0}^{k_0+l_2^0-1}(\widetilde{u}_{1,t+1}-\widetilde{u}_{2,t+1})^2,\\
&\widetilde{\phi}(\tau_0,\lambda_1^0,\lambda_2^0)=\frac{1}{\widetilde{\omega}}_4\frac{1}{n(1-\tau_0)}\sum_{l_2=\lfloor n\tau_0\rfloor+1}^n\frac{n}{l_2}\frac{1}{\sqrt{n}}\sum_{t=k_0}^{k_0+l_2-1}(\widetilde{u}_{1,t+1}-\widetilde{u}_{2,t+1})^2,
\end{align}
where the dependence on $\lambda_1^0, \lambda_2^0$ follows through the variance estimators. \footnote{Note that strictly we have $\widetilde{\psi}(\tau_0,\lambda_1^0,\lambda_2^0)=\frac{1}{\widetilde{\omega}_3}\frac{1}{\lambda_2^0}\frac{1}{n(1-\tau_0)}\sum_{l_1=\lfloor n\tau_0\rfloor+1}^n\frac{1}{\sqrt{n}}\sum_{t=k_0}^{k_0+l_2^0-1}(\widetilde{u}_{1,t+1}-\widetilde{u}_{2,t+1})^2$. However, 
\begin{align*}
   \widetilde{\psi}(\tau_0,\lambda_1^0,\lambda_2^0)&=\frac{1}{\widetilde{\omega}_3}\frac{1}{\lambda_2^0}\frac{1}{n(1-\tau_0)}\sum_{l_1=\lfloor n\tau_0\rfloor+1}^n\frac{1}{\sqrt{n}}\sum_{t=k_0}^{k_0+l_2^0-1}(\widetilde{u}_{1,t+1}-\widetilde{u}_{2,t+1})^2\\
   &=\frac{1}{\widetilde{\omega}_3}\frac{1}{\lambda_2^0}\frac{n-\lfloor n \tau_0 \rfloor}{n(1-\tau_0)}\frac{1}{\sqrt{n}}\sum_{t=k_0}^{k_0+l_2^0-1}(\widetilde{u}_{1,t+1}-\widetilde{u}_{2,t+1})^2=\frac{1}{\widetilde{\omega}_3}\frac{1}{\lambda_2^0}\frac{1}{\sqrt{n}}\sum_{t=k_0}^{k_0+l_2^0-1}(\widetilde{u}_{1,t+1}-\widetilde{u}_{2,t+1})^2 +O(n^{-1}),
\end{align*}
therefore they are equivalent.} Clearly, the feasible versions based on the estimated factors are given by
\begin{align}
    &\widehat{\zeta}(\lambda_1^0,\lambda_2^0)=\frac{1}{\widehat{\omega}_2}\frac{1}{\lambda_2^0}\frac{1}{\sqrt{n}}\sum_{t=k_0}^{k_0+l_2^0-1}(\widetilde{u}_{1,t+1}-\widehat{u}_{2,t+1})^2,\\
&\widehat{\psi}(\tau_0,\lambda_1^0,\lambda_2^0)=\frac{1}{\widehat{\omega}}_{3}\frac{1}{\lambda_2^0}\frac{1}{\sqrt{n}}\sum_{t=k_0}^{k_0+l_2^0-1}(\widetilde{u}_{1,t+1}-\widehat{u}_{2,t+1})^2,\\
&\widehat{\phi}(\tau_0,\lambda_1^0,\lambda_2^0)=\frac{1}{\widehat{\omega}}_4\frac{1}{n(1-\tau_0)}\sum_{l_2=\lfloor n\tau_0\rfloor+1}^n\frac{n}{l_2}\frac{1}{\sqrt{n}}\sum_{t=k_0}^{k_0+l_2-1}(\widetilde{u}_{1,t+1}-\widehat{u}_{2,t+1})^2
\end{align}
We can approximate them with the feasible adjustments. 
\begin{proposition} \label{Prop3} Under Assumptions A.1 - A.4 with either homogeneous or heterogeneous loading weakness we have \begin{align*}
  &\widehat{\zeta}(\lambda_1^0,\lambda_2^0)=\widetilde{\zeta}(\lambda_1^0,\lambda_2^0) + o_p(1), \\
  &\widehat{\psi}(\tau_0,\lambda_1^0,\lambda_2^0)=\widetilde{\psi}(\tau_0,\lambda_1^0,\lambda_2^0) + o_p(1),\\
  &\widehat{\phi}(\tau_0,\lambda_1^0,\lambda_2^0)=\widetilde{\phi}(\tau_0,\lambda_1^0,\lambda_2^0)+o_p(1).
    \end{align*}
\end{proposition}

\noindent \textbf{Proof}. Given part $\mathrm{(iv.)}$ of Lemma \ref{Lemma7}, we can demonstrate that 
\begin{align}
    \frac{1}{\sqrt{n}}\sum_{t=k_0}^{k_0+l_2^0-1}(\widetilde{u}_{1,t+1}-\widehat{u}_{2,t+1})^2&=\frac{1}{\sqrt{n}}\sum_{t=k_0}^{k_0+l_2^0-1}(\widetilde{u}_{1,t+1}-\widetilde{u}_{2,t+1}+\widetilde{u}_{2,t+1}-\widehat{u}_{2,t+1})^2\notag\\
    &=\frac{1}{\sqrt{n}}\sum_{t=k_0}^{k_0+l_2^0-1}(\widetilde{u}_{1,t+1}-\widetilde{u}_{2,t+1})^2+\frac{2}{\sqrt{n}}\sum_{t=k_0}^{k_0+l_2^0-1}(\widetilde{u}_{1,t+1}-\widetilde{u}_{2,t+1})(\widetilde{u}_{2,t+1}-\widehat{u}_{2,t+1})\notag\\
    &+\frac{1}{\sqrt{n}}\sum_{t=k_0}^{k_0+l_2^0-1}(\widetilde{u}_{1,t+1}-\widehat{u}_{2,t+1})^2\notag\\
    &=\frac{1}{\sqrt{n}}\sum_{t=k_0}^{k_0+l_2^0-1}(\widetilde{u}_{1,t+1}-\widetilde{u}_{2,t+1})^2+o_p(1),
\end{align}
because the last term is negligible based on Lemma \ref{Lemma7} (i.). Moreover, by Cauchy-Schwarz inequality,
\begin{align}
    \Bigg|\frac{1}{\sqrt{n}}\sum_{t=k_0}^{k_0+l_2^0-1}(\widetilde{u}_{1,t+1}-\widetilde{u}_{2,t+1})(\widetilde{u}_{1,t+1}&-\widehat{u}_{2,t+1}) \Bigg| \leq \frac{1}{\sqrt{n}}\sum_{t=k_0}^{k_0+l_2^0-1}|\widetilde{u}_{1,t+1}-\widetilde{u}_{2,t+1}||\widetilde{u}_{2,t+1}-\widehat{u}_{2,t+1}|\notag\\
    &\leq \left(\frac{1}{\sqrt{n}}\sum_{t=k_0}^{k_0+l_2^0-1}(\widetilde{u}_{1,t+1}-\widetilde{u}_{2,t+1})^2\right)^{1/2}\left(\frac{1}{\sqrt{n}}\sum_{t=k_0}^{k_0+l_2^0-1}(\widetilde{u}_{2,t+1}-\widehat{u}_{2,t+1})^2 \right)^{1/2}\notag\\
    &=o_p(1),
\end{align}
due to Lemma \ref{Lemma7}, and the first component coming from the infeasible adjustment term, which is well-behaved under the local alternative. In particular,
\begin{align}
    \frac{1}{\sqrt{n}}\sum_{t=k_0}^{k_0+l_2^0-1}(\widetilde{u}_{1,t+1}-\widetilde{u}_{2,t+1})^2&=\frac{1}{\sqrt{n}}\sum_{t=k_0}^{k_0+l_2^0-1}\widetilde{u}_{1,t+1}^2 - \frac{2}{\sqrt{n}}\sum_{t=k_0}^{k_0+l_2^0-1}\widetilde{u}_{1,t+1}\widetilde{u}_{2,t+1} + \frac{1}{\sqrt{n}}\sum_{t=k_0}^{k_0+l_2^0-1}\widetilde{u}_{2,t+1}^2 \notag\\
    &= \frac{1}{\sqrt{n}}\sum_{t=k_0}^{k_0+l_2^0-1}(\widetilde{u}_{1,t+1}^2-u_{t+1}^2) -\frac{2}{\sqrt{n}}\sum_{t=k_0}^{k_0+l_2^0-1}(\widetilde{u}_{1,t+1}\widetilde{u}_{2,t+1}-u_{t+1}^2)\notag\\
    &+\frac{1}{\sqrt{n}}\sum_{t=k_0}^{k_0+l_2^0-1}(\widetilde{u}_{2,t+1}^2-u_{t+1}^2)\notag\\
    &=\frac{1}{\sqrt{n}}\sum_{t=k_0}^{k_0+l_2^0-1}(\widetilde{u}_{1,t+1}^2-u_{t+1}^2) + O_p(T^{-1/4})\notag\\
    &\to_p \sqrt{1-\pi_0}\+\beta^{0\prime}(\*I_r-\+\Sigma_{\*w\*f}'\+\Sigma^{-1}_\*w\+\Sigma_{\*w\*f})\+\beta^{0},
\end{align}
as $T\to \infty$, because of the arguments in the proof of Proposition \ref{Prop2}. The proof regarding the $\widehat{\phi}(\tau_0,\lambda_1^0,\lambda_2^0)$ follows the steps in (\ref{gf4_power}), and it goes 
\begin{align}
    &\frac{1}{n(1-\tau_0)}\sum_{l_2=\lfloor n\tau_0\rfloor+1}^n\frac{n}{l_2}\frac{1}{\sqrt{n}}\sum_{t=k_0}^{k_0+l_2-1}(\widetilde{u}_{1,t+1}-\widehat{u}_{2,t+1})^2\notag\\
    &= \frac{1}{n(1-\tau_0)}\sum_{l_2=\lfloor n\tau_0\rfloor+1}^n\frac{n}{l_2}\frac{1}{\sqrt{n}}\sum_{t=k_0}^{k_0+l_2-1}(\widetilde{u}_{1,t+1}-\widetilde{u}_{2,t+1})^2\notag\\
    &+\frac{2}{n(1-\tau_0)}\sum_{l_2=\lfloor n\tau_0\rfloor+1}^n\frac{n}{l_2}\frac{1}{\sqrt{n}}\sum_{t=k_0}^{k_0+l_2-1}(\widetilde{u}_{1,t+1}-\widetilde{u}_{2,t+1})(\widetilde{u}_{2,t+1}-\widehat{u}_{2,t+1})\notag\\
    &+\frac{1}{n(1-\tau_0)}\sum_{l_2=\lfloor n\tau_0\rfloor+1}^n\frac{n}{l_2}\frac{1}{\sqrt{n}}\sum_{t=k_0}^{k_0+l_2-1}(\widetilde{u}_{1,t+1}-\widehat{u}_{2,t+1})^2 \notag\\
    &=\frac{1}{n(1-\tau_0)}\sum_{l_2=\lfloor n\tau_0\rfloor+1}^n\frac{n}{l_2}\frac{1}{\sqrt{n}}\sum_{t=k_0}^{k_0+l_2-1}(\widetilde{u}_{1,t+1}-\widetilde{u}_{2,t+1})^2 + o_p(1)
\end{align}
under our conditions, because the last term is exactly (\ref{negligible_integral}) and the middle term follows
\begin{align}
  &\left|\frac{1}{n(1-\tau_0)}\sum_{l_2=\lfloor n\tau_0\rfloor+1}^n\frac{n}{l_2}\frac{1}{\sqrt{n}}\sum_{t=k_0}^{k_0+l_2-1}(\widetilde{u}_{1,t+1}-\widetilde{u}_{2,t+1})(\widetilde{u}_{2,t+1}-\widehat{u}_{2,t+1})\right|\notag\\
  &\leq \frac{1}{n(1-\tau_0)}\sum_{l_2=\lfloor n\tau_0\rfloor+1}^n\frac{n}{l_2}\left(\frac{1}{\sqrt{n}}\sum_{t=k_0}^{k_0+l_2-1}(\widetilde{u}_{1,t+1}-\widetilde{u}_{2,t+1})^2\right)^{1/2}\left(\frac{1}{\sqrt{n}}\sum_{t=k_0}^{k_0+l_2-1}(\widetilde{u}_{2,t+1}-\widehat{u}_{2,t+1})^2 \right)^{1/2}\notag\\
  &\leq \left(\frac{1}{\sqrt{n}}\sum_{t=k_0}^{T-1} (\widetilde{u}_{2,t+1}-\widehat{u}_{2,t+1})^2 \right)^{1/2} \times  \underbrace{\frac{1}{n(1-\tau_0)}\sum_{l_2=\lfloor n\tau_0\rfloor+1}^n\frac{n}{l_2}\left(\frac{1}{\sqrt{n}}\sum_{t=k_0}^{k_0+l_2-1}(\widetilde{u}_{1,t+1}-\widetilde{u}_{2,t+1})^2\right)^{1/2}}_{O_p(1)}\notag\\
  &= \left(\frac{1}{\sqrt{n}}\sum_{t=k_0}^{T-1} (\widetilde{u}_{2,t+1}-\widehat{u}_{2,t+1})^2 \right)^{1/2} \times \left(\sqrt{1-\pi_0}\+\beta^{0\prime}(\*I_r-\+\Sigma_{\*w\*f}'\+\Sigma^{-1}_\*w\+\Sigma_{\*w\*f})\+\beta^{0}\right)^{1/2} \left(\frac{1}{n(1-\tau_0)}\sum_{l_2=\lfloor n\tau_0\rfloor+1}^n\frac{n}{l_2}\right) \notag\\
  &+o_p(1) \notag\\ 
   &=\left(\frac{1}{\sqrt{n}}\sum_{t=k_0}^{T-1} (\widetilde{u}_{2,t+1}-\widehat{u}_{2,t+1})^2 \right)^{1/2} \times \left(\sqrt{1-\pi_0}\+\beta^{0\prime}(\*I_r-\+\Sigma_{\*w\*f}'\+\Sigma^{-1}_\*w\+\Sigma_{\*w\*f})\+\beta^{0}\right)^{1/2} \left(\frac{1}{(1-\tau_0)}\int_{s=\tau_0}^1s^{-1}ds\right)\notag\\
   &+o_p(1)\notag\\
  &=o_p(1).
\end{align}

\section{Additional Monte Carlo Results}
\subsection{Design, Size and Power}
We design a DGP similar to DGP2 in \citet{pitarakis2025novel} but where factors are specified in the same way as DGP2 of \citet{bai2023approximate} with strong/ weak homogeneous/ heterogeneous loadings. Throughout, we set the number of factors $r=3$.
\begin{align*}
  &y_{t+1}=c+\theta_1 y_t+\+\beta'\*f_t+u_{t+1}, \quad \*f_t\in\mathbb{R}^{r=3}, \\ &u_{t+1}=\begin{cases}
      NID(0,1) \hspace{2mm} (\textit{Baseline}),\\
\sigma_{t+1}\varepsilon_{t+1},\quad \varepsilon_{t+1}\sim NID(0,1),\quad  \sigma^2_{t+1}=\omega+\alpha u_t^2+\eta \sigma^2_t,\quad \omega,\alpha=0.1,\quad  \eta=0.2,  
  \end{cases} 
  \\ 
&x_{i,t}=\+\lambda_i'\*f_t+e_{i,t},\\
  &e_{i,t}=\rho_ie_{i,t-1}+\sqrt{1-\rho_i^2}v_{i,t},\quad   v_{i,t} =\begin{cases}
  NID(0,1)\hspace{2mm} (\textit{Baseline}),\\
      \epsilon_{i,t}+ \sum_{k = i+1}^{K} \xi( \epsilon_{i-k,t} + \epsilon_{i+k,t}), \quad \epsilon_{i,0} =0, \quad K = 5,\quad \xi=0.4,
  \end{cases}\\
&\*f_t\sim N(0,\*I_3),\quad \+\lambda_i\sim \*G_i\*D\*B_N/\sqrt{N}+\*G_i\pi/\sqrt{N}, \quad  \*F'\*F/T\approx\*I_3, \quad \*B_N^{-1}\*\Lambda'\*\Lambda\*B_N^{-1}\approx \*D^2,
\end{align*}
where: $\*G_i\sim N(0,\*I_3)$, $\*D^2=\operatorname{diag}(3\;2\;1);$ $ \*B_N=\operatorname{diag}(N^{\alpha_1/2}\;N^{\alpha_2/2}\;N^{\alpha_3/2})$, $(\alpha_1,\alpha_2,\alpha_3)=(1, 1, 1)$ for strong homogeneous loadings, $(0.51, 0.51, 0.51)$ for weak homogeneous loadings, $(1, 0.7, 0.51)$ for mixed strong/weak heterogeneous loadings. 

\begin{remark}
    Note that $\+\lambda_i$ is simulated differently than in \cite{bai2023approximate}. The reason is that the rate of $\|\*B_N^{-1}\+\Lambda'\+\Lambda\*B_N^{-1}- \+\Sigma_{\+\Lambda} \|$ plays an important role in our asymptotic analysis, whereas it did not matter for \cite{bai2023approximate}. This simulation method mimics A.3 i), because we can show that under such design $\*B_N^{-1}\+\Lambda'\+\Lambda\*B_N^{-1}= \*D^2 + O_p(N^{-\alpha_r/2})$ as desired. Consider $\+\lambda_i\sim \*G_{i}\*D\*B_N/\sqrt{N}+\*G_{i}\pi/\sqrt{N}$, where $\pi>0$ regulates the strength of the random perturbation:
    \begin{align*}
\*B_N^{-1} \*\Lambda' \*\Lambda \*B_N^{-1} &= \*B_N^{-1} \left( \*B_N \*D \frac{1}{N} \sum_{i=1}^{N} \*G_{i} \*G'_{i} \*D \*B_N \right) \*B_N^{-1} 
+ \*B_N^{-1} \left( \*B_N \*D \frac{1}{N} \sum_{i=1}^{N} \*G_{i} \*G'_{i} \right) \*B_N^{-1}\\
&+ \*B_N^{-1} \left( \*B_N \*D \frac{1}{N} \sum_{i=1}^{N} \*G_{i} \*G'_{i} \right)' \*B_N^{-1}
+ \*B_N^{-1} \frac{1}{N} \sum_{i=1}^{N} \*G_{i} \*G'_{i} \*B_N^{-1}\\ 
&= \*D \frac{1}{N} \sum_{i=1}^{N} \*G_i \*G'_i \*D + \*D \pi \frac{1}{N} \sum_{i=1}^{N} \*G_{i} \*G'_{i} \*B_N^{-1}
+ \*B_N^{-1}\left( \*D \pi\frac{1}{N} \sum_{i=1}^{N} \*G_{i} \*G'_{i} \right)' \\
&+ \*B_N^{-1} \pi^2\frac{1}{N} \sum_{i=1}^{N} \*G_{i} \*G'_{i} \*B_N^{-1} 
= 1 + 2 + 3 + 4.
\end{align*}

\noindent Clearly, 1 is $O_p(1)$. Next, 2 (and hence 3) is $O_p(N^{-\alpha_r/2})$:
\begin{align*}
    \left\|2 \right\|=\left\|3 \right\|\leq \pi\left\|\*D \right\| \left\|  \frac{1}{N} \sum_{i=1}^{N} \*G_{i} \*G'_{i}\right\|\left\|\*B^{-1} \right\|=O_p(N^{-\alpha_r/2}).
\end{align*}
Lastly,
\begin{align*}
    \| 4 \| \leq \pi^2\| \*B_N^{-1} \|^2 \left\| \frac{1}{N} \sum_{i=1}^{N} \*G_{i} \*G'_{i} \right\| = O_p(N^{-\alpha_r}).
\end{align*}
\noindent Therefore, the perturbation $\*G_{i}\pi/\sqrt{N}$ induces the reaction to the weakest loading. If the factors are strong, we obtain $O_p(N^{-1/2})$ rate as it should be. Of course, the functional form of the perturbation can be altered, as $\*G_i$ can be substituted for another random matrix, which may be correlated or uncorrelated to $\*G_i$. 

\end{remark}

\noindent Cross-sectional and time dimensions are as follows: $(N,T)=(800,500)$. For the practical implementations, following \citet{pitarakis2025novel}, we set $c=1.25$, $\theta_1=0.5$, $\rho_i=0.3+N(0,1)_i\times 0.5$, $\+\beta=(0,0,0)'$, $\pi=24$, for size, and $\+\beta=(j,j,j)'$ for $j\in \{0.1, 0.2, 0.3, 0.35, 0.4,\allowbreak 0.45, 0.5, 0.55, 0.6\}$ for power.

\subsection{Size and Power Tables} Here we display the full Monte Carlo simulation tables for all the tests --and different choices of the tuning parameters-- in the baseline scenario (see also Section 3.1 of the main paper). 
    \begin{table}[ht]
\centering
\caption{$g_{\widehat{f},1}$, DGP Size/Power (nom 5\%),  $\alpha_i=[1,1,1]$}\label{tab_size_test_mu}
\begin{tabular}{l c c c c}
\toprule
 & \multicolumn{4}{c}{$\mu_0$} \\
\cmidrule(lr){2-5}
$g_{\widehat{f},1}$ & 0.30 & 0.35 & 0.40 & 0.45 \\
\midrule
$\beta_j=0$ & 0.036 & 0.052 & 0.054 & 0.098 \\
$\beta_j=0.10$ & 0.156 & 0.240 & 0.382 & 0.620 \\
$\beta_j=0.20$ & 0.654 & 0.800 & 0.892 & 0.938 \\
$\beta_j=0.30$ & 0.888 & 0.924 & 0.948 & 0.974 \\
$\beta_j=0.35$ & 0.920 & 0.946 & 0.962 & 0.978 \\
$\beta_j=0.40$ & 0.934 & 0.958 & 0.972 & 0.980 \\
$\beta_j=0.45$ & 0.944 & 0.962 & 0.976 & 0.980 \\
$\beta_j=0.50$ & 0.954 & 0.966 & 0.980 & 0.984 \\
$\beta_j=0.55$ & 0.956 & 0.972 & 0.980 & 0.986 \\
$\beta_j=0.60$ & 0.962 & 0.974 & 0.982 & 0.986 \\
\bottomrule
\end{tabular}
\end{table}

\begin{table}[ht]
\centering
\caption{$g_{\widehat{f},1}$, DGP Size/Power (nom 5\%),  $\alpha_i=[0.51, 0.51, 0.51]$}\label{tab_size_test_mu}
\begin{tabular}{l c c c c}
\toprule
 & \multicolumn{4}{c}{$\mu_0$} \\
\cmidrule(lr){2-5}
$g_{\widehat{f},1}$ & 0.30 & 0.35 & 0.40 & 0.45 \\
\midrule
$\beta_j=0$ & 0.030 & 0.052 & 0.054 & 0.088 \\
$\beta_j=0.10$ & 0.122 & 0.172 & 0.278 & 0.508 \\
$\beta_j=0.20$ & 0.492 & 0.654 & 0.756 & 0.846 \\
$\beta_j=0.30$ & 0.770 & 0.832 & 0.878 & 0.922 \\
$\beta_j=0.35$ & 0.814 & 0.870 & 0.902 & 0.934 \\
$\beta_j=0.40$ & 0.854 & 0.890 & 0.920 & 0.944 \\
$\beta_j=0.45$ & 0.872 & 0.902 & 0.922 & 0.948 \\
$\beta_j=0.50$ & 0.886 & 0.910 & 0.924 & 0.954 \\
$\beta_j=0.55$ & 0.892 & 0.918 & 0.928 & 0.954 \\
$\beta_j=0.60$ & 0.896 & 0.918 & 0.932 & 0.956 \\
\bottomrule
\end{tabular}
\end{table}

\begin{table}[ht]
\centering
\caption{$g_{\widehat{f},1}$, DGP Size/Power (nom 5\%),  $\alpha_i=[0.1, 0.1, 0.1]$}\label{tab_size_test_mu}
\begin{tabular}{l c c c c}
\toprule
 & \multicolumn{4}{c}{$\mu_0$} \\
\cmidrule(lr){2-5}
$g_{\widehat{f},1}$ & 0.30 & 0.35 & 0.40 & 0.45 \\
\midrule
$\beta_j=0$ & 0.032 & 0.038 & 0.044 & 0.088 \\
$\beta_j=0.10$ & 0.118 & 0.158 & 0.234 & 0.416 \\
$\beta_j=0.20$ & 0.404 & 0.552 & 0.664 & 0.804 \\
$\beta_j=0.30$ & 0.690 & 0.762 & 0.820 & 0.860 \\
$\beta_j=0.35$ & 0.734 & 0.796 & 0.850 & 0.882 \\
$\beta_j=0.40$ & 0.758 & 0.826 & 0.864 & 0.894 \\
$\beta_j=0.45$ & 0.786 & 0.840 & 0.868 & 0.900 \\
$\beta_j=0.50$ & 0.812 & 0.850 & 0.880 & 0.904 \\
$\beta_j=0.55$ & 0.824 & 0.858 & 0.886 & 0.912 \\
$\beta_j=0.60$ & 0.834 & 0.866 & 0.892 & 0.918 \\
\bottomrule
\end{tabular}
\end{table}

\begin{table}[ht]
\centering
\caption{$g_{\widehat{f},1}$, DGP Size/Power (nom 5\%),  $\alpha_i=[0.51, 0.7, 1]$}\label{tab_size_test_mu}
\begin{tabular}{l c c c c}
\toprule
 & \multicolumn{4}{c}{$\mu_0$} \\
\cmidrule(lr){2-5}
$g_{\widehat{f},1}$ & 0.30 & 0.35 & 0.40 & 0.45 \\
\midrule
$\beta_j=0$ & 0.032 & 0.042 & 0.056 & 0.092 \\
$\beta_j=0.10$ & 0.158 & 0.258 & 0.400 & 0.660 \\
$\beta_j=0.20$ & 0.730 & 0.850 & 0.906 & 0.968 \\
$\beta_j=0.30$ & 0.928 & 0.954 & 0.978 & 0.986 \\
$\beta_j=0.35$ & 0.954 & 0.972 & 0.982 & 0.986 \\
$\beta_j=0.40$ & 0.966 & 0.978 & 0.984 & 0.992 \\
$\beta_j=0.45$ & 0.970 & 0.980 & 0.986 & 0.992 \\
$\beta_j=0.50$ & 0.978 & 0.982 & 0.986 & 0.994 \\
$\beta_j=0.55$ & 0.982 & 0.984 & 0.988 & 0.994 \\
$\beta_j=0.60$ & 0.982 & 0.984 & 0.990 & 0.994 \\
\bottomrule
\end{tabular}
\end{table}
\begin{table}[ht]
\centering
\caption{$g_{\widehat{f},2}$ and $g_{\widehat{f},2}^{adj}$, DGP Size/Power (nom 5\%),  $\tau_0=0.8$, $\alpha_i=[1,1,1]$}\label{tab_size_test2}
\begin{tabular}{l l c c c c c c c c c c}
\toprule
 &$\lambda_1^0=1$& \multicolumn{10}{c}{$(N,T)=(800,500)$} \\
\cmidrule(lr){3-12}
 &$\lambda_2^0$& 0.5 & 0.55 & 0.6 & 0.65 & 0.7 & 0.75 & 0.8 & 0.85 & 0.9 & 0.95 \\
\midrule
&$g_{\widehat{f},2}$ \\
&$\beta_j=0$ & 0.032 & 0.036 & 0.026 & 0.030 & 0.044 & 0.038 & 0.040 & 0.034 & 0.030 & 0.030 \\
&$\beta_j=0.10$ & 0.060 & 0.058 & 0.062 & 0.060 & 0.082 & 0.084 & 0.130 & 0.124 & 0.152 & 0.220 \\
&$\beta_j=0.20$ & 0.258 & 0.290 & 0.316 & 0.360 & 0.404 & 0.436 & 0.518 & 0.568 & 0.640 & 0.722 \\
&$\beta_j=0.30$ & 0.546 & 0.600 & 0.638 & 0.668 & 0.690 & 0.732 & 0.750 & 0.782 & 0.820 & 0.838 \\
&$\beta_j=0.35$ & 0.670 & 0.690 & 0.720 & 0.732 & 0.756 & 0.768 & 0.802 & 0.820 & 0.830 & 0.852 \\
\bottomrule
\end{tabular}
\end{table}

\begin{table}[ht]
\centering
\caption{$g_{\widehat{f},2}$ and $g_{\widehat{f},2}^{adj}$, DGP Size/Power (nom 5\%),  $\tau_0=0.8$, $\alpha_i=[0.51,0.51,0.51]$}\label{tab_size_test2}
\begin{tabular}{l l c c c c c c c c c c}
\toprule
 &$\lambda_1^0=1$& \multicolumn{10}{c}{$(N,T)=(800,500)$} \\
\cmidrule(lr){3-12}
 &$\lambda_2^0$& 0.5 & 0.55 & 0.6 & 0.65 & 0.7 & 0.75 & 0.8 & 0.85 & 0.9 & 0.95 \\
\midrule
&$g_{\widehat{f},2}$ \\
&$\beta_j=0$ & 0.028 & 0.034 & 0.024 & 0.028 & 0.048 & 0.046 & 0.046 & 0.040 & 0.028 & 0.032 \\
&$\beta_j=0.10$ & 0.036 & 0.044 & 0.052 & 0.050 & 0.078 & 0.072 & 0.080 & 0.092 & 0.098 & 0.134 \\
&$\beta_j=0.20$ & 0.144 & 0.170 & 0.192 & 0.210 & 0.250 & 0.280 & 0.320 & 0.378 & 0.432 & 0.526 \\
&$\beta_j=0.30$ & 0.368 & 0.404 & 0.422 & 0.444 & 0.464 & 0.508 & 0.550 & 0.578 & 0.604 & 0.636 \\
&$\beta_j=0.35$ & 0.442 & 0.460 & 0.512 & 0.526 & 0.544 & 0.560 & 0.606 & 0.626 & 0.644 & 0.680 \\
\bottomrule
\end{tabular}
\end{table}

\begin{table}[ht]
\centering
\caption{$g_{\widehat{f},2}$ and $g_{\widehat{f},2}^{adj}$, DGP Size/Power (nom 5\%),  $\tau_0=0.8$, $\alpha_i=[0.1,0.1,0.1]$}\label{tab_size_test2}
\begin{tabular}{l l c c c c c c c c c c}
\toprule
 &$\lambda_1^0=1$& \multicolumn{10}{c}{$(N,T)=(800,500)$} \\
\cmidrule(lr){3-12}
 &$\lambda_2^0$& 0.5 & 0.55 & 0.6 & 0.65 & 0.7 & 0.75 & 0.8 & 0.85 & 0.9 & 0.95 \\
\midrule
&$g_{\widehat{f},2}$ \\
&$\beta_j=0$ & 0.028 & 0.032 & 0.022 & 0.028 & 0.038 & 0.044 & 0.040 & 0.026 & 0.032 & 0.032 \\
&$\beta_j=0.10$ & 0.030 & 0.040 & 0.048 & 0.050 & 0.066 & 0.056 & 0.078 & 0.070 & 0.098 & 0.110 \\
&$\beta_j=0.20$ & 0.110 & 0.124 & 0.120 & 0.142 & 0.192 & 0.220 & 0.248 & 0.294 & 0.348 & 0.400 \\
&$\beta_j=0.30$ & 0.278 & 0.298 & 0.332 & 0.354 & 0.370 & 0.390 & 0.432 & 0.454 & 0.486 & 0.518 \\
&$\beta_j=0.35$ & 0.354 & 0.376 & 0.400 & 0.420 & 0.420 & 0.454 & 0.482 & 0.498 & 0.514 & 0.564 \\
\bottomrule
\end{tabular}
\end{table}

\begin{table}[ht]
\centering
\caption{$g_{\widehat{f},2}$ and $g_{\widehat{f},2}^{adj}$, DGP Size/Power (nom 5\%),  $\tau_0=0.8$, $\alpha_i=[0.51,0.7,1]$}\label{tab_size_test2}
\begin{tabular}{l l c c c c c c c c c c}
\toprule
 &$\lambda_1^0=1$& \multicolumn{10}{c}{$(N,T)=(800,500)$} \\
\cmidrule(lr){3-12}
 &$\lambda_2^0$& 0.5 & 0.55 & 0.6 & 0.65 & 0.7 & 0.75 & 0.8 & 0.85 & 0.9 & 0.95 \\
\midrule
&$g_{\widehat{f},2}$ \\
&$\beta_j=0$    & 0.030 & 0.034 & 0.028 & 0.030 & 0.038 & 0.036 & 0.042 & 0.038 & 0.030 & 0.026 \\
&$\beta_j=0.10$ & 0.058 & 0.056 & 0.058 & 0.066 & 0.088 & 0.100 & 0.130 & 0.144 & 0.166 & 0.232 \\
&$\beta_j=0.20$ & 0.272 & 0.284 & 0.324 & 0.358 & 0.428 & 0.458 & 0.542 & 0.602 & 0.678 & 0.776 \\
&$\beta_j=0.30$ & 0.636 & 0.676 & 0.700 & 0.724 & 0.746 & 0.780 & 0.800 & 0.826 & 0.878 & 0.902 \\
&$\beta_j=0.35$ & 0.732 & 0.750 & 0.768 & 0.770 & 0.796 & 0.814 & 0.848 & 0.880 & 0.904 & 0.918 \\
\bottomrule
\end{tabular}
\end{table}

   \begin{table}[ht]
\centering
\caption{$g_{\widehat{f},3}$ and $g_{\widehat{f},3}^{adj}$, DGP Size/Power (nom 5\%),  $\tau_0=0.8$, $\alpha_i=[1,1,1]$}\label{tab_size_test3}
\begin{tabular}{l l c c c c c c c c c c c}
\toprule
 && \multicolumn{11}{c}{$(N,T)=(800,500)$} \\
\cmidrule(lr){3-13}
 &$\lambda_2^0$& 0.5 & 0.55 & 0.6 & 0.65 & 0.7 & 0.75 & 0.8 & 0.85 & 0.9 & 0.95 & 1.0 \\
\midrule
&$g_{\widehat{f},3}$ \\
&$\beta_j=0$ & 0.022 & 0.030 & 0.028 & 0.030 & 0.030 & 0.030 & 0.046 & 0.022 & 0.020 & 0.012 & 0.020 \\
&$\beta_j=0.10$ & 0.062 & 0.052 & 0.068 & 0.068 & 0.094 & 0.106 & 0.180 & 0.242 & 0.326 & 0.264 & 0.174 \\
&$\beta_j=0.20$ & 0.274 & 0.304 & 0.354 & 0.396 & 0.476 & 0.558 & 0.644 & 0.724 & 0.772 & 0.746 & 0.704 \\
&$\beta_j=0.30$ & 0.576 & 0.638 & 0.678 & 0.696 & 0.722 & 0.770 & 0.812 & 0.844 & 0.848 & 0.836 & 0.818 \\
&$\beta_j=0.35$ & 0.684 & 0.706 & 0.736 & 0.764 & 0.784 & 0.802 & 0.846 & 0.854 & 0.862 & 0.852 & 0.840 \\
\bottomrule
\end{tabular}
\end{table}

    \begin{table}[ht]
\centering
\caption{$g_{\widehat{f},3}$ and $g_{\widehat{f},3}^{adj}$, DGP Size/Power (nom 5\%),  $\tau_0=0.8$, $\alpha_i=[0.51,0.51,0.51]$}\label{tab_size_test3}
\begin{tabular}{l l c c c c c c c c c c c}
\toprule
 && \multicolumn{11}{c}{$(N,T)=(800,500)$} \\
\cmidrule(lr){3-13}
 &$\lambda_2^0$& 0.5 & 0.55 & 0.6 & 0.65 & 0.7 & 0.75 & 0.8 & 0.85 & 0.9 & 0.95 & 1.0 \\
\midrule
&$g_{\widehat{f},3}$ \\
&$\beta_j=0$ & 0.022 & 0.024 & 0.026 & 0.020 & 0.030 & 0.040 & 0.052 & 0.024 & 0.024 & 0.016 & 0.020 \\
&$\beta_j=0.10$ & 0.042 & 0.042 & 0.046 & 0.050 & 0.066 & 0.080 & 0.118 & 0.180 & 0.216 & 0.174 & 0.124 \\
&$\beta_j=0.20$ & 0.158 & 0.180 & 0.200 & 0.244 & 0.300 & 0.368 & 0.454 & 0.546 & 0.572 & 0.552 & 0.494 \\
&$\beta_j=0.30$ & 0.396 & 0.430 & 0.468 & 0.492 & 0.528 & 0.564 & 0.606 & 0.646 & 0.678 & 0.648 & 0.618 \\
&$\beta_j=0.35$ & 0.480 & 0.496 & 0.540 & 0.556 & 0.584 & 0.614 & 0.652 & 0.676 & 0.706 & 0.696 & 0.666 \\
\bottomrule
\end{tabular}
\end{table}

  \begin{table}[ht]
\centering
\caption{$g_{\widehat{f},3}$ and $g_{\widehat{f},3}^{adj}$, DGP Size/Power (nom 5\%),  $\tau_0=0.8$, $\alpha_i=[0.1,0.1,0.1]$}\label{tab_size_test4}
\begin{tabular}{l l c c c c c c c c c c c}
\toprule
 && \multicolumn{11}{c}{$(N,T)=(800,500)$} \\
\cmidrule(lr){3-13}
 &$\lambda_2^0$& 0.5 & 0.55 & 0.6 & 0.65 & 0.7 & 0.75 & 0.8 & 0.85 & 0.9 & 0.95 & 1.0 \\
\midrule
&$g_{\widehat{f},3}$ \\
&$\beta_j=0$ & 0.020 & 0.024 & 0.022 & 0.026 & 0.030 & 0.030 & 0.038 & 0.020 & 0.026 & 0.016 & 0.022 \\
&$\beta_j=0.10$ & 0.034 & 0.040 & 0.048 & 0.048 & 0.060 & 0.060 & 0.102 & 0.126 & 0.156 & 0.132 & 0.098 \\
&$\beta_j=0.20$ & 0.118 & 0.132 & 0.150 & 0.174 & 0.240 & 0.288 & 0.364 & 0.432 & 0.472 & 0.436 & 0.392 \\
&$\beta_j=0.30$ & 0.306 & 0.340 & 0.372 & 0.400 & 0.414 & 0.458 & 0.514 & 0.542 & 0.568 & 0.552 & 0.522 \\
&$\beta_j=0.35$ & 0.378 & 0.408 & 0.420 & 0.444 & 0.482 & 0.520 & 0.548 & 0.568 & 0.588 & 0.578 & 0.556 \\
\bottomrule
\end{tabular}
\end{table}

\begin{table}[ht]
\centering
\caption{$g_{\widehat{f},3}$ and $g_{\widehat{f},3}^{adj}$, DGP Size/Power (nom 5\%),  $\tau_0=0.8$, $\alpha_i=[0.51,0.7,1]$}\label{tab_size_test3}
\begin{tabular}{l l c c c c c c c c c c c}
\toprule
 && \multicolumn{11}{c}{$(N,T)=(800,500)$} \\
\cmidrule(lr){3-13}
 &$\lambda_2^0$& 0.5 & 0.55 & 0.6 & 0.65 & 0.7 & 0.75 & 0.8 & 0.85 & 0.9 & 0.95 & 1.0 \\
\midrule
&$g_{\widehat{f},3}$ \\
&$\beta_j=0$ & 0.018 & 0.026 & 0.030 & 0.024 & 0.028 & 0.032 & 0.044 & 0.022 & 0.018 & 0.022 & 0.024 \\
&$\beta_j=0.10$ & 0.054 & 0.056 & 0.064 & 0.070 & 0.108 & 0.126 & 0.180 & 0.262 & 0.342 & 0.286 & 0.208 \\
&$\beta_j=0.20$ & 0.308 & 0.326 & 0.384 & 0.440 & 0.522 & 0.578 & 0.694 & 0.788 & 0.834 & 0.798 & 0.740 \\
&$\beta_j=0.30$ & 0.668 & 0.706 & 0.722 & 0.754 & 0.774 & 0.806 & 0.860 & 0.904 & 0.910 & 0.906 & 0.884 \\
&$\beta_j=0.35$ & 0.742 & 0.760 & 0.776 & 0.790 & 0.818 & 0.854 & 0.900 & 0.918 & 0.922 & 0.922 & 0.904 \\
\bottomrule
\end{tabular}
\end{table}

    \begin{table}[ht]
\centering
\caption{$g_{\widehat{f},4}$ and $g_{\widehat{f},4}^{adj}$, DGP Size/Power (nom 5\%),  $\tau_0=0.8$}\label{tab_size_test5}
\begin{tabular}{l c c c c c c c c c c}
\toprule
 &$\lambda_1^0=1$& \multicolumn{9}{c}{$(N,T)=(800,500)$} \\
\cmidrule(lr){2-11}
 &$\alpha_i$& 1 & 0.51 & 0.1 \\
\midrule
$g_{\widehat{f},4}$ \\
$\beta_j=0$ & &0.026 & 0.024  & 0.024 \\
$\beta_j=0.10$ && 0.176 & 0.100  & 0.098 \\
$\beta_j=0.20$ && 0.686 & 0.488 & 0.380 \\
$\beta_j=0.30$ && 0.830 & 0.630  & 0.514 \\
$\beta_j=0.35$ && 0.844 & 0.666  & 0.546 \\
$\beta_j=0.40$ && 0.858 & 0.698  & 0.570 \\
$\beta_j=0.45$ && 0.870 & 0.706  & 0.584 \\
$\beta_j=0.50$ && 0.876 & 0.724  & 0.598 \\
$\beta_j=0.55$ && 0.882 & 0.732  & 0.610 \\
$\beta_j=0.60$ && 0.880 & 0.740  & 0.622 \\
\midrule
$g_{\widehat{f},4}^{adj}$ \\
$\beta_j=0$ && 0.092 & 0.098 & 0.088 \\
$\beta_j=0.10$ && 0.546 & 0.430 & 0.372  \\
$\beta_j=0.20$ && 0.924 & 0.826 & 0.752  \\
$\beta_j=0.30$ && 0.962 & 0.912 & 0.860  \\
$\beta_j=0.35$ && 0.970 & 0.920 & 0.874  \\
$\beta_j=0.40$ && 0.972 & 0.932 & 0.884  \\
$\beta_j=0.45$ && 0.980 & 0.934 & 0.890  \\
$\beta_j=0.50$ && 0.984 & 0.936 & 0.894  \\
$\beta_j=0.55$ && 0.984 & 0.938 & 0.900  \\
$\beta_j=0.60$ && 0.984 & 0.946 & 0.904  \\
\bottomrule
\end{tabular}
\end{table}
   
\begin{table}[ht]
\centering
\caption{$g_{\widehat{f},4}$ and $g_{\widehat{f},4}^{adj}$, DGP Size/Power (nom 5\%),  $\tau_0=0.8$, $\alpha_i=[0.51, 0.7, 1]$}\label{tab_size_test5}
\begin{tabular}{l c c c}
\toprule
  & $\lambda_1^0=1$ & \multicolumn{2}{c}{$(N,T)=(800,500)$} \\
\cmidrule(lr){3-4}
 &  & $g_{\widehat{f},4}$ & $g_{\widehat{f},4}^{adj}$ \\
\midrule
$\beta_j=0$ & & 0.022 & 0.092 \\
$\beta_j=0.10$ & & 0.200 & 0.550 \\
$\beta_j=0.20$ & & 0.730 & 0.954 \\
$\beta_j=0.30$ & & 0.894 & 0.982 \\
$\beta_j=0.35$ & & 0.916 & 0.984 \\
$\beta_j=0.40$ & & 0.920 & 0.990 \\
$\beta_j=0.45$ & & 0.926 & 0.992 \\
$\beta_j=0.50$ & & 0.934 & 0.992 \\
$\beta_j=0.55$ & & 0.936 & 0.992 \\
$\beta_j=0.60$ & & 0.940 & 0.992 \\
\bottomrule
\end{tabular}
\end{table}
 
\begin{table}[ht]
\centering
\caption{$g_{\widehat{f},4}$ and $g_{\widehat{f},4}^{adj}$, DGP Size/Power (nom 5\%),  $\tau_0=0.8$}\label{tab_size_test5}
\begin{tabular}{l c c c c c c c c c c}
\toprule
 &$\lambda_1^0=0.6$& \multicolumn{9}{c}{$(N,T)=(800,500)$} \\
\cmidrule(lr){2-11}
 &$\alpha_i$& 1 & 0.51 & 0.1 \\
\midrule
$g_{\widehat{f},4}$ \\
$\beta_j=0$ & & 0.036 & 0.030  & 0.030 \\
$\beta_j=0.10$ && 0.088 & 0.072  & 0.062 \\
$\beta_j=0.20$ && 0.388 & 0.242 & 0.194 \\
$\beta_j=0.30$ && 0.716 & 0.484  & 0.376 \\
$\beta_j=0.35$ && 0.766 & 0.558  & 0.420 \\
$\beta_j=0.40$ && 0.780 & 0.580  & 0.460 \\
$\beta_j=0.45$ && 0.806 & 0.612  & 0.484 \\
$\beta_j=0.50$ && 0.820 & 0.626  & 0.498 \\
$\beta_j=0.55$ && 0.832 & 0.632  & 0.522 \\
$\beta_j=0.60$ && 0.834 & 0.652  & 0.540 \\
\midrule
$g_{\widehat{f},4}^{adj}$ \\
$\beta_j=0$ && 0.054 & 0.054 & 0.050 \\
$\beta_j=0.10$ && 0.248 & 0.196 & 0.178  \\
$\beta_j=0.20$ && 0.788 & 0.608 & 0.518  \\
$\beta_j=0.30$ && 0.904 & 0.780 & 0.708  \\
$\beta_j=0.35$ && 0.930 & 0.824 & 0.742  \\
$\beta_j=0.40$ && 0.942 & 0.864 & 0.764  \\
$\beta_j=0.45$ && 0.950 & 0.878 & 0.790  \\
$\beta_j=0.50$ && 0.956 & 0.892 & 0.810  \\
$\beta_j=0.55$ && 0.956 & 0.908 & 0.826  \\
$\beta_j=0.60$ && 0.960 & 0.910 & 0.832  \\
\bottomrule
\end{tabular}
\end{table}

\begin{table}[ht]
\centering
\caption{$g_{\widehat{f},4}$ and $g_{\widehat{f},4}^{adj}$, DGP Size/Power (nom 5\%),  $\tau_0=0.8$, $\alpha_i=[0.51, 0.7, 1]$}\label{tab_size_test5}
\begin{tabular}{l c c c}
\toprule
  & $\lambda_1^0=0.6$ & \multicolumn{2}{c}{$(N,T)=(800,500)$} \\
\cmidrule(lr){3-4}
 &  & $g_{\widehat{f},4}$ & $g_{\widehat{f},4}^{adj}$ \\
\midrule
$\beta_j=0$ & & 0.034 & 0.054 \\
$\beta_j=0.10$ & & 0.090 & 0.278 \\
$\beta_j=0.20$ & & 0.442 & 0.814 \\
$\beta_j=0.30$ & & 0.748 & 0.956 \\
$\beta_j=0.35$ & & 0.820 & 0.970 \\
$\beta_j=0.40$ & & 0.852 & 0.976 \\
$\beta_j=0.45$ & & 0.880 & 0.982 \\
$\beta_j=0.50$ & & 0.890 & 0.986 \\
$\beta_j=0.55$ & & 0.902 & 0.986 \\
$\beta_j=0.60$ & & 0.910 & 0.988 \\
\bottomrule
\end{tabular}
\end{table}
\clearpage
\subsection{Sensitivity to Smaller Sample Sizes}
In this subsection we display the baseline Monte Carlo simulations of $g_{\widehat{f},1},g_{\widehat{f},2}^{adj}, g_{\widehat{f},3}^{adj},g_{\widehat{f},4}^{adj}$ for the same DGP as in Section 5, for the cases where $N=100$ with $T=200, 350$. This allow us to have a better look at the finite sample behavior of the proposed tests when the samples are sensibly smaller than those considered in the main Monte Carlo section.
\begin{table}[ht]
\centering
\caption{$g_{\widehat{f},1}$, DGP Size/Power (nom 5\%), $\alpha_i=[1,1,1]$, $N=100, T=200$}
\label{tab_size_test_beta2}
\begin{tabular}{l c c c c}
\toprule
 & \multicolumn{4}{c}{$\mu_0$} \\
\cmidrule(lr){2-5}
$g_{\widehat{f},1}$ & 0.30 & 0.35 & 0.40 & 0.45 \\
\midrule
$\beta_j=0.00$ & 0.074 & 0.080 & 0.100 & 0.136 \\
$\beta_j=0.10$ & 0.132 & 0.168 & 0.282 & 0.408 \\
$\beta_j=0.20$ & 0.394 & 0.494 & 0.628 & 0.774 \\
$\beta_j=0.30$ & 0.664 & 0.730 & 0.816 & 0.882 \\
$\beta_j=0.35$ & 0.736 & 0.780 & 0.856 & 0.908 \\
$\beta_j=0.40$ & 0.782 & 0.824 & 0.876 & 0.916 \\
$\beta_j=0.45$ & 0.814 & 0.852 & 0.896 & 0.922 \\
$\beta_j=0.50$ & 0.836 & 0.866 & 0.904 & 0.930 \\
$\beta_j=0.55$ & 0.842 & 0.876 & 0.910 & 0.936 \\
$\beta_j=0.60$ & 0.858 & 0.882 & 0.912 & 0.936 \\
\bottomrule
\end{tabular}
\end{table}

\begin{table}[ht]
\centering
\caption{$g_{\widehat{f},1}$, DGP Size/Power (nom 5\%), $\alpha_i=[1,1,1]$, $N=100, T=350$}
\label{tab_size_test_beta2_T350}
\begin{tabular}{l c c c c}
\toprule
 & \multicolumn{4}{c}{$\mu_0$} \\
\cmidrule(lr){2-5}
$g_{\widehat{f},1}$ & 0.30 & 0.35 & 0.40 & 0.45 \\
\midrule
$\beta_j=0.00$ & 0.048 & 0.046 & 0.060 & 0.130 \\
$\beta_j=0.10$ & 0.128 & 0.170 & 0.300 & 0.512 \\
$\beta_j=0.20$ & 0.542 & 0.646 & 0.776 & 0.860 \\
$\beta_j=0.30$ & 0.788 & 0.830 & 0.880 & 0.918 \\
$\beta_j=0.35$ & 0.826 & 0.854 & 0.896 & 0.922 \\
$\beta_j=0.40$ & 0.852 & 0.876 & 0.906 & 0.932 \\
$\beta_j=0.45$ & 0.862 & 0.888 & 0.908 & 0.940 \\
$\beta_j=0.50$ & 0.872 & 0.892 & 0.912 & 0.946 \\
$\beta_j=0.55$ & 0.874 & 0.900 & 0.916 & 0.948 \\
$\beta_j=0.60$ & 0.880 & 0.904 & 0.922 & 0.950 \\
\bottomrule
\end{tabular}
\end{table}


\begin{table}[ht]
\centering
\caption{$g_{\widehat{f},1}$, DGP Size/Power (nom 5\%), $\alpha_i=[0.51, 0.51, 0.51]$, $N=100, T=200$}
\label{tab_size_test_beta2_alpha051}
\begin{tabular}{l c c c c}
\toprule
 & \multicolumn{4}{c}{$\mu_0$} \\
\cmidrule(lr){2-5}
$g_{\widehat{f},1}$ & 0.30 & 0.35 & 0.40 & 0.45 \\
\midrule
$\beta_j=0.00$ & 0.072 & 0.080 & 0.102 & 0.158 \\
$\beta_j=0.10$ & 0.124 & 0.184 & 0.240 & 0.390 \\
$\beta_j=0.20$ & 0.366 & 0.450 & 0.610 & 0.746 \\
$\beta_j=0.30$ & 0.602 & 0.698 & 0.780 & 0.846 \\
$\beta_j=0.35$ & 0.684 & 0.750 & 0.810 & 0.876 \\
$\beta_j=0.40$ & 0.726 & 0.780 & 0.844 & 0.904 \\
$\beta_j=0.45$ & 0.748 & 0.800 & 0.864 & 0.914 \\
$\beta_j=0.50$ & 0.782 & 0.828 & 0.872 & 0.926 \\
$\beta_j=0.55$ & 0.806 & 0.844 & 0.878 & 0.924 \\
$\beta_j=0.60$ & 0.818 & 0.856 & 0.884 & 0.922 \\
\bottomrule
\end{tabular}
\end{table}

\begin{table}[ht]
\centering
\caption{$g_{\widehat{f},1}$, DGP Size/Power (nom 5\%), $\alpha_i=[0.51, 0.51, 0.51]$, $N=100, T=350$}
\label{tab_size_test_beta2_alpha051_T350}
\begin{tabular}{l c c c c}
\toprule
 & \multicolumn{4}{c}{$\mu_0$} \\
\cmidrule(lr){2-5}
$g_{\widehat{f},1}$ & 0.30 & 0.35 & 0.40 & 0.45 \\
\midrule
$\beta_j=0.00$ & 0.042 & 0.036 & 0.056 & 0.118 \\
$\beta_j=0.10$ & 0.106 & 0.158 & 0.266 & 0.468 \\
$\beta_j=0.20$ & 0.486 & 0.594 & 0.702 & 0.810 \\
$\beta_j=0.30$ & 0.712 & 0.780 & 0.828 & 0.888 \\
$\beta_j=0.35$ & 0.776 & 0.814 & 0.852 & 0.906 \\
$\beta_j=0.40$ & 0.806 & 0.834 & 0.876 & 0.914 \\
$\beta_j=0.45$ & 0.824 & 0.846 & 0.888 & 0.922 \\
$\beta_j=0.50$ & 0.838 & 0.856 & 0.896 & 0.924 \\
$\beta_j=0.55$ & 0.842 & 0.868 & 0.902 & 0.924 \\
$\beta_j=0.60$ & 0.852 & 0.878 & 0.906 & 0.930 \\
\bottomrule
\end{tabular}
\end{table}


\begin{table}[ht]
\centering
\caption{$g_{\widehat{f},1}$, DGP Size/Power (nom 5\%), $\alpha_i=[0.1, 0.1, 0.1]$, $N=100$, $T=200$}
\label{tab_size_test_beta2_alpha010}
\begin{tabular}{l c c c c}
\toprule
 & \multicolumn{4}{c}{$\mu_0$} \\
\cmidrule(lr){2-5}
$g_{\widehat{f},1}$ & 0.30 & 0.35 & 0.40 & 0.45 \\
\midrule
$\beta_j=0.00$ & 0.076 & 0.076 & 0.108 & 0.176 \\
$\beta_j=0.10$ & 0.132 & 0.174 & 0.246 & 0.400 \\
$\beta_j=0.20$ & 0.364 & 0.478 & 0.610 & 0.756 \\
$\beta_j=0.30$ & 0.630 & 0.722 & 0.794 & 0.854 \\
$\beta_j=0.35$ & 0.706 & 0.764 & 0.826 & 0.878 \\
$\beta_j=0.40$ & 0.740 & 0.798 & 0.848 & 0.892 \\
$\beta_j=0.45$ & 0.770 & 0.816 & 0.856 & 0.906 \\
$\beta_j=0.50$ & 0.788 & 0.832 & 0.874 & 0.914 \\
$\beta_j=0.55$ & 0.810 & 0.840 & 0.882 & 0.916 \\
$\beta_j=0.60$ & 0.826 & 0.850 & 0.882 & 0.922 \\
\bottomrule
\end{tabular}
\end{table}

\begin{table}[ht]
\centering
\caption{$g_{\widehat{f},1}$, DGP Size/Power (nom 5\%), $\alpha_i=[0.1, 0.1, 0.1]$, $N=100$, $T=350$}
\label{tab_size_test_beta2_alpha010_T350}
\begin{tabular}{l c c c c}
\toprule
 & \multicolumn{4}{c}{$\mu_0$} \\
\cmidrule(lr){2-5}
$g_{\widehat{f},1}$ & 0.30 & 0.35 & 0.40 & 0.45 \\
\midrule
$\beta_2=0.00$ & 0.046 & 0.040 & 0.068 & 0.110 \\
$\beta_2=0.10$ & 0.112 & 0.152 & 0.278 & 0.450 \\
$\beta_2=0.20$ & 0.492 & 0.564 & 0.676 & 0.798 \\
$\beta_2=0.30$ & 0.706 & 0.778 & 0.838 & 0.890 \\
$\beta_2=0.35$ & 0.762 & 0.808 & 0.860 & 0.914 \\
$\beta_2=0.40$ & 0.802 & 0.830 & 0.878 & 0.928 \\
$\beta_2=0.45$ & 0.818 & 0.862 & 0.900 & 0.932 \\
$\beta_2=0.50$ & 0.828 & 0.870 & 0.908 & 0.938 \\
$\beta_2=0.55$ & 0.842 & 0.876 & 0.916 & 0.940 \\
$\beta_2=0.60$ & 0.856 & 0.886 & 0.918 & 0.944 \\
\bottomrule
\end{tabular}
\end{table}
\begin{table}[]
\centering
\caption{$g_{\widehat{f},2}^{adj}$, DGP Size/Power (nom 5\%),  $\tau_0=0.8$, $\alpha_i=[1,1,1]$}
\label{tab_size_test2_adj_only}
\begin{tabular}{l l c c c c c c c c c c}
\toprule
 &$\lambda_1^0=1$& \multicolumn{10}{c}{$(N,T)=(100,200)$} \\
\cmidrule(lr){3-12}
 &$\lambda_2^0$& 0.5 & 0.55 & 0.6 & 0.65 & 0.7 & 0.75 & 0.8 & 0.85 & 0.9 & 0.95 \\
\midrule
&$g_{\widehat{f},2}^{adj}$ \\
&$\beta_j=0$ & 0.066 & 0.060 & 0.052 & 0.062 & 0.078 & 0.088 & 0.078 & 0.086 & 0.104 & 0.152 \\
&$\beta_j=0.10$ & 0.124 & 0.132 & 0.146 & 0.172 & 0.186 & 0.198 & 0.214 & 0.250 & 0.312 & 0.412 \\
&$\beta_j=0.20$ & 0.368 & 0.388 & 0.406 & 0.470 & 0.510 & 0.534 & 0.582 & 0.648 & 0.692 & 0.770 \\
&$\beta_j=0.30$ & 0.632 & 0.646 & 0.684 & 0.706 & 0.716 & 0.748 & 0.784 & 0.808 & 0.854 & 0.882 \\
&$\beta_j=0.35$ & 0.710 & 0.718 & 0.748 & 0.758 & 0.782 & 0.798 & 0.814 & 0.852 & 0.878 & 0.902 \\
&$\beta_j=0.40$ & 0.748 & 0.768 & 0.784 & 0.806 & 0.812 & 0.828 & 0.846 & 0.872 & 0.900 & 0.916 \\
&$\beta_j=0.45$ & 0.780 & 0.790 & 0.796 & 0.818 & 0.834 & 0.854 & 0.864 & 0.882 & 0.906 & 0.922 \\
&$\beta_j=0.50$ & 0.792 & 0.806 & 0.814 & 0.832 & 0.844 & 0.862 & 0.874 & 0.884 & 0.910 & 0.928 \\
&$\beta_j=0.55$ & 0.818 & 0.814 & 0.828 & 0.836 & 0.858 & 0.868 & 0.884 & 0.894 & 0.912 & 0.932 \\
&$\beta_j=0.60$ & 0.826 & 0.820 & 0.838 & 0.850 & 0.860 & 0.878 & 0.886 & 0.898 & 0.920 & 0.938 \\
\bottomrule
\end{tabular}
\end{table}

\begin{table}[ht]
\centering
\caption{$g_{\widehat{f},2}^{adj}$, DGP Size/Power (nom 5\%),  $\tau_0=0.8$, $\alpha_i=[1,1,1]$}
\label{tab_size_test2_adj_only_N100_T350}
\begin{tabular}{l l c c c c c c c c c c}
\toprule
 &$\lambda_1^0=1$& \multicolumn{10}{c}{$(N,T)=(100,350)$} \\
\cmidrule(lr){3-12}
 &$\lambda_2^0$& 0.5 & 0.55 & 0.6 & 0.65 & 0.7 & 0.75 & 0.8 & 0.85 & 0.9 & 0.95 \\
\midrule
&$g_{\widehat{f},2}^{adj}$ \\
&$\beta_j=0$ & 0.044 & 0.036 & 0.032 & 0.056 & 0.054 & 0.056 & 0.078 & 0.084 & 0.082 & 0.122 \\
&$\beta_j=0.10$ & 0.130 & 0.146 & 0.154 & 0.198 & 0.206 & 0.226 & 0.272 & 0.290 & 0.370 & 0.484 \\
&$\beta_j=0.20$ & 0.478 & 0.518 & 0.540 & 0.600 & 0.628 & 0.684 & 0.714 & 0.760 & 0.790 & 0.846 \\
&$\beta_j=0.30$ & 0.746 & 0.770 & 0.792 & 0.796 & 0.812 & 0.824 & 0.844 & 0.866 & 0.886 & 0.900 \\
&$\beta_j=0.35$ & 0.792 & 0.802 & 0.820 & 0.820 & 0.836 & 0.856 & 0.870 & 0.882 & 0.898 & 0.926 \\
&$\beta_j=0.40$ & 0.820 & 0.830 & 0.832 & 0.856 & 0.860 & 0.874 & 0.884 & 0.896 & 0.916 & 0.932 \\
&$\beta_j=0.45$ & 0.834 & 0.840 & 0.854 & 0.866 & 0.876 & 0.880 & 0.892 & 0.908 & 0.920 & 0.934 \\
&$\beta_j=0.50$ & 0.842 & 0.850 & 0.862 & 0.876 & 0.888 & 0.884 & 0.900 & 0.916 & 0.924 & 0.940 \\
&$\beta_j=0.55$ & 0.850 & 0.854 & 0.872 & 0.890 & 0.892 & 0.892 & 0.904 & 0.916 & 0.932 & 0.948 \\
&$\beta_j=0.60$ & 0.858 & 0.858 & 0.880 & 0.892 & 0.892 & 0.896 & 0.906 & 0.920 & 0.932 & 0.950 \\
\bottomrule
\end{tabular}
\end{table}
\begin{table}[ht]
\centering
\caption{$g_{\widehat{f},2}^{adj}$, DGP Size/Power (nom 5\%),  $\tau_0=0.8$, $\alpha_i=[0.51,0.51,0.51]$}
\label{tab_size_test2_adj_only_N100_T200}
\begin{tabular}{l l c c c c c c c c c c}
\toprule
 &$\lambda_1^0=1$& \multicolumn{10}{c}{$(N,T)=(100,200)$} \\
\cmidrule(lr){3-12}
 &$\lambda_2^0$& 0.5 & 0.55 & 0.6 & 0.65 & 0.7 & 0.75 & 0.8 & 0.85 & 0.9 & 0.95 \\
\midrule
&$g_{\widehat{f},2}^{adj}$ \\
&$\beta_j=0$ & 0.076 & 0.060 & 0.056 & 0.074 & 0.084 & 0.094 & 0.092 & 0.090 & 0.108 & 0.156 \\
&$\beta_j=0.10$ & 0.136 & 0.138 & 0.138 & 0.158 & 0.184 & 0.192 & 0.212 & 0.232 & 0.290 & 0.364 \\
&$\beta_j=0.20$ & 0.344 & 0.350 & 0.384 & 0.430 & 0.478 & 0.506 & 0.524 & 0.598 & 0.656 & 0.716 \\
&$\beta_j=0.30$ & 0.606 & 0.616 & 0.644 & 0.670 & 0.692 & 0.734 & 0.752 & 0.766 & 0.802 & 0.846 \\
&$\beta_j=0.35$ & 0.674 & 0.690 & 0.706 & 0.716 & 0.748 & 0.782 & 0.792 & 0.812 & 0.846 & 0.876 \\
&$\beta_j=0.40$ & 0.712 & 0.740 & 0.750 & 0.770 & 0.788 & 0.804 & 0.824 & 0.838 & 0.868 & 0.890 \\
&$\beta_j=0.45$ & 0.756 & 0.760 & 0.768 & 0.794 & 0.820 & 0.830 & 0.840 & 0.854 & 0.878 & 0.898 \\
&$\beta_j=0.50$ & 0.766 & 0.770 & 0.790 & 0.806 & 0.834 & 0.840 & 0.850 & 0.866 & 0.892 & 0.914 \\
&$\beta_j=0.55$ & 0.776 & 0.790 & 0.802 & 0.828 & 0.842 & 0.848 & 0.852 & 0.878 & 0.900 & 0.920 \\
&$\beta_j=0.60$ & 0.790 & 0.804 & 0.812 & 0.836 & 0.848 & 0.850 & 0.860 & 0.886 & 0.902 & 0.924 \\
\bottomrule
\end{tabular}
\end{table}

\begin{table}[ht]
\centering
\caption{$g_{\widehat{f},2}^{adj}$, DGP Size/Power (nom 5\%),  $\tau_0=0.8$, $\alpha_i=[0.51,0.51,0.51]$}
\label{tab_size_test2_adj_only_N100_T350}
\begin{tabular}{l l c c c c c c c c c c}
\toprule
 &$\lambda_1^0=1$& \multicolumn{10}{c}{$(N,T)=(100,350)$} \\
\cmidrule(lr){3-12}
 &$\lambda_2^0$& 0.5 & 0.55 & 0.6 & 0.65 & 0.7 & 0.75 & 0.8 & 0.85 & 0.9 & 0.95 \\
\midrule
&$g_{\widehat{f},2}^{adj}$ \\
&$\beta_j=0$ & 0.052 & 0.042 & 0.040 & 0.050 & 0.054 & 0.058 & 0.070 & 0.078 & 0.078 & 0.116 \\
&$\beta_j=0.10$ & 0.112 & 0.130 & 0.138 & 0.170 & 0.186 & 0.196 & 0.244 & 0.278 & 0.358 & 0.454 \\
&$\beta_j=0.20$ & 0.452 & 0.494 & 0.528 & 0.570 & 0.598 & 0.644 & 0.668 & 0.698 & 0.748 & 0.786 \\
&$\beta_j=0.30$ & 0.718 & 0.730 & 0.746 & 0.754 & 0.776 & 0.780 & 0.794 & 0.810 & 0.832 & 0.872 \\
&$\beta_j=0.35$ & 0.754 & 0.760 & 0.768 & 0.784 & 0.786 & 0.800 & 0.822 & 0.828 & 0.862 & 0.898 \\
&$\beta_j=0.40$ & 0.778 & 0.774 & 0.784 & 0.798 & 0.804 & 0.828 & 0.842 & 0.854 & 0.874 & 0.908 \\
&$\beta_j=0.45$ & 0.790 & 0.790 & 0.800 & 0.812 & 0.824 & 0.850 & 0.854 & 0.866 & 0.892 & 0.914 \\
&$\beta_j=0.50$ & 0.808 & 0.802 & 0.820 & 0.826 & 0.842 & 0.858 & 0.862 & 0.868 & 0.896 & 0.922 \\
&$\beta_j=0.55$ & 0.818 & 0.816 & 0.824 & 0.834 & 0.852 & 0.866 & 0.864 & 0.878 & 0.906 & 0.924 \\
&$\beta_j=0.60$ & 0.824 & 0.826 & 0.836 & 0.834 & 0.854 & 0.864 & 0.872 & 0.884 & 0.908 & 0.922 \\
\bottomrule
\end{tabular}
\end{table}


\begin{table}[ht]
\centering
\caption{$g_{\widehat{f},2}^{adj}$, DGP Size/Power (nom 5\%),  $\tau_0=0.8$, $\alpha_i=[0.1,0.1,0.1]$}
\label{tab_size_test2_adj_only_N100_T200_alpha01}
\begin{tabular}{l l c c c c c c c c c c}
\toprule
 &$\lambda_1^0=1$& \multicolumn{10}{c}{$(N,T)=(100,200)$} \\
\cmidrule(lr){3-12}
 &$\lambda_2^0$& 0.5 & 0.55 & 0.6 & 0.65 & 0.7 & 0.75 & 0.8 & 0.85 & 0.9 & 0.95 \\
\midrule
&$g_{\widehat{f},2}^{adj}$ \\
&$\beta_j=0$ & 0.074 & 0.068 & 0.066 & 0.064 & 0.088 & 0.096 & 0.084 & 0.092 & 0.106 & 0.146 \\
&$\beta_j=0.10$ & 0.142 & 0.144 & 0.140 & 0.154 & 0.186 & 0.188 & 0.208 & 0.252 & 0.294 & 0.364 \\
&$\beta_j=0.20$ & 0.348 & 0.360 & 0.390 & 0.414 & 0.464 & 0.508 & 0.532 & 0.598 & 0.658 & 0.712 \\
&$\beta_j=0.30$ & 0.584 & 0.602 & 0.640 & 0.670 & 0.690 & 0.714 & 0.754 & 0.778 & 0.818 & 0.840 \\
&$\beta_j=0.35$ & 0.666 & 0.670 & 0.706 & 0.722 & 0.742 & 0.774 & 0.798 & 0.818 & 0.846 & 0.870 \\
&$\beta_j=0.40$ & 0.706 & 0.708 & 0.742 & 0.758 & 0.778 & 0.802 & 0.820 & 0.838 & 0.860 & 0.882 \\
&$\beta_j=0.45$ & 0.744 & 0.754 & 0.770 & 0.790 & 0.804 & 0.820 & 0.830 & 0.850 & 0.866 & 0.894 \\
&$\beta_j=0.50$ & 0.766 & 0.774 & 0.794 & 0.804 & 0.826 & 0.834 & 0.846 & 0.864 & 0.886 & 0.904 \\
&$\beta_j=0.55$ & 0.774 & 0.794 & 0.810 & 0.820 & 0.836 & 0.846 & 0.858 & 0.876 & 0.892 & 0.912 \\
&$\beta_j=0.60$ & 0.784 & 0.808 & 0.822 & 0.836 & 0.844 & 0.856 & 0.862 & 0.884 & 0.902 & 0.912 \\
\bottomrule
\end{tabular}
\end{table}

\begin{table}[ht]
\centering
\caption{$g_{\widehat{f},2}^{adj}$, DGP Size/Power (nom 5\%), $\tau_0=0.8$, $\alpha_i=[0.1,0.1,0.1]$}
\label{tab_size_test2_adj_only_N100_T350_alpha01}
\begin{tabular}{l l c c c c c c c c c c}
\toprule
 &$\lambda_1^0=1$& \multicolumn{10}{c}{$(N,T)=(100,350)$} \\
\cmidrule(lr){3-12}
 &$\lambda_2^0$& 0.5 & 0.55 & 0.6 & 0.65 & 0.7 & 0.75 & 0.8 & 0.85 & 0.9 & 0.95 \\
\midrule
&$g_{\widehat{f},2}^{adj}$ \\
&$\beta_j=0$     & 0.052 & 0.038 & 0.040 & 0.050 & 0.046 & 0.062 & 0.068 & 0.078 & 0.084 & 0.130 \\
&$\beta_j=0.10$  & 0.114 & 0.118 & 0.134 & 0.164 & 0.180 & 0.194 & 0.244 & 0.264 & 0.344 & 0.426 \\
&$\beta_j=0.20$  & 0.440 & 0.466 & 0.504 & 0.550 & 0.584 & 0.624 & 0.660 & 0.692 & 0.732 & 0.776 \\
&$\beta_j=0.30$  & 0.700 & 0.692 & 0.718 & 0.756 & 0.762 & 0.790 & 0.804 & 0.814 & 0.840 & 0.880 \\
&$\beta_j=0.35$  & 0.730 & 0.734 & 0.764 & 0.780 & 0.800 & 0.812 & 0.832 & 0.844 & 0.860 & 0.894 \\
&$\beta_j=0.40$  & 0.758 & 0.774 & 0.786 & 0.808 & 0.822 & 0.834 & 0.840 & 0.858 & 0.880 & 0.906 \\
&$\beta_j=0.45$  & 0.790 & 0.806 & 0.814 & 0.830 & 0.836 & 0.850 & 0.860 & 0.874 & 0.892 & 0.916 \\
&$\beta_j=0.50$  & 0.808 & 0.822 & 0.828 & 0.838 & 0.848 & 0.860 & 0.878 & 0.884 & 0.906 & 0.918 \\
&$\beta_j=0.55$  & 0.830 & 0.832 & 0.834 & 0.846 & 0.856 & 0.868 & 0.888 & 0.892 & 0.910 & 0.924 \\
&$\beta_j=0.60$  & 0.844 & 0.840 & 0.842 & 0.856 & 0.864 & 0.876 & 0.888 & 0.898 & 0.912 & 0.924 \\
\bottomrule
\end{tabular}
\end{table}
\begin{table}[ht]
\centering
\caption{$g_{\widehat{f},3}^{adj}$, DGP Size/Power (nom 5\%), $\tau_0=0.8$, $\alpha_i=[1,1,1]$}
\label{tab_size_test3_hat_N100_T200_alpha1}
\begin{tabular}{l l c c c c c c c c c c}
\toprule
 &$\lambda_1^0=1$& \multicolumn{10}{c}{$(N,T)=(100,200)$} \\
\cmidrule(lr){3-12}
 &$\lambda_2^0$& 0.5 & 0.55 & 0.6 & 0.65 & 0.7 & 0.75 & 0.8 & 0.85 & 0.9 & 0.95 \\
\midrule
&$g_{\widehat{f},3}^{adj}$ \\
&$\beta_j=0$     & 0.078 & 0.070 & 0.062 & 0.076 & 0.088 & 0.100 & 0.130 & 0.168 & 0.192 & 0.150 \\
&$\beta_j=0.10$  & 0.142 & 0.160 & 0.168 & 0.194 & 0.234 & 0.266 & 0.328 & 0.420 & 0.476 & 0.430 \\
&$\beta_j=0.20$  & 0.384 & 0.422 & 0.452 & 0.518 & 0.552 & 0.626 & 0.688 & 0.768 & 0.808 & 0.780 \\
&$\beta_j=0.30$  & 0.650 & 0.668 & 0.700 & 0.724 & 0.768 & 0.798 & 0.836 & 0.874 & 0.898 & 0.874 \\
&$\beta_j=0.35$  & 0.722 & 0.744 & 0.758 & 0.790 & 0.816 & 0.832 & 0.868 & 0.898 & 0.912 & 0.900 \\
&$\beta_j=0.40$  & 0.750 & 0.776 & 0.786 & 0.820 & 0.838 & 0.856 & 0.888 & 0.908 & 0.922 & 0.912 \\
&$\beta_j=0.45$  & 0.780 & 0.788 & 0.814 & 0.828 & 0.854 & 0.868 & 0.900 & 0.910 & 0.932 & 0.920 \\
&$\beta_j=0.50$  & 0.792 & 0.804 & 0.818 & 0.842 & 0.868 & 0.886 & 0.902 & 0.916 & 0.930 & 0.932 \\
&$\beta_j=0.55$  & 0.804 & 0.816 & 0.836 & 0.854 & 0.870 & 0.888 & 0.904 & 0.920 & 0.932 & 0.934 \\
&$\beta_j=0.60$  & 0.816 & 0.828 & 0.852 & 0.856 & 0.878 & 0.890 & 0.904 & 0.924 & 0.932 & 0.938 \\
\bottomrule
\end{tabular}
\end{table}

\begin{table}[ht]
\centering
\caption{$g_{\widehat{f},3}^{adj}$, DGP Size/Power (nom 5\%), $\tau_0=0.8$, $\alpha_i=[1,1,1]$}
\label{tab_size_test3_hat_N100_T350_alpha1}
\begin{tabular}{l l c c c c c c c c c c}
\toprule
 &$\lambda_1^0=1$& \multicolumn{10}{c}{$(N,T)=(100,350)$} \\
\cmidrule(lr){3-12}
 &$\lambda_2^0$& 0.5 & 0.55 & 0.6 & 0.65 & 0.7 & 0.75 & 0.8 & 0.85 & 0.9 & 0.95 \\
\midrule
&$g_{\widehat{f},3}^{adj}$ \\
&$\beta_j=0$     & 0.042 & 0.038 & 0.044 & 0.056 & 0.062 & 0.072 & 0.130 & 0.146 & 0.186 & 0.122 \\
&$\beta_j=0.10$  & 0.148 & 0.172 & 0.190 & 0.232 & 0.246 & 0.306 & 0.430 & 0.540 & 0.614 & 0.550 \\
&$\beta_j=0.20$  & 0.514 & 0.546 & 0.612 & 0.666 & 0.690 & 0.756 & 0.798 & 0.852 & 0.876 & 0.864 \\
&$\beta_j=0.30$  & 0.754 & 0.790 & 0.802 & 0.814 & 0.830 & 0.852 & 0.884 & 0.908 & 0.922 & 0.912 \\
&$\beta_j=0.35$  & 0.802 & 0.808 & 0.820 & 0.844 & 0.858 & 0.870 & 0.894 & 0.914 & 0.924 & 0.928 \\
&$\beta_j=0.40$  & 0.826 & 0.840 & 0.848 & 0.864 & 0.874 & 0.886 & 0.906 & 0.916 & 0.932 & 0.932 \\
&$\beta_j=0.45$  & 0.838 & 0.848 & 0.862 & 0.880 & 0.890 & 0.894 & 0.912 & 0.916 & 0.938 & 0.938 \\
&$\beta_j=0.50$  & 0.854 & 0.856 & 0.872 & 0.888 & 0.894 & 0.898 & 0.912 & 0.922 & 0.940 & 0.940 \\
&$\beta_j=0.55$  & 0.856 & 0.858 & 0.884 & 0.890 & 0.896 & 0.902 & 0.914 & 0.930 & 0.946 & 0.946 \\
&$\beta_j=0.60$  & 0.860 & 0.866 & 0.888 & 0.896 & 0.900 & 0.904 & 0.914 & 0.934 & 0.948 & 0.950 \\
\bottomrule
\end{tabular}
\end{table}
\begin{table}[ht]
\centering
\caption{$g_{\widehat{f},3}^{adj}$, DGP Size/Power (nom 5\%), $\tau_0=0.8$, $\alpha_i=[0.51,0.51,0.51]$}
\label{tab_size_test3_hat_N100_T200_alpha051}
\begin{tabular}{l l c c c c c c c c c c}
\toprule
 &$\lambda_1^0=1$& \multicolumn{10}{c}{$(N,T)=(100,200)$} \\
\cmidrule(lr){3-12}
 &$\lambda_2^0$& 0.5 & 0.55 & 0.6 & 0.65 & 0.7 & 0.75 & 0.8 & 0.85 & 0.9 & 0.95 \\
\midrule
&$g_{\widehat{f},3}^{adj}$ \\
&$\beta_j=0$     & 0.072 & 0.076 & 0.068 & 0.078 & 0.098 & 0.108 & 0.116 & 0.162 & 0.184 & 0.170 \\
&$\beta_j=0.10$  & 0.142 & 0.158 & 0.168 & 0.188 & 0.236 & 0.256 & 0.310 & 0.382 & 0.450 & 0.408 \\
&$\beta_j=0.20$  & 0.362 & 0.382 & 0.432 & 0.482 & 0.520 & 0.576 & 0.652 & 0.734 & 0.772 & 0.734 \\
&$\beta_j=0.30$  & 0.630 & 0.638 & 0.670 & 0.694 & 0.742 & 0.766 & 0.812 & 0.846 & 0.876 & 0.858 \\
&$\beta_j=0.35$  & 0.688 & 0.710 & 0.730 & 0.762 & 0.782 & 0.796 & 0.838 & 0.872 & 0.892 & 0.874 \\
&$\beta_j=0.40$  & 0.716 & 0.756 & 0.762 & 0.792 & 0.808 & 0.834 & 0.862 & 0.890 & 0.902 & 0.892 \\
&$\beta_j=0.45$  & 0.756 & 0.774 & 0.788 & 0.804 & 0.832 & 0.848 & 0.870 & 0.900 & 0.914 & 0.902 \\
&$\beta_j=0.50$  & 0.770 & 0.790 & 0.806 & 0.822 & 0.844 & 0.860 & 0.880 & 0.906 & 0.920 & 0.916 \\
&$\beta_j=0.55$  & 0.780 & 0.806 & 0.820 & 0.834 & 0.852 & 0.866 & 0.888 & 0.910 & 0.926 & 0.920 \\
&$\beta_j=0.60$  & 0.794 & 0.814 & 0.828 & 0.842 & 0.858 & 0.874 & 0.888 & 0.912 & 0.924 & 0.924 \\
\bottomrule
\end{tabular}
\end{table}

\begin{table}[ht]
\centering
\caption{$g_{\widehat{f},3}^{adj}$, DGP Size/Power (nom 5\%), $\tau_0=0.8$, $\alpha_i=[0.51,0.51,0.51]$}
\label{tab_size_test3_hat_N100_T350_alpha051}
\begin{tabular}{l l c c c c c c c c c c}
\toprule
 &$\lambda_1^0=1$& \multicolumn{10}{c}{$(N,T)=(100,350)$} \\
\cmidrule(lr){3-12}
 &$\lambda_2^0$& 0.5 & 0.55 & 0.6 & 0.65 & 0.7 & 0.75 & 0.8 & 0.85 & 0.9 & 0.95 \\
\midrule
&$g_{\widehat{f},3}^{adj}$ \\
&$\beta_j=0$     & 0.062 & 0.046 & 0.038 & 0.056 & 0.062 & 0.078 & 0.134 & 0.138 & 0.172 & 0.126 \\
&$\beta_j=0.10$  & 0.128 & 0.150 & 0.172 & 0.208 & 0.240 & 0.296 & 0.400 & 0.514 & 0.566 & 0.520 \\
&$\beta_j=0.20$  & 0.478 & 0.528 & 0.586 & 0.624 & 0.678 & 0.718 & 0.760 & 0.802 & 0.838 & 0.828 \\
&$\beta_j=0.30$  & 0.730 & 0.746 & 0.762 & 0.778 & 0.792 & 0.818 & 0.846 & 0.878 & 0.894 & 0.882 \\
&$\beta_j=0.35$  & 0.766 & 0.766 & 0.778 & 0.796 & 0.816 & 0.840 & 0.870 & 0.890 & 0.912 & 0.896 \\
&$\beta_j=0.40$  & 0.796 & 0.790 & 0.804 & 0.814 & 0.838 & 0.850 & 0.888 & 0.900 & 0.920 & 0.910 \\
&$\beta_j=0.45$  & 0.808 & 0.804 & 0.816 & 0.836 & 0.848 & 0.868 & 0.892 & 0.904 & 0.920 & 0.914 \\
&$\beta_j=0.50$  & 0.816 & 0.818 & 0.832 & 0.844 & 0.856 & 0.868 & 0.894 & 0.908 & 0.924 & 0.922 \\
&$\beta_j=0.55$  & 0.828 & 0.832 & 0.840 & 0.852 & 0.868 & 0.874 & 0.898 & 0.910 & 0.930 & 0.930 \\
&$\beta_j=0.60$  & 0.834 & 0.840 & 0.844 & 0.856 & 0.866 & 0.878 & 0.902 & 0.912 & 0.928 & 0.932 \\
\bottomrule
\end{tabular}
\end{table}
\begin{table}[ht]
\centering
\caption{$g_{\widehat{f},3}^{adj}$, DGP Size/Power (nom 5\%), $\tau_0=0.8$, $\alpha_i=[0.1,0.1,0.1]$}
\label{tab_size_test3_hat_N100_T200_alpha01}
\begin{tabular}{l l c c c c c c c c c c}
\toprule
 &$\lambda_1^0=1$& \multicolumn{10}{c}{$(N,T)=(100,200)$} \\
\cmidrule(lr){3-12}
 &$\lambda_2^0$& 0.5 & 0.55 & 0.6 & 0.65 & 0.7 & 0.75 & 0.8 & 0.85 & 0.9 & 0.95 \\
\midrule
&$g_{\widehat{f},3}^{adj}$ \\
&$\beta_j=0$     & 0.070 & 0.074 & 0.082 & 0.070 & 0.100 & 0.096 & 0.140 & 0.156 & 0.182 & 0.158 \\
&$\beta_j=0.10$  & 0.140 & 0.162 & 0.168 & 0.182 & 0.234 & 0.264 & 0.316 & 0.400 & 0.450 & 0.408 \\
&$\beta_j=0.20$  & 0.360 & 0.388 & 0.428 & 0.480 & 0.532 & 0.592 & 0.660 & 0.738 & 0.768 & 0.736 \\
&$\beta_j=0.30$  & 0.608 & 0.626 & 0.668 & 0.700 & 0.734 & 0.776 & 0.802 & 0.846 & 0.874 & 0.856 \\
&$\beta_j=0.35$  & 0.690 & 0.688 & 0.728 & 0.758 & 0.778 & 0.804 & 0.834 & 0.866 & 0.886 & 0.874 \\
&$\beta_j=0.40$  & 0.726 & 0.738 & 0.764 & 0.792 & 0.798 & 0.830 & 0.858 & 0.898 & 0.898 & 0.890 \\
&$\beta_j=0.45$  & 0.758 & 0.778 & 0.794 & 0.816 & 0.826 & 0.846 & 0.882 & 0.898 & 0.910 & 0.898 \\
&$\beta_j=0.50$  & 0.776 & 0.792 & 0.812 & 0.832 & 0.838 & 0.856 & 0.894 & 0.906 & 0.918 & 0.912 \\
&$\beta_j=0.55$  & 0.792 & 0.808 & 0.818 & 0.842 & 0.850 & 0.868 & 0.894 & 0.912 & 0.922 & 0.914 \\
&$\beta_j=0.60$  & 0.802 & 0.818 & 0.826 & 0.844 & 0.858 & 0.874 & 0.896 & 0.914 & 0.926 & 0.918 \\
\bottomrule
\end{tabular}
\end{table}

\begin{table}[ht]
\centering
\caption{$g_{\widehat{f},3}$, DGP Size/Power (nom 5\%), $\tau_0=0.8$, $\alpha_i=[0.1,0.1,0.1]$}
\label{tab_size_test3_hat_N100_T350_alpha01}
\begin{tabular}{l l c c c c c c c c c c}
\toprule
 &$\lambda_1^0=1$& \multicolumn{10}{c}{$(N,T)=(100,350)$} \\
\cmidrule(lr){3-12}
 &$\lambda_2^0$& 0.5 & 0.55 & 0.6 & 0.65 & 0.7 & 0.75 & 0.8 & 0.85 & 0.9 & 0.95 \\
\midrule
&$g_{\widehat{f},3}$ \\
&$\beta_j=0$     & 0.054 & 0.040 & 0.040 & 0.054 & 0.052 & 0.074 & 0.108 & 0.126 & 0.164 & 0.110 \\
&$\beta_j=0.10$  & 0.130 & 0.132 & 0.160 & 0.200 & 0.222 & 0.296 & 0.380 & 0.482 & 0.554 & 0.484 \\
&$\beta_j=0.20$  & 0.456 & 0.518 & 0.552 & 0.590 & 0.648 & 0.708 & 0.744 & 0.802 & 0.824 & 0.794 \\
&$\beta_j=0.30$  & 0.706 & 0.714 & 0.744 & 0.768 & 0.798 & 0.818 & 0.852 & 0.874 & 0.892 & 0.880 \\
&$\beta_j=0.35$  & 0.738 & 0.758 & 0.768 & 0.808 & 0.828 & 0.838 & 0.872 & 0.898 & 0.904 & 0.904 \\
&$\beta_j=0.40$  & 0.776 & 0.784 & 0.798 & 0.834 & 0.846 & 0.864 & 0.888 & 0.906 & 0.912 & 0.916 \\
&$\beta_j=0.45$  & 0.792 & 0.810 & 0.828 & 0.848 & 0.854 & 0.874 & 0.892 & 0.912 & 0.920 & 0.926 \\
&$\beta_j=0.50$  & 0.824 & 0.824 & 0.842 & 0.862 & 0.868 & 0.880 & 0.906 & 0.916 & 0.926 & 0.930 \\
&$\beta_j=0.55$  & 0.838 & 0.838 & 0.848 & 0.864 & 0.866 & 0.884 & 0.910 & 0.922 & 0.928 & 0.936 \\
&$\beta_j=0.60$  & 0.846 & 0.842 & 0.858 & 0.872 & 0.874 & 0.890 & 0.912 & 0.930 & 0.934 & 0.938 \\
\bottomrule
\end{tabular}
\end{table}
\begin{table}[ht]
\centering
\caption{$g_{\widehat{f},4}^{adj}$, DGP Size/Power (nom 5\%),  $\tau_0=0.8$, $\alpha_i=[1,1,1]$}
\label{tab_size_test5_adj_N100_T200_alpha1}
\begin{tabular}{l c }
\toprule
  $\lambda_1^0=0.6$ & $(N,T)=(100,200)$ \\
\midrule
$\beta_j=0$     & 0.064 \\
$\beta_j=0.10$  & 0.172 \\
$\beta_j=0.20$  & 0.472 \\
$\beta_j=0.30$  & 0.722 \\
$\beta_j=0.35$  & 0.762 \\
$\beta_j=0.40$  & 0.808 \\
$\beta_j=0.45$  & 0.834 \\
$\beta_j=0.50$  & 0.848 \\
$\beta_j=0.55$  & 0.860 \\
$\beta_j=0.60$  & 0.868 \\
\bottomrule
\end{tabular}
\end{table}

\begin{table}[ht]
\centering
\caption{$g_{\widehat{f},4}^{adj}$, DGP Size/Power (nom 5\%),  $\tau_0=0.8$, $\alpha_i=[1,1,1]$}
\label{tab_size_test5_adj_N100_T350_alpha1}
\begin{tabular}{l c }
\toprule
  $\lambda_1^0=0.6$ & $(N,T)=(100,350)$ \\
\midrule
$\beta_j=0$     & 0.048 \\
$\beta_j=0.10$  & 0.198 \\
$\beta_j=0.20$  & 0.586 \\
$\beta_j=0.30$  & 0.796 \\
$\beta_j=0.35$  & 0.816 \\
$\beta_j=0.40$  & 0.842 \\
$\beta_j=0.45$  & 0.862 \\
$\beta_j=0.50$  & 0.872 \\
$\beta_j=0.55$  & 0.888 \\
$\beta_j=0.60$  & 0.894 \\
\bottomrule
\end{tabular}
\end{table}
\begin{table}[ht]
\centering
\caption{$g_{\widehat{f},4}^{adj}$, DGP Size/Power (nom 5\%),  $\tau_0=0.8$, $\alpha_i=[0.51,0.51,0.51]$}
\label{tab_size_test5_adj_N100_T200_alpha051}
\begin{tabular}{l c }
\toprule
  $\lambda_1^0=0.6$ & $(N,T)=(100,200)$ \\
\midrule
$\beta_j=0$     & 0.056 \\
$\beta_j=0.10$  & 0.170 \\
$\beta_j=0.20$  & 0.448 \\
$\beta_j=0.30$  & 0.662 \\
$\beta_j=0.35$  & 0.732 \\
$\beta_j=0.40$  & 0.774 \\
$\beta_j=0.45$  & 0.782 \\
$\beta_j=0.50$  & 0.808 \\
$\beta_j=0.55$  & 0.820 \\
$\beta_j=0.60$  & 0.832 \\
\bottomrule
\end{tabular}
\end{table}

\begin{table}[ht]
\centering
\caption{$g_{\widehat{f},4}^{adj}$, DGP Size/Power (nom 5\%),  $\tau_0=0.8$, $\alpha_i=[0.51,0.51,0.51]$}
\label{tab_size_test5_adj_N100_T350_alpha051}
\begin{tabular}{l c }
\toprule
  $\lambda_1^0=0.6$ & $(N,T)=(100,350)$ \\
\midrule
$\beta_j=0$     & 0.054 \\
$\beta_j=0.10$  & 0.196 \\
$\beta_j=0.20$  & 0.578 \\
$\beta_j=0.30$  & 0.762 \\
$\beta_j=0.35$  & 0.792 \\
$\beta_j=0.40$  & 0.814 \\
$\beta_j=0.45$  & 0.830 \\
$\beta_j=0.50$  & 0.842 \\
$\beta_j=0.55$  & 0.856 \\
$\beta_j=0.60$  & 0.862 \\
\bottomrule
\end{tabular}
\end{table}
\begin{table}[ht]
\centering
\caption{$g_{\widehat{f},4}^{adj}$, DGP Size/Power (nom 5\%),  $\tau_0=0.8$, $\alpha_i=[0.1,0.1,0.1]$}
\label{tab_size_test5_adj_N100_T200_alpha01}
\begin{tabular}{l c }
\toprule
  $\lambda_1^0=0.6$ & $(N,T)=(100,200)$ \\
\midrule
$\beta_j=0$     & 0.050 \\
$\beta_j=0.10$  & 0.176 \\
$\beta_j=0.20$  & 0.436 \\
$\beta_j=0.30$  & 0.650 \\
$\beta_j=0.35$  & 0.724 \\
$\beta_j=0.40$  & 0.774 \\
$\beta_j=0.45$  & 0.786 \\
$\beta_j=0.50$  & 0.810 \\
$\beta_j=0.55$  & 0.822 \\
$\beta_j=0.60$  & 0.828 \\
\bottomrule
\end{tabular}
\end{table}

\begin{table}[ht]
\centering
\caption{$g_{\widehat{f},4}^{adj}$, DGP Size/Power (nom 5\%),  $\tau_0=0.8$, $\alpha_i=[0.1,0.1,0.1]$}
\label{tab_size_test5_adj_N100_T350_alpha01}
\begin{tabular}{l c }
\toprule
  $\lambda_1^0=0.6$ & $(N,T)=(100,350)$ \\
\midrule
$\beta_j=0$     & 0.056 \\
$\beta_j=0.10$  & 0.184 \\
$\beta_j=0.20$  & 0.542 \\
$\beta_j=0.30$  & 0.756 \\
$\beta_j=0.35$  & 0.782 \\
$\beta_j=0.40$  & 0.812 \\
$\beta_j=0.45$  & 0.838 \\
$\beta_j=0.50$  & 0.848 \\
$\beta_j=0.55$  & 0.852 \\
$\beta_j=0.60$  & 0.862 \\
\bottomrule
\end{tabular}
\end{table}
\clearpage 
\subsection{Cross-sectional dependent idiosyncratics and Garch errors for homogeneous and heterogeneous loadings}
Here we display the Monte Carlo simulations of $g_{\widehat{f},1},g_{\widehat{f},2}^{adj}, g_{\widehat{f},3}^{adj},g_{\widehat{f},4}^{adj}$ for the same DGP as in Section 5, $N=800, T=500$, with either cross-sectionally dependent idiosyncratics (see Section 3.2 of the main paper) or both cross-sectionally dependent idiosyncratics and GARCH$(1,1)$ forecast errors (see Section 3.3 of the main paper), when the loadings are both homogeneous and heterogeneous.

\subsubsection{Cross-sectional dependent idiosyncratics}
\begin{table}[ht]
\centering
\caption{$g_{\widehat{f},1}$, DGP Size/Power (nom 5\%), $\alpha_i=[1,1,1]$}
\label{tab_size_test_mu_new}
\begin{tabular}{lcccc}
\toprule
 & \multicolumn{4}{c}{$\mu_0$} \\
\cmidrule(lr){2-5}
$g_{\widehat{f},1}$ & 0.30 & 0.35 & 0.40 & 0.45 \\
\midrule
$\beta_{2}=0.00$ & 0.032 & 0.042 & 0.048 & 0.082 \\
$\beta_{2}=0.10$ & 0.158 & 0.246 & 0.392 & 0.652 \\
$\beta_{2}=0.20$ & 0.704 & 0.822 & 0.898 & 0.952 \\
$\beta_{2}=0.30$ & 0.902 & 0.942 & 0.966 & 0.980 \\
$\beta_{2}=0.35$ & 0.936 & 0.964 & 0.976 & 0.988 \\
$\beta_{2}=0.40$ & 0.954 & 0.974 & 0.982 & 0.992 \\
$\beta_{2}=0.45$ & 0.960 & 0.982 & 0.984 & 0.994 \\
$\beta_{2}=0.50$ & 0.966 & 0.982 & 0.990 & 0.994 \\
$\beta_{2}=0.55$ & 0.970 & 0.984 & 0.992 & 0.994 \\
$\beta_{2}=0.60$ & 0.970 & 0.986 & 0.992 & 0.994 \\
\bottomrule
\end{tabular}
\end{table}

\begin{table}[ht]
\centering
\caption{$g_{\widehat{f},1}$, DGP Size/Power (nom 5\%), $\alpha_i=[0.51,0.51,0.51]$}
\label{tab_size_test_mu_adjG1}
\begin{tabular}{lcccc}
\toprule
 & \multicolumn{4}{c}{$\mu_0$} \\
\cmidrule(lr){2-5}
$g_{\widehat{f},1}$ & 0.30 & 0.35 & 0.40 & 0.45 \\
\midrule
$\beta_{2}=0.00$ & 0.040 & 0.046 & 0.056 & 0.082 \\
$\beta_{2}=0.10$ & 0.136 & 0.214 & 0.316 & 0.516 \\
$\beta_{2}=0.20$ & 0.558 & 0.684 & 0.810 & 0.910 \\
$\beta_{2}=0.30$ & 0.816 & 0.886 & 0.916 & 0.946 \\
$\beta_{2}=0.35$ & 0.854 & 0.914 & 0.940 & 0.960 \\
$\beta_{2}=0.40$ & 0.878 & 0.916 & 0.950 & 0.968 \\
$\beta_{2}=0.45$ & 0.894 & 0.928 & 0.956 & 0.970 \\
$\beta_{2}=0.50$ & 0.902 & 0.936 & 0.958 & 0.970 \\
$\beta_{2}=0.55$ & 0.916 & 0.940 & 0.962 & 0.974 \\
$\beta_{2}=0.60$ & 0.922 & 0.946 & 0.964 & 0.974 \\
\bottomrule
\end{tabular}
\end{table}

\begin{table}[ht]
\centering
\caption{$g_{\widehat{f},1}$, DGP Size/Power (nom 5\%), $\alpha_i=[0.1,0.1,0.1]$}
\label{tab_size_test_mu_alpha_0_1}
\begin{tabular}{lcccc}
\toprule
 & \multicolumn{4}{c}{$\mu_0$} \\
\cmidrule(lr){2-5}
$g_{\widehat{f},1}$ & 0.30 & 0.35 & 0.40 & 0.45 \\
\midrule
$\beta_{2}=0.00$ & 0.030 & 0.042 & 0.052 & 0.082 \\
$\beta_{2}=0.10$ & 0.110 & 0.162 & 0.248 & 0.460 \\
$\beta_{2}=0.20$ & 0.476 & 0.588 & 0.718 & 0.856 \\
$\beta_{2}=0.30$ & 0.726 & 0.800 & 0.858 & 0.898 \\
$\beta_{2}=0.35$ & 0.780 & 0.844 & 0.880 & 0.912 \\
$\beta_{2}=0.40$ & 0.810 & 0.870 & 0.892 & 0.920 \\
$\beta_{2}=0.45$ & 0.834 & 0.880 & 0.902 & 0.928 \\
$\beta_{2}=0.50$ & 0.850 & 0.884 & 0.908 & 0.930 \\
$\beta_{2}=0.55$ & 0.868 & 0.890 & 0.910 & 0.932 \\
$\beta_{2}=0.60$ & 0.876 & 0.894 & 0.914 & 0.934 \\
\bottomrule
\end{tabular}
\end{table}

\begin{table}[ht]
\centering
\caption{$g_{\widehat{f},1}$, DGP Size/Power (nom 5\%),  $\alpha_i=[0.51,0.7,1]$}\label{tab_size_test_mu_new}
\begin{tabular}{l c c c c}
\toprule
 & \multicolumn{4}{c}{$\mu_0$} \\
\cmidrule(lr){2-5}
$g_{\widehat{f},1}$ & 0.30 & 0.35 & 0.40 & 0.45 \\
\midrule
$\beta_j=0$   & 0.034 & 0.044 & 0.062 & 0.096 \\
$\beta_j=0.10$& 0.170 & 0.278 & 0.426 & 0.680 \\
$\beta_j=0.20$& 0.736 & 0.864 & 0.924 & 0.970 \\
$\beta_j=0.30$& 0.934 & 0.970 & 0.978 & 0.986 \\
$\beta_j=0.35$& 0.954 & 0.976 & 0.984 & 0.990 \\
$\beta_j=0.40$& 0.974 & 0.982 & 0.988 & 0.990 \\
$\beta_j=0.45$& 0.978 & 0.982 & 0.988 & 0.992 \\
$\beta_j=0.50$& 0.982 & 0.986 & 0.988 & 0.992 \\
$\beta_j=0.55$& 0.982 & 0.988 & 0.990 & 0.992 \\
$\beta_j=0.60$& 0.982 & 0.988 & 0.990 & 0.992 \\
\bottomrule
\end{tabular}
\end{table}

\begin{table}[ht]
\centering
\caption{$g_{\widehat{f},2}^{adj}$, DGP Size/Power (nom 5\%), $\tau_0=0.8$, $\alpha_i=[1,1,1]$}
\label{tab_size_test2_adj_alpha111}
\begin{tabular}{l l c c c c c c c c c c}
\toprule
 &$\lambda_1^0=1$& \multicolumn{10}{c}{$(N,T)=(800,500)$} \\
\cmidrule(lr){3-12}
 &$\lambda_2^0$& 0.5 & 0.55 & 0.6 & 0.65 & 0.7 & 0.75 & 0.8 & 0.85 & 0.9 & 0.95 \\
\midrule
&$g_{\widehat{f},2}^{adj}$ \\
&$\beta_j=0$    & 0.050 & 0.056 & 0.042 & 0.054 & 0.084 & 0.080 & 0.096 & 0.104 & 0.116 & 0.130 \\
&$\beta_j=0.10$ & 0.148 & 0.164 & 0.192 & 0.204 & 0.276 & 0.306 & 0.352 & 0.410 & 0.486 & 0.598 \\
&$\beta_j=0.20$ & 0.646 & 0.698 & 0.724 & 0.766 & 0.826 & 0.860 & 0.878 & 0.902 & 0.920 & 0.956 \\
&$\beta_j=0.30$ & 0.888 & 0.904 & 0.916 & 0.924 & 0.938 & 0.948 & 0.954 & 0.970 & 0.978 & 0.984 \\
&$\beta_j=0.35$ & 0.918 & 0.928 & 0.934 & 0.940 & 0.954 & 0.964 & 0.964 & 0.976 & 0.980 & 0.986 \\
&$\beta_j=0.40$ & 0.932 & 0.938 & 0.940 & 0.952 & 0.966 & 0.968 & 0.972 & 0.978 & 0.984 & 0.990 \\
&$\beta_j=0.45$ & 0.936 & 0.946 & 0.954 & 0.962 & 0.972 & 0.974 & 0.972 & 0.980 & 0.984 & 0.994 \\
&$\beta_j=0.50$ & 0.948 & 0.954 & 0.958 & 0.966 & 0.974 & 0.974 & 0.976 & 0.980 & 0.986 & 0.994 \\
&$\beta_j=0.55$ & 0.950 & 0.956 & 0.962 & 0.968 & 0.974 & 0.974 & 0.978 & 0.984 & 0.988 & 0.994 \\
&$\beta_j=0.60$ & 0.950 & 0.958 & 0.966 & 0.968 & 0.976 & 0.974 & 0.982 & 0.988 & 0.990 & 0.994 \\
\bottomrule
\end{tabular}
\end{table}

\begin{table}[ht]
\centering
\caption{$g_{\widehat{f},2}^{adj}$, DGP Size/Power (nom 5\%), $\tau_0=0.8$, $\alpha_i=[0.51,0.51,0.51]$}
\label{tab_size_test2_adj_alpha051}
\begin{tabular}{l l c c c c c c c c c c}
\toprule
 &$\lambda_1^0=1$& \multicolumn{10}{c}{$(N,T)=(800,500)$} \\
\cmidrule(lr){3-12}
 &$\lambda_2^0$& 0.5 & 0.55 & 0.6 & 0.65 & 0.7 & 0.75 & 0.8 & 0.85 & 0.9 & 0.95 \\
\midrule
&$g_{\widehat{f},2}^{adj}$ \\
&$\beta_j=0$    & 0.042 & 0.058 & 0.046 & 0.044 & 0.068 & 0.074 & 0.086 & 0.090 & 0.104 & 0.122 \\
&$\beta_j=0.10$ & 0.122 & 0.136 & 0.140 & 0.164 & 0.212 & 0.232 & 0.276 & 0.324 & 0.378 & 0.496 \\
&$\beta_j=0.20$ & 0.488 & 0.552 & 0.588 & 0.622 & 0.662 & 0.698 & 0.762 & 0.792 & 0.824 & 0.888 \\
&$\beta_j=0.30$ & 0.790 & 0.798 & 0.800 & 0.820 & 0.838 & 0.864 & 0.884 & 0.898 & 0.912 & 0.934 \\
&$\beta_j=0.35$ & 0.826 & 0.842 & 0.838 & 0.862 & 0.882 & 0.892 & 0.902 & 0.918 & 0.928 & 0.952 \\
&$\beta_j=0.40$ & 0.846 & 0.862 & 0.870 & 0.880 & 0.898 & 0.904 & 0.916 & 0.926 & 0.948 & 0.958 \\
&$\beta_j=0.45$ & 0.872 & 0.884 & 0.886 & 0.898 & 0.912 & 0.916 & 0.928 & 0.940 & 0.954 & 0.960 \\
&$\beta_j=0.50$ & 0.884 & 0.890 & 0.898 & 0.904 & 0.916 & 0.922 & 0.930 & 0.950 & 0.960 & 0.964 \\
&$\beta_j=0.55$ & 0.892 & 0.898 & 0.906 & 0.912 & 0.920 & 0.926 & 0.936 & 0.952 & 0.962 & 0.964 \\
&$\beta_j=0.60$ & 0.900 & 0.906 & 0.910 & 0.914 & 0.928 & 0.930 & 0.946 & 0.952 & 0.962 & 0.968 \\
\bottomrule
\end{tabular}
\end{table}

\begin{table}[ht]
\centering
\caption{$g_{\widehat{f},2}^{adj}$, DGP Size/Power (nom 5\%), $\tau_0=0.8$, $\alpha_i=[0.1,0.1,0.1]$}
\label{tab_size_test2_adj_alpha010}
\begin{tabular}{l l c c c c c c c c c c}
\toprule
 &$\lambda_1^0=1$& \multicolumn{10}{c}{$(N,T)=(800,500)$} \\
\cmidrule(lr){3-12}
 &$\lambda_2^0$& 0.5 & 0.55 & 0.6 & 0.65 & 0.7 & 0.75 & 0.8 & 0.85 & 0.9 & 0.95 \\
\midrule
&$g_{\widehat{f},2}^{adj}$ \\
&$\beta_j=0$    & 0.040 & 0.056 & 0.054 & 0.056 & 0.068 & 0.066 & 0.088 & 0.084 & 0.108 & 0.150 \\
&$\beta_j=0.10$ & 0.104 & 0.114 & 0.144 & 0.160 & 0.178 & 0.202 & 0.258 & 0.304 & 0.362 & 0.450 \\
&$\beta_j=0.20$ & 0.446 & 0.472 & 0.504 & 0.538 & 0.568 & 0.630 & 0.672 & 0.726 & 0.760 & 0.814 \\
&$\beta_j=0.30$ & 0.698 & 0.708 & 0.730 & 0.740 & 0.774 & 0.796 & 0.816 & 0.834 & 0.864 & 0.888 \\
&$\beta_j=0.35$ & 0.738 & 0.772 & 0.782 & 0.794 & 0.812 & 0.836 & 0.854 & 0.866 & 0.880 & 0.900 \\
&$\beta_j=0.40$ & 0.782 & 0.792 & 0.804 & 0.808 & 0.838 & 0.854 & 0.866 & 0.874 & 0.890 & 0.904 \\
&$\beta_j=0.45$ & 0.800 & 0.820 & 0.816 & 0.840 & 0.854 & 0.864 & 0.870 & 0.874 & 0.890 & 0.908 \\
&$\beta_j=0.50$ & 0.812 & 0.826 & 0.834 & 0.854 & 0.870 & 0.872 & 0.872 & 0.882 & 0.896 & 0.912 \\
&$\beta_j=0.55$ & 0.832 & 0.832 & 0.844 & 0.870 & 0.876 & 0.874 & 0.874 & 0.888 & 0.898 & 0.916 \\
&$\beta_j=0.60$ & 0.840 & 0.844 & 0.852 & 0.874 & 0.882 & 0.876 & 0.878 & 0.890 & 0.900 & 0.920 \\
\bottomrule
\end{tabular}
\end{table}

\begin{table}[ht]
\centering
\caption{$g_{\widehat{f},2}^{adj}$, DGP Size/Power (nom 5\%),  $\tau_0=0.8$, $\alpha_i=[0.51,0.7,1]$}
\label{tab_size_test2_adj_large}
\begin{tabular}{l l c c c c c c c c c c}
\toprule
 &$\lambda_1^0=1$& \multicolumn{10}{c}{$(N,T)=(800,500)$} \\
\cmidrule(lr){3-12}
 &$\lambda_2^0$& 0.5 & 0.55 & 0.6 & 0.65 & 0.7 & 0.75 & 0.8 & 0.85 & 0.9 & 0.95 \\
\midrule
&$g_{\widehat{f},2}^{adj}$ \\
&$\beta_j=0$    & 0.050 & 0.060 & 0.056 & 0.054 & 0.072 & 0.066 & 0.088 & 0.100 & 0.106 & 0.142 \\
&$\beta_j=0.10$ & 0.158 & 0.190 & 0.236 & 0.232 & 0.274 & 0.312 & 0.364 & 0.430 & 0.518 & 0.654 \\
&$\beta_j=0.20$ & 0.682 & 0.724 & 0.766 & 0.804 & 0.828 & 0.870 & 0.904 & 0.924 & 0.950 & 0.970 \\
&$\beta_j=0.30$ & 0.904 & 0.922 & 0.934 & 0.940 & 0.948 & 0.956 & 0.970 & 0.978 & 0.978 & 0.988 \\
&$\beta_j=0.35$ & 0.934 & 0.948 & 0.948 & 0.954 & 0.970 & 0.972 & 0.978 & 0.978 & 0.982 & 0.990 \\
&$\beta_j=0.40$ & 0.950 & 0.958 & 0.962 & 0.968 & 0.976 & 0.980 & 0.980 & 0.984 & 0.986 & 0.992 \\
&$\beta_j=0.45$ & 0.958 & 0.966 & 0.968 & 0.970 & 0.980 & 0.980 & 0.986 & 0.986 & 0.990 & 0.994 \\
&$\beta_j=0.50$ & 0.962 & 0.966 & 0.970 & 0.976 & 0.980 & 0.986 & 0.988 & 0.990 & 0.990 & 0.992 \\
&$\beta_j=0.55$ & 0.966 & 0.974 & 0.976 & 0.980 & 0.986 & 0.986 & 0.988 & 0.990 & 0.992 & 0.992 \\
&$\beta_j=0.60$ & 0.972 & 0.976 & 0.980 & 0.982 & 0.986 & 0.986 & 0.990 & 0.990 & 0.992 & 0.994 \\
\bottomrule
\end{tabular}
\end{table}
\begin{table}[ht]
\centering
\caption{$g_{\widehat{f},3}^{adj}$, DGP Size/Power (nom 5\%), $\tau_0=0.8$, $\alpha_i=[1,1,1]$}
\label{tab_size_test4_adj_alpha111}
\begin{tabular}{l l c c c c c c c c c c }
\toprule
 &$\lambda_1^0=1$& \multicolumn{10}{c}{$(N,T)=(800,500)$} \\
\cmidrule(lr){3-12}
 &$\lambda_2^0$& 0.5 & 0.55 & 0.6 & 0.65 & 0.7 & 0.75 & 0.8 & 0.85 & 0.9 & 0.95 \\
\midrule
&$g_{\widehat{f},4}^{adj}$ \\
&$\beta_j=0$    & 0.058 & 0.054 & 0.050 & 0.062 & 0.072 & 0.092 & 0.126 & 0.176 & 0.218 & 0.168  \\
&$\beta_j=0.10$ & 0.176 & 0.210 & 0.224 & 0.258 & 0.328 & 0.376 & 0.516 & 0.634 & 0.726 & 0.672  \\
&$\beta_j=0.20$ & 0.670 & 0.738 & 0.768 & 0.814 & 0.870 & 0.902 & 0.926 & 0.956 & 0.962 & 0.954 \\
&$\beta_j=0.30$ & 0.894 & 0.908 & 0.922 & 0.932 & 0.954 & 0.960 & 0.974 & 0.980 & 0.984 & 0.980  \\
&$\beta_j=0.35$ & 0.918 & 0.934 & 0.940 & 0.950 & 0.962 & 0.968 & 0.980 & 0.980 & 0.988 & 0.982 \\
&$\beta_j=0.40$ & 0.936 & 0.942 & 0.950 & 0.956 & 0.970 & 0.974 & 0.980 & 0.984 & 0.988 & 0.988  \\
&$\beta_j=0.45$ & 0.944 & 0.952 & 0.954 & 0.964 & 0.972 & 0.974 & 0.980 & 0.988 & 0.990 & 0.992  \\
&$\beta_j=0.50$ & 0.952 & 0.954 & 0.962 & 0.968 & 0.974 & 0.976 & 0.982 & 0.988 & 0.994 & 0.994 \\
&$\beta_j=0.55$ & 0.954 & 0.960 & 0.966 & 0.970 & 0.976 & 0.976 & 0.986 & 0.990 & 0.994 & 0.994 \\
&$\beta_j=0.60$ & 0.954 & 0.960 & 0.966 & 0.970 & 0.976 & 0.982 & 0.988 & 0.990 & 0.994 & 0.994  \\
\bottomrule
\end{tabular}
\end{table}
\begin{table}[ht]
\centering
\caption{$g_{\widehat{f},3}^{adj}$, DGP Size/Power (nom 5\%), $\tau_0=0.8$, $\alpha_i=[0.51,0.51,0.51]$}
\label{tab_size_test3_adj_alpha051}
\begin{tabular}{l l c c c c c c c c c c }
\toprule
 &$\lambda_1^0=1$& \multicolumn{10}{c}{$(N,T)=(800,500)$} \\
\cmidrule(lr){3-12}
 &$\lambda_2^0$& 0.5 & 0.55 & 0.6 & 0.65 & 0.7 & 0.75 & 0.8 & 0.85 & 0.9 & 0.95 \\
\midrule
&$g_{\widehat{f},3}^{adj}$ \\
&$\beta_j=0$    & 0.040 & 0.048 & 0.050 & 0.058 & 0.078 & 0.080 & 0.132 & 0.158 & 0.208 & 0.144  \\
&$\beta_j=0.10$ & 0.124 & 0.152 & 0.176 & 0.204 & 0.256 & 0.324 & 0.408 & 0.558 & 0.612 & 0.544  \\
&$\beta_j=0.20$ & 0.528 & 0.608 & 0.632 & 0.668 & 0.708 & 0.784 & 0.836 & 0.886 & 0.918 & 0.890  \\
&$\beta_j=0.30$ & 0.800 & 0.802 & 0.820 & 0.840 & 0.866 & 0.896 & 0.916 & 0.944 & 0.954 & 0.946  \\
&$\beta_j=0.35$ & 0.832 & 0.848 & 0.858 & 0.876 & 0.896 & 0.918 & 0.928 & 0.950 & 0.964 & 0.952  \\
&$\beta_j=0.40$ & 0.850 & 0.868 & 0.882 & 0.900 & 0.914 & 0.926 & 0.938 & 0.964 & 0.964 & 0.964  \\
&$\beta_j=0.45$ & 0.874 & 0.890 & 0.900 & 0.906 & 0.918 & 0.940 & 0.946 & 0.966 & 0.968 & 0.962  \\
&$\beta_j=0.50$ & 0.890 & 0.898 & 0.912 & 0.910 & 0.930 & 0.948 & 0.950 & 0.968 & 0.970 & 0.962  \\
&$\beta_j=0.55$ & 0.898 & 0.900 & 0.914 & 0.916 & 0.934 & 0.950 & 0.954 & 0.972 & 0.972 & 0.962  \\
&$\beta_j=0.60$ & 0.902 & 0.908 & 0.916 & 0.916 & 0.934 & 0.950 & 0.954 & 0.972 & 0.974 & 0.962  \\
\bottomrule
\end{tabular}
\end{table}
\begin{table}[ht]
\centering
\caption{$g_{\widehat{f},3}^{adj}$, DGP Size/Power (nom 5\%), $\tau_0=0.8$, $\alpha_i=[0.1,0.1,0.1]$}
\label{tab_size_test3_adj_alpha111}
\begin{tabular}{l l c c c c c c c c c c }
\toprule
 &$\lambda_1^0=1$& \multicolumn{10}{c}{$(N,T)=(800,500)$} \\
\cmidrule(lr){3-12}
 &$\lambda_2^0$& 0.5 & 0.55 & 0.6 & 0.65 & 0.7 & 0.75 & 0.8 & 0.85 & 0.9 & 0.95  \\
\midrule
&$g_{\widehat{f},3}^{adj}$ \\
&$\beta_j=0$    & 0.044 & 0.040 & 0.042 & 0.052 & 0.080 & 0.086 & 0.136 & 0.172 & 0.220 & 0.164  \\
&$\beta_j=0.10$ & 0.134 & 0.140 & 0.170 & 0.194 & 0.246 & 0.296 & 0.382 & 0.534 & 0.580 & 0.498  \\
&$\beta_j=0.20$ & 0.480 & 0.496 & 0.564 & 0.584 & 0.632 & 0.714 & 0.770 & 0.836 & 0.858 & 0.832  \\
&$\beta_j=0.30$ & 0.704 & 0.728 & 0.762 & 0.782 & 0.808 & 0.842 & 0.860 & 0.888 & 0.896 & 0.884  \\
&$\beta_j=0.35$ & 0.754 & 0.774 & 0.800 & 0.816 & 0.838 & 0.854 & 0.872 & 0.892 & 0.898 & 0.898  \\
&$\beta_j=0.40$ & 0.784 & 0.806 & 0.818 & 0.836 & 0.856 & 0.872 & 0.874 & 0.894 & 0.914 & 0.908  \\
&$\beta_j=0.45$ & 0.796 & 0.818 & 0.832 & 0.852 & 0.868 & 0.876 & 0.880 & 0.900 & 0.914 & 0.912  \\
&$\beta_j=0.50$ & 0.830 & 0.832 & 0.846 & 0.870 & 0.874 & 0.878 & 0.896 & 0.910 & 0.918 & 0.914  \\
&$\beta_j=0.55$ & 0.842 & 0.844 & 0.856 & 0.874 & 0.878 & 0.884 & 0.900 & 0.912 & 0.918 & 0.920 \\
&$\beta_j=0.60$ & 0.848 & 0.856 & 0.862 & 0.874 & 0.882 & 0.886 & 0.902 & 0.916 & 0.920 & 0.924  \\
\bottomrule
\end{tabular}
\end{table}
\begin{table}[ht]
\centering
\caption{$g_{\widehat{f},3}^{adj}$, DGP Size/Power (nom 5\%),  $\tau_0=0.8$, $\alpha_i=[0.51,0.7,1]$}
\label{tab_size_test3_adj_large_2}
\begin{tabular}{l l c c c c c c c c c c}
\toprule
 &$\lambda_1^0=1$& \multicolumn{10}{c}{$(N,T)=(800,500)$} \\
\cmidrule(lr){3-12}
 &$\lambda_2^0$& 0.5 & 0.55 & 0.6 & 0.65 & 0.7 & 0.75 & 0.8 & 0.85 & 0.9 & 0.95 \\
\midrule
&$g_{\widehat{f},3}^{adj}$ \\
&$\beta_j=0$    & 0.048 & 0.060 & 0.060 & 0.064 & 0.080 & 0.086 & 0.142 & 0.172 & 0.186 & 0.180 \\
&$\beta_j=0.10$ & 0.184 & 0.228 & 0.274 & 0.274 & 0.328 & 0.400 & 0.552 & 0.682 & 0.750 & 0.688 \\
&$\beta_j=0.20$ & 0.704 & 0.754 & 0.802 & 0.848 & 0.874 & 0.908 & 0.946 & 0.966 & 0.978 & 0.978 \\
&$\beta_j=0.30$ & 0.910 & 0.938 & 0.938 & 0.948 & 0.964 & 0.968 & 0.978 & 0.984 & 0.988 & 0.990 \\
&$\beta_j=0.35$ & 0.944 & 0.952 & 0.954 & 0.964 & 0.974 & 0.982 & 0.982 & 0.986 & 0.992 & 0.990 \\
&$\beta_j=0.40$ & 0.952 & 0.964 & 0.966 & 0.974 & 0.980 & 0.984 & 0.986 & 0.990 & 0.994 & 0.994 \\
&$\beta_j=0.45$ & 0.958 & 0.966 & 0.970 & 0.980 & 0.982 & 0.986 & 0.988 & 0.990 & 0.994 & 0.994 \\
&$\beta_j=0.50$ & 0.966 & 0.974 & 0.976 & 0.982 & 0.984 & 0.990 & 0.990 & 0.990 & 0.994 & 0.996 \\
&$\beta_j=0.55$ & 0.968 & 0.976 & 0.980 & 0.984 & 0.988 & 0.990 & 0.990 & 0.990 & 0.996 & 0.996 \\
&$\beta_j=0.60$ & 0.970 & 0.980 & 0.982 & 0.986 & 0.988 & 0.990 & 0.990 & 0.992 & 0.996 & 0.996 \\
\bottomrule
\end{tabular}
\end{table}
\begin{table}[ht]
\centering
\caption{$g_{\widehat{f},4}^{adj}$, DGP Size/Power (nom 5\%), $\tau_0=0.8$, $\alpha_i=[1,1,1]$}
\label{tab_size_test5_adj_N800_T500_alpha111}
\begin{tabular}{l c}
\toprule
$\lambda_2^0=0.6$ & $(N,T)=(800,500)$ \\
\midrule
$\beta_{2}=0.00$ & 0.052 \\
$\beta_{2}=0.10$ & 0.266 \\
$\beta_{2}=0.20$ & 0.834 \\
$\beta_{2}=0.30$ & 0.930 \\
$\beta_{2}=0.35$ & 0.946 \\
$\beta_{2}=0.40$ & 0.960 \\
$\beta_{2}=0.45$ & 0.966 \\
$\beta_{2}=0.50$ & 0.968 \\
$\beta_{2}=0.55$ & 0.974 \\
$\beta_{2}=0.60$ & 0.976 \\
\bottomrule
\end{tabular}
\end{table}
\begin{table}[ht]
\centering
\caption{$g_{\widehat{f},4}^{adj}$, DGP Size/Power (nom 5\%), $\tau_0=0.8$, $\alpha_i=[0.51,0.51,0.51]$}
\label{tab_size_test5_adj_N800_T500_alpha051}
\begin{tabular}{l c}
\toprule
$\lambda_2^0=0.6$ & $(N,T)=(800,500)$ \\
\midrule
$\beta_{2}=0.00$ & 0.056 \\
$\beta_{2}=0.10$ & 0.206 \\
$\beta_{2}=0.20$ & 0.658 \\
$\beta_{2}=0.30$ & 0.834 \\
$\beta_{2}=0.35$ & 0.866 \\
$\beta_{2}=0.40$ & 0.888 \\
$\beta_{2}=0.45$ & 0.902 \\
$\beta_{2}=0.50$ & 0.906 \\
$\beta_{2}=0.55$ & 0.910 \\
$\beta_{2}=0.60$ & 0.912 \\
\bottomrule
\end{tabular}
\end{table}
\begin{table}[ht]
\centering
\caption{$g_{\widehat{f},4}^{adj}$, DGP Size/Power (nom 5\%), $\tau_0=0.8$, $\alpha_i=[0.1,0.1,0.1]$}
\label{tab_size_test5_adj_N800_T500_alpha010}
\begin{tabular}{l c}
\toprule
$\lambda_2^0=0.6$ & $(N,T)=(800,500)$ \\
\midrule
$\beta_{2}=0.00$ & 0.048 \\
$\beta_{2}=0.10$ & 0.180 \\
$\beta_{2}=0.20$ & 0.576 \\
$\beta_{2}=0.30$ & 0.768 \\
$\beta_{2}=0.35$ & 0.812 \\
$\beta_{2}=0.40$ & 0.826 \\
$\beta_{2}=0.45$ & 0.852 \\
$\beta_{2}=0.50$ & 0.858 \\
$\beta_{2}=0.55$ & 0.864 \\
$\beta_{2}=0.60$ & 0.872 \\
\bottomrule
\end{tabular}
\end{table}
\begin{table}[ht]
\centering
\caption{$g_{\widehat{f},4}^{adj}$, DGP Size/Power (nom 5\%),  $\tau_0=0.8$, $\alpha_i=[0.51,0.7,1]$}
\label{tab_size_test5_adj_N800_T500_alpha_hetero}
\begin{tabular}{l c }
\toprule
  $\lambda_2^0=0.6$ & $(N,T)=(800,500)$ \\
\midrule
$\beta_j=0$     & 0.054 \\
$\beta_j=0.10$  & 0.286 \\
$\beta_j=0.20$  & 0.842 \\
$\beta_j=0.30$  & 0.956 \\
$\beta_j=0.35$  & 0.976 \\
$\beta_j=0.40$  & 0.982 \\
$\beta_j=0.45$  & 0.986 \\
$\beta_j=0.50$  & 0.986 \\
$\beta_j=0.55$  & 0.986 \\
$\beta_j=0.60$  & 0.986 \\
\bottomrule
\end{tabular}
\end{table}

\clearpage
\subsubsection{Cross-sectional dependent idiosyncratics and GARCH$(1,1)$ forecast errors}

\begin{table}[ht]
\centering
\caption{$g_{\widehat{f},1}$, DGP Size/Power (nom 5\%), $\alpha_i=[1,1,1]$}
\label{tab_size_test_mu_high_alpha111}
\begin{tabular}{lcccc}
\toprule
 & \multicolumn{4}{c}{$\mu_0$} \\
\cmidrule(lr){2-5}
$g_{\widehat{f},1}$ & 0.30 & 0.35 & 0.40 & 0.45 \\
\midrule
$\beta_{2}=0.00$ & 0.078 & 0.088 & 0.100 & 0.130 \\
$\beta_{2}=0.10$ & 0.890 & 0.952 & 0.968 & 0.986 \\
$\beta_{2}=0.20$ & 0.974 & 0.984 & 0.992 & 0.994 \\
$\beta_{2}=0.30$ & 0.980 & 0.990 & 0.994 & 0.994 \\
$\beta_{2}=0.35$ & 0.984 & 0.990 & 0.994 & 0.994 \\
$\beta_{2}=0.40$ & 0.984 & 0.990 & 0.994 & 0.994 \\
$\beta_{2}=0.45$ & 0.988 & 0.990 & 0.994 & 0.994 \\
$\beta_{2}=0.50$ & 0.988 & 0.990 & 0.994 & 0.994 \\
$\beta_{2}=0.55$ & 0.988 & 0.990 & 0.994 & 0.994 \\
$\beta_{2}=0.60$ & 0.988 & 0.990 & 0.994 & 0.994 \\
\bottomrule
\end{tabular}
\end{table}
\begin{table}[ht]
\centering
\caption{$g_{\widehat{f},1}$, DGP Size/Power (nom 5\%), $\alpha_i=[0.51,0.51,0.51]$}
\label{tab_size_test_mu_high_alpha051}
%
\end{table}
\begin{table}[ht]
\centering
\caption{$g_{\widehat{f},1}$, DGP Size/Power (nom 5\%), $\alpha_i=[0.1,0.1,0.1]$}
\label{tab_size_test_mu_high_alpha010}
%
\end{table}
\begin{table}[ht]
\centering
\caption{$g_{\widehat{f},1}$, DGP Size/Power (nom 5\%),  $\alpha_i=[0.51,0.7,1]$}\label{tab_size_test_mu_high}
%
\end{table}
\begin{table}[ht]
\centering
\caption{$g_{\widehat{f},2}^{adj}$, DGP Size/Power (nom 5\%), $\tau_0=0.8$, $\alpha_i=[1,1,1]$}
\label{tab_size_test2_adj_large_alpha111}
%
\end{table}
\begin{table}[ht]
\centering
\caption{$g_{\widehat{f},2}^{adj}$, DGP Size/Power (nom 5\%), $\tau_0=0.8$, $\alpha_i=[0.51,0.51,0.51]$}
\label{tab_size_test2_adj_large_alpha051}
%
\end{table}
\begin{table}[ht]
\centering
\caption{$g_{\widehat{f},2}^{adj}$, DGP Size/Power (nom 5\%), $\tau_0=0.8$, $\alpha_i=[0.1,0.1,0.1]$}
\label{tab_size_test2_adj_large_alpha010}
%
\end{table}
\begin{table}[ht]
\centering
\caption{$g_{\widehat{f},2}^{adj}$, DGP Size/Power (nom 5\%),  $\tau_0=0.8$, $\alpha_i=[0.51,0.7,1]$}
\label{tab_size_test2_adj_large_2}
%
\end{table}
\begin{table}[ht]
\centering
\caption{$g_{\widehat{f},3}^{adj}$, DGP Size/Power (nom 5\%), $\tau_0=0.8$, $\alpha_i=[1,1,1]$}
\label{tab_size_test3_adj_large_alpha111}
%
\end{table}
\begin{table}[ht]
\centering
\caption{$g_{\widehat{f},3}^{adj}$, DGP Size/Power (nom 5\%), $\tau_0=0.8$, $\alpha_i=[0.51,0.51,0.51]$}
\label{tab_size_test3_adj_large_alpha051}
%
\end{table}
\begin{table}[ht]
\centering
\caption{$g_{\widehat{f},3}^{adj}$, DGP Size/Power (nom 5\%),  $\tau_0=0.8$, $\alpha_i=[0.51,0.7,1]$}
\label{tab_size_test3_adj_large_2}
%
\end{table}
\begin{table}[ht]
\centering
\caption{$g_{\widehat{f},4}^{adj}$, DGP Size/Power (nom 5\%),  $\tau_0=0.8$, $\alpha_i=[1,1,1]$}
\label{tab_size_test5_adj_N800_T500_alpha111}
\begin{tabular}{l c }
\toprule
  $\lambda_2^0=0.6$ & $(N,T)=(800,500)$ \\
\midrule
$\beta_2=0$     & 0.064 \\
$\beta_2=0.10$  & 0.904 \\
$\beta_2=0.20$  & 0.970 \\
$\beta_2=0.30$  & 0.986 \\
$\beta_2=0.35$  & 0.988 \\
$\beta_2=0.40$  & 0.988 \\
$\beta_2=0.45$  & 0.988 \\
$\beta_2=0.50$  & 0.988 \\
$\beta_2=0.55$  & 0.988 \\
$\beta_2=0.60$  & 0.988 \\
\bottomrule
\end{tabular}
\end{table}
\begin{table}[ht]
\centering
\caption{$g_{\widehat{f},4}^{adj}$, DGP Size/Power (nom 5\%),  $\tau_0=0.8$, $\alpha_i=[0.51,0.51,0.51]$}
\label{tab_size_test5_adj_N800_T500_alpha051}
%
\end{table}
\begin{table}[ht]
\centering
\caption{$g_{\widehat{f},4}^{adj}$, DGP Size/Power (nom 5\%),  $\tau_0=0.8$, $\alpha_i=[0.51, 0.7, 1]$}
\label{tab_size_test5_adj_N800_T500_alpha051_07_1}
%
\end{table}
\clearpage 
\section{Additional Tables and Figures}\label{sec_additions}
Here, we replicate the same exercise in \cite{pitarakis2023direct}, where the grand average is employed to proxy the global inflation. 
\begin{table}[ht]
\centering
\caption{$p$-values: 1-quarter-ahead}
\begin{threeparttable}
%
\begin{tablenotes}
\footnotesize
\item Notes: $AR(1)$ vs grand-average augmented $AR(1)$; $\mu_0=0.45$, $\tau_0=0.8$.
\end{tablenotes}
\end{threeparttable}
\end{table}

Table \ref{tab:fredmd_variables} collects the full names of the FRED-MD time series selected in our second application in Section 4.2

\begin{sidewaystable}[ht]
\centering
\caption{Full names of selected FRED-MD macroeconomic variables}
\label{tab:fredmd_variables}
%
\end{sidewaystable}
\clearpage

\bibliography{literature}
\end{document}